\pdfoutput=1
\documentclass[11pt]{article}
\usepackage{dsfont}

\usepackage{amsmath,amssymb,graphicx}
\usepackage{diagbox}
\usepackage{hyperref}
\usepackage[nosort]{cite}

\usepackage{tikz}
\usetikzlibrary{calc} \usetikzlibrary{patterns} \usetikzlibrary{decorations.pathreplacing} \usetikzlibrary{decorations.markings} \usetikzlibrary{decorations.pathmorphing} \usetikzlibrary{positioning}

\usepackage{color}
\definecolor{darkred}{rgb}{0.65,0.15,0}
\definecolor{newgreen}{rgb}{0.2,0.62,0.14}
\hypersetup{pdfborder={0 0 0},colorlinks=true,urlcolor=blue,citecolor=blue,linkcolor=darkred,linktocpage=true}

\DeclareFontFamily{U}{mathx}{}
\DeclareFontShape{U}{mathx}{m}{n}{ <-> mathx10 }{}
\DeclareSymbolFont{mathx}{U}{mathx}{m}{n}
\DeclareFontSubstitution{U}{mathx}{m}{n}

\DeclareMathAccent{\widecheck}{0}{mathx}{"71}

\numberwithin{equation}{section}

\def\nn{\nonumber}

\def\spa#1.#2{\left\langle#1\,#2\right\rangle}
\def\spb#1.#2{\left[#1\,#2\right]}

\def\ep{\epsilon}

\newcommand{\bsv}{\beta^{\rm sv}}
\newcommand{\betasv}[1]{\beta^{\rm sv}\! \left[\begin{smallmatrix}#1\end{smallmatrix}\right]}
\newcommand{\hatbetasv}[1]{\widehat \beta^{\rm sv}\! \left[\begin{smallmatrix}#1\end{smallmatrix}\right]}
\newcommand{\bsvBR}[3]{\beta^{\rm sv} \! \left[\begin{smallmatrix}#1\\#2\end{smallmatrix};#3\right]}
\newcommand{\bsvBRno}[2]{\beta^{\rm sv}\! \left[\begin{smallmatrix}#1\\#2\end{smallmatrix}\right]}
\newcommand{\hatbsvBR}[3]{\widehat \beta^{\rm sv} \! \left[\begin{smallmatrix}#1\\#2\end{smallmatrix};#3\right]}
\newcommand{\alphaBR}[3]{\alpha\! \left[\begin{smallmatrix}#1\\#2\end{smallmatrix};#3\right]}
\newcommand{\alphaBRno}[2]{\alpha\! \left[\begin{smallmatrix}#1\\#2\end{smallmatrix}\right]}

\newcommand{\cc}{\text{c.c.}}

\newcommand{\rcross}{{\color{red}\times}}

\def\beq{\begin{equation}}
\def\eeq{\end{equation}}
\let\Re\relax
\let\Im\relax
\DeclareMathOperator{\Re}{Re}
\DeclareMathOperator{\Im}{Im}

\newcommand{\eq}{\begin{equation}}
\newcommand{\eqe}{\end{equation}}
\newcommand{\eqa}{\begin{eqnarray}}
\newcommand{\eqae}{\end{eqnarray}}

\newcommand{\bea}{\begin{eqnarray}}
\newcommand{\eea}{\end{eqnarray}}
\newcommand{\dd}{\mathrm{d}}

\newcommand{\CC}{\mathbb C}
\newcommand{\NN}{\mathbb N}
\newcommand{\ZZ}{\mathbb Z}

\newcommand{\EEodd}{{\rm E}^{(-)}}

\newcommand{\Jp}[3]{{\rm J}^{+[#1]}_{#2,#3}}
\newcommand{\Jm}[3]{{\rm J}^{-[#1]}_{#2,#3}}
\newcommand{\Jpm}[3]{{\rm J}^{\pm[#1]}_{#2,#3}}
\newcommand{\FFpm}[3]{{\rm F}^{\pm(#1)}_{#2,#3}}
\newcommand{\FFp}[3]{{\rm F}^{+(#1)}_{#2,#3}}
\newcommand{\FFm}[3]{{\rm F}^{-(#1)}_{#2,#3}}

\newcommand{\seedp}[3]{f^{+(#1)}_{#2,#3}}
\newcommand{\seedm}[3]{f^{-(#1)}_{#2,#3}}
\newcommand{\Jseedp}[3]{j^{+[#1]}_{#2,#3}}
\newcommand{\Jseedm}[3]{j^{-[#1]}_{#2,#3}}
\newcommand{\cFFpm}[3]{\widecheck{\rm F}^{\pm(#1)}_{#2,#3}}
\newcommand{\cFFp}[3]{\widecheck{\rm F}^{+(#1)}_{#2,#3}}
\newcommand{\cFFm}[3]{\widecheck{\rm F}^{-(#1)}_{#2,#3}}

\newcommand{\seedpalt}[3]{\tilde f^{+(#1)}_{#2,#3}}
\newcommand{\Jseedpalt}[3]{\tilde{\jmath}^{+[#1]}_{#2,#3}}
\newcommand{\seedmalt}[3]{\tilde f^{-(#1)}_{#2,#3}}
\newcommand{\Jseedmalt}[3]{\tilde{\jmath}^{-[#1]}_{#2,#3}}
\newcommand{\seedpmalt}[3]{\tilde f^{\pm(#1)}_{#2,#3}}
\newcommand{\Jseedpmalt}[3]{\tilde{\jmath}^{\pm[#1]}_{#2,#3}}

\newbox\charbox
\newbox\slabox
\def\s#1{{      % Feynman slash
        \setbox\charbox=\hbox{$#1$}
        \setbox\slabox=\hbox{$/$}
        \dimen\charbox=\ht\slabox
        \advance\dimen\charbox by -\dp\slabox
        \advance\dimen\charbox by -\ht\charbox
        \advance\dimen\charbox by \dp\charbox
        \divide\dimen\charbox by 2
        \raise-\dimen\charbox\hbox to \wd\charbox{\hss/\hss}
        \llap{$#1$}
}}

\reversemarginpar
\newcounter{todocounter}

\colorlet{ddcolor}{green!40!white}

\newcommand{\ddinline}[2][]{
  \ifthenelse { \equal {#1} {} }
    { \def\temp {#2} }  % if #1 == blank
    { \def\temp {#1} }   % else (not blank)
  \refstepcounter{todocounter}\todo[color=ddcolor,inline,caption={\textbf{\thetodocounter. DD} \temp}]{\textbf{\thetodocounter. DD:} #2}{}}

\colorlet{oscolor}{blue!20!white}

\newcommand{\osinline}[2][]{
  \ifthenelse { \equal {#1} {} }
    { \def\temp {#2} }  % if #1 == blank
    { \def\temp {#1} }   % else (not blank)
  \refstepcounter{todocounter}\todo[color=oscolor,inline,caption={\textbf{\thetodocounter. OS} \temp}]{\textbf{\thetodocounter. OS:} #2}{}}

\colorlet{akcolor}{yellow!40!white}

\newcommand{\akinline}[2][]{
  \ifthenelse { \equal {#1} {} }
    { \def\temp {#2} }  % if #1 == blank
    { \def\temp {#1} }   % else (not blank)
  \refstepcounter{todocounter}\todo[color=akcolor,inline,caption={\textbf{\thetodocounter. AK} \temp}]{\textbf{\thetodocounter. AK:} #2}{}}

\newcommand{\betalagpm}[1]{\beta^{{\rm sv}\pm,\, #1}_{m,k}}
\newcommand{\betalagp}[1]{\beta^{{\rm sv}+,\, #1}_{m,k}}
\newcommand{\betalagm}[1]{\beta^{{\rm sv}-,\, #1}_{m,k}}
\newcommand{\hatbetalagp}[1]{\widehat \beta^{{\rm sv}+,\, #1}_{m,k}}  
\newcommand{\hatbetalagm}[1]{\widehat \beta^{{\rm sv}-,\, #1}_{m,k}}  
\newcommand{\hatbetalagpm}[1]{\widehat \beta^{{\rm sv}\pm,\, #1}_{m,k}}

\newcommand{\GG}{ {\rm G} }
\newcommand{\EE}{ {\rm E} }

\newcommand{\MLD}{\text{ mod lower depth}}

\newcommand{\PS}{\sum_{\gamma \in B(\ZZ)\backslash {\rm SL}(2,\ZZ)}}
\newcommand{\seeed}{\varphi}
\newcommand{\summ}{\Phi}

\begin{document}

 {\flushright UUITP-41/21\\
 DCPT-21/13\\[15mm]}

\begin{center}

{\bf {\LARGE \sc  Poincar\'e series for \\[2mm] modular graph forms at depth two}\\[4mm]
{\Large I. Seeds and Laplace systems}}\\[5mm]

\vspace{6mm}
\normalsize
{\large  Daniele Dorigoni${}^{1}$, Axel Kleinschmidt${}^{2,3}$ and Oliver Schlotterer${}^4$}

\vspace{10mm}
${}^1${\it Centre for Particle Theory \& Department of Mathematical Sciences\\
Durham University, Lower Mountjoy, Stockton Road, Durham DH1 3LE, UK}
\vskip 1 em
${}^2${\it Max-Planck-Institut f\"{u}r Gravitationsphysik (Albert-Einstein-Institut)\\
Am M\"{u}hlenberg 1, DE-14476 Potsdam, Germany}
\vskip 1 em
${}^3${\it International Solvay Institutes\\
ULB-Campus Plaine CP231, BE-1050 Brussels, Belgium}
\vskip 1 em
${}^4${\it  Department of Physics and Astronomy, Uppsala University, 75108 Uppsala, Sweden}

\vspace{10mm}

\hrule

\vspace{5mm}

 \begin{tabular}{p{14cm}}
We derive new Poincar\'e-series representations for infinite families of non-holomorphic modular invariant functions that include modular graph forms as they appear in the low-energy expansion of closed-string scattering amplitudes at genus one. The Poincar\'e series are constructed from
iterated integrals over single holomorphic Eisenstein series and their complex conjugates, decorated by suitable
combinations of zeta values. We evaluate the Poincar\'e sums over these
iterated Eisenstein integrals of depth one and deduce new representations for all 
modular graph forms built from iterated Eisenstein integrals at depth two. In a companion
paper, some of the Poincar\'e sums over depth-one integrals going beyond modular graph forms 
will be described in terms of iterated integrals over holomorphic cusp forms and their L-values.
\end{tabular}

\vspace{6mm}
\hrule
\end{center}

\thispagestyle{empty}

\newpage
\setcounter{page}{1}

\setcounter{tocdepth}{2}
\tableofcontents

\bigskip

%%%%%%%%%%%%%%%%%%%%%%%%%%%%%%%%%%%%%%%%%%%%%%%%%%%%%%%%%%%
\section{Introduction}
\label{sec:1}
%%%%%%%%%%%%%%%%%%%%%%%%%%%%%%%%%%%%%%%%%%%%%%%%%%%%%%%%%%%

The low-energy expansion of string scattering amplitudes at genus one introduced infinite classes of
non-holomorphic so-called {\it modular graph forms} (MGFs) \cite{DHoker:2015gmr, DHoker:2015wxz, DHoker:2016mwo}.
The fascinating properties of modular graph forms include multiple zeta values in their expansion around 
the cusp $\tau \rightarrow i \infty$, with $\tau$ the modular parameter of the torus, and intricate networks of algebraic and differential relations. Accordingly,
the study of MGFs has received considerable attention in both the physics
\cite{Green:1999pv, Green:2008uj, Green:2013bza, DHoker:2015gmr, DHoker:2015sve, Basu:2015ayg, DHoker:2015wxz, DHoker:2016mwo, Basu:2016xrt, Basu:2016kli, Basu:2016mmk, DHoker:2016quv, Kleinschmidt:2017ege, Basu:2017nhs, Broedel:2018izr, Ahlen:2018wng,Gerken:2018zcy, Gerken:2018jrq, DHoker:2019txf, Dorigoni:2019yoq, DHoker:2019xef, DHoker:2019mib, DHoker:2019blr, Basu:2019idd,  Gerken:2019cxz, Hohenegger:2019tii, Gerken:2020yii, Basu:2020kka, Vanhove:2020qtt, Basu:2020pey, Basu:2020iok , Hohenegger:2020slq}
and mathematics literature \cite{Brown:mmv, Zerbini:2015rss, Brown:I, Brown:II, DHoker:2017zhq,Zerbini:2018sox, Zerbini:2018hgs, Zagier:2019eus, Berg:2019jhh}, also see
\cite{Gerken:review} for a review, \cite{Gerken:2020aju} for a {\tt Mathematica} implementation and \cite{DHoker:2013fcx, DHoker:2014oxd, Pioline:2015qha, DHoker:2017pvk, DHoker:2018mys, Basu:2018bde, DHoker:2020tcq, DHoker:2020uid, Basu:2021xdt} for generalisations to higher genus.

The direct evaluation of world-sheet integrals in closed-string genus-one amplitudes yields lattice-sum representations of MGFs \cite{Green:1999pv, Green:2008uj, DHoker:2015gmr, DHoker:2015wxz, Gerken:2018jrq}. Their differential equations \cite{DHoker:2016mwo} by contrast suggest to represent MGFs via iterated integrals over holomorphic Eisenstein series ${\rm G}_k(\tau)$ and their complex conjugates \cite{DHoker:2015wxz, Broedel:2018izr, Gerken:2020yii}. The lattice sums refer to the discrete momenta $p=m\tau + n$ on a torus world-sheet with $m,n \in \mathbb Z$, manifest the modular properties and lead to the interpretation of MGFs as discretised Feynman integrals for a scalar field on the torus. However, the lattice sum representation of an MGF is not unique and there are many non-manifest relations among lattice sums. By contrast, iterated-Eisenstein-integral representations are unique~\cite{Nilsnewarticle}, expose the entirety of algebraic and differential relations of MGFs and make the detailed form of their Fourier-expansion accessible. Hence, from their implications for different properties of MGFs, it is desirable to swiftly translate between the lattice-sum and iterated-Eisenstein-integral viewpoints.

In this work, we investigate Poincar\'e-series representations of MGFs and restrict to the modular-invariant case, i.e.\ modular graph functions and modular-invariant combinations of forms. With a Poincar\'e-series representation we mean a rewriting of a modular-invariant function in terms of a sum over images under the modular group ${\rm SL}(2,\mathbb{Z})$ of a simpler function that we call its (Poincar\'e) seed function. As we shall see, our actual space of functions transcends the space of MGFs in a controlled way related to iterated integrals of holomorphic cusp forms, a theme that will be explored in great detail in the companion Part~II~\cite{PartII}.

MGFs associated with one-loop graphs with $k\geq 2$ links are known to be given by non-holomorphic Eisenstein series $\EE_k(\tau)$. These cases can be expressed as (iterated) integrals of a single holomorphic Eisenstein series and hence they all are said to be of depth one. The non-holomorphic Eisenstein series $\EE_k(\tau)$ is known to be proportional to the sum over images of the simple monomial seed $(\Im \tau)^k$, see e.g.~\cite{Iwaniec:2002,Fleig:2015vky}.

Our key results
advance and apply the dictionary between lattice sums and iterated Eisenstein integrals
to the depth-two case, i.e.\ to iterated integrals of two holomorphic Eisenstein series.
In particular, we
generalise the studies of Poincar\'e-series representations of depth-two MGFs
in \cite{DHoker:2015gmr, Ahlen:2018wng, DHoker:2019txf, Dorigoni:2019yoq,Dorigoni:2020oon, Basu:2020kka} to arbitrary weight. On the one
hand, since Poincar\'e series add up images of simpler functions (their Poincar\'e seeds) under the 
modular group ${\rm SL}(2,\mathbb Z)$, they manifest the modular properties of the resulting
MGFs.
On the other hand, we will see from comparison between the iterated-Eisenstein-integral representations of 
MGFs and our representatives of their seed functions that Poincar\'e sums effectively 
add one unit of depth. The choice of seed function for a given Poincar\'e series is of course far from unique and we will discuss different choices of seed functions. 

In this work we  argue that it is always possible to choose a seed for depth-two iterated Eisenstein integrals that is built from depth-one iterated Eisenstein integrals. From this class of seed functions one can obtain the Fourier expansion of the modular-invariant function in a systematic way.  In particular, with our choice of depth-one iterated Eisenstein integrals as seeds, we deduce the precise zeta values appearing in the Laurent polynomials. In fact, iterated Eisenstein integrals will also play a crucial role for finding the
system of Laplace equations obeyed by both the modular invariant MGFs
at depth two and by their seed functions.

We emphasise that the notion of depth of an MGF is in general different from the loop 
order of the graph defining it. While MGFs corresponding to one-loop Feynman graphs
can be represented by iterated Eisenstein integrals of depth one,
the two-loop MGFs do not exhaust all depth-two modular invariant objects.
It was already known that several two-loop MGFs can be reduced to one-loop ones and
odd zeta values \cite{Green:2008uj, DHoker:2015gmr}, which illustrates that the notion of 
depth and loop order are not always lined up. 
In those cases of both depth and loop order two, Poincar\'e-series can be viewed as interpolating 
between double integrals over holomorphic Eisenstein series and double 
sums over lattice momenta: Our choices of seed functions then involve a single sum over 
${\rm SL}(2,\mathbb Z)$ transformations (akin to one lattice momentum) of a depth-one integral.

Poincar\'e seeds for all two-loop MGFs were already derived in \cite{DHoker:2019txf}, and the results of the present work extend those in the reference in several ways. 
Firstly, we give a streamlined ${\rm SL}(2,\ZZ)$-representative of the seed functions which does not contain any powers of $q\bar q$, with $q=\exp(2\pi i\tau)$, in the Fourier expansion and without having to rely on a lattice-sum formulation.
Secondly, we also spell out seed functions for those imaginary combinations of iterated Eisenstein integrals 
of depth two which necessitate three- or higher-loop MGFs, see \cite{Gerken:2020yii} for a simple weight-five example.
Thirdly, although we will not pursue this in detail in the present work, with our methods we could exploit resurgent analysis, as explained in \cite{Dorigoni:2019yoq,Dorigoni:2020oon}, to reconstruct the complete non-perturbative completion of the MGFs, i.e.\ the $q\bar q$ terms, from their perturbative expansion around the cusp $\tau \rightarrow i \infty$. However, the iterated-Eisenstein-integral representation that we shall derive gives a complementary way of obtaining the $q\bar{q}$-terms, see section~\ref{sec:q-series} for further comments. 

In this work and its companion paper \cite{PartII}, we will give a complete description of the
Poincar\'e sums of iterated Eisenstein integrals at depth one with Brown's integration kernels
$\tau^j {\rm G}_k(\tau)$ subject to $k\geq 4$ and $0\leq j\leq k{-}2$ \cite{Brown:mmv}. In this way, we recover 
all MGFs which are expressible in terms of Brown's iterated Eisenstein integrals of depth $\leq 2$. Moreover, 
certain Poincar\'e series turn out
to augment double integrals over holomorphic Eisenstein series by depth-one integrals over holomorphic
cusp forms. These real-analytic modular invariants go beyond MGFs and will
be discussed in the companion Part~II. Iterated integrals of holomorphic 
cusp forms do not admit the lattice-sum representations with integer exponents
characteristic for MGFs. This can be seen from the fact that repeated $\tau$-derivatives of MGFs
give rise to holomorphic Eisenstein series but not to holomorphic cusp forms 
\cite{DHoker:2016mwo, Gerken:2019cxz}. Hence, we identify modular invariant combinations of iterated integrals of holomorphic modular forms and their complex conjugates (conjecturally examples of Brown’s equivariant iterated Eisenstein integrals~\cite{Brown:mmv,Brown:I,Brown:II}) that cannot be represented in terms of MGFs.

Among the real MGFs of depth two, the most prominent instances are the two-loop 
lattice sums $C_{a,b,c}(\tau)$ \cite{DHoker:2015gmr} built from integrals over $a{+}b{+}c$ closed-string 
Green functions. At transcendental weight $a{+}b{+}c=8$, Poincar\'e sums over real seed functions 
similar to those of $C_{a,b,c}(\tau)$ also generate a modular invariant that involves iterated integrals
over the $\Delta_{12}(\tau)$ discriminant, the holomorphic cusp form of modular weight 12. 
Poincar\'e sums over imaginary seed functions already generate iterated integrals over 
$\Delta_{12}(\tau)$ at transcendental weight 7, see sections \ref{sec:FcF} and \ref{sec:odd.dpt1} for further
comments and Part~II for a detailed analysis.

\subsection{Laplace systems}
\label{sec:1.1summ}

The main focus of this pair of papers will be the real and imaginary modular invariant functions that will be denoted by $\FFp{s}{m}{k}$ and $\FFm{s}{m}{k}$, respectively, and labelled by positive integers $s,m,k$ to be explained below. 
More precisely, the $\FFp{s}{m}{k}$ ($\FFm{s}{m}{k}$) are even (odd) under the involution $\tau \rightarrow - \bar \tau$ of the upper half-plane.
 On the one hand, these modular invariant functions $\FFpm{s}{m}{k}$ determine all MGFs of depth two. On the other hand, not all instances of the $\FFpm{s}{m}{k}$ can be expressed as lattice sums and so they transcend the space of MGFs but still will be expressible in terms of iterated integrals over holomorphic modular forms (including cuspidal ones) of depth at most two. For this reason we will be referring to $\FFpm{s}{m}{k}$ as depth-two modular invariant functions. 

The even modular invariant functions $\FFp{s}{m}{k}$ are characterised by inhomogeneous Laplace eigenvalue 
equations similar to those of the two-loop lattice sums MGFs $C_{a,b,c}$ in \cite{DHoker:2015gmr} namely
\begin{equation}
\label{int:Fmk}
\big(\Delta - s(s{-}1) \big) \FFp{s}{m}{k} = {\rm E}_m {\rm E}_k \, , \ \ \ \ \ \
s \in \left\lbrace k{-}m{+}2,k{-}m{+}4,\ldots,k{+}m{-}4,k{+}m{-}2 \right\rbrace \, ,
\end{equation}
where $2\leq m \leq k$ and $\EE_m$, $\EE_k$ are non-holomorphic Eisenstein series and $\Delta=4 (\Im \tau)^2 \partial_\tau \partial_{\bar \tau}$ the ${\rm SL}(2,\mathbb Z)$ invariant Laplacian. This differential equation fixes the asymptotics of $\FFp{s}{m}{k}$ at
the cusp $\tau \rightarrow i\infty$ up to two integration constants. The latter will be inferred from 
Poincar\'e-series representations whose seed functions enjoy shift symmetry 
under $\tau \rightarrow \tau{+}1$. We note that the differential equation is invariant under the swap of the numbers $s$ and $1{-}s$ and we always take $s$ to be the larger one of them. Furthermore, the equation is invariant under the swap of $m$ and $k$ and we label the function $\FFp{s}{m}{k}$ with $m\leq k$.

Moreover, the seed functions of $\FFp{s}{m}{k}$ are systematically reduced to real parts of convergent 
iterated Eisenstein integrals at depth one, multiplied by a positive integer power of $\Im \tau$. If the 
resulting modular invariant function involves a double integral over
holomorphic Eisenstein series $({\rm G}_{2m} ,{\rm G}_{2k})$ with $m\neq k$, we
find that two types of depth-one seed functions lead to the same Poincar\'e 
sums, even if they cannot be related directly by an ${\rm SL}(2,\mathbb Z)$ transformation. This can be thought of as a depth-two generalisation of the standard functional relation between non-holomorphic Eisenstein series $\Gamma(k) \EE_k = \Gamma(1{-}k) \EE_{1-k}$ which superficially relates the seed functions $(\Im \tau)^k$ and $(\Im\tau)^{1-k}$ (up to proportionality) even though they are not related by an ${\rm SL}(2,\mathbb{Z})$ transformation. However, the Poincar\'e sum of the seed $(\Im\tau)^{1-k}$ is not convergent for $k\in\NN$ but typically its Poincar\'e sum is defined by analytic continuation using the functional relation. In the same way, we shall find different ways of expressing the same modular invariant function through different seeds, however, only one of the seeds will have a convergent Poincar\'e sum. 

In a similar fashion we will also introduce odd modular invariant functions of depth two, denoted by $\FFm{s}{m}{k}$, and characterised by inhomogeneous Laplace eigenvalue equations 
\begin{align}
\label{int:Fmkm}
\big(\Delta-s(s{-}1) \big)\FFm{s}{m}{k} &= \frac{ 
 (\nabla \EE_m )(\overline{\nabla} \EE_k)
 - ( \nabla \EE_k) (\overline{\nabla} \EE_m)}{2(\Im \tau)^2}  \, \\
s &\in  \left\lbrace k{-}m{+}1,k{-}m{+}3,\ldots,k{+}m{-}3,k{+}m{-}1 \right\rbrace\,,\nn
\end{align}
with $2\leq m < k$ and
Cauchy--Riemann derivatives $\nabla = 2 i (\Im \tau)^2 \partial_\tau $ 
and $\overline{\nabla}=- 2 i (\Im \tau)^2 \partial_{\bar{\tau}} $.
Under the swap of $m$ and $k$ the right-hand side of the equation changes sign and we always assume $m\leq k$ for $\FFm{s}{m}{k}$ (and in fact $\FFm{s}{m}{m}=0$ for our choice of boundary conditions).

Unlike the $\FFp{s}{m}{k}$, the modular objects $\FFm{s}{m}{k}$ will be cusp forms which do not allow
for integration constants proportional to powers of $\Im \tau$.
Their corresponding seed functions will reduce to imaginary parts of convergent iterated Eisenstein integrals at depth one, multiplied by positive integer powers of $\Im \tau$. However, yet again, the resulting odd modular invariant functions $\FFm{s}{m}{k}$ involve double integrals over holomorphic Eisenstein series $({\rm G}_{2m}, {\rm G}_{2k})$ 
\cite{DHoker:2019txf, Gerken:2020yii}. The Laplace system (\ref{int:Fmkm}) will provide a generalisation of the 
cusp forms first discussed in the references.

The objects ${\rm F}^{\pm(s)}_{m,k}$ together with products of two non-holomorphic Eisenstein series and their Cauchy--Riemann derivatives generate all modular invariant MGFs of depth $\leq 2$. This can be seen for instance from the generating series of MGFs in \cite{Gerken:2020yii} which also features certain dropouts among the iterated Eisenstein integrals at depth $\geq 2$ and cannot contain all the ${\rm F}^{\pm(s)}_{m,k}$. These dropouts can be traced back to Tsunogai's derivation algebra \cite{Tsunogai,Pollack} which also governs Brown's construction of real-analytic modular forms \cite{Brown:II}. As will be detailed in Part~II, those $\FFp{s}{m}{k},\FFm{s}{m}{k}$ beyond MGFs contain iterated integrals of holomorphic cusp forms with ratios of L-values in their coefficients.

\medskip

We note that an equation very similar to~\eqref{int:Fmk} appeared for the first time in the context of higher-derivative corrections to the type IIB low-energy effective action where now ${\rm SL}(2,\mathbb{Z})$ plays the role of U-duality acting on the axio-dilaton~\cite{Green:2005ba}. In this case the indices $m$ and $k$ on the inhomogeneity $\EE_m\EE_k$ are half-integers, see also~\cite{Green:1997tv,Green:1998by,Green:2005ba,Green:2014yxa,Pioline:2015yea,Bossard:2015uga,Bossard:2020xod} for further developments and~\cite{Chester:2020vyz} for recent work in the context of $\mathcal{N}{=}4$ super Yang--Mills theory.
Our focus is on Poincar\'e-seed representations of the solutions and their relation to (single-valued) iterated integrals. One remarkable outcome of our work is that iterated integrals of cusp forms also play a central role for modular-invariant solutions to the Laplace equations. One consequence of this is that the Fourier expansion of the solutions $\FFpm{s}{m}{k}$ can also contain terms that are associated, mode-by-mode, with homogeneous solutions of the Laplace equations, a behaviour that has not been encountered in the U-duality context yet~\cite{Green:2014yxa,Klinger:2018,Chester:2020vyz,Minprogress}. 

Laplace systems akin to~\eqref{int:Fmk} and~\eqref{int:Fmkm} of special type at depth three have been recently investigated in~\cite{Drewitt:2021}.
At depth two in turn, iterated integrals of cusp forms have been studied in relation to so-called higher modular forms in~\cite{Diamantis:2020}. 
More specifically, Poincar\'e seeds built from depth-one iterated integrals of holomorphic cusp forms have been considered in this reference, which can be viewed as the cuspidal counterparts of the seeds in this work.

As further motivation for our work, we stress that 
MGFs appear in the $\alpha'$-expansion of closed-string scattering amplitudes at genus one. As such they are crucial ingredients of the non-perturbatively completed couplings in the Type-IIB low-energy effective action in flat space, multiplying the higher-curvature corrections ${\cal R}^4, D^4 {\cal R}^4$ and $D^6 {\cal R}^4$  known from \cite{Green:1997tv, Green:1997as, Green:2014yxa}. Besides playing a central role for checks of U-duality in Type-IIB superstrings, these couplings are also relevant for precisions tests of the AdS/CFT correspondence.
On the AdS/CFT side, recent developments on the flat-space limit of Type-IIB effective actions on $AdS_5 \times S^5$ involving localisation and conformal-bootstrap methods include \cite{Binder:2019jwn, Binder:2019mpb, Chester:2019pvm, Chester:2019jas, Chester:2020vyz, Green:2020eyj,Dorigoni:2021bvj,Dorigoni:2021guq}, and the interplay with correlation functions in ${\cal N}=4$ super Yang--Mills has for instance been investigated in \cite{Okuda:2010ym, Penedones:2010ue, Alday:2018pdi, Alday:2018kkw, Drummond:2019odu, Aprile:2019rep, Drummond:2019hel, Bissi:2020wtv, Bissi:2020woe}.

%%%%%%%%
\subsection{Preview example}
\label{sec:1.example}
%%%%%%%%

The real-analytic modular invariant functions $\FFpm{s}{m}{k}$ can be represented in 
various ways and the different viewpoints are combined in this work.
As an example, the depth-two function $\FFp{3}{2}{3}$ can be written in the 
following three equivalent forms:
\begin{align}
\label{eq:F233intro}
\FFp{3}{2}{3} &= - \frac14 C_{3,1,1} + \frac{43}{140} \EE_5 -\frac{\zeta_5}{240}\nn\\
&= \PS  \left[ \frac{y^5}{297675} -\frac{y^2\zeta_3}{1890} + \frac{y^2}{315} \Re [\mathcal{E}_0(4,0^2)]\right]_\gamma\\
&=  30 \betasv{2& 1\\4& 6} + 30\betasv{3& 0\\6& 4}
 -20  \zeta_{3} \betasv{1\\6} - \frac{  3  \zeta_{5}}{y}  \betasv{0\\4} + \frac{  \zeta_{5}}{360 } + \frac{  \zeta_{3} \zeta_{5}}{8 y^3} - \frac{  7 \zeta_{7}}{64 y^2}\,,\nn
\end{align}
where $y= \pi \Im \tau$.
The first form is the representation in terms of MGFs that, in particular, involves the double lattice sum $C_{3,1,1}$, see~\eqref{rev.05} for the detailed definition. This form connects directly to expressions arising in perturbative string calculations and has manifest modular invariance in each term. There are many relations among MGFs that are not directly manifest from such a representation and the relation between the number of lattice summations and the depth of the modular invariant is not always one to one.

The second representation of $\FFp{3}{2}{3}$ given in~\eqref{eq:F233intro} is the Poincar\'e-series form in terms of translates under the action of an ${\rm SL}(2,\ZZ)$ element $\gamma$ on a simpler building block. This building block is constructed out of iterated integrals $\mathcal{E}_0(4,0^2)$ of depth one that are discussed in detail in section~\ref{sec:dep1E0}, and the Poincar\'e series are in this sense constructed from lower-depth objects. The (non-unique) Poincar\'e-series form of $\FFp{3}{2}{3}$ makes modular invariance manifest and also gives direct access to certain parts of the Fourier expansion, see~\eqref{eq:FLP}.

The final expression in~\eqref{eq:F233intro} is in terms of single and double integrals $\bsvBRno{j}{k}$ and $\bsvBRno{j_1 &j_2}{k_1 &k_2}$ over Brown's kernels $\tau^{j_i} {\rm G}_{k_i}(\tau)$ along with their complex conjugates \cite{Gerken:2020yii}, see \eqref{eq:bsv1} and \eqref{eq:bsv2} for the precise definitions. 
Their advantage is that they make all relations manifest, meaning that different labels $j_i, k_i$ correspond to independent objects~\cite{Nilsnewarticle}. However, the individual terms do not have good modular properties and only specific combinations are modular invariant.

This pair of articles revolves around the interplay of the different representations of modular-invariant functions $\FFpm{s}{m}{k}$ of depth two defined by the Laplace equations~\eqref{int:Fmk} and~\eqref{int:Fmkm}. In particular, the Poincar\'e-series and iterated-integral representation in terms of $\bsv$ will be combined to explore the space of functions $\FFpm{s}{m}{k}$. While the Poincar\'e-series representation in~\eqref{eq:F233intro} exists for all $\FFpm{s}{m}{k}$, the lattice-sum and the $\bsv$ representation do not always exist. It is the central theme of Part~II to clarify this point and to explain how the space of $\bsv$ has to be extended by iterated integrals of cusp forms in order to represent the $\FFpm{s}{m}{k}$.

%%%%%%%%
\subsection{Outline}
\label{sec:1.2out}
%%%%%%%%

Our work is structured as follows. In section~\ref{sec:2}, we review the basic notions of modular graph functions, iterated Eisenstein integrals and Poincar\'e-series representations. In section~\ref{sec:3}, we then show how the central Laplace equations~\eqref{int:Fmk} and \eqref{int:Fmkm} arise from iterated Eisenstein integrals and how they can be solved using the method of Poincar\'e series. We shall also consider different bases of Poincar\'e seeds where the Laplace equations arrange in a step-form system. The considerations of section~\ref{sec:3} are concerned with the leading-depth contributions to modular-invariant functions, and we explain how to add their lower-depth tails in section~\ref{sec:4}. 
While sections~\ref{sec:3} and~\ref{sec:4} are mostly focussed on the even $\FFp{s}{m}{k}$,
the analogous discussions for the odd $\FFm{s}{m}{k}$ can be found in section~\ref{oddsection}.
Section~\ref{sec:4alt} contains a discussion of alternative seed functions for the depth-two modular invariant functions and how they are related to the ones derived in section~\ref{sec:3}.
In section~\ref{sec:5}, we present possible further directions of investigation. Section~\ref{sec:last} contains concluding remarks.
Several appendices collect additional technical details and more involved examples. 
In an ancillary file appended to the arXiv submission and journal publication 
we enclose a large collection of data and examples.

%%%%%%%%%%%%%%%%%%%%%%%%%%%%%%%%%%%%%%%%%%%%%%%%%%%%%%%%%%%
\section{Review}
\label{sec:2}
%%%%%%%%%%%%%%%%%%%%%%%%%%%%%%%%%%%%%%%%%%%%%%%%%%%%%%%%%%%

In this section, we recall the salient features of MGFs and Poincar\'e sums in order to set the notation and terminology for our results.

%%%%%%%%%%%%%%%%%%%%%%%%%%%%%%%%%%%%%%%%%%%%%%%%%%%%%%%%%%%
\subsection{Modular graph functions}
\label{sec:2.1}
%%%%%%%%%%%%%%%%%%%%%%%%%%%%%%%%%%%%%%%%%%%%%%%%%%%%%%%%%%%

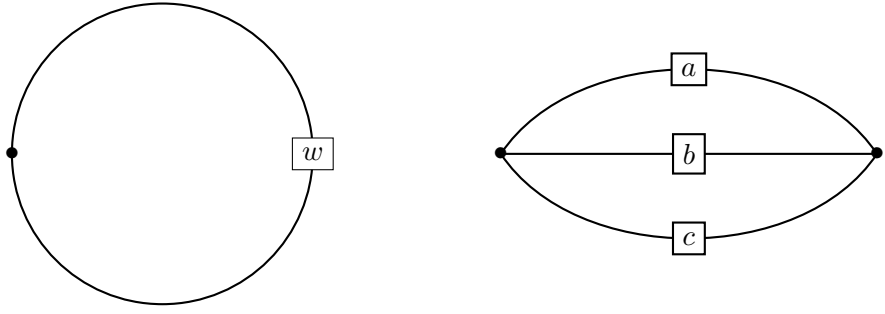
\begin{figure}[t!]
\centering
\begin{tikzpicture}
\draw [thick] (-2,0) circle (2);
\draw (-4,0)node{$\bullet$};
\draw (0,0) node [rectangle,fill=white,draw=black] {$w$};
%%%
\scope[xshift=1.5cm]
    \draw[thick,postaction={decorate}] (1,0)node{$\bullet$} .. controls (2,1.5) and (5,1.5) .. node[fill=white,draw=black]{$a$} (6,0)node{$\bullet$};
    \draw[thick,postaction={decorate}] (1,0) .. controls (2,0) and (5,0) .. node[fill=white,draw=black]{$b$} (6,0);
    \draw[thick,postaction={decorate}] (1,0) .. controls (2,-1.5) and (5,-1.5) .. node[fill=white,draw=black]{$c$} (6,0);
    \endscope
\end{tikzpicture}
\caption{\label{fig:12}\textit{The graphs corresponding to the one-loop and two-loop modular graph functions $\EE_w$ and $C_{a,b,c}$ where a link with a boxed number $w$ indicates $w$ concatenated Green functions.}}
\end{figure}

The central objects in this work are modular graph functions \cite{DHoker:2015wxz} that are generated from
the low-energy expansion of the following configuration-space integral
relevant to closed-string amplitudes\footnote{More precisely, the four-point one-loop
amplitude of type-II superstrings is proportional to the $\tau$-integral of ${\cal M}_4(s_{ij},\tau)$
over the fundamental domain of ${\rm SL}(2,\mathbb Z)$ \cite{Green:1982sw}.
The five-point type-II amplitude in turn involves both ${\cal M}_5(s_{ij},\tau)$ and additional 
configuration-space integrals with singularities as $z_i \rightarrow z_j$ in the integrand 
\cite{Richards:2008jg, Green:2013bza}.} at genus one \cite{Green:1999pv, Green:2008uj}: 
\beq
{\cal M}_n(s_{ij},\tau) = \bigg( \prod_{j=2}^n \int_{\mathfrak T} \frac{\dd^2 z_j}{\Im \tau} \bigg) \prod_{1\leq i<j}^n \exp\bigg( \sum_{1\leq i<j}^n s_{ij} G(z_i{-}z_j,\tau) \bigg)\,.
\label{rev.01}
\eeq
Each puncture $z_j$ for $j\geq 2$ is independently integrated over the torus ${\mathfrak T} = \mathbb C/(\mathbb Z + \tau \mathbb Z)$ with complex modular parameter $\tau$ subject to $\Im \tau>0$, and one can use translation invariance on the torus to set $z_1$ to an arbitrary value. Upon Taylor-expanding the integrand in
the dimensionless Mandelstam invariants $s_{ij} \in \mathbb C$, it remains
to integrate monomials in closed-string Green functions
\beq
G(z,\tau) = - \log \left|  \frac{ \theta_1(z,\tau) }{\eta(\tau)}\right|^2 + \frac{ 2\pi (\Im z)^2 }{\Im \tau} = \frac{ \Im \tau}{\pi} \sum_{(m,n) \neq (0,0)} \frac{ e^{2\pi i (mv-nu)}}{|m\tau {+}n|^2}\,,
\label{rev.02}
\eeq
with the standard Dedekind eta function $\eta(\tau)$ and odd Jacobi theta function $\theta_1(z,\tau)$.\footnote{These are given by the $q$-series, where $q=e^{2\pi i \tau}$,
\begin{equation*}
\eta(\tau) = q^{1/24} \prod_{n= 1}^{\infty} (1-q^n)\,,\quad \quad \theta_1(z,\tau) = i \sum_{n=-\infty}^\infty (-1)^n q^{(n-1/2)^2} e^{2\pi i z(n-1/2)}\,.
\end{equation*}}
The Fourier sum in the last step is only conditionally convergent and is understood using the Eisenstein summation convention~\cite{ApostolTomM1976MfaD}. Such integrals over degree-$w$ monomials in $G(z_i{-}z_j,\tau)$ are referred to as {\it modular graph functions}
of weight $w$, and they are modular invariant since both the Green function (\ref{rev.02}) and the 
measure in (\ref{rev.01}) are. The integrals over the torus punctures $z_j = u_j \tau + v_j$ with $u_j,v_j \in \mathbb R$
and $\frac{\dd^2 z_j}{\Im \tau} = \dd u_j \, \dd v_j$ are
particularly convenient to perform with the lattice-sum representation of the Green function in (\ref{rev.02}).
When visualizing $G(z_i{-}z_j,\tau)$ as an edge between vertices $i$ and $j$, each modular graph function
corresponds to a Feynman graph on the torus. The integrals over $z_j$ impose conservation
of the lattice momenta
\beq
p= m \tau + n \in \Lambda' \, , \ \ \ \ \ \ \Lambda' =( \mathbb Z + \tau \mathbb Z) \setminus \{0\}
\label{rev.03}
\eeq
at each vertex and lead to vanishing modular graph functions for one-particle reducible graphs.
Hence, the simplest non-vanishing modular graph functions are non-holomorphic Eisenstein
series of weight $w \neq 1$ associated with closed one-loop graphs
\beq
{\rm E}_w(\tau) = \bigg( \frac{ \Im \tau}{\pi} \bigg)^w \sum_{p \in \Lambda'} \frac{1}{|p|^{2w}}\,,
\label{rev.04}
\eeq
followed by two-loop modular graph functions of weight $w=a{+}b{+}c$:
\beq
C_{a,b,c}(\tau) =  \bigg( \frac{ \Im \tau}{\pi} \bigg)^{a+b+c} \sum_{p_1,p_2,p_3 \in \Lambda'}
\frac{\delta(p_1{+}p_2{+}p_3)}{|p_1|^{2a}|p_2|^{2b}|p_3|^{2c}} \, .
\label{rev.05}
\eeq
The graphs corresponding to the MGFs $\EE_w$ and $C_{a,b,c}$ are depicted in figure~\ref{fig:12}.

It is straightforward to represent arbitrary modular graph functions as nested lattice sums,
but it requires more effort to find their algebraic relations. Already the one- and two-loop
modular graph functions in (\ref{rev.04}) and (\ref{rev.05}) obey intricate relations over
$\mathbb Q$-linear combinations of multiple zeta values (MZVs)
\beq
\zeta_{n_1,n_2,\ldots,n_r} = \sum_{0< k_1 <k_2<\ldots <k_r} k_1^{-n_1}k_2^{-n_2}\ldots k_r^{-n_r} \, , \ \ \ \ \ \
n_i \in \mathbb N \, , \ \
n_r \geq 2
\label{rev.06}
\eeq
starting with~\cite{DHoker:2015gmr}
\beq
C_{1,1,1}(\tau) = {\rm E}_3(\tau) + \zeta_3 \, , \ \ \ \ \ \
C_{2,2,1}(\tau) = \frac{2}{5} {\rm E}_5(\tau) + \frac{ \zeta_5}{30} \, .
\label{rev.07}
\eeq
Another important property of 
modular graph functions that is not yet readily available from their lattice-sum representations
is their asymptotic expansion around the cusp. In the variables
\beq
y  = \pi \Im \tau \, , \ \ \    \ \ \ q = e^{2\pi i \tau}\, , \ \ \    \ \ \ \bar q = e^{-2\pi i  \bar \tau} \, ,
\label{rev.08}
\eeq
the non-holomorphic Eisenstein series (\ref{rev.04}) can be written as the Fourier series
\begin{align}
\label{eq:FEk}
\EE_w(\tau) &=(-1)^{w-1} \frac{ {\rm B}_{2w} }{(2w)!} (4y)^w + 
\frac{4(2w{-}3)!  \zeta_{2w-1}}{(w{-}2)!(w{-}1)!} (4y)^{1-w}
\\
&\quad + \frac{2}{\Gamma(w)} \sum_{n=1}^{\infty} n^{w-1} \sigma_{1-2w}(n) \left[\sum_{a=0}^{w-1} (4ny)^{-a} \frac{\Gamma(w{+}a)}{a!\, \Gamma(w{-}a)}\right] (q^n+\bar{q}^n)
\notag
\end{align}
with Bernoulli numbers ${\rm B}_{2w}$ that are related to even Riemann zeta values by
\begin{equation}
\label{eq:zB}
2\zeta_{2w} = (-1)^{w+1} \frac{4^w \pi^{2w}}{(2w)!} {\rm B}_{2w}\,, \quad\quad w=1,2,3,\ldots\,.
\end{equation}
%}
Here, we have assumed that $w$ is a positive integer $w>1$ to replace the usual Bessel function $K_{w-1/2}(2\pi |n| \Im\tau)$ appearing in the non-zero Fourier mode by its exact functional form
\begin{equation}
\label{eq:Besselexp}
K_{w-1/2} (z) = \sqrt{\frac{\pi}{2z}} e^{-z} \sum_{\ell=0}^{w-1} (2z)^{-\ell} \frac{\Gamma(w+\ell)}{\ell! \Gamma(w-\ell)}\,.
\end{equation}
Moreover,
\beq
\sigma_{s}(n) = \sum_{d | n} d^s
\label{divsum}
\eeq
denotes a divisor sum over positive divisors of $n$. 

The general form of the expansion of modular graph functions around the cusp follows a structure similar to (\ref{eq:FEk}): 
The coefficients of $q^m \bar q^n$ with $m,n \geq 0$ are Laurent polynomials in $y$ whose
coefficients are $\mathbb Q$-linear combinations of MZVs.\footnote{This follows from the
method of Panzer outlined in \cite{Panzertalk} to express modular graph functions in terms of elliptic multiple zeta values \cite{Enriquez:Emzv} and their complex conjugates. While Zerbini proved the weaker statement that the coefficients in the
Laurent polynomials are cyclotomic multiple zeta values \cite{Zerbini:2015rss}, it is conjectured
\cite{Zerbini:2015rss, DHoker:2015wxz} and supported by a growing body of evidence \cite{DHoker:2019xef, Zagier:2019eus}
that the coefficients in the Laurent polynomials are single-valued MZVs \cite{Schnetz:2013hqa, Brown:2013gia} such as $\zeta_{2w+1}$ with $w\in \mathbb N$. Our results show that up to depth two only single-valued MZVs occur.} The simplest Laurent polynomials of irreducible
two-loop modular graph functions are \cite{Green:2008uj, DHoker:2015gmr}
\begin{align}
C_{2,1,1}(\tau) &= \frac{ 2y^4 }{14175} + \frac{ \zeta_3 y}{45} + \frac{5 \zeta_5}{12y} - \frac{ \zeta_3^2 }{4y^2}+ \frac{ 9 \zeta_7}{16y^3} +O(q,\bar q)\,,
\label{rev.10}
\\
C_{3,1,1}(\tau) &= \frac{ 2y^5 }{155925} + \frac{2 \zeta_3 y^2}{945} - \frac{ \zeta_5}{180} + \frac{7 \zeta_7 }{16y^2}
- \frac{ \zeta_3 \zeta_5 }{2y^3}+ \frac{ 43 \zeta_9}{64y^4} +O(q,\bar q)\, ,
\notag
\end{align}
and the Laurent polynomials for arbitrary $C_{a,b,c}$ are explicitly known \cite{DHoker:2017zhq} in terms of $\zeta_{2w-1}$ and products thereof.
Notice that the first non-trivial single-valued MZV $\zeta_{3,5,3}$ appears in an MGF of trihedral topology of weight seven~\cite{Zerbini:2015rss}. 

In contrast to (\ref{eq:FEk}), the $q$-series of $C_{a,b,c}$ also involves terms $q^m \bar q^n$ with both of $m,n\neq 0$, see \cite{Broedel:2018izr} for their explicit form at $a{+}b{+}c\leq 6$ and \cite{DHoker:2019txf} at general weight. We note that there can also be terms of the form $(q\bar{q})^n$ that are independent of $\Re\tau$ and behave as $e^{-4n y}$. These arise as non-perturbative terms already in the zero mode and are subsumed in the $O(q,\bar{q})$ symbol in~\eqref{rev.10}. 
These $(q\bar{q})^n$ non-perturbative terms can actually be entirely reconstructed from the purely perturbative Laurent polynomials, or rather a suitable deformation thereof, using resurgence analysis, see \cite{Dorigoni:2019yoq,Dorigoni:2020oon}, we will however not discuss such construction in the present work.

%%%%%%%%%%%%%%%%%%%%%%%%%%%%%%%%%%%%%%%%%%%%%%%%%%%%%%%%%%%
\subsection{Iterated Eisenstein integrals}
\label{sec:2.2}
%%%%%%%%%%%%%%%%%%%%%%%%%%%%%%%%%%%%%%%%%%%%%%%%%%%%%%%%%%%

In this section, we briefly review the formalism of iterated integrals over holomorphic Eisenstein
series, leaving the analogous discussion of iterated integrals over holomorphic cusp forms to Part~II.

Modular graph functions can be represented via iterated integrals over holomorphic
Eisenstein series and their complex conjugates whose coefficients are $\mathbb Q$-linear
combinations of MZVs \cite{DHoker:2015wxz, DHoker:2016mwo, Broedel:2018izr, Panzertalk, DHoker:2019txf, Gerken:2019cxz, Gerken:2020yii}.
This follows from their differential equations with respect to the Cauchy--Riemann operator
\beq
\nabla = 2 i (\Im \tau)^2 \partial_\tau \, , \ \ \ \ \ \ \overline \nabla =  - 2 i (\Im \tau)^2 \partial_{\bar \tau}\,,
\label{rev.11}
\eeq
which maps modular graph functions to non-holomorphic modular forms dubbed 
{\it modular graph forms} \cite{DHoker:2016mwo}, possibly accompanied by
holomorphic Eisenstein series
\beq
{\rm G}_k(\tau) = \sum_{p \in \Lambda'} \frac{1}{p^k} 
= 2\zeta_{k} + \frac{2 (2\pi i)^k}{(k{-}1)!} \sum_{n>0} \sigma_{k-1}(n) q^n
\, , \ \ \ \ \ \ k \geq 4 \, ,
\label{rev.12}
\eeq
see (\ref{divsum}) for the divisor sum $\sigma_{k-1}(n)$.

The Cauchy--Riemann equations of modular graph functions with known asymptotics at the cusp
can be solved via iterated Eisenstein integrals. This exposes the entirety of their algebraic relations. 
There are different ways of defining iterated Eisenstein integrals~\cite{Brown:mmv, Broedel:2015hia, Broedel:2018izr}. In the present work, we require only a subset of the general case and therefore restrict to presenting the relevant definitions.

\subsubsection{Depth-one iterated Eisenstein integrals}
\label{sec:dep1E0}

The integral for $k>0$
\begin{align}
\label{eq:E0depth1}
\mathcal{E}_0(k,0^p;\tau) &= \frac{(2\pi i)^{p+1-k}}{p!}  \int_\tau^{i \infty} \dd\tau_1 (\tau{-}\tau_1)^p \GG_{k}^0(\tau_1)
\end{align}
is said to be an iterated integral of depth one and the notation $0^p$ is a short-hand of $p$ successive zeros. Higher-depth versions, where the iterated integral structure becomes more evident, can be found in~\cite{Broedel:2015hia,Broedel:2018izr}. The holomorphic Eisenstein series $\GG_{k}^0$ appearing in the integrand has its zero mode removed compared to~\eqref{rev.12} and so is defined as
\begin{align}
{\rm G}_k^0(\tau) &= \left\{ \begin{array}{cl}
{\rm G}_k(\tau) - 2 \zeta_k &: \ k>0\, , \ {\rm even} \\[1mm]
0&: \ k>0  \, ,\ {\rm odd} 
%\\
\end{array} \right.
\end{align}
and we additionally define ${\rm G}_0^0=-1$.\footnote{The conditionally convergent and non-modular form $\GG_2(\tau)$ does not play any role in our analysis.}
The integral in (\ref{eq:E0depth1}) converges for $p \geq 0$ 
and from the $q$-expansion of~\eqref{rev.12} one can deduce~\cite{Broedel:2015hia, Dorigoni:2020oon}
\begin{align}
\mathcal{E}_0(k,0^p;\tau) &= -\frac{2}{(k{-}1)!}\sum_{m,n=1}^{\infty} \frac{m^{k-1}}{(mn)^{p+1}} q^{mn} 
=-\frac{2}{(k{-}1)!}\sum_{m=1}^{\infty} m^{k-p-2}\sigma_{1-k}(m) q^m  \label{eq:E0sigma} \\
&  =-\frac{2}{(k{-}1)!}\sum_{m=1}^{\infty} m^{-p-1}\sigma_{k-1}(m) q^m\,.
\notag
\end{align}
This expression can also be considered for arbitrary $p$.
The way we shall use~\eqref{eq:E0sigma} is mainly in the other direction, namely such that we can translate a term involving divisor sums (\ref{divsum}) into iterated integrals.

The non-holomorphic Eisenstein series $\EE_k$ from~\eqref{eq:FEk} can be recast in terms of~\eqref{eq:E0depth1} by using the expansion of the Bessel function~\eqref{eq:Besselexp}
as \cite{Ganglzagier, DHoker:2015wxz}
\begin{align}
\EE_k (\tau) &= (-1)^{k-1} \frac{ {\rm B}_{2k} }{(2k)!} (4y)^k + 
\frac{4(2k{-}3)!  \zeta_{2k-1}}{(k{-}2)!(k{-}1)!} (4y)^{1-k}
\nn\\
&\quad\quad
- 2 \frac{\Gamma(2k)}{\Gamma(k)} \sum_{a=0}^{k-1}  (4y)^{-a} \frac{\Gamma(k{+}a)}{a!\Gamma(k{-}a)} \Re \mathcal{E}_0\big(2k, 0^{k-1+a};\tau\big)\nn\\
&= (-1)^{k-1} \frac{ {\rm B}_{2k} }{(2k)!} (4y)^k + 
\frac{4(2k{-}3)!  \zeta_{2k-1}}{(k{-}2)!(k{-}1)!} (4y)^{1-k}
\label{eq:EII} \\
&\quad\quad +
\left[ -\frac{1}{2\pi i} \frac{\Gamma(2k)}{[\Gamma(k)]^2} (4y)^{1-k} \int_\tau^{i \infty} \dd\tau_1 (\tau-\tau_1)^{k-1}(\bar\tau-\tau_1)^{k-1} \GG_{2k}^0(\tau_1) + \cc \right]\,. \notag
\end{align}
From this formula or the lattice-sum representation (\ref{rev.04}) one can check that \cite{DHoker:2016mwo} 
\begin{align}
(\pi \nabla)^k \EE_k (\tau) &= 
\frac{\Gamma(2k)}{\Gamma(k)}   (\Im \tau)^{2k} \left[ 2\zeta_{2k} + \GG_{2k}^0(\tau)\right] = \frac{\Gamma(2k)}{\Gamma(k)}   (\Im \tau)^{2k}  \GG_{2k}(\tau)\, ,
\label{eq:Ediff}
\end{align}
where the Cauchy--Riemann derivative $\nabla$ was introduced in~\eqref{rev.11} and we record the following useful identity
\begin{align}
\label{eq:CRII}
(\pi \nabla)^s \left[ i y^{1-s} \int_{\tau}^{i \infty} \dd\tau_1 (\tau-\tau_1)^{s-1} (\bar\tau-\tau_1)^{s-1} f(\tau_1)\right]  =  2\pi  \Gamma(s) 4^{s-1} (\Im\tau)^{2s} f(\tau) \, .
\end{align}
The integral converges for any function $f$ exponentially decaying for $\tau\to i\infty$ (without any assumption on the modular properties), and (\ref{eq:CRII}) will also be applied to more general $f$ based on tangential-base-point regularisation of endpoint divergences \cite{Brown:mmv}.

\subsubsection{More general iterated Eisenstein integrals}

Another key role will be played by certain iterated integrals denoted by $\bsv$ defined in \cite{Gerken:2020yii}\footnote{The superscript `${\rm sv}$' indicates that these real-analytic functions are conjecturally the single-valued versions of holomorphic iterated integral $\beta$ defined in~\cite{Gerken:2020xfv}. We shall often suppress the argument $\tau$ in order not to clutter the notation.}
\begin{align}
\label{eq:bsv1}
\bsvBR{j}{k}{\tau} =  \frac{(2\pi i)^{-1}}{(4y)^{k-2-j}} \bigg\{ \int\limits_{\tau}^{i \infty} \dd \tau_1 (\tau{-}\tau_1)^{k-2-j} (\bar\tau{-}\tau_1)^{j} \GG_k(\tau_1) - \! \int\limits_{\bar\tau}^{-i\infty} \! \dd\bar\tau_1 (\tau{-}\bar\tau_1)^{k-2-j} (\bar\tau{-}\bar\tau_1)^{j} \overline{\GG_k(\tau_1)} \bigg\}
\end{align}
as well as their depth-two generalisations
\begin{align}
\label{eq:bsv2}
\bsvBR{j_1 &j_2}{k_1 &k_2}{\tau} &=
\sum_{p_1=0}^{k_1{-}2{-}j_1} \sum_{p_2=0}^{k_2{-}2{-}j_2} \frac{\binom{k_1{-}2{-}j_1}{p_1}\binom{k_2{-}2{-}j_2}{p_2}}{(4y)^{p_1+p_2}} \overline{\alphaBR{j_1 +p_1&j_2+p_2}{k_1 &k_2}{\tau}}
 + \frac{(2\pi i)^{-2}}{(4y)^{k_1+k_2-j_1-j_2-4}} \\
&\! \! \!  \! \! \quad \times \bigg\{
\int\limits^{i\infty}_\tau \dd\tau_2  (\tau{-}\tau_2)^{k_2-j_2-2} (\bar\tau{-}\tau_2)^{j_2} \GG_{k_2}(\tau_2) \int\limits^{i\infty}_{\tau_2} \dd\tau_1 (\tau{-}\tau_1)^{k_1-j_1-2}(\bar\tau{-}\tau_1)^{j_1}  \GG_{k_1}(\tau_1) \nn\\
&\! \! \! \! \! \quad\quad -  \int\limits^{i\infty}_\tau \dd\tau_2(\tau{-}\tau_2)^{k_2-j_2-2} (\bar\tau{-}\tau_2)^{j_2}  \GG_{k_2}(\tau_2) \! \int\limits^{-i\infty}_{\bar\tau} \! \dd\bar\tau_1  (\tau{-}\bar\tau_1)^{k_1-j_1-2}(\bar\tau{-}\bar\tau_1)^{j_1}\overline{\GG_{k_1}(\tau_1)}\nn\\
&\! \! \! \! \! \quad\quad + \! \int\limits^{-i\infty}_{\bar\tau} \! \dd\bar\tau_1(\tau{-}\bar\tau_1)^{k_1-j_1-2}(\bar\tau{-}\bar\tau_1)^{j_1} \overline{\GG_{k_1}(\tau_1)} \! \int\limits^{-i\infty}_{\bar\tau_1} \! \dd\bar\tau_2  (\tau{-}\bar\tau_2)^{k_2-j_2-2} (\bar\tau{-}\bar\tau_2)^{j_2} \overline{\GG_{k_2}(\tau_2)}\bigg\}\,.\nn
\end{align}
The objects $\overline{\alpha[\begin{smallmatrix} j_1 &j_2\\ k_1 &k_2 \end{smallmatrix}]}$ appearing for depth two are purely antiholomorphic functions which are determined on a case-by-case basis in \cite{Gerken:2020yii, Gerken:2020xfv} and preserve
the differential equations\footnote{The higher-depth version of these equations can be found in~\cite{Gerken:2020yii}.}
\begin{subequations}
\begin{align}
2\pi i (\tau{-}\bar \tau)^2 \partial_\tau
\bsvBR{j}{k}{\tau} &= (k{-}2{-}j) \bsvBR{j+1}{k}{\tau}  -\delta_{j,k-2} (\tau{-}\bar\tau)^k \GG_k(\tau) \,,\\
2\pi i (\tau{-}\bar \tau)^2 \partial_\tau
\bsvBR{j_1 &j_2}{k_1 &k_2}{\tau} &=
 (k_1{-}j_1{-}2) \bsvBR{j_1+1 &j_2}{k_1 &k_2}{\tau} +  (k_2{-}j_2{-}2) \bsvBR{j_1 &j_2+1}{k_1 &k_2}{\tau}   \notag \\
 &\quad - \delta_{j_2,k_2-2} (\tau{-}\bar \tau)^{k_2} {\rm G}_{k_2}(\tau) \bsvBR{j_1}{k_1}{\tau} 
 \label{rev.20}
\end{align}
\end{subequations}
manifested by the integral definitions above.
The antiholomorphic integration constants 
$\overline{\alpha[\begin{smallmatrix} j_1 &j_2\\ k_1 &k_2 \end{smallmatrix}]}$  are 
invariant under $\tau \rightarrow \tau{+}1$ and believed to be expressible via $\zeta_{2w+1}$
multiplying antiholomorphic iterated Eisenstein integrals (\ref{eq:E0depth1}) at depth $\leq 1$ \cite{Gerken:2020yii}. The absence of similar integration constants at depth one follows from an analysis of the limiting behaviour $\tau\to i\infty$. The simplest explicit examples include \cite{Gerken:2020yii}
\begin{align}
\alpha[\begin{smallmatrix}  1 &0 \\ 4 &4 \end{smallmatrix}] &= 0 \, ,
&\alpha[\begin{smallmatrix}  2 &0 \\ 6 &4 \end{smallmatrix}] &= - \frac{ \zeta_3 }{630} {\cal E}_0(4)\, ,
\notag \\
\alpha[\begin{smallmatrix}  2 &0 \\ 4 &4 \end{smallmatrix}] &= \frac{2 \zeta_3}{3}  {\cal E}_0(4)\, ,
&\alpha[\begin{smallmatrix}  1 &2 \\ 6 &4 \end{smallmatrix}] &= - \frac{ \zeta_3 }{210} {\cal E}_0(4,0)
-  \frac{2 \zeta_3}{3}  {\cal E}_0(6,0) \, ,
\label{exsalpha} \\
\alpha[\begin{smallmatrix}  2 &1 \\ 4 &4 \end{smallmatrix}] &= \frac{2 \zeta_3}{3}  {\cal E}_0(4,0) \, ,
&\alpha[\begin{smallmatrix}  4 &0 \\ 6 &4 \end{smallmatrix}] &= - \frac{ 2\zeta_3 }{105} {\cal E}_0(4,0^2)
+ \frac{2 \zeta_5}{5}  {\cal E}_0(4)  \, ,
\notag
\end{align}
and we will discuss them in more detail in sections~\ref{sec:4.3} and \ref{sec:odd.bsvrep}.
In fact, the methods of this work lead to new ways of determining the
$\overline{\alpha[\begin{smallmatrix} j_1 &j_2\\ k_1 &k_2 \end{smallmatrix}]}$
and we will provide a large number of examples in an ancillary file.
The shuffle property
\begin{align}
\label{eq:betashuffle}
\betasv{j_1\\k_1}\betasv{j_2\\k_2} = \betasv{j_1&j_2\\k_1&k_2} + \betasv{j_2&j_1\\k_2&k_1}
\end{align}
places constraints on combinations of the integration constants in the form
\begin{align}
\alphaBRno{j_1 &j_2}{k_1 &k_2} = - \alphaBRno{j_2&j_1}{k_2 &k_1}
\end{align}
since there are no integration constants at depth one.

Note that we can use the integral representation \eqref{eq:E0depth1} to rewrite \eqref{eq:bsv1} in the suggestive form
\begin{align}
\betasv{j \\k} &\nn=  \frac{ {\rm B}_{k}  j! (k{-}2{-}j)! (-4y)^{j+1} }{k! \, (k{-}1)!}  +\sum_{a=0}^{j} (k{-}j{-}2{+}a)! \binom{j}{a} (4y)^{2+2j-k-a} {\cal E}_0(k,0^{k-j-2+a})\\
&\quad\label{eq:betasvE0}+\sum_{b=0}^{k-j-2} (j{+}b)! \binom{k{-}2{-}j}{b} (4y)^{-b}\overline{ {\cal E}_0(k,0^{j+b})} \,,
\end{align}
where a tangential base-point regularisation has been used to compute the $y^{j+1}$ term of depth zero, see \cite{Brown:mmv, Broedel:2018izr, Gerken:2020yii}.
A similar representation, although involving also depth-two iterated integrals, can be derived for \eqref{eq:bsv2} as well, modulo the presence of these antiholomorphic objects $\overline{\alpha[\begin{smallmatrix} j_1 &j_2\\ k_1 &k_2 \end{smallmatrix}]}$. The explicit form of their exponentially suppressed terms will be given in section \ref{sec:q-series}, i.e.\ modes of the form $q^m \bar q^n$ with both of $m,n> 0$ due to crossterms ${\cal E}_0 \overline{ {\cal E}_0}$.

As a major result of this work, we shall extend the iterated-Eisenstein-integral representations~\cite[Eq.~(4.9)]{Gerken:2020yii}
\begin{align}
(\pi \nabla)^m {\rm E}_k &= \Big( {-}\frac{1}{4} \Big)^{m} \frac{ (2k{-}1)! }{(k{-}1)! (k{-}1{-}m)!} 
\bigg\{
{-} \betasv{ k-1+m\\ 2k} + \frac{ 2 \zeta_{2k-1} }{(2k{-}1) (4y)^{k-1-m} }
\bigg\} \,,\notag \\
\frac{(\pi \overline{\nabla})^m {\rm E}_k}{y^{2m}} &=  \frac{({-}4)^{m} (2k{-}1)! }{(k{-}1)! (k{-}1{-}m)!} 
\bigg\{
{-} \betasv{ k-1-m\\ 2k} + \frac{ 2 \zeta_{2k-1} }{(2k{-}1) (4y)^{k-1+m} }
\bigg\} \, ,
\label{oct24.2}
\end{align}
valid for $0{\leq}m{<}k$,
to general depth-two MGFs and modular invariant functions. For instance, the two-loop modular 
graph functions $C_{a,b,c}$ in (\ref{rev.05}) will be related to 
$\beta^{\rm sv}[\begin{smallmatrix} j_1 &j_2\\ k_1 &k_2 \end{smallmatrix}]$
 subject to the modular-invariance condition $j_1{+}j_2 = \frac{1}{2}(k_1{+}k_2{-}4)$.

{}From the definitions~\eqref{eq:bsv1} and~\eqref{eq:bsv2} one can check the following reality and modularity properties of the $\bsv$ at leading order in depth~\cite{Gerken:2020yii}:
\begin{align}
\overline{\bsvBR{j}{k}{\tau}} &= (4y)^{2+2j-k} \bsvBR{k-2-j}{k}{\tau} \,,  \notag \\[2mm]
\bsvBR{j}{k}{-\tfrac{1}{\tau}} &= \bar\tau^{k-2-2j}  \bsvBR{j}{k}{\tau} \MLD
\end{align}
and
\begin{align}
\label{eq:bsv2prop}
\overline{\bsvBR{j_1 &j_2}{k_1 &k_2}{\tau}} &= (4y)^{4+2j_1+2j_2-k_1-k_2} \bsvBR{k_2-2-j_2&k_1-2-j_1}{k_2 &k_1}{\tau} \MLD \,,\nn\\[2mm]
\bsvBR{j_1 &j_2}{k_1 &k_2}{-\tfrac{1}{\tau}} &= \bar\tau^{k_1+k_2-4-2j_1-2j_2} \bsvBR{j_1 &j_2}{k_1 &k_2}{\tau} \MLD\,,
\end{align}
where we recall $y=\pi \Im\tau$.

The lower-depth terms in the complex-conjugation properties of 
$\beta^{\rm sv}[\begin{smallmatrix} j_1 &j_2\\ k_1 &k_2 \end{smallmatrix}]$ can be entirely
attributed to the $\overline{\alpha[\begin{smallmatrix} j_1 &j_2\\ k_1 &k_2 \end{smallmatrix}]}$ in (\ref{eq:bsv2})
\begin{align}
\overline{\betasv{j_1 &j_2\\ k_1 &k_2}} &= (4y)^{4+2j_1+2j_2-k_1-k_2} \betasv{k_2-2-j_2&k_1-2-j_1\\k_2 &k_1}\nn\\
&\quad + \sum_{p_1=0}^{k_1{-}2{-}j_1} \sum_{p_2=0}^{k_2{-}2{-}j_2} (4y)^{-p_1-p_2}\binom{k_1{-}2{-}j_1}{p_1}\binom{k_2{-}2{-}j_2}{p_2} \alphaBRno{j_1 +p_1&j_2+p_2}{k_1 &k_2} \label{eq:ccbsv2}\\
&\quad -  \sum_{p_1=0}^{j_1} \sum_{p_2=0}^{j_2} (4y)^{4+2j_1+2j_2-k_1-k_2-p_1-p_2} \binom{j_1}{p_1}\binom{j_2}{p_2} \overline{\alphaBRno{k_2-2-j_2+p_2&k_1-2-j_1+p_1}{k_2 &k_1}}\,. \nn
\end{align}
Even though this is not manifest from the above equation, the $\alpha$ and $\bar{\alpha}$ on the right-hand side always conspire to produce $\bsv$ of depth one, plus possibly depth-zero terms. This follows from the fact that all iterated integrals in the generating series for MGFs \cite{Gerken:2020yii} can be expressed exclusively through $\bsv$, and this property is preserved by complex conjugation.

The examples of $\overline{\alphaBRno{j_1 &j_2}{4 &4}}$ in (\ref{exsalpha}) for instance lead to
\begin{equation}
\overline{\bsvBR{1 &0}{4 &4}{\tau}}  
= \frac{  \bsvBR{2 &1}{4 &4}{\tau} }{16y^2} - \frac{ \zeta_3}{24y^2} \bsvBR{1}{4}{\tau} + \frac{ \zeta_3}{96 y^3}
\bsvBR{2}{4}{\tau} - \frac{ \zeta_3}{2160}
\, ,
\end{equation}
and a complete list of $\overline{\beta^{\rm sv}[\begin{smallmatrix} j_1 &j_2\\ k_1 &k_2 \end{smallmatrix}]}$  
with $k_1{+}k_2\leq 28$ can be found in the ancillary file.

\medskip

As in this work we are interested in modular invariant functions, most of the $\beta^{\rm sv}$ of depth two appearing in the remainder of the paper satisfy $j_1{+}j_2=\frac{1}{2} (k_1{+}k_2{-}4)$. For these values of the parameters we define the following depth-two combinations that shall feature prominently in the subsequent analysis
\begin{equation}
\label{eq:betalag}
\betalagpm{j} =
\bsvBRno{2m-2-j& k-m+j}{2m& 2k} \pm
\bsvBRno{k+m-2-j &j}{2k& 2m}
\end{equation}
with $ 0\leq j \leq 2m{-}2$. These combinations are modular invariant modulo terms of lower depth by~\eqref{eq:bsv2prop}. 
The $\beta^{{\rm sv}\pm}$ have eigenvalue $\pm1$ under complex conjugation modulo lower depth and we shall refer to them as \textit{even} and \textit{odd} combinations, respectively. In (\ref{eq:betalag}), we have introduced the integers $m=\frac{k_1}{2}$ and $k=\frac{k_2}{2}$ and we shall assume, without loss of generality, that $m\leq k$ throughout sections~\ref{sec:3} to~\ref{oddsection}. 
Note that the odd combinations can vanish for some values of parameters, e.g.
\beq
\beta^{{\rm sv}-,j}_{k,k}=0 \, , \ \ \ \ \ \ 0\leq j\leq 2k{-}2\, .
\label{nooddkk}
\eeq
Most $\bsv$ appear in the generating series of all MGFs~\cite{Gerken:2020yii} and therefore possess representations as (nested) lattice sums over discrete loop momenta of $\alpha'$-expanded genus-one string amplitudes. The reality properties of the MGFs can be used to determine most of the integration constants $\overline{\alpha[\begin{smallmatrix} j_1 &j_2\\ k_1 &k_2 \end{smallmatrix}]}$; however, there are some cases, controlled by Tsunogai's derivation algebra~\cite{Tsunogai,Pollack}, where this is not possible with the methods of~\cite{Gerken:2020yii}. One of our new results here and in Part~II is a determination of \textit{all} integration constants at depth two even beyond the constraints of the derivation algebra since the class of functions we are reaching via Poincar\'e series is larger than that of MGFs.

The $\bsv$ defined here are expected to be equivalent to the single-valued iterated Eisenstein integrals defined
for arbitrary depth by Brown in~\cite{Brown:mmv, Brown:I, Brown:II}. 
The tentative analogues of the integration constants $\overline{\alpha[\begin{smallmatrix} j_1 &j_2\\ k_1 &k_2 \end{smallmatrix}]}$ in Brown's setup 
are determined by the modular properties of holomorphic iterated Eisenstein integrals.

%%%%%%%%%%%%%%%%%%%%%%%%%%%%%%%%%%%%%%%%%%%%%%%%%%%%%%%%%%%
\subsection{Poincar\'e series}
\label{sec:2.3}
%%%%%%%%%%%%%%%%%%%%%%%%%%%%%%%%%%%%%%%%%%%%%%%%%%%%%%%%%%%

The modular invariant functions appearing in this paper can be expressed as Poincar\'e series, i.e.\ as a sum over images of a so-called seed function under the action of ${\rm SL}(2,\ZZ)$~\cite{Iwaniec:2002,Fleig:2015vky}. Denoting a modular invariant function by $\summ(\tau)$ and its seed by $\seeed(\tau)$, the Poincar\'e series is
\begin{align}
\label{eq:Psum}
\summ (\tau) = \PS \seeed(\gamma\cdot \tau)\,,
\end{align}
where
\begin{align}
\label{eq:SLact}
\gamma = \begin{pmatrix} a& b\\ c& d\end{pmatrix} \in {\rm SL}(2,\ZZ)\,,
\quad\quad \gamma\cdot \tau = \frac{a\tau+b}{c\tau+d}\,,
\end{align}
and we assumed that the seed function is periodic in the real direction, $\seeed(\tau{+}n)=\seeed(\tau)$ for all $n\in\ZZ$, which explains the (Borel) stabiliser
\begin{align}
B(\ZZ) = \left\{ \begin{pmatrix} \pm 1 & n\\0 & \pm1\end{pmatrix} \,\middle|\, n\in\ZZ\right\} \subset {\rm SL}(2,\ZZ)
\end{align}
in~\eqref{eq:Psum}. The Poincar\'e sum~\eqref{eq:Psum} is only absolutely convergent for appropriate seeds but can often be defined in other cases by analytic continuation when $\seeed$ depends on a complex parameter.

The simplest instance of a Poincar\'e series is
\begin{align}
\label{eq:PSEk}
\EE_k(\tau) &= \frac{2\zeta_{2k}}{\pi^k}   \PS (\Im \gamma\cdot \tau)^k 
\end{align}
that converges absolutely for $\Re(k)>1$. The seed here is given by $\seeed(\tau) = \frac{2\zeta_{2k}}{\pi^k} (\Im \tau)^k =\frac{2\zeta_{2k}}{\pi^{2k}} y^k$ and for integer $k$ the prefactor of $y^k$ becomes the rational number $(-1)^{k-1} \frac{4^k {\rm B}_{2k} }{(2k)!}$.  For $\Re(k)<1$, the non-holomorphic Eisenstein series can be defined by analytic continuation and one has the functional relation
\begin{align}
\label{eq:FR}
\Gamma(k) \EE_k(\tau) =  \Gamma(1{-}k) \EE_{1-k}(\tau)\,.
\end{align}
In our convention $\EE_0(\tau)=-1$ whereas $\EE_1(\tau)$ is infinite.

As an MGF, the Eisenstein series $\EE_k$ is of depth one while its seed in (\ref{eq:PSEk}) is a pure power of $y=\pi \Im \tau$ which is of depth zero. This exemplifies that the transition to the Poincar\'e seed reduces the functional complexity, and this viewpoint was exploited in~\cite{DHoker:2015gmr,Ahlen:2018wng,Dorigoni:2019yoq,Basu:2020kka} to obtain Poincar\'e-series representations of depth-two MGFs. As an example we recall the Poincar\'e-series representation of the two-loop modular graph function $C_{2,1,1}$ in (\ref{rev.05}) from~\cite{DHoker:2015gmr}
\begin{align}
C_{2,1,1}(\tau) &= \sum_{\gamma \in B(\ZZ)\backslash {\rm SL}(2,\ZZ)}  \left[
\frac{2y^4}{14175}+ \frac{ y \zeta_3}{90} + \frac{ y}{90} \sum_{m=1}^{\infty} \sigma_{-3}(m) (q^m + \bar q^m) 
 \right]_{\gamma}
  \notag\\
 &= \sum_{\gamma \in B(\ZZ)\backslash {\rm SL}(2,\ZZ)}  \left[
\frac{2y^4}{14175}+ \frac{ y \zeta_3}{90} - \frac{ y}{15} \Re  {\cal E}_0(4, 0^2;\tau) 
 \right]_{\gamma} \, , \label{fristcase}
\end{align}
and that of $C_{3,1,1}$ from~\cite{Ahlen:2018wng}
\begin{align}
C_{3,1,1}(\tau) &= \sum_{\gamma \in B(\ZZ)\backslash {\rm SL}(2,\ZZ)}  \left[ \frac{2 y^5}{155\,925}  + \frac{2\zeta_3}{945} y^2 - \frac{\zeta_5}{60} \bigg(\frac{y}{\pi} \bigg)^\epsilon +  \frac{2y^2}{945} \sum_{m=1}^{\infty} \sigma_{-3}(m) ( q^m + \bar{q}^m)\right]_{\gamma}\nn\\
&=\sum_{\gamma \in B(\ZZ)\backslash {\rm SL}(2,\ZZ)}  \left[ \frac{2 y^5}{155\,925}  + \frac{2\zeta_3}{945} y^2 - \frac{\zeta_5}{60} \bigg(\frac{y}{\pi} \bigg)^\epsilon -    \frac{2y^2}{315}\Re \mathcal{E}_0(4,0^2;\tau )\right]_{\gamma}\,,
\label{secondcase}
\end{align}
where the notation $[\cdots]_{\gamma}$ means that $\gamma$ acts on all occurrences of $\tau$ inside the bracket using the fractional linear action~\eqref{eq:SLact}. 
In the second lines, we have rewritten the $q$-series in terms of the real part of an iterated Eisenstein integral~\eqref{eq:E0depth1} with $q$-expansion~\eqref{eq:E0sigma}. 

We note that, as discussed in~\cite{Ahlen:2018wng,Dorigoni:2019yoq}, both of the cases (\ref{fristcase}) and (\ref{secondcase}) require some care: The $C_{2,1,1}$ seed function contains a linear term in $y$ that would lead to a divergence upon Poincar\'e summation, see~\eqref{eq:FR}. By contrast, the $C_{3,1,1}$ example has, as written, a term $y^\epsilon$ whose Poincar\'e sum goes to a constant after Poincar\'e summation and use of~\eqref{eq:FR}. Both of these cases have to be dealt with using analytic continuation. For the case of $C_{3,1,1}$ we have shown this explicitly here using the $y^\epsilon$ with $\epsilon\to 0$ after the Poincar\'e sum. In the case of $C_{2,1,1}$ this is slightly more subtle but can be done to arrive at finite Fourier expansions \cite{Dorigoni:2019yoq}. This is reviewed in more detail in appendix~\ref{app:Poincare}.

\subsubsection{Laurent polynomials from Poincar\'e series}

While the Poincar\'e-series representation \eqref{eq:Psum} in terms of the seed reduces the depth of the modular invariant function by one unit, it makes extracting some properties of the modular invariant function $\summ$ more cumbersome. For instance, extracting the Laurent polynomial of the zeroth Fourier mode of $\summ$ involves now additional steps. For Eisenstein series going from~\eqref{eq:PSEk} to~\eqref{eq:FEk} is standard~\cite{Iwaniec:2002,Fleig:2015vky}, but for general Poincar\'e series the analysis is more involved and relies on certain Kloosterman sums. These were studied in~\cite{Ahlen:2018wng,Dorigoni:2019yoq}, where it was shown how to determine the Laurent polynomial (LP) of the Poincar\'e sum of seeds that are of the form $\sigma_a(\ell) (4\pi \ell )^b (\Im \tau)^r (q^\ell+\bar{q}^\ell)$, where $\ell>0$, such as the terms above, where we recall $y=\pi\Im\tau$. This is reviewed in appendix~\ref{app:Poincare}, where a quantity $I(a,b,r)=I(-a,a+b,r)$ for the Laurent polynomial of such seeds is given in~\eqref{eq:Iabr} that converges for $\Re(r)>1$ and can be analytically continued to (almost) all $a,b,r\in\mathbb{C}$.

We here note for future reference that
\begin{align}
\text{seed:}\quad\quad y^\alpha  \Re \mathcal{E}_0(k,0^p) 
\quad &= \quad
-\frac{2}{(k-1)!} y^\alpha \sum_{m\geq 1} m^{-p-1} \sigma_{k-1}(m) (q^m+\bar{q}^m) 
\notag \\
&\underset{\text{LP}}{\rightarrow} \quad
-\frac{2^{2p+3}\pi^{p+\alpha+1}}{(k-1)!}  I(k-1,-p-1,\alpha) \,,
\label{conrul.1}
\end{align}
where $I(k{-}1,{-}p{-}1,\alpha)$ reduces to odd zeta values in the cases of interest to this work
with $\pi,\Im \tau$ dependent coefficients.
The contribution of a pure power $y^\alpha$  to the Laurent polynomial is simple since this is the case of Eisenstein series with $\Re\alpha {>}1$, which according to~\eqref{eq:FEk} lead to 
\begin{align}
\text{seed:}\quad\quad  y^\alpha\quad &\underset{\text{LP}}{\rightarrow} \quad
y^\alpha +
\frac{\pi^{2\alpha-1/2} \Gamma(\alpha-1/2)\zeta_{2\alpha-1}}{\Gamma(\alpha)\zeta_{2\alpha}} y^{1-\alpha} 
 \notag \\
&\quad= y^\alpha+
\frac{ (-1)^{\alpha - 1} (2 \alpha)! (  2 \alpha - 3)! \zeta_{   2 \alpha - 1}}{  4^{2 \alpha - 2}  (\alpha - 2)! 
(     \alpha - 1)! {\rm B}_{2 \alpha}}   y^{1-\alpha}  
 \,.  \label{conrul.2}
\end{align}
For integer $\alpha\geq 2$, all explicit factors of $\pi$ in the second term disappear, leaving 
a rational number times $\zeta_{2\alpha-1} y^{1-\alpha}$ as shown explicitly.

\subsubsection{Examples of Poincar\'e series}

We close this section by recording a few more Poincar\'e series that will be used in this paper. The first one expresses the modular invariant $(\Im \tau)^k {\rm G}_k \overline{{\rm G}_k}$ as a Poincar\'e sum according to
\begin{align}
(\Im \tau)^k {\rm G}_k(\tau) \overline{{\rm G}_k(\tau)} = 2\,\zeta_{k} \PS \left[   (\Im \tau)^k {\rm G}_k(\tau)
\right]_\gamma \, .
\label{oct21.0}
\end{align}
We stress that the ${\rm SL}(2,\ZZ)$ action here is just on $\tau$, there is no extra factor of automorphy as one sometimes uses in the `slash operator' for Poincar\'e-series representations of $\GG_k$ alone~\cite{Fleig:2015vky}. The sum converges absolutely for $k>2$. Similarly, we will use the generalisations of (\ref{oct21.0}) to 
\begin{subequations}
\label{ppeis.all}
\begin{align}
\PS &\left[y^{k+m} \GG_{2m}(\tau)  \right]_\gamma
=  -   \frac{(2k)! (k{-}1)!}{(-4)^{k} {\rm B}_{2k} (k{+}m{-}1)!} \GG_{2m} (\pi  \overline \nabla)^m \EE_k 
 \label{ppeis.1} \\
& =   (-4)^{m-k} \frac{(2k)!(2k{-}1)! y^{2m}\GG_{2m}}{{\rm B}_{2k} (k{+}m{-}1)! (k{-}m{-}1)!}  
\bigg\{ \betasv{k-m-1\\2k} - \frac{ 2 \zeta_{2k-1} }{(2k{-}1) (4y)^{k+m-1}} \bigg\}
\notag \,,\\
\PS &\left[y^{k+m} \overline{\GG_{2m}(\tau)}  \right]_\gamma
=  -   \frac{(2k)! (k{-}1)!}{(-4)^{k} {\rm B}_{2k} (k{+}m{-}1)!}\overline{\GG_{2m}} (\pi \nabla)^m \EE_k 
 \label{ppeis.2} \\
& =   (-4)^{-k-m} \frac{(2k)!(2k{-}1)! \overline{\GG_{2m}}}{{\rm B}_{2k} (k{+}m{-}1)! (k{-}m{-}1)!}  
\bigg\{ \betasv{k+m-1\\2k} - \frac{ 2 \zeta_{2k-1} }{(2k{-}1) (4y)^{k-m-1}} \bigg\}\, ,
\notag
\end{align}
\end{subequations}
where $y=\pi \Im\tau$. The Poincar\'e sums in the respective first steps converge absolutely for $k{+}m>1$ and the rewritings in the respective second lines are obtained using~\eqref{oct24.2} and require $0\leq m <k$.

Another Poincar\'e sum that we shall use was given in~\cite[Eq.~(3.10)]{Bossard:2017kfv}:
\begin{align}
\label{Besselsum}
\PS\left[ \sqrt{|n| \Im \tau} K_{s-1/2}(2\pi|n| \Im \tau) e^{2\pi i n \Re \tau} \right]_{\gamma} 
= \frac{\pi^{2s+1/2}  \sigma_{2s-1}(|n|) {\rm E}_s(\tau) }{4 |n|^{s-1} \cos(\pi s) \Gamma(s{+}1/2)\zeta_{2s-1} \zeta_{2s}}  \,.
\end{align}
This Poincar\'e sum is over any given non-zero Fourier mode of $\EE_s$ and is therefore expected to be proportional to $\EE_s$ again. The sum itself is divergent but the result on the right-hand side was argued for in~\cite{Bossard:2017kfv} by analytic continuation. 

%%%%%%%%%%%%%%%%%%%%%%%%%%%%%%%%%%%%%%%%%%%%%%%%%%%%%%%%%%%
\section{Laplace equations for even seed functions, MGFs and beyond}
\label{sec:3}
%%%%%%%%%%%%%%%%%%%%%%%%%%%%%%%%%%%%%%%%%%%%%%%%%%%%%%%%%%%

In this section, we introduce a method to determine the Laplace equations of modular 
invariant functions of depth two, expressed through the $\betalagpm{j}$ defined in~\eqref{eq:betalag}
along with iterated integrals of depth $\leq 1$. In particular, we will focus on the even case in this
section and infer the associated seed 
functions of the schematic form $y^a  \Re {\cal E}_0(2m,0^b)$ with $a\geq1$ and $0\leq b\leq 2m{-2}$. In the later section \ref{oddsection}, similar methods will be applied to infer seed functions $y^a \Im {\cal E}_0(2m,0^b)$ for odd modular invariant functions and in section~\ref{sec:4alt} we consider non-convergent seeds for other ranges of $a$ and $b$.

%%%%%%%%%%%%%%%%%%%%%%%%%%%%%%%%%%%%%%%%%%%%%%%%%%%%%%%%%%%
\subsection{Laplacian of modular graph functions and iterated Eisenstein integrals}
\label{sec:3.0}
%%%%%%%%%%%%%%%%%%%%%%%%%%%%%%%%%%%%%%%%%%%%%%%%%%%%%%%%%%%

A prominent feature of MGFs is that they satisfy (possibly inhomogeneous) eigenvalue equations
with respect to the ${\rm SL}(2,\mathbb Z)$-invariant Laplace operator \cite{DHoker:2015gmr, Basu:2016xrt, Kleinschmidt:2017ege, Basu:2019idd}
\beq
\Delta = 4 (\Im \tau)^2 \frac{ \partial^2 }{\partial \tau \partial \bar \tau} = \pi \overline \nabla\left[ \frac{1}{y^2} \pi \nabla\right] \,,
\label{lapsec.1}
\eeq
where we recall $y=\pi \Im\tau$. At depth one, these are the homogeneous eigenvalue equations of the non-holomorphic Eisenstein
series (\ref{rev.04}), 
\begin{align}
\big(\Delta - s(s{-}1) \big) {\rm E}_s  = 0 \, ,
\label{lapsec.2}
\end{align}
while the Laplace action on the two-loop modular graph functions $C_{a,b,c}$
in (\ref{rev.05}) is known from \cite{DHoker:2015gmr} to be given by
\begin{align}
&\big( \Delta - a(a{-}1)  - b(b{-}1) - c(c{-}1)  \big) C_{a,b,c}  \label{lapsec.3} \\
& \ \ = ab \big(C_{a-1,b+1,c} + C_{a+1,b-1,c} + C_{a+1,b+1,c-2}
-2 C_{a,b+1,c-1} - 2 C_{a+1,b,c-1} \big) + {\rm cyc}(a,b,c)\,,
\notag
\end{align}
with~\cite{DHoker:2015gmr}
\begin{align}
C_{a,b,0} = {\rm E}_a {\rm E}_b - {\rm E}_{a+b} \, , \ \ \ \ \ \ 
C_{a,b,-1} = {\rm E}_{a-1} {\rm E}_b+{\rm E}_a {\rm E}_{b-1}
\label{lapsec.3a}
\end{align}
and $+{\rm cyc}(a,b,c)$ instructs us to add the remaining two cyclic permutations of $(a,b,c)$.

A major goal of this work is to relate the Laplace equations of $C_{a,b,c}$ to their representations
in terms of Poincar\'e series and iterated Eisenstein integrals. Poincar\'e-series representations for the $C_{a,b,c}$ were found in~\cite{DHoker:2019txf} but we shall cover the more general space of depth-two (as opposed to two-loop) functions $\betalagpm{j}$ and furthermore seek alternative representatives of the seed functions without the powers of
$q\bar q$ in those of the reference. 

Our results on the depth-one seed functions for $C_{a,b,c}$ are based on the central claim:
\begin{align}
&\text{Any $C_{a,b,c}(\tau)$ is expressible in terms of the objects $\betalagp{j}$ in~\eqref{eq:betalag} subject to} \label{lapsec.4} \\
&\text{$m{+}k=a{+}b{+}c$ and modulo combinations of $\zeta_{2w+1}$, powers of $y$ and $\beta^{\rm sv}$ of depth $\leq 1$}
\notag 
\end{align}
which follows from the following considerations:
\begin{itemize}
\item Any $C_{a,b,c}(\tau)$ is contained in the generating series $Y^\tau_{\vec{\eta}}$ of MGFs \cite{Gerken:2020yii} 
whose $\alpha'$-expansion comprises $\beta^{\rm sv}$ combined with powers of $y^{-1}$ and a constant series
capturing the $\tau \rightarrow i\infty$ degenerations of MGFs. For instance, all the $C_{a,b,c}(\tau)$
are contained in the five-point instance of $Y^\tau_{\vec{\eta}}$ at the order of $s^0_{ij}$, when isolating the component integrals over Kronecker--Eisenstein coefficients $f^{(a)}_{12} f^{(b)}_{24} f^{(c)}_{25}
\overline{f^{(a)}_{13} f^{(b)}_{34} f^{(c)}_{35}}$ (with $f^{(a)}_{ij} = f^{(a)}(z_{i}{-}z_j,\tau)$ and $f^{(a)}(z,\tau)
=(-1)^{a-1} \sum_{p \in \Lambda'} p^{-a}$), where we refer to~\cite{Gerken:2020yii} for the notation.
\item The $\beta^{\rm sv}[\begin{smallmatrix} j_1 & j_2 &\ldots &j_\ell\\ k_1 &k_2 &\ldots &k_\ell \end{smallmatrix}]$ 
of depth $\ell\leq 2$ are sufficient to represent the $C_{a,b,c}$: Both of 
$\beta^{\rm sv}[\begin{smallmatrix} j_1 &j_2\\ k_1 &k_2 \end{smallmatrix}]$ 
and $C_{a,b,c}$ reduce to products of holomorphic
Eisenstein series and depth-one objects under repeated Cauchy--Riemann derivatives (\ref{rev.11}). 
This can be seen from the differential equation (\ref{rev.20}) of the 
$\beta^{\rm sv}[\begin{smallmatrix} j_1 &j_2\\ k_1 &k_2 \end{smallmatrix}]$ and the action of
$\nabla$ on lattice sums \cite{DHoker:2016mwo}. In particular, repeated $\nabla$-derivatives
eventually reduce the lattice-sum representations (\ref{rev.05}) of $C_{a,b,c}$
to single lattice sums after applying holomorphic subgraph reduction.
\item As can be anticipated from the dictionary (\ref{oct24.2}) between ${\rm E}_k$ and $\beta^{\rm sv}$ at depth one,
the leading-depth terms $\beta^{\rm sv}[\begin{smallmatrix} j_1 &j_2\\ k_1 &k_2 \end{smallmatrix}]$ entering $C_{a,b,c}$ are accompanied by a tail
of terms $y^{-\alpha}\betasv{j \\ k}$ and $y^{-\beta}$ accompanied by odd zeta values. This is necessary to attain exact
modular invariance beyond the leading-depth term in~\eqref{eq:bsv2prop}.
\end{itemize}
As a consequence of (\ref{lapsec.4}), the Laplace equations of ${\rm E}_k$ and 
$C_{a,b,c}$ can be studied at the level of the $\beta^{\rm sv}$. 
The complex-conjugation properties (\ref{eq:bsv2prop}) together with the differential equations
(\ref{rev.20}) of the $\beta^{\rm sv}$ fix the leading-depth term in the Laplacian of 
$\beta^{\rm sv}[\begin{smallmatrix} j_1 &j_2\\ k_1 &k_2 \end{smallmatrix}]$. 
In particular, for the entries $j_1{+}j_2 = \frac{1}{2}(k_1{+}k_2{-}4)$ relevant to $C_{a,b,c}$, we find
\begin{align}
\Delta  \betasv{j_1& j_2\\ k_1& k_2} \, \Big|_{ j_1{+}j_2 = \frac{1}{2}(k_1{+}k_2{-}4)}&=
\big( (k_1{-}j_1{-}2)(j_1{+}1) + (k_2{-}j_2{-}2)(j_2{+}1) \big) \betasv{j_1& j_2\\ k_1& k_2} \notag \\
&\hspace{-1.6cm}+ j_2(k_1{-}j_1{-}2)  \betasv{j_1+1& j_2-1\\ k_1& k_2}
+ j_1(k_2{-}j_2{-}2)  \betasv{j_1-1& j_2+1\\ k_1& k_2} \label{ccbsv.4} \\
&\hspace{-1.6cm}- j_1 \delta_{j_2,k_2-2} {\rm G}_{k_2}(\tau{-}\bar \tau)^{k_2}  \betasv{j_1-1\\ k_1}
-(k_2{-}j_2{-}2) \delta_{j_1,0} \frac{  \overline{ {\rm G}_{k_1} } }{(2\pi i)^{k_1}} \betasv{j_2+1\\  k_2} \notag \\
&\hspace{-1.6cm}+ \delta_{j_2,k_2-2}  \delta_{j_1,0} \frac{  \overline{ {\rm G}_{k_1} } }{(2\pi i)^{k_1}}{\rm G}_{k_2}(\tau{-}\bar \tau)^{k_2} \ {\rm mod} \ {\rm lower} \ {\rm depth} \, ,
\notag
\end{align}
where the third line vanishes for $k_1=k_2$ and the last one vanishes for $k_1\neq k_2$.
One can reformulate (\ref{ccbsv.4}) in terms of the combinations $\betalagpm{j}$ introduced in~\eqref{eq:betalag} for $0\leq j\leq 2m{-}2$ as
\begin{align}
\Delta \betalagpm{j} &=\notag \Big[j \Big(2m-j-1\Big)+\Big(m+k -2-j\Big) \Big(k-m+j+1\Big)\Big]\betalagpm{j} \\
&\ \ \notag+j\Big(k-m+j\Big)\betalagpm{j-1}+\Big(2m-2-j\Big)\Big( m+k-j-2\Big) \betalagpm{j+1}\\
&\ \  +\delta_{j,2m-2} \Big(m-k\Big) \bigg[ \frac{\overline{\GG_{2m}}}{(2\pi i)^{2m}} \betasv{k+m-1 \\ 2k}\pm \GG_{2m}(\tau{-}\bar{\tau})^{2m} \betasv{k-m-1 \\ 2k}\bigg] \label{eq:LapLag} \\
&\ \ +(1\pm 1) \delta_{j,2m-2}\,\delta_{m,k} \frac{  \overline{ {\rm G}_{2k}}  }{(2\pi i)^{2k}}{\rm G}_{2k}(\tau{-}\bar \tau)^{2k} \ {\rm mod} \ {\rm lower} \ {\rm depth}\,.
\notag
\end{align}
In the following we shall analyse Laplace systems associated with the $\betalagpm{j}$ in more detail and find Poincar\'e-series solutions to them. We recall that we assume $m\leq k$ throughout without loss of generality. While the even cases will be discussed in the rest of this section, the odd ones are relegated to
section \ref{oddsection}.

%%%%%%%%%%%%%%%%%%%%%%%%%%%%%%%%%%%%%%%%%%%%%%%%%%%%%%%%%%%
\subsection{\texorpdfstring{Laplacian of even combinations of $\beta^{\rm  sv}$}{Laplacian of even combinations of betasv}}
\label{sec:3.1}
%%%%%%%%%%%%%%%%%%%%%%%%%%%%%%%%%%%%%%%%%%%%%%%%%%%%%%%%%%%

The action of the Laplacian on the depth-two  $\betalagpm{j}$ in~\eqref{eq:LapLag} is given by a tri-diagonal matrix of size $(2m{-}1)\times (2m{-}1)$. For the even combinations $\betalagp{j}$, we shall focus on the top left corner of size $(m{-}1)\times (m{-}1)$ corresponding to $0\leq j \leq m{-}2$. The reason being that for $j=m{-}1$, the expression~\eqref{eq:betalag} is a pure shuffle according to~\eqref{eq:betashuffle}  that reduces to a product of Eisenstein series at leading depth according to~\eqref{oct24.2}:
\begin{align}
\label{eq:shuffle}
\betalagp{m-1} &=\betasv{m-1& k-1\\ 2m& 2k}  + \betasv{k-1 & m-1\\ 2k& 2m}  = \betasv{m-1 \\ 2m}\betasv{k-1 \\2 k} \\
&\notag =\frac{\big[(m{-}1)!(k{-}1)!\big]^2}{(2m{-}1)!(2k{-}1)!} \, \EE_{m}\EE_{k} \ {\rm mod} \ {\rm lower} \ {\rm depth}\,.
\end{align}
Similarly, the even $\betalagp{j}$ in the range $m \leq j \leq 2m{-}2$ are determined by those with 
$0 \leq j \leq m{-}2$ by the following shuffle relations
\beq
\betalagp{j} + \betalagp{2m-2-j} = 
\betasv{j\\ 2m} \betasv{k+m-j-2\\ 2k}
+ \betasv{j+k-m \\ 2k } \betasv{2m-j-2\\ 2m}\, .
\label{moreshuf}
\eeq
For $j$ in the range $0 \leq j\leq m{-}2$ we notice moreover that the Kronecker deltas in (\ref{eq:LapLag}) do not contribute and the Laplace equation can be rewritten as
\begin{equation}\label{eq:LapMatrix}
\Delta \betalagp{j} = \sum_{i=0}^{m-2} M^+_{ji} \, \betalagp{i} +   \delta_{j,m-2}
\, \frac{4\big[m!k!\big]^2}{(2m)!(2k)!} \, \EE_{m}\EE_{k} \ {\rm mod} \ {\rm lower} \ {\rm depth}\,,
\end{equation}
where the $(m{-}1)\times (m{-}1)$ matrix $M^+_{ji}$ is given by
\begin{equation}\label{eq:MatrixForm}
M^+_{ji} = 
\left\lbrace \begin{array}{ll} j \big(2m-j-1\big)+\big(m+k -2-j\big) \big(k-m+j+1\big) & \text{for $i=j$}\,, \\[1mm]
j\big(k-m+j\big) &\text{for $i=j-1$}\,,\\[1mm]
\big(2m-2-j\big)\big( m+k-j-2\big) & \text{for $i=j+1$}\,,\\[1mm]
0 & \text{otherwise}\,.
\end{array}\right.
\end{equation}

\subsubsection{\texorpdfstring{Completing the $\beta^{\rm sv}$ at depth two}{Completing the betasv at depth two}}
\label{sec:3.2.1}

From the structure of the Laplace system (\ref{eq:LapLag}) we can easily see that 
the shuffle irreducible $\betalagp{j}$ with $0 \leq j\leq m{-}2$ together with 
repeated Laplace action on the pure shuffle (\ref{eq:shuffle}) at $j=m{-}1$
determine the remaining $\betalagp{j}$ with $ m\leq j \leq 2m{-}2$.
For example, the Laplace equation (\ref{eq:LapLag}) for the pure shuffle $\betalagp{j=m-1}$ will determine the leading-depth part of $\betalagp{j=m}$ in terms of $\betalagp{m-2}$,
\begin{equation}\label{eq:shufflelap}
\big( \Delta   - m(m{-}1) - k (k{-}1)\big) \betalagp{ m-1 } =
 (m{-}1) (k{-}1)\Big(\betalagp{m-2 }+\betalagp{m }\Big) \ {\rm mod} \ {\rm lower} \ {\rm depth}\,.
\end{equation}
This equation can be solved for $\betalagp{m}$ and one obtains an explicit expression if one uses the Laplacian of the shuffle that can be calculated as
\begin{align}
\Delta\Big(\EE_m \EE_k\Big)
&= {\rm E}_k \, \Delta {\rm E}_m +  {\rm E}_m \,\Delta {\rm E}_k   
+ \frac{(\pi \nabla {\rm E}_m) (\pi \overline{ \nabla} {\rm E}_k) +  (\pi \overline{ \nabla} {\rm E}_m) (\pi \nabla {\rm E}_k) }{y^2} \, .
\label{altrec.3}
\end{align}
The same procedure will allow us to determine all the leading-depth terms for $m\leq j \leq 2m{-}2$ in terms of the previous $\betalagp{j}$ and higher powers of the Laplacians of the shuffle. The terms in the higher Laplacians can be expressed in terms of the auxiliary modular invariant objects
\beq
\Jp{\ell}{m}{k} =  \Re \bigg[  \frac{\big((\pi \nabla)^\ell {\rm E}_m \big) (\pi  \overline \nabla)^\ell {\rm E}_k}{y^{2\ell} } \bigg]  \quad\quad \text{for $\ell\ge0 $} \, .
\label{prodd1}
\eeq
For instance,~\eqref{altrec.3} in this language is 
\begin{align}
\label{eq:PPP0}
\big(\Delta- m(m{-}1) - k(k{-}1) \big) \Jp{0}{m}{k} =  2 \Jp{1}{m}{k}\,,
\end{align}
where we made use of the well-known Laplacian $\Delta {\rm E}_k = k(k{-}1) {\rm E}_k$.
The Laplacians on the other $\Jp{\ell}{m}{k}$ can be worked out as a recurrence relation for $\ell\geq1$:
\begin{align}
\label{altrec.10}
\big( \Delta - m(m{-}1) - k(k{-}1) + 2\ell^2 \big) \Jp{\ell}{m}{k} =
 \Jp{\ell+1}{m}{k} +  (m{+}\ell{-}1)(m{-}\ell) (k{+}\ell{-}1)(k{-}\ell) \Jp{\ell-1}{m}{k} 
\end{align}
that complements~\eqref{eq:PPP0}. In order to derive the recurrence one uses
\beq
\pi \nabla \bigg( \frac{ ( \pi \overline{\nabla} )^p {\rm E}_k}{y^{2p}} \bigg) = (k{-}p)(k{+}p{-}1) 
\frac{ ( \pi \overline{\nabla} )^{p-1} {\rm E}_k}{y^{2p-2}} 
\label{altrec.6}
\eeq
and its complex conjugate. We note that the Laplace equation~\eqref{altrec.10} does not close on a finite set as it always generates $\Jp{\ell}{m}{k}$ for increasing $\ell$. 

The modular functions $\Jp{\ell}{m}{k}$ are directly related to $\betalagp{j}$ with $j\geq m{-}1$.
From equation (\ref{eq:shuffle}) we see that
\beq
\Jp{0}{m}{k} = \mathcal{N}_{m,k} \betalagp{m-1} \ {\rm mod} \ {\rm lower} \ {\rm depth}\,,
\eeq
where 
\beq
\mathcal{N}_{m,k}
= \frac{(2m{-}1)!(2k{-}1)!}{[(m{-}1)!(k{-}1)!]^2} \,.
\eeq

Using then (\ref{eq:shufflelap}) together with (\ref{eq:PPP0}) we obtain
\beq
\Jp{1}{m}{k} = \frac{ \mathcal{N}_{m,k}}{2}(m{-}1)(k{-}1) \Big(  \betalagp{m}+\betalagp{m-2}\Big) \ {\rm mod} \ {\rm lower} \ {\rm depth}\,.
\eeq

We can then keep on exploiting the Laplace system (\ref{eq:LapLag}) combined with (\ref{altrec.10}) to find the general expression for $0\leq \ell\leq m{-}1$
\beq
\Jp{\ell}{m}{k} = \frac{\mathcal{N}_{m,k}}{2}\Big[ \prod_{j=1}^{\ell} (m{-}j)(k{-}j)\Big]\Big(  \betalagp{m+\ell-1}+\betalagp{m-\ell-1}\Big) \ {\rm mod} \ {\rm lower} \ {\rm depth}\,.
\label{bsvjplus}
\eeq
Note that here we are assuming that $\ell <m$ since for $\ell \geq m$ the expression (\ref{prodd1})
can no longer be written in terms of $\beta^{\rm sv}$ but involves holomorphic Eisenstein series by
(\ref{eq:Ediff}).

\subsubsection{Spectrum of the Laplacian}
\label{sec:3.2.2}

In order to understand the space of modular-invariant even combinations $\betalagp{j}$ it is then crucial to characterise the space of solutions to the system (\ref{eq:LapMatrix}).
As a concrete example we can rewrite (\ref{eq:LapMatrix}) for the cases $(m,k)=(2,2)$,
$(m,k)=(3,4)$ and $(m,k)=(4,4)$
\begin{align}
\Delta \beta^{{\rm sv}+,\, 0}_{2,2} &= 2 \beta^{{\rm sv}+,\, 0}_{2,2} + \frac{1}{9} \EE_2^2  \ {\rm mod} \ {\rm lower} \ {\rm depth}\,,
\nn \\
\Delta \left(\begin{matrix}\beta^{{\rm sv}+,\, 0}_{3,4} \\ \beta^{{\rm sv}+,\, 1}_{3,4}  \end{matrix}\right) &= \left(\begin{matrix} 10 & 20 \\
2 &16
  \end{matrix}\right)\left(\begin{matrix}\beta^{{\rm sv}+,\, 0}_{3,4} \\ \beta^{{\rm sv}+,\, 1}_{3,4}  \end{matrix}\right) +\frac{1}{350}\left(\begin{matrix}0\\ \EE_3\EE_4 \end{matrix}\right) \ {\rm mod} \ {\rm lower} \ {\rm depth}\,,
\\
\Delta \left(\begin{matrix}\beta^{{\rm sv}+,\, 0}_{4,4} \\ \beta^{{\rm sv}+,\, 1}_{4,4} \\ \beta^{{\rm sv}+,\, 2}_{4,4} \end{matrix}\right) &= \left(\begin{matrix} 6 & 36 & 0 \\
 1 & 16 &25 \\
  0 &4 &22
  \end{matrix}\right)\left(\begin{matrix}\beta^{{\rm sv}+,\, 0}_{4,4} \\ \beta^{{\rm sv}+,\, 1}_{4,4} \\ \beta^{{\rm sv}+,\, 2}_{4,4} \end{matrix}\right) +\frac{1}{1225}\left(\begin{matrix}0\\0\\ \EE_4^2 \end{matrix}\right) \ {\rm mod} \ {\rm lower} \ {\rm depth}\nn \,.
\end{align}
The matrices above can be diagonalised and have eigenvalues $\{6,20\}$
in case of $(m,k)=(3,4)$ and $\{2,12,30\}$ in case of $(m,k)=(4,4)$. 

More generally, we find that the tridiagonal matrix $M^+_{ji}$ in (\ref{eq:MatrixForm}) has eigenvalues given by the spectrum (recall $m\leq k$)
\begin{equation}\label{eq:Spectrum}
s(s{-}1) \,\,\mbox{with}\,\,s \in \left\lbrace k{-}m{+}2,k{-}m{+}4,\ldots,k{+}m{-}4,k{+}m{-}2 \right\rbrace
\end{equation}
and each eigenvalue has multiplicity one. We only have even values for $s$ when the integers $m$ and $k$ have the same parity and odd values otherwise when $m$ and $k$ have opposite parity, and none of the eigenvalues vanishes since $s\geq 2$. This spectrum was found on the basis of a large number of examples including all $m,k$ with $m{+}k \leq 28$ and is in general conjectural\footnote{A promising strategy for a proof could be to clarify the relation between the $\beta^{\rm sv}$ and the single-valued iterated Eisenstein integrals of Brown \cite{Brown:mmv, Brown:I, Brown:II} and to then exploit their properties in the references.}. We shall next describe the eigenvalue problem more concretely which also yields the proof of the spectrum for some infinite families of $m$ and $k$.

The diagonalisation of $M^+$ proceeds by writing the linear combination
\begin{equation}
\label{eq:tildeb}
\tilde{\beta}_{(s)}^+ = \sum_{i=0}^{m-2} v^i_{(s)} \betalagp{i}\,,
\end{equation}
expressed in terms of the eigenvector $\mathbf{v}_{(s)}=\big(v_{(s)}^0,\ldots, v_{(s)}^{m-2}\big)^T$ of $(M^+)^T$ with eigenvalues $s(s{-}1)$, where $(M^+)^T$ denotes the transpose of $M^+$ given in (\ref{eq:MatrixForm}). With~\eqref{eq:tildeb} the Laplace equation (\ref{eq:LapMatrix}) reduces to
\begin{equation}\label{eq:LaplacianDiag}
\Big(\Delta -s(s{-}1) \Big) \tilde{\beta}_{(s)}^+= \alpha_{(s)}\, \EE_{m}\EE_{k} \ {\rm mod} \ {\rm lower} \ {\rm depth}
\end{equation}
for some rational coefficients $\alpha_{(s)}$.

Since $M^+$ is tridiagonal, the eigenvalue equation
$(M^+)^T \mathbf{v}_{(s)}  = s(s{-}1)\mathbf{v}_{(s)}$ 
translates into a three-term recurrence given by
\begin{align}
M^+_{i-1,i}\,v^{i-1}_{(s)} + \big(M_{ii}^+ -s(s{-}1) \big)\,v^i_{(s)} + M_{i+1,i}^+ \,v^{i+1}_{(s)} = 0\,,
\label{oct20.1}
\end{align}
with the boundary conditions $v^{-1}_{(s)} = v^{m-1}_{(s)}= 0$.
In order to have a non-zero solution for $\mathbf{v}_{(s)}$ we see in particular that both $v_{(s)}^0$ and $v^{m-2}_{(s)}$ must be non-zero so in particular the constants $\alpha_{(s)}$ in equation (\ref{eq:LaplacianDiag}) in front of the source term will never vanish.

For generic $m\leq k$ and $s$ in (\ref{eq:Spectrum}) we do not have a closed-form solution. However, for $m\leq k$ and $s=k{-}m{+}2$, corresponding to the lowest possible eigenvalue in~\eqref{eq:Spectrum}, one can prove that
\begin{equation}
v^i_{(s)} =  \frac{ (-1)^i (2m {-} 2 i {-}2) \Gamma(2m{-}2)}{i!\, \Gamma(2m{-}i {-} 1)}\,,
\label{oct20.2}
\end{equation}
where we normalised $v^0_{(s)}=1$.

For other configurations of $m$, $k$ and $s$ one can find closed expressions in a few instances and we have used them to perform large scans over matrices $M^+_{ji}$ given in~\eqref{eq:LapMatrix} to test that the claimed spectrum~\eqref{eq:Spectrum} seems indeed correct.  

As we argued above, our general considerations show that the coefficient $\alpha_{(s)}$ in~\eqref{eq:LaplacianDiag} is always non-zero. For this reason we shall in the following study the equation
\begin{equation}
\label{eq:Fmk}
\big(\Delta - s(s{-}1) \big) \FFp{s}{m}{k} = {\rm E}_m {\rm E}_k \, , \ \ \ \ \ \
s \in \left\lbrace k{-}m{+}2,k{-}m{+}4,\ldots,k{+}m{-}4,k{+}m{-}2 \right\rbrace \, ,
\end{equation}
with integers $2\leq m \leq k$, ignoring lower-depth terms in the underlying Laplace equation
(\ref{eq:LapLag}) of the $\beta^{\rm sv}$ which will be re-instated in section~\ref{sec:4} below.
Given the modular invariance of the Laplacian and the non-holomorphic Eisenstein series,
the ultimate goal of this work is to construct modular invariant solutions to (\ref{eq:Fmk}) as well as to its odd
counterpart (\ref{eq:Fmkm}).
The $\FFp{s}{m}{k}$ in (\ref{eq:Fmk}) exhaust and transcend the real MGFs at depth two and zero modular weights, and the discussion of their iterated-integral and Poincar\'e-series
representations in the next sections is a key result of this work.

Assigning transcendental weights $m$ and $k$ to $\EE_m$ and $\EE_k$, respectively, we deduce that $\FFp{s}{m}{k}$ should have transcendental weight $m{+}k$, a fact that will be supported by its Laurent polynomial~\eqref{eq:FLP} below. Equation~\eqref{eq:Fmk} allows for the modular-invariant homogeneous solution $\EE_s$ whose coefficient must be a rational multiple of $\zeta_{m+k-s}$ by uniform transcendentality and being at depth two. However, the allowed spectrum for $s$ in~\eqref{eq:Fmk} shows that $m{+}k{-}s>0$ is always an even integer. As there are no single-valued zeta values with this property, the homogeneous solution $\EE_s$ is therefore disallowed by our assumptions: uniform transcendentality and maximum depth two.

We note that the spectrum in~\eqref{eq:Fmk} excludes the cases $s=0$ and $s=1$ for which the Laplace eigenvalue would vanish. The cases of vanishing Laplace eigenvalue are the ones where constants can appear as homogeneous solutions.
In the case of MGFs such homogeneous solutions do arise, see for instance~\eqref{rev.07}, where these constants are odd zeta values and can be determined from the lattice-sum representations \cite{DHoker:2017zhq}.

In the even sector, equation~\eqref{eq:Fmk} is the most general one to consider when restricting to sources built out of $\Jp{\ell}{m}{k}$, since a source with $\ell>0$ can always be reduced to $\Jp{0}{m}{k}=\EE_m\EE_k$ using the recursion in \eqref{eq:PPP0} and \eqref{altrec.10} at the price of redefining the function $\FFp{s}{m}{k}$.

\subsubsection{\texorpdfstring{Examples at $m=k$}{Examples at m=k}}
\label{sec:3.2.3}

For the Laplace equations (\ref{eq:Fmk}) with $m=k$, the simplest solutions $\FFp{s}{k}{k} $ with $k=2,3$ have already 
been studied from the perspective of both modular graph functions \cite{DHoker:2015gmr} and 
iterated Eisenstein integrals \cite{Broedel:2018izr}. The eigenvectors of the relevant
$(k{-}1)\times (k{-}1)$ matrices $M^+_{ji}$ (\ref{eq:MatrixForm}) in the normalisation of (\ref{eq:Fmk}) are given by
\begin{align}
 \FFp{2}{2}{2} &
   = 18     \betasv{ 2 & 0\\4& 4}  \ {\rm mod} \ {\rm lower} \ {\rm depth}\,,
 \notag\\
 \FFp{2}{3}{3} &
 = 100  (2 \betasv{3& 1\\6& 6} - \betasv{4& 0\\6& 6})
  \ {\rm mod} \ {\rm lower} \ {\rm depth}\,,
  \label{oct20.3}
 \\
 \FFp{4}{3}{3} &
 = 25  (8 \betasv{3& 1\\6& 6} + \betasv{4& 0\\6& 6}) 
  \ {\rm mod} \ {\rm lower} \ {\rm depth}\, .
 \notag
\end{align}
The combinations of $\beta^{\rm sv}$ on the right-hand sides appeared in
\cite{Gerken:2020yii} as the leading-depth terms of the modular graph functions ${\rm E}_{2,2},{\rm E}_{3,3},
{\rm E}_{3,3}'$ introduced in \cite{Broedel:2018izr}. On these grounds, a modular invariant completion of
(\ref{oct20.3}) by lower-depth terms is furnished by
\begin{align}
 \FFp{2}{2}{2} &= - {\rm E}_{2,2}  = - C_{2,1,1} + \frac{9}{10} {\rm E}_4 \,, \notag\\
 \FFp{2}{3}{3} &= \frac{2}{9} {\rm E}_{3,3} - \frac{5}{3}{\rm E}_{3,3}' = - \frac{1}{4} C_{2,2,2} - C_{3,2,1} + \frac{13}{28} {\rm E}_6 \label{oct20.4}\,,
 \\
 \FFp{4}{3}{3} &=  \frac{1}{2} {\rm E}_{3,3} - \frac{5}{3} {\rm E}_{3,3}' = \frac{1}{36} C_{2,2,2} - \frac{1}{6} C_{3,2,1} + \frac{1}{6} {\rm E}_6 \, , \notag
\end{align}
and one can confirm from the Laplace equations (\ref{lapsec.3}) of the $C_{a,b,c}$ \cite{DHoker:2015gmr}
that these examples indeed satisfy $(\Delta - s(s{-}1))\FFp{s}{k}{k} = {\rm E}_k^2$.

\subsubsection{\texorpdfstring{Examples at $m<k$}{Examples at m<k}}
\label{sec:3.2.4}

For the simplest examples of the Laplace equations (\ref{eq:Fmk}) with $m< k$, 
the appropriately normalised eigenvectors of the relevant
$(m{-}1)\times (m{-}1)$ matrices $M^+_{ji}$ (\ref{eq:MatrixForm}) are
\begin{align}
 \FFp{3}{2}{ 3}& =   30 (\betasv{2& 1\\4& 6} + \betasv{3& 0\\6& 4})
  \ {\rm mod} \ {\rm lower} \ {\rm depth}\,,  \label{oct20.5} \\
 \FFp{4}{ 2}{ 4}& =   105 (\betasv{2& 2\\4& 8} + \betasv{4& 0\\8& 4})
  \ {\rm mod} \ {\rm lower} \ {\rm depth}  \, .\notag
  \end{align}
The $\beta^{\rm sv}$ can again be identified as leading-depth terms 
\cite{Gerken:2020yii} of the modular graph functions ${\rm E}_{2,3}$ and ${\rm E}_{2,4}$ \cite{Broedel:2018izr}.
Hence, a modular invariant completion of (\ref{oct20.5}) by lower-depth terms is given by
\begin{align}
 \FFp{3}{2}{ 3}&=  - \frac{1}{4} {\rm E}_{2,3} - \frac{ \zeta_5  }{240} =  - \frac{1}{4} C_{3,1,1} + \frac{43}{140} {\rm E}_5 - \frac{ \zeta_5  }{240} \label{oct20.6} \,, \\
 \FFp{4}{ 2}{ 4}&= -\frac{1}{54} {\rm E}_{2,4} =   - \frac{1}{54} C_{2,2,2} - \frac{1}{18} C_{3,2,1} - \frac{1}{6} C_{4,1,1} + \frac{13}{54} {\rm E}_6 \, ,
 \notag
  \end{align}  
where the subtraction of $- \frac{ \zeta_5  }{240} $ ensures that the zeta 
constant in $(\Delta- 6) {\rm E}_{2,3} = \frac{ \zeta_5}{10} - 4 {\rm E}_2 {\rm E}_3$
does not appear in $(\Delta- 6)  \FFp{3}{2}{ 3} =  {\rm E}_2 {\rm E}_3$.

%%%%%%%%%%%%%%%%%%%%%%%%%%%%%%%%%%%%%%%%%%%%%%%%%%%%%%%%%%%
\subsection{Solution to even Laplace equations via Poincar\'e series}
\label{sec:3.2new}
%%%%%%%%%%%%%%%%%%%%%%%%%%%%%%%%%%%%%%%%%%%%%%%%%%%%%%%%%%%

In the examples (\ref{oct20.3}) and (\ref{oct20.5}) of $\FFp{s}{m}{ k} $ with $m{+}k\leq 6$, the basis of modular graph 
functions is known from \cite{DHoker:2016quv}, and the lower-depth terms could be inferred from 
\cite{Gerken:2020yii, Gerken:2020aju}. As a main result of this work, we shall now introduce an 
alternative method to determine the modular invariant $\FFp{s}{m}{ k} $ at arbitrary $m,k$ without any recourse to 
earlier expressions for the MGFs at these weights.
Our method relies on Poincar\'e-series representations of $\FFp{s}{m}{k}$ to be derived in this
section from the methods of~\cite{Ahlen:2018wng,Dorigoni:2019yoq}. We make the ansatz
\begin{align}
\FFp{s}{m}{k}(\tau)  = \PS \seedp{s}{m}{k}(\gamma\cdot \tau)
 \label{oct20.7}
\end{align}
in terms of an even (under $\tau\to-\bar\tau$) and periodic (under $\tau\to\tau{+}1$) seed function $\seedp{s}{m}{k}$ and also replace $\EE_k$ on the right-hand side of~\eqref{eq:Fmk} by its Poincar\'e series~\eqref{eq:PSEk}, usually dubbed as \textit{folding} $\EE_k$. We shall assume without loss of generality here that $k\geq m$, and we replace $\EE_k$ rather than $\EE_m$ by its Poincar\'e series in order to obtain an $\seedp{s}{m}{k}$ whose Poincar\'e sum is absolutely convergent
for $m<k$ and hence modular invariant by construction. 
For $m=k$, the resulting Poincar\'e seed produces a divergent sum (due to the presence of a term linear in $y$ in the seed, similar to $\EE_1$). However, as explained in~\cite{Ahlen:2018wng,Dorigoni:2019yoq}, this case can be treated by considering $k\to k+\epsilon$ in the final expressions (e.g.\ for the Laurent polynomial), and taking the limit $\epsilon\to 0$ at the end, see also appendix~\ref{app:Poincare}, thereby reaching the modular-invariant diagonal case $\FFp{s}{k}{k}$. Alternative seeds where the Eisenstein series  $\EE_m$ with $m<k$ is folded will be discussed in section~\ref{sec:fold.2}.

\subsubsection{Deriving the seed function}
\label{sec:3.3.1}

With the above Poincar\'e-series ansatz, we reduce~\eqref{eq:Fmk} to
\begin{align}
\label{eq:seedeq}
\big(\Delta - s(s{-}1) \big) \seedp{s}{m}{k}  &=
(-1)^{k-1} \frac{ {\rm B}_{2k} }{(2k)!} (4y)^k \EE_m\\
&=
(-1)^{k+m} \frac{ {\rm B}_{2k} {\rm B}_{2m} }{(2k)! (2m)!} (4y)^{k+m}
-(-1)^{k} \frac{4 {\rm B}_{2k} (2m{-}3)! \zeta_{2m-1} }{(2k)!(m{-}2)!(m{-}1)!} (4y)^{k+1-m}
\nn \\
&\quad - (-1)^{k} \frac{ 2{\rm B}_{2k}   (4y)^k}{(2k)! \Gamma(m)} 
 \sum_{n=1}^{\infty} n^{m-1} \sigma_{1-2m}(n)  (q^n +\bar{q}^n) \sum_{a=0}^{m-1} (4 n y)^{-a} \frac{\Gamma(m{+}a)}{a! \Gamma(m{-}a)}\,, \nn
\end{align}
where we have used the truncating Fourier expansion~\eqref{eq:FEk} for $\EE_m$
and emphasise that the coefficients of $y^{m+k}, \ \zeta_{2m-1}y^{1+k-m}$ and $y^{k-a} (q^n+\bar{q}^n)$ are rational.
Since $k\geq m$, all powers of $y$ that occur are positive. We shall solve this equation by Fourier decomposing the periodic function 
\begin{align}
\label{eq:seeed}
\seedp{s}{m}{k}(\tau) = c_0(y) + \sum_{n=1}^{\infty} c_n(y) (q^n+\bar{q}^n)
\end{align}
and assuming that $c_n(y)$ for $n>0$ is a Laurent polynomial
\begin{align}
c_n(y)  = \sum_{a=k-m+1}^{k-1} c_{n,a} y^a
\end{align}
of the same form as the right-hand side of the differential equation (\ref{eq:seedeq}). Note that the power $y^k$ is absent in the ansatz compared to the right-hand side of the differential equation. The ansatz~\eqref{eq:seeed} makes a choice of boundary conditions and, by the relative coefficient of $q^n$ and $\bar q^n$, has a built-in evenness under $\tau \to -\bar\tau$. We furthermore take real coefficients $c_{n,a}$, so that $\seedp{s}{m}{k}$ is real under complex conjugation.

Substituting the ansatz for the seed into the differential equation~\eqref{eq:seedeq} then leads to a second-order differential equation for $c_0(y)$ and to recurrence relations for the coefficients $c_{n,a}$. We solve these equations by
\begin{align}
\label{eq:C0CNsol}
c_0(y) &= (-1)^{k+m} \frac{ {\rm B}_{2k} {\rm B}_{2m}  (4y)^{k+m} }{(2k)! (2m)! (\mu_{k+m}-\mu_s) } 
-(-1)^{k} \frac{4 {\rm B}_{2k} (2m{-}3)! \zeta_{2m-1}  (4y)^{k+1-m} }{(2k)!(m{-}2)!(m{-}1)! (\mu_{k-m+1}-\mu_s)} 
\,,\nn\\
c_n(y) &= (-1)^k \frac{ 2  {\rm B}_{2k} }{(2k)! \Gamma(m)}
\sigma_{1-2m}(n) n^{m-k-1} \sum_{\ell=k-m+1}^{k-1} g^+_{m,k,\ell,s} (4ny)^\ell\,,
\end{align}
with $\mu_s=s(s{-}1)$ and rational coefficients
\begin{align}
\label{eq:fkm}
g^+_{m,k,\ell,s} = \frac{\Gamma(\ell)}{\Gamma(\ell{+}s)} \sum_{i=\ell}^{k-1}  \frac{ (\ell{+}1{-}s)_{i-\ell}\Gamma(s{+}i)\Gamma(m{+}k{-}i{-}1)}{\Gamma(k{-}i)\Gamma(i{+}1)\Gamma(m{-}k{+}i{+}1)}\,,
\end{align}
where $(a)_n=a (a{+}1) \cdots (a{+}n{-}1)=\frac{\Gamma(a+n)}{\Gamma(a)}$ is the (ascending) Pochhammer symbol.

In the zero mode $c_0(y)$ we have only used the powers of the right-hand side of the differential equation and set the homogeneous powers $y^s$ and $y^{1-s}$ to zero, which again is our choice of boundary conditions. Given that $s\geq 2$ by (\ref{eq:Fmk}), the second homogeneous solution $y^{1-s}$ clearly leads to a divergent Poincar\'e sum, but one that is formally related to $\EE_s$ by~\eqref{eq:FR} and that corresponds to the Poincar\'e sum of the  first homogeneous solution $y^s$~\eqref{eq:PSEk}. These corrrespond to $\beta^{\rm sv}$ at depth one and which we would like to disentangle from the $\FFp{s}{m}{k}$ at depth two. For this reason we impose that they vanish in the seed.

Using the relation~\eqref{eq:E0sigma} between $q$-series over divisor sums and iterated integrals, we can therefore write the Poincar\'e seed of $\FFp{s}{m}{k}$ as 
\begin{align}
\label{eq:genseed}
\seedp{s}{m}{k}(\tau) = c_0(y)
- (-1)^{k} \frac{ 2  {\rm B}_{2k} \Gamma(2m)
}{(2k)! \Gamma(m)} \sum_{\ell=k-m+1}^{k-1} g^+_{m,k,\ell,s} (4y)^\ell \Re \mathcal{E}_0 (2m,0^{k+m-\ell-1})
\end{align}
with $c_0(y)$ given by (\ref{eq:C0CNsol}) and the constants $g^+_{m,k,\ell,s}$ in (\ref{eq:fkm}).
Given that this result for $k\geq m$ has been obtained from folding ${\rm E}_k$ instead of ${\rm E}_m$
for the sake of a convergent Poincar\'e sum, the seed function only involves 
iterated Eisenstein integrals over $\GG_{2m}$. In section~\ref{sec:fold.2} we shall consider divergent Poincar\'e sums whose seeds feature iterated integrals over $\GG_{2k}$ with $k>m$ instead.
The evenness of $\seedp{s}{m}{k}$ under $\tau\to-\bar\tau$ here is reflected in the reality of the coefficients and the occurrence of $\Re {\cal E}_0$. 

From the general expression (\ref{eq:genseed}) we notice that there are only two terms accompanied by the power $y^{k-m+1}$. One is coming from the unique odd zeta $\zeta_{2m-1}$ due to $c_0(y)$, as in the first line of (\ref{eq:C0CNsol}). The other is coming from the non-zero Fourier modes' contribution with $\ell = k{-}m{+}1$ which leads to the iterated integral $ \Re \mathcal{E}_0 (2m,0^{2m-2})$. 

As we shall be explaining later on from a different perspective, the unique odd zeta value and the maximal iterated integral are interlocked and always appear in a very specific combination.
Isolating the two terms with $y^{k-m+1}$ in  (\ref{eq:genseed})  we have 
\beq\label{eq:MaxEOddZeta}
\seedp{s}{m}{k}(\tau) \, \Big|_{y^{k-m+1}} = \frac{ (-1)^k 4^{k-m+2}  {\rm B}_{2k} (2m{-}3)! (2m{-}1)! }{(2k)!
 (m{-}2)! (m{-}1)! (\mu_{k-m+1}-\mu_s)}
\Big[  \Re \mathcal{E}_0 (2m,0^{2m-2})- \frac{\zeta_{2m-1}}{(2m{-}1)!}\Big]\,.
\eeq

\subsubsection{\texorpdfstring{Examples at $m=k$}{Examples at m=k}}
\label{sec:3.3.3}

For the examples of $ \FFp{s}{k}{k}$ in (\ref{oct20.3}) and (\ref{oct20.4}) with $k\leq 3$,
the general formula (\ref{eq:genseed}) for the seed functions yields
\begin{align}
\seedp{2}{ 2 }{ 2} &=
\frac{ y^4}{20250}  - \frac{ y \zeta_3}{90 }  +\frac{ y}{15 } \Re[{\cal E}_0(4, 0^2)] \,,\notag \\
\seedp{2}{ 3 }{ 3} &= \frac{y^6}{6251175}  - \frac{ y \zeta_5}{1260} + \frac{2 y^2}{63 } \Re[{\cal E}_0(6, 0^{3})] + \frac{2y}{21}  \Re[{\cal E}_0(6, 0^{4})]\,,
 \label{oct20.9} \\
\seedp{4}{ 3 }{ 3} &= \frac{2 y^6}{8037225} - \frac{  y \zeta_5}{7560} +\frac{ 2 y^2}{63}  \Re[{\cal E}_0(6, 0^{3})] + \frac{y}{63}  \Re[{\cal E}_0(6, 0^{4})] \, ,
 \notag
\end{align}
also see \cite{DHoker:2015gmr,Dorigoni:2019yoq} for $\seedp{2}{ 2 }{ 2}$ and \cite{Ahlen:2018wng,Dorigoni:2019yoq,Basu:2020kka} for $\seedp{2}{ 3 }{3},\seedp{4}{ 3 }{ 3}$. The simplest Poincar\'e seeds for $ \FFp{s}{k}{k}$ beyond the state of the art read
\begin{align} 
\seedp{2}{ 4 }{ 4} &= \frac{y^8}{1205583750} - \frac{ y \zeta_7}{15120} +\frac{ 4 y^3}{135}  \Re[{\cal E}_0(8, 0^{4})] 
+\frac{ 4 y^2}{27}  \Re[{\cal E}_0(8, 0^{5})] + \frac{y}{3}  \Re[{\cal E}_0(8, 0^{6})]\,,  \notag \\
\seedp{4}{ 4 }{ 4} &= \frac{y^8}{982327500} - \frac{ y \zeta_7}{90720} + \frac{ 4 y^3}{135 } \Re[{\cal E}_0(8, 0^{4})] + \frac{y^2}{9}  \Re[{\cal E}_0(8, 0^{5})] + 
\frac{ y}{18}  \Re[{\cal E}_0(8, 0^{6})]\,,   \label{oct20.10} \\
\seedp{6}{ 4 }{ 4} &= \frac{y^8}{580466250} - \frac{ y \zeta_7}{226800} + \frac{4 y^3}{135}  \Re[{\cal E}_0(8, 0^{4})] + \frac{2 y^2}{45}  \Re[{\cal E}_0(8, 0^{5})] + 
 \frac{y}{45}  \Re[{\cal E}_0(8, 0^{6})] \, .
 \notag
 \end{align}
A seed related to $\seedp{2}{ 4 }{ 4}$ was given in~\cite{Basu:2020kka}.

\subsubsection{\texorpdfstring{Examples at $m<k$}{Examples at m<k}}
\label{sec:3.3.4}

For the examples of $ \FFp{s}{m}{k}$ in (\ref{oct20.5}) and (\ref{oct20.6}) with $(m,k)=(2,3)$ and $(2,4)$,
the general formula (\ref{eq:genseed}) for the seed function yields
\begin{align}
\seedp{3}{ 2 }{ 3} &= \frac{ y^5}{297675} - \frac{  y^2 \zeta_3}{1890} + \frac{ y^2}{315}  \Re[{\cal E}_0(4, 0^2)] \,,   \label{oct20.11} \\
\seedp{4}{ 2 }{ 4} &= \frac{ y^6}{3827250} - \frac{  y^3 \zeta_3}{28350} + \frac{ y^3}{4725  }  \Re[{\cal E}_0(4, 0^2)]\,,\notag
\end{align}
see also \cite{Ahlen:2018wng,Dorigoni:2019yoq}. We shall give the complete set of seed functions $\seedp{s}{ m }{ k}$ for the modular invariant functions of weight $m{+}k\leq 8$,
\begin{align}
\seedp{5}{ 2 }{ 5} &= \frac{ y^7}{46309725} - \frac{  y^4 \zeta_3}{374220} + \frac{  y^4 }{62370  } \Re[{\cal E}_0(4, 0^2)]\,, \label{oct20.12} \\
\seedp{6}{ 2 }{ 6} &= \frac{ 691 y^8}{373530031875}  - \frac{   691 y^5 \zeta_3}{3192564375} + \frac{  1382 y^5 }{1064188125} \Re[{\cal E}_0(4, 0^2)]\,,
\notag
\end{align}
as well as
\begin{align}
\seedp{5}{ 3 }{ 4} &= \frac{ y^7}{49116375} - \frac{   y^2 \zeta_5}{113400} + \frac{  2 y^3}{945}  \Re[{\cal E}_0(6, 0^{3})] + \frac{ y^2}{945}  \Re[{\cal E}_0(6, 0^{4})] \,,
 \notag \\
\seedp{3}{ 3 }{ 4} &= \frac{ y^7}{80372250} - \frac{   y^2 \zeta_5}{25200} +\frac{  2  y^3}{945} \Re[{\cal E}_0(6, 0^{3})] + \frac{ y^2}{210 } \Re[{\cal E}_0(6, 0^{4})] \,,
 \label{oct20.13} \\
\seedp{6}{ 3 }{ 5} &= \frac{ 2 y^8}{1149323175 }- \frac{ y^3 \zeta_5}{1496880 }+ \frac{  y^4 }{6237} \Re[{\cal E}_0(6, 0^{3})]+ \frac{  y^3 }{12474}  \Re[{\cal E}_0(6, 0^{4})]\,, \notag \\
\seedp{4}{ 3 }{ 5} &= \frac{ y^8}{972504225 } - \frac{   y^3 \zeta_5}{374220 } + \frac{  y^4 }{6237 }  \Re[{\cal E}_0(6, 0^{3})]+ \frac{  2 y^3 }{6237 } \Re[{\cal E}_0(6, 0^{4})]\, .
 \notag
 \end{align}
 Performing the Poincar\'e sum of these seeds produces $\FFp{s}{m}{k}$ which are modular invariant and even under $\tau\to -\bar\tau$ by construction.

%%%%%%%%%%%%%%%%%%%%%%%%%%%%%%%%%%%%%%%%%%%%%%%%%%%%%%%%%%%
\subsection{Seed functions for even shuffles}
\label{sec:3.3}
%%%%%%%%%%%%%%%%%%%%%%%%%%%%%%%%%%%%%%%%%%%%%%%%%%%%%%%%%%%

In~\eqref{eq:genseed} we saw that the seed functions of the  $\FFp{s}{m}{k}$   can be expressed in terms of iterated integrals $y^\ell \Re \mathcal{E}_0(2m,0^{k+m-\ell-1})$ with $k{-}m{+}1\leq \ell \leq k{-}1$ such that we only span the iterated integrals $\Re \mathcal{E}_0(2m,0^{p})$ with $m\leq p\leq 2m{-}2$. In order to exhaust the remaining cases with $0\leq p<m$, we recall that the $(m,k)$ sector of real modular invariants also
contains the (sums of) shuffles $\Jp{\ell}{m}{k}$ in~\eqref{prodd1} built from $\EE_m,\EE_k$ and their derivatives.
We shall now determine the corresponding seed functions for $0\leq \ell <m$ and see that they contain the missing iterated integrals.
This turns out to be simpler as the $\Jp{\ell}{m}{k}$ are sums of shuffles unlike the $\FFp{s}{m}{k}$.
 
As we shall see, the seed functions of the modular invariant functions $\FFp{s}{m}{k}$, $\Jp{\ell}{m}{k}$
at depth two are both built from rational combinations of $y^{m+k}$ or $y^{k-m+1} \zeta_{2m-1}$ and  
$y^{k+m-1-p}   \Re[{\cal E}_0({2m},0^{p}) ] $ with $0\leq p\leq 2m{-}2$. Hence, the transition from 
the modular invariant functions to their Poincar\'e seeds once more reduces the depth of the contributing iterated Eisenstein 
integrals by one unit just like in (\ref{eq:PSEk}).

\subsubsection{\texorpdfstring{Seed functions of $\Jp{\ell}{m}{k}$}{Seed functions of Jplus(l,m,k)}}
\label{sec:3.4.1}

We want to derive a seed function $\Jseedp{\ell}{m}{k}$ in
\begin{align}
\Jp{\ell}{m}{k}(\tau)  = \PS \Jseedp{\ell}{m}{k} (\gamma\cdot \tau)\,,
 \label{oct20.14}
\end{align}
for the objects $\Jp{\ell}{m}{k} (\tau)$ with $\ell\ge0 $
defined in equation (\ref{prodd1}). We recall that we are working under the assumptions $m\leq k$ without loss of generality. Starting from $\ell=\min(m,k)=m$ we then generate holomorphic Eisenstein series $\GG_{2m}$ according to~\eqref{eq:Ediff}. Hence, the $\Jp{\ell}{m}{k}$ with $\ell \geq m$ are going beyond the $\mathbb Q[y^{\pm 1},{\rm MZV}]$ combinations of $\beta^{\rm sv} [ \begin{smallmatrix}  j_1 &j_2 &\ldots &j_\ell \\ k_1 &k_2 &\ldots &k_\ell
\end{smallmatrix}]$ with $j_i\leq k_i{-}2$ that form the backbone of MGFs and the modular invariant functions $\FFpm{s}{m}{k}$ here studied.

Using the seed representation (\ref{eq:PSEk}) for $ {\rm E}_k$, we can easily obtain
\begin{align}
\Jseedp{\ell}{m}{k}
&=
-  \frac{ (-4)^k  (k)_\ell {\rm B}_{2k}}{2 (2k)!}
y^{k-\ell}  \big( (\pi \nabla)^\ell   +   (\pi \overline \nabla)^\ell \big)  {\rm E}_m\,.
\end{align}
For the regime $\ell<m$ that we are considering, this seed is indeed convergent.

Then, using the explicit expression (\ref{eq:FEk}) for the non-holomorphic Eisenstein series, we have to compute
\begin{align}
(\pi \nabla)^\ell \Big( y^a (q^n+\bar{q}^n) \Big) & = y^{a+\ell}\Big[(a)_\ell \bar{q}^n +q^n \sum_{s=0}^\ell (-1)^s  \binom{\ell}{s}(a+s)_{\ell-s} (4 n y)^s \Big]\,, \notag\\
(\pi {\overline \nabla})^\ell \Big( y^a (q^n+\bar{q}^n) \Big) & = y^{a+\ell}\Big[(a)_\ell q^n +{\bar{q}}^n \sum_{s=0}^\ell (-1)^s \binom{\ell}{s}(a+s)_{\ell-s} (4 n y)^s \Big]\,,
\end{align}
which one can prove by induction. We also note $(\pi \nabla)^\ell y^a= (\pi \overline{\nabla})^\ell y^a = (a)_\ell y^{a+\ell}$\,.

Putting everything together we obtain
\begin{align}
\Jseedp{\ell}{m}{k}
&\notag = 
(-1)^{k + m} \frac{ {\rm B}_{2 k}   {\rm B}_{2 m} }{ (2 k)!  (2 m)!} (m)_\ell (k)_\ell   (4 y)^{k + m}
-4 (-1)^{k+\ell} \frac{ {\rm B}_{2 k}  (  2 m {-} 3)!  \zeta_{   2 m - 1} }{  (2 k)! (m {-} 2)!  (m {-} 1{-}\ell)!}
(k)_{\ell}  (4 y)^{k + 1 - m}
 \\
&\quad  -(-1)^{k+\ell } \frac{  {\rm B}_{2k} (k)_\ell }{(2 k)!   \Gamma(m)} (4 y)^k
\sum_{n=1}^{\infty}  n^{m - 1} \sigma_{1-2m}(n) (q^n+\bar{q}^n)
\sum_{a=0}^{m-1}  (4 n y)^{-a}  \frac{ \Gamma(   m {+} a)}{a! \Gamma(m {-} a) } \label{eq:seedRFull} \\
& \hspace{20mm}  \times \Big[(a{+}1{-}\ell)_\ell +\sum_{s=0}^\ell \binom{\ell}{s} (a{+}1{-}\ell)_{\ell-s} (4 n y)^s\Big]\,.
\notag
\end{align}
The $q$-series over the divisor sums can again be written in terms of iterated Eisenstein integrals over $\GG_{2m}$ of different lengths and multiplied by different powers of $y$ by using~\eqref{eq:E0sigma}:
\begin{align}
\label{eqseedR}
\Jseedp{\ell}{m}{k}
& =
(-1)^{k + m} \frac{ {\rm B}_{2 k}   {\rm B}_{2 m} }{ (2 k)!  (2 m)!} (m)_\ell (k)_\ell   (4 y)^{k + m}
-4 (-1)^{k+\ell} \frac{ {\rm B}_{2 k}  (  2 m {-} 3)!  \zeta_{   2 m - 1} }{  (2 k)! (m {-} 2)!  (m {-} 1{-}\ell)!}
(k)_{\ell}  (4 y)^{k + 1 - m}
 \nn\\
&\quad +(-1)^{k+\ell}  \frac{  {\rm B}_{2k}(k)_\ell \Gamma(2m)}{ (2k)! \Gamma(m)} (4y)^{k}\sum_{a=0}^{m-1}  \frac{\Gamma(m{+}a)}{a! \Gamma(m{-}a)}  \Bigg[(a{+}1{-}\ell)_\ell\, (4y)^{-a}\Re[{\cal E}_0(2m, 0^{m+a-1}) ]  \nn\\
&\hspace{45mm} +\sum_{s=0}^\ell \binom{\ell}{s} (a{+}1{-}\ell)_{\ell-s} \,(4 y)^{s-a} \Re[{\cal E}_0(2m, 0^{m+a-s-1}) ]\Bigg]\,.
\end{align}
When combined with (\ref{eq:genseed}), this shows that the expressions 
\begin{align}
\label{eq:E0bas}
y^{k+m-1-b} \Re \mathcal{E}_0(2m,0^b) \quad\quad \text{for $0\leq b\leq 2m{-}2$ }\, ,
\end{align}
together with $y^{k+m}$ and $y^{k-m+1}\zeta_{2m-1}$ from the Fourier zero mode, provide a basis of the Poincar\'e seeds for the modular invariants built from $\betalagp{j}$. Note that $0\leq j \leq 2m{-}2$ so that the counting agrees. The Fourier zero-mode contributions $y^{k+m}$ and $y^{k-m+1}$ appear only in such a way to produce the correct Laplace equations and are lower depth than the iterated integrals.

As in (\ref{eq:genseed}), the odd zeta value in (\ref{eqseedR}) always occurs with the same relative coefficient $\Re[{\cal E}_0({2m},0^{2m-2}) ] - \frac{ \zeta_{2m-1}}{(2m-1)!} $.
This can be easily seen from (\ref{eqseedR}) where the only instances of the maximally integrated $\Re[{\cal E}_0({2m},0^{2m-2}) ]$ occur for $a=m{-}1$ and $s=0$, so isolating the unique odd zeta value and the two maximal iterated integrals we have 
\beq
\Jseedp{\ell}{m}{k} \, \Big|_{ y^{k-m+1}} =
\frac{ (-1)^{k + \ell}   4^{k + 2 - m} {\rm B}_{2 k} ( 2 m {-} 3)!  ( 2 m {-} 1)! (k)_\ell}{
(2 k)! (m {-} 2)! (m {-} 1 {-} \ell)!}
   \Big[\Re[{\cal E}_0({2m},0^{2m-2}) ] - \frac{ \zeta_{2m-1}}{(2m{-}1)!}\Big]\,,
\label{noted2}
\eeq
upon using the Legendre duplication formula. 
This very same combination, with a different rational prefactor, was found in (\ref{eq:MaxEOddZeta}) when we discussed the seed functions for the $ \FFp{s}{m}{k}$. We shall come back to this observation in section~\ref{sec:3.5}.

\subsubsection{\texorpdfstring{Examples at $m=k$}{Examples at m=k}}
\label{sec:3.4.2}

The simplest examples of the seed functions (\ref{eqseedR}) related to bilinears in
${\rm E}_2$ and ${\rm E}_3$ are
\begin{align}
\Jseedp{0}{2}{2}  &= \frac{ y^4}{2025} +   \frac{ y \zeta_3}{45} -  \frac{4y^2 }{15 } \Re[{\cal E}_0(4, 0)] - \frac{ 2y}{15}  \Re[{\cal E}_0(4, 0^2)] \,,
\\
\Jseedp{1}{2}{2}  &=
\frac{4 y^4}{2025} - \frac{ 2y \zeta_3 }{45 } +  \frac{16y^3}{15}  \Re[{\cal E}_0(4)] + \frac{ 8y^2}{15}  \Re[{\cal E}_0(4, 0)] + 
\frac{ 4y}{15 } \Re[{\cal E}_0(4, 0^2)] \,,
 \notag
\end{align}
and 
\begin{align}
\Jseedp{0}{3}{3}  &=\frac{4 y^6}{893025} + \frac{y \zeta_5}{630}   - \frac{ 16 y^3}{63}  \Re[{\cal E}_0(6, 0^2)]
 - \frac{ 8y^2}{21}  \Re[{\cal E}_0(6, 0^3)] - \frac{ 4 y}{21}  \Re[{\cal E}_0(6, 0^4)] \,,
 \notag \\
\Jseedp{1}{3}{3}  &= \frac{4 y^6}{99225}  - \frac{ y  \zeta_5 }{105} + \frac{ 32 y^4}{21 } \Re[{\cal E}_0(6, 0)] + \frac{16y^3}{7}  \Re[{\cal E}_0(6, 0^2)] \notag \\
&\ \ \ \ \ \ \ \ + 
 \frac{ 16y^2}{7}  \Re[{\cal E}_0(6, 0^3)] + \frac{8y}{7} \Re[{\cal E}_0(6, 0^4)]\,,
\\
\Jseedp{2}{3}{3}  &=\frac{64 y^6}{99225 } + \frac{ 4y \zeta_5}{105 }  - \frac{ 512 y^5}{21 } \Re[{\cal E}_0(6)] 
- \frac{ 512 y^4 }{21}  \Re[{\cal E}_0(6, 0)] - 
\frac{ 128 y^3 }{7} \Re[{\cal E}_0(6, 0^2)] \notag \\
&\ \ \ \ \ \ \ \  - \frac{ 64 y^2 }{7}  \Re[{\cal E}_0(6, 0^3)] - \frac{32y }{7}  \Re[{\cal E}_0(6, 0^4)] \, ,
 \notag
\end{align}
respectively. By combining these seeds with those for $ \FFp{s}{k}{k}$ in (\ref{oct20.9}),
we can isolate the iterated Eisenstein integrals in the linear combinations
{\allowdisplaybreaks
\begin{align}
\frac{y^4}{1350}+ y \Big( \Re[{\cal E}_0({4, 0^2}) ]  - \frac{ \zeta_3}{6} \Big) &= 15  \seedp{2}{ 2}{ 2} \,, \notag \\
-\frac{y^4}{450}+ y^2 \Re[{\cal E}_0({4, 0}) ]  &= - \frac{ 15}{4} (2 \seedp{2}{ 2}{ 2} + \Jseedp{0}{ 2}{ 2})
 \label{oct20.15} \,, \\
\frac{y^4}{360}+ y^3  \Re[{\cal E}_0({4}) ]  &=\frac{15}{16} (2 \Jseedp{0}{ 2}{ 2} + \Jseedp{1}{ 2}{ 2}) \, ,
\notag 
\end{align}}%
and
\begin{align}
-\frac{y^6}{893025}+y \Big( \Re[{\cal E}_0({6, 0^4}) ]  - \frac{ \zeta_5}{5!} \Big) &=
\frac{63}{5} (\seedp{2}{ 3}{ 3} - \seedp{4}{ 3}{ 3})\,,
\notag \\
\frac{y^6}{119070}+y^2 \Re[{\cal E}_0({6,0^3}) ]  &=  -\frac{63}{10} (\seedp{2}{ 3}{ 3} - 6 \seedp{4}{ 3}{ 3})\,,
\notag \\
-\frac{y^6}{34020}+y^3  \Re[{\cal E}_0({6,0^2}) ]  &= -\frac{63}{16} (12 \seedp{4}{ 3}{ 3} + \Jseedp{0}{ 3}{ 3})
 \label{oct20.16}\,,\\
\frac{y^6}{17010}+y^4  \Re[{\cal E}_0({6,0}) ]  &=  \frac{21}{32} (36 \seedp{4}{ 3}{ 3} + 9 \Jseedp{0}{ 3}{ 3} + \Jseedp{1}{ 3}{ 3})\,,
\notag 
\\
-\frac{y^6}{15120}+y^5  \Re[{\cal E}_0({6}) ]  &=  -\frac{21}{512} (72 \Jseedp{0}{ 3}{ 3} + 16 \Jseedp{1}{ 3}{ 3} + \Jseedp{2}{ 3}{ 3})  \, .
\notag
\end{align}
Similar relations between Poincar\'e sums over $\Re[{\cal E}_0({8, 0^p}) ]$ and $ \FFp{s}{4}{4}$
can be found in appendix~\ref{app:ex.1}.

\subsubsection{\texorpdfstring{Examples at $m<k$}{Examples at m<k}}
\label{sec:3.4.3}

One can similarly combine the seeds for $ \FFp{3}{2}{3}$, $\FFp{4}{2}{4}$ in (\ref{oct20.11}) 
with the following $\Jseedp{\ell}{m}{k} $ from~(\ref{eqseedR}),
\begin{align}
\Jseedp{0}{2}{3}  &= \frac{2 y^5}{42525 }+ \frac{ 2y^2 \zeta_3}{945}   -  \frac{8y^3}{ 315}  \Re[{\cal E}_0(4, 0)] - 
  \frac{ 4y^2 }{315}  \Re[{\cal E}_0(4, 0^2)] \,, \notag \\
\Jseedp{1}{2}{3}  &= \frac{4 y^5}{14175 } - \frac{ 2y^2 \zeta_3 }{315 } + \frac{ 16y^4}{105 } \Re[{\cal E}_0(4)] + 
  \frac{ 8y^3 }{105 } \Re[{\cal E}_0(4, 0)] +  \frac{4y^2 }{105 } \Re[{\cal E}_0(4, 0^2)] \,,
\notag \\
\Jseedp{0}{2}{4}  &=
 \frac{ y^6}{212625} + \frac{   y^3 \zeta_3}{4725} - \frac{4 y^4 }{1575} \Re[{\cal E}_0(4, 0)] - \frac{2 y^3 }{1575} \Re[{\cal E}_0(4, 0^2)] \,, \\
\Jseedp{1}{2}{4}  &= \frac{8 y^6}{212625} - \frac{   4 y^3 \zeta_3}{4725} 
+ \frac{32 y^5 }{1575} \Re[{\cal E}_0(4)] + \frac{   16 y^4 }{1575} \Re[{\cal E}_0(4, 0)] + \frac{8 y^3 }{1575} \Re[{\cal E}_0(4, 0^2)]\,,  \notag
\end{align}
and thereby isolate the iterated Eisenstein integrals in 
the linear combinations
\begin{align}
 \frac{y^5}{945}+ y^2 \Big( \Re[{\cal E}_0({4, 0^2}) ]  - \frac{ \zeta_3}{6} \Big)  &= 
315 \seedp{3}{ 2}{ 3} \,, \notag \\
-\frac{ y^5}{420}  +y^3 \Re[{\cal E}_0({4, 0}) ]  &= 
-\frac{315}{8} (4 \seedp{3}{ 2}{ 3} + \Jseedp{0}{ 2}{ 3})
 \label{oct20.17}\,,\\
 \frac{y^5}{360} +y^4 \Re[{\cal E}_0({4}) ] &= 
\frac{105}{16} (3 \Jseedp{0}{ 2}{ 3} + \Jseedp{1}{ 2}{ 3})
\notag 
\end{align}
and
\begin{align}
\frac{ y^6}{810}+y^3 \Big( \Re[{\cal E}_0({4, 0^2}) ]  - \frac{ \zeta_3}{6} \Big)  &=
4725 \seedp{4}{ 2}{ 4} \,,
 \notag \\
  - \frac{y^6}{405}  +y^4 \Re[{\cal E}_0({4, 0}) ]&= 
-\frac{1575}{4}(6 \seedp{4}{ 2}{ 4} + \Jseedp{0}{ 2}{ 4})\,,
 \label{oct20.18} \\
\frac{y^6}{360} +y^5 \Re[{\cal E}_0({4}) ]  &=
\frac{1575}{32 }(4 \Jseedp{0}{ 2}{ 4} + \Jseedp{1}{ 2}{ 4})\,.
\notag 
\end{align}
Similar relations between Poincar\'e sums over $\Re[{\cal E}_0({2k, 0^p}) ]$ and $ \FFp{s}{2}{5}, \FFp{s}{2}{6}, \FFp{s}{3}{4}, \FFp{s}{3}{5}$ can be found in appendix \ref{app:ex.2}. Comparing (\ref{oct20.15}) with (\ref{oct20.17}) and (\ref{oct20.18}) illustrates once more that the seed functions $\seedp{s}{ m}{ k}$ and $\Jseedp{\ell}{ m}{ k}$
with $m\leq k$ derived above only feature iterated Eisenstein integrals over ${\rm G}_{2m}$ rather than ${\rm G}_{2k}$. Still, $k$ leaves
a fingerprint in the structure of the seed function through the power of $y$ in the terms $y^{k+m-p-1} \Re[{\cal E}_0({2m},0^p) ]$.

%%%%%%%%%%%%%%%%%%%%%%%%%%%%%%%%%%%%%%%%%%%%%%%%%%%%%%%%%%%
\subsection{Laplace equations of even combinations in step form}
\label{sec:3.5}
%%%%%%%%%%%%%%%%%%%%%%%%%%%%%%%%%%%%%%%%%%%%%%%%%%%%%%%%%%%

The above results motivate an alternative organisation of the system of Laplace equations at each $(m,k)$ where seeds of the form in~\eqref{eq:E0bas} take center stage. We will now describe a procedure to directly construct the combinations of $\FFp{s}{m}{k}$ in the Poincar\'e sums over a given $y^{k+m-p-1} \Re[{\cal E}_0({2m},0^p) ]$.

 As noted in (\ref{eq:MaxEOddZeta}) and (\ref{noted2}), all the seed functions $\seedp{s}{m}{k}$ and $\Jseedp{\ell}{ m}{ k}$ feature a term
involving $y^{k-m+1} ( \Re[{\cal E}_0({2m},0^{2m-2}) ] - \frac{ \zeta_{2m-1}}{(2m-1)!} )$, i.e.\ the iterated Eisenstein integral $ \Re[{\cal E}_0({2m},0^{2m-2}) ]$ with the maximal number of zeros. Both the seed $y^{k-m+1}  \Re[{\cal E}_0({2m},0^{2m-2}) ]$ and its Poincar\'e sum require the maximum number $2m{-}1$ of Laplace actions until a holomorphic Eisenstein series is generated. For given $(m,k)$, there is a unique real depth-two combination $\betalagp{j}$ which 
shares this property, as can be seen from the Laplace system (\ref{eq:LapLag}), namely $\betalagp{0}$. Therefore we conclude
\beq
\sum_{\gamma \in B(\ZZ)\backslash {\rm SL}(2,\ZZ)}
\bigg[ y^{k-m+1} \Big( \Re[{\cal E}_0({2m},0^{2m-2}) ] - \frac{ \zeta_{2m-1}}{(2m{-}1)!} \Big)  \bigg]_\gamma = \rho_{m,k}\,
\betalagp{0} \ \text{mod lower depth}
\label{oct21.1}
\eeq
with rational prefactor
\begin{align}
\rho_{m,k} =   \frac{  (-4)^{m-k-1}  (2k)!(2k{-}1)!}{2 {\rm B}_{2k} (k{+}m{-}2)! (k{-}m)!(2m{-}2)!} \,,
\label{rhoexpr}
\end{align}
such that in particular $\rho_{k,k} = -\frac{k (2k-1)^2}{4 {\rm B}_{2k}} $.
The combination~\eqref{oct21.1} is the only choice, where the occurrence of the holomorphic Eisenstein series is maximally delayed to the $(2m{-}1)^{\rm th}$ 
power of the Laplacian, i.e.\ to $\Delta^{2m-1}\betalagp{j=0}$. In order to generate the Poincar\'e sums over shorter iterated Eisenstein integrals $\sim 
 \Re[{\cal E}_0({2m},0^{2m-2-r}) ]$, we apply combinations of Laplace operators  
 to both sides of~(\ref{oct21.1}). From $\Delta y^n = n(n{-}1) y^n $ and
\begin{align}
\Delta \Big( y^n   \Re[{\cal E}_0({2m}) ]  \Big) &= \frac{ 4n  y^{n+1} }{(2\pi i)^{2m}} \Re[{\rm G}_{2m}^0 ]+ n(n{-}1) y^n  \Re[{\cal E}_0({2m}) ]\,,
\label{d1lemma}\\
\Delta \Big( y^n   \Re[{\cal E}_0({2m,0^p}) ]  \Big) &= -  4n   y^{n+1}  \Re[{\cal E}_0({2m,0^{p-1}}) ] + n(n{-}1) y^n  \Re[{\cal E}_0({2m,0^p}) ] \, , \ \ \ \ \ \ p \neq 0\,,\nn 
\end{align}
we deduce that the operator (for $\ell>0$)
\begin{align}
\mathcal{O}_\ell = -\frac{1}{4\ell} \big( \Delta - \ell (\ell{-}1)\big)
\label{oct21.3}
\end{align}
has the property 
\begin{subequations}
\begin{align}
\label{eq:step}
\mathcal{O}_{k+m-b-1}\Big(y^{k+m-b-1}  \Re \mathcal{E}_0(2m,0^b) \Big) &= y^{k+m-b}  \Re \mathcal{E}_0(2m,0^{b-1}) \, ,\quad \quad (b>0)\,,\\
\label{eq:laststep}
\mathcal{O}_{k+m-1}\Big(y^{k+m-1}  \Re \mathcal{E}_0(2m) \Big) &= -\frac{y^{k+m}}{(2\pi i)^{2m}}  \Re  \GG_{2m}^0 \,.
\end{align}
\end{subequations}
It therefore can be used to reduce the number of zeros of an iterated integral while increasing the power of $y$ in exactly the same way as they appear in the Poincar\'e seeds studied in the previous sections.

However, the seeds (\ref{oct21.0}) of the holomorphic Eisenstein series generated from $\Delta \betalagp{2m-2}$
involve the full $\GG_{2m}=\GG_{2m}^0 + 2 \zeta_{2m}$ rather than $\GG_{2m}^0$ seen in \eqref{eq:laststep}. 
Hence, we have to combine the $\Re \mathcal{E}_0(2m,0^p)$ term with a power of $y$ as in
\begin{align}
\mathcal{O}_{k+m-1}\Big(y^{k+m-1} \Re \mathcal{E}_0(2m) -\frac{2 {\rm B}_{2m}}{(2m)!} y^{k+m}
\Big) = -\frac{y^{k+m}}{(2\pi i)^{2m}}  \Re  \GG_{2m}\,,
\label{oct21.4}
\end{align}
where we used the 
relation between the even $\zeta$-value in terms of Bernoulli numbers given in~\eqref{eq:zB}. This fixes the coefficients of the pure powers of $y$ in the seeds of the last lines in (\ref{oct20.15}) to (\ref{oct20.18}) -- they ensure that no additional ${\rm E}_{m+k}$ are generated in the respective Poincar\'e sums.

From the above arguments, the Poincar\'e sums over individual iterated Eisenstein integrals yield
modular forms with the following leading-depth terms ($r=1,2,\ldots,2m{-}2$)
\begin{align}
&\sum_{\gamma \in B(\ZZ)\backslash {\rm SL}(2,\ZZ)}
\Big[  a_{m,k}^{r} y^{k+m} + y^{k-m+1+r}   \Re[{\cal E}_0({2m},0^{2m-2-r}) ]  \Big]_\gamma  \label{oct21.5}\\
& \ \ \ \ \ \
\sim  {\cal O}_{k-m+r} {\cal O}_{k-m+r-1} \ldots {\cal O}_{k-m+2} {\cal O}_{k-m+1} \betalagp{0}  \ \text{mod lower depth} \, .
\notag
\end{align}
The product of operators (\ref{oct21.3}) can be straightforwardly evaluated via 
(\ref{eq:LapLag}) and results in a combination of $\betalagp{0},\betalagp{1},\betalagp{2},\ldots,\betalagp{r}$.
 The coefficient of $y^{k+m}$ in (\ref{oct21.5}) is given by
\beq
a_{m,k}^{r} = - \frac{ 2 (-4)^{2m-2-r} {\rm B}_{2m} (k{+}m{-}2)! (2k{+}r{-}1)!}{(2m)! (k{-}m{+}r)! (2m{-}1{-}r)!(2k{+}2m{-}3)!}
\label{oct21.6}
\eeq
and ensures that the terms of lower depth do not include any ${\rm E}_{m+k}$.
Note in particular that the $2m{-}1$ possibilities of inserting $j=0,1,2,\ldots,2m{-}2$ zeros
into the seed $\Re[{\cal E}_0({2m},0^{j}) ]$ precisely match the number of leading-depth-two terms $\betalagp{j}$
for real modular invariants, cf.\ (\ref{eq:betalag}).

Based on (\ref{ppeis.all}), we deduce the following Poincar\'e sum over (\ref{oct21.4}), 
\begin{align}
&\quad \frac{1}{(2\pi i)^{2m}} \PS \left[y^{k+m} \Re(\GG_{2m}(\tau))  \right]_\gamma   \label{notthis} \\
& =   - \frac{(2k)! (k{-}1)!  }{2 (-4)^k {\rm B}_{2k} \Gamma(k{+}m)} \bigg\{
\frac{ {\rm G}_{2m} }{(2\pi i)^{2m}} (\pi \overline \nabla)^m {\rm E}_k
+ \frac{\overline{ {\rm G}_{2m} }}{(2\pi i)^{2m}} (\pi  \nabla)^m {\rm E}_k
\bigg\}\, ,
\notag
\end{align}
or alternatively
\begin{align}
\frac{1}{(2\pi i)^{2m}} &\PS \left[y^{k+m} \Re(\GG_{2m}(\tau))  \right]_\gamma 
=  -\delta_{k,m} \frac{(2k)!}{4^{2k} {\rm B}_{2k}} \frac{ \overline{\GG_{2k}}}{(2\pi i)^{2k}} (\tau{-}\bar\tau)^{2k}\GG_{2k} \notag \\
&+ \frac{(2k)!(2k{-}1)!(k{-}m)}{2 (-4)^{k+m}{\rm B}_{2k} \Gamma(k{+}m) (k{-}m)!} \left\{
(\tau{-}\bar\tau)^{2m}  \GG_{2m} \betasv{ k-m-1\\2k}
+ \frac{ \overline{\GG_{2m}}}{(2\pi i)^{2m}}\betasv{k+m-1\\2k}  \right\} \notag \\
&\ \ \ \ \ \ \ \  \MLD \, . \label{notthis2}
\end{align}
The Poincar\'e sum converges for $k+m>1$ and the rewriting using~\eqref{oct24.2} in terms of $\bsv$ requires $0\leq m< k$.

\subsection{\texorpdfstring{Comparison to $C_{a,b,c}$ MGFs}{Comparison to C(a,b,c) MGFs}}
\label{sec:3.2.5}

In the examples~\eqref{oct20.4} and \eqref{oct20.6} we expressed some of the $\FFp{s}{m}{k}$ that are determined by~\eqref{eq:Fmk} in terms of the two-loop modular graph functions $C_{a,b,c}$ defined in~\eqref{rev.05}, together with $\EE_{m+k}$ and possibly $\zeta_{m+k}$. This was possible because the $C_{a,b,c}$ that appeared contained the same inhomogeneous terms $\EE_m\EE_k$  in their Laplace equations. We shall take  the appearance of $\EE_m\EE_k$ as the defining feature of what we call the $(m,k)$ depth-two sector obtained from double integrals of holomorphic Eisenstein series $({\rm G}_{2m},{\rm G}_{2k})$ as in section \ref{sec:2.2}. 
Thinking of the $\FFp{s}{m}{k}$ as the most general real and shuffle-irreducible depth-two objects in the $(m,k)$ sector, a natural question is how they relate in general to the $C_{a,b,c}$. In the following discussion we restrict to only even $\FFp{s}{m}{k}$; a similar analysis can be done for the odd ones which relate to cuspidal MGFs.

From the general analysis of \cite{DHoker:2015gmr}, we know that the $C_{a,b,c}$ are closed under the action of the Laplacian at fixed weight $w=a{+}b{+}c$ up to source terms of the form ${\rm E}_{m}{\rm E}_k$ with $w=m{+}k$ and ${\rm E}_w$, as was recalled in (\ref{lapsec.3}) and (\ref{lapsec.3a}). This closure condition is not met by the modular invariants $\Jp{\ell}{m}{k}$ with $\ell \leq m$
defined in (\ref{prodd1}), so the $\FFp{s}{m}{k}$ are the appropriate choice of modular invariant functions at
depth two to represent the $C_{a,b,c}$. Moreover, the dimension of the vector space, $\mathcal{V}_{C}(w,s)$, of $C_{a,b,c}$ at a given weight $w=a{+}b{+}c$ and given eigenvalue $s(s{-}1)$ was determined in~\cite{DHoker:2015gmr} to be
\begin{equation}
\label{eq:dimC}
\dim \mathcal{V}_C(w,s) = \left\{\begin{array}{cl} \left\lfloor \frac{s+2}{3}\right\rfloor &\text{for $1\leq s\leq w{-}2$ and $s,w$ of same parity,} \\[2mm]
0 &\text{otherwise.}
\end{array} \right. 
\end{equation}
We can perform a similar counting of the number of independent $\FFp{s}{m}{k}$ using the spectrum in~\eqref{eq:Fmk} and find
\begin{equation}
\label{eq:dimF}
\dim \mathcal{V}_{{\rm F}^+}(w,s) = \left\{\begin{array}{cl} \left\lfloor \frac{s}{2}\right\rfloor &\text{for $2\leq s\leq w{-}2$ and $s,w$ of same parity,}\\[2mm]
0 &\text{otherwise,}\end{array}\right.
\end{equation}
where $w=m{+}k$. We note that the only difference in the allowed values of $s$ occurs when $w$ is odd and $s=1$ which corresponds to vanishing Laplace eigenvalue. The corresponding modular invariant solution is a constant. For $C_{a,b,c}$ it is known that this must be $\zeta_w$ times a rational number~\cite{DHoker:2017zhq}. 

Comparing the dimensions~\eqref{eq:dimC} and~\eqref{eq:dimF} we see that the $\FFp{s}{m}{k}$ are in general more numerous than the $C_{a,b,c}$. The first deviations occur at
\beq
\dim \mathcal{V}_C(w,s)  < \dim \mathcal{V}_{{\rm F}^+}(w,s) \, , \ \ \ \ \ \text{at} \ s= 6,8,9,10,\ldots
 \ \text{i.e.} \ w= 8,10,11,12,\ldots \, ,
 \label{cabc.0}
\eeq
i.e.\ at weight $w=8$ and at any weight $w\geq 10$, the number of independent $\FFp{s}{m}{k}$ is 
strictly larger than that of $C_{a,b,c}$. Hence, there exist even and modular invariant combinations 
involving $\beta^{\rm sv}$ at
depth two which cannot be represented in terms of two-loop MGFs $C_{a,b,c}$. 
However, we are not claiming that the ``missing'' modular invariant functions are built from genuine MGFs in that they possess a lattice-sum representation. We will come back to this point at the end of this section.

Given that the $C_{a,b,c}$ span a subspace of the depth-two objects in the $(m,k)$ sector, we can expand
\begin{equation}
\label{eq:decomp} C_{a,b,c} = \sum_{m=2}^{\lfloor{\frac{w}{2}\rfloor}} \sum_{s=w-2m+2}^{w-2} \alpha_{m}^{(s)}(a,b,c) \FFp{s}{m}{w-m} +\gamma(a,b,c) {\rm E}_{w}+\lambda(a,b,c) \zeta_w\,,
\end{equation}
for some rational\footnote{The coefficients appearing in the Laplace system (\ref{lapsec.3}) and (\ref{lapsec.3a}) for the $C_{a,b,c}$ are all integers, furthermore from (\ref{eq:Fmk}) we know that the action of the Laplacian on $\FFp{s}{m}{w-m}$ produces linear combinations of $\FFp{s}{m}{w-m}$ and  $ {\rm E}_m {\rm E}_{w-m}$ with integer coefficients, hence this problem reduces to a linear system for the coefficients $\alpha_{m}^{(s)}$, $\gamma$, $\lambda$ with integer coefficients. If a solution exists it must be rational.} coefficients $\gamma,\lambda,\alpha_{m}^{(s)}$, with $s$ running in steps of $2$, and the $\zeta_w$ only occur for odd values of $w=a{+}b{+}c$. The Laplace system of the $C_{a,b,c}$ given in~\eqref{lapsec.3} and~\eqref{lapsec.3a} shows that the only possible source term linear in Eisenstein series is $\EE_w$ and therefore this term can arise in~\eqref{eq:decomp}, with a rational coefficient due to uniform transcendentality. 
An argument similar to the one below~\eqref{eq:Fmk} can be used to exclude any other term linear in Eisenstein series.
From the spectrum~\eqref{eq:dimC}, $\EE_s$ could also arise from a homogeneous solution; uniform transcendentality and depth two would require its coefficient to be a rational multiple of $\zeta_{w-s}$. Since $w$ and $s$ have the same parity by~\eqref{eq:dimC}, no such single-valued zeta exists for $s>1$. For $s=1$ (which is possible only for odd $w$) we can use the Eisenstein functional equation for $s\to 1{-}s$ to obtain the term  $\zeta_w \EE_0= - \zeta_w$ indeed present in the decomposition~\eqref{eq:decomp}.

At fixed weight $w=a{+}b{+}c$, we can substitute this ansatz in the Laplace system for the $C_{a,b,c}$ in~\eqref{lapsec.3} and, using the defining equation (\ref{eq:Fmk}), solve for the unknown rational coefficients $\alpha_{m}^{(s)}$, $\gamma$. By comparing the Laurent polynomials of $C_{a,b,c}$ known from \cite{DHoker:2017zhq} with those of the $\FFp{s}{m}{k}$ to be determined in section \ref{sec:4.1}, it will also be possible to determine the coefficient $\lambda$ of $\zeta_w$.
 An inverse change of basis is in general not possible. 
 
This procedure reproduces the examples in sections \ref{sec:3.2.3} and \ref{sec:3.2.4}, e.g.
\begin{align}
C_{1,1,1} &= {\rm E}_3 + \zeta_3\,, & C_{3,1,1} &= - 4 \FFp{3}{2}{3} + \frac{43}{35} {\rm E}_5 - \frac{ \zeta_5}{60} \,, \notag \\
C_{2,1,1} &= - \FFp{2}{2}{2} + \frac{9}{10} {\rm E}_4\,, & C_{2,2,1} &= \frac{2}{5} {\rm E}_5 + \frac{ \zeta_5}{30}\,,
\label{cabc.1}
\end{align}
as well as
\begin{align}
C_{2,2,2} &=  - \frac{12 }{5} \FFp{2}{3}{3} + \frac{ 72 }{5}\FFp{4}{3}{3} - \frac{ 9}{7} {\rm E}_6\,,
\notag \\
C_{3,2,1} &=   - \frac{ 2 }{5}\FFp{2}{3}{3} - \frac{ 18}{5} \FFp{4}{3}{3} + \frac{ 11}{14} {\rm E}_6\,,
\label{cabc.2} \\
C_{4,1,1} &=  \frac{ 2}{5} \FFp{2}{3}{3}  - \frac{ 2}{5}\FFp{4}{3}{3} - 6 \FFp{4}{2}{4} + \frac{ 167 }{126}{\rm E}_6  \, .
\notag
\end{align}
Moreover, we obtain the following new decompositions at weight seven
\begin{align}
C_{3,2,2} &= - \frac{ 24}{7} \FFp{3}{ 3}{ 4} + \frac{ 108}{7} \FFp{5}{ 3}{ 4}  - \frac{  23 }{21}{\rm E}_7 + \frac{ \zeta_7}{630}\,,
\notag \\
C_{3,3,1} &=  \frac{  24}{7} \FFp{3}{ 3}{ 4} - \frac{ 108}{7} \FFp{5}{ 3}{ 4} + \frac{ 32 }{21}{\rm E}_7+ \frac{ \zeta_7}{420}\,,
\notag \\
C_{4,2,1} &=  - \frac{  24}{7} \FFp{3}{ 3}{ 4} - \frac{ 18}{7} \FFp{5}{ 3}{ 4} + \frac{ 16}{21} {\rm E}_7- \frac{ \zeta_7 }{630}\,,
\label{cabc.3} \\
C_{5,1,1} &=   \frac{ 12}{7} \FFp{3}{ 3}{ 4} - \frac{ 12}{7} \FFp{5}{ 3}{ 4} - 8 \FFp{5}{ 2}{ 5}  + \frac{ 661 }{462} {\rm E}_7+ \frac{ \zeta_7 }{2520}\, ,
\notag 
\end{align}
where the coefficients of $\zeta_7$ are based on the results of \cite{DHoker:2017zhq} (see section \ref{sec:4.1}) and
\beq
C_{3,3,1}+C_{3,2,2} = \frac{3}{7} {\rm E}_7 + \frac{ \zeta_7}{252} \, .
\label{z7rel}
\eeq
Weight eight is then the first instance where the five linearly independent $C_{a,b,c}$ modulo ${\rm E}_8$ do not suffice to span the space of six $\FFp{s}{m}{k}$,
\beq
\left. \begin{array}{c}
C_{3,3,2}, \, C_{4,2,2} , \, C_{4,3,1} \\
C_{5,2,1} , \, C_{6,1,1}
\end{array}\right\} \ \ \longrightarrow \ \ 
\left\{ \begin{array}{c}
\FFp{6}{4}{4}, \, \FFp{6}{3}{5}, \, \FFp{6}{2}{6}\, \phantom{.}  \\
\FFp{4}{4}{4}, \, \FFp{4}{3}{5}, \, \FFp{2}{4}{4}\, .
\end{array}\right. 
\label{cabc.4}
\eeq
From~\cite{DHoker:2015gmr} it is known that the Laplacian on the five $C_{a,b,c}$ of weight eight can be diagonalised with eigenvalues $\{2,4,4,6,6\}$, so that only a co-dimension one subspace of combinations of $ \FFp{6}{ 2}{ 6}$, $\FFp{6}{ 3}{ 5}$  and $\FFp{6}{ 4}{ 4}$ can be expressed through the $C_{a,b,c}$ at leading depth. 
The representations of $C_{a,b,c}$ at weight eight in terms of $\FFp{s}{m}{8-m}$ and ${\rm E}_8$
analogous to (\ref{cabc.1}) to (\ref{cabc.3}) can be found in appendix \ref{app:cabc.1}.

These first instances of discrepancy in the counting occur for eigenvalue $s=6$ consistent with~\eqref{cabc.0}. Our results in Part II imply that all these discrepancies in counting are due to relations in Tsunogai's derivation algebra.  We know that the generating series of MGFs introduced in~\cite{Gerken:2020yii} relates MGFs to expressions in terms of the $\bsv$; furthermore this generating series contains a conjectural matrix realisation of Tsunogai's derivation algebra~\cite{Tsunogai,Pollack} and therefore there are combinations of the $\bsv$ that do not arise in the generating series due to relations in the algebra. This means that there are modular invariant completions of the $\bsv$ for which no lattice-sum representation is currently available and none is expected. 

As explained in detail in Part~II, the modular invariant $\FFpm{s}{m}{k}$ beyond MGFs cannot be represented solely in terms of $\mathbb Q[y^{\pm 1},{\rm MZV}]$ combinations of $\bsv$. The $\bsv$ are iterated Eisenstein integrals and it turns out that the representation of all modular invariant $\FFpm{s}{m}{k}$ also requires the inclusion of iterated integrals of holomorphic cusp forms. From~\eqref{cabc.0} we see that discrepancies arise whenever $s$ is half the weight of a holomorphic cusp form. In particular, we can prove that the difference
\beq
\dim \mathcal{V}_{{\rm F}^+}(w,s)- \dim \mathcal{V}_C(w,s) = \dim \mathcal{S}_{2s} = \left\lbrace \begin{array}{ll}
\left\lfloor \frac{2s}{12}\right\rfloor -1 &: \ 2s = 2 \!\!\!\mod 12\,,\\[2mm]
 \left\lfloor \frac{2s}{12}\right\rfloor \phantom{-1}&: \ \mbox{otherwise}
\end{array}\right.
\label{dimsmatch}
\eeq
matches the dimension of the space $\mathcal{S}_{2s}$ of holomorphic cusp forms with even modular weights $(2s,0)$~\cite{ApostolTomM1976MfaD}. The second part of this paper~\cite{PartII} is dedicated to clarifying these points by explicitly constructing the necessary additions of iterated integrals of cusp forms.

%%%%%%%%%%%%%%%%%%%%%%%%%%%%%%%%%%%%%%%%%%%%%%%%%%%%%%%%%%%
\section{Reinstating lower depth}
\label{sec:4}
%%%%%%%%%%%%%%%%%%%%%%%%%%%%%%%%%%%%%%%%%%%%%%%%%%%%%%%%%%%

In the previous section we found a map between the $\FFp{s}{m}{k}$ and $\betalagp{j}$ that is based on diagonalising the Laplace equation~\eqref{eq:LapLag} at leading depth. While the $\FFp{s}{m}{k}$ constructed from Poincar\'e series are modular invariant by construction, this is not the case for the pure depth-two $\betalagp{j}$. We shall now describe a procedure to add lower-depth $\bsv$ to the leading-depth expression of the $\FFp{s}{m}{k}$ in terms of the depth-two $\betalagp{j}$ that is based on Cauchy--Riemann equations. In many cases this leads to modular-invariant expressions for the $\FFp{s}{m}{k}$ through $\bsv$. The analogous map between $\FFm{s}{m}{k}$ and $\betalagm{j}$
along with the lower-depth terms in the odd case can be found in section \ref{oddsection}.

As already indicated at the end of the previous section, cases where a completion only in terms of $\bsv$ is not possible are tied to iterated integrals of holomorphic cusp forms and will be treated in detail in Part~II. Once this is achieved, we have the full Fourier expansions of the $\FFpm{s}{m}{k}$ at our disposal, see section~\ref{sec:q-series} for comments on this. An alternative route would be to explore the Fourier expansion from the Poincar\'e series using resurgence~\cite{Dorigoni:2019yoq,Dorigoni:2020oon} but we shall not follow this approach here.

%%%%%%%%%%%%%%%%%%%%%%%%%%%%%%%%%%%%%%%%%%%%%%%%%%%%%%%%%%%
\subsection{Cauchy--Riemann equations}
\label{sec:4.1CR}
%%%%%%%%%%%%%%%%%%%%%%%%%%%%%%%%%%%%%%%%%%%%%%%%%%%%%%%%%%%

The Laplace equation~\eqref{eq:Fmk} was studied in detail in the previous section and now we will look at Cauchy--Riemann equations that are compatible with it. On any function $F$ we have for any $p\geq 0$ that
\begin{align}
\label{eq:CRcom}
(\pi \overline{\nabla}) \left[ y^{-2p} (\pi \nabla)^p F \right] =  y^{-2(p-1)} (\pi \nabla)^{p-1} \left[\left(\Delta - p(p{-}1)\right) F\right]\,,
\end{align}
with $\nabla$ defined in (\ref{rev.11}). We can use this for $p=s$ together with the Laplace equation~\eqref{eq:Fmk} to determine a Cauchy--Riemann equation (of order $s$) compatible with the Laplace equation for $\FFp{s}{m}{k}$. The value $p=s$ is the lowest value where genuine depth-two terms (that are not products of depth-one terms) disappear and this case is the generalisation of the condition~\eqref{eq:Ediff} for depth one. 
We take the ansatz\footnote{Contributions $i=0$ and $i=s$ to the sum in (\ref{eq:CRFp}) are absent
since $\overline \nabla$ acting on this equation would be inconsistent with the Laplace equation of $\FFp{s}{m}{k} $.}
\begin{align}
\label{eq:CRFp}
(\pi \nabla)^s \FFp{s}{m}{k} = \sum_{i=1}^{s-1} c^+_i (\pi\nabla)^i \EE_m (\pi \nabla)^{s-i} \EE_k + (\pi \nabla)^s H^+
\end{align}
in terms of products of Cauchy--Riemann derivatives of the constituent Eisenstein series and possible homogeneous solutions $H^+$ of the Laplace equation. A consistent modular transformation of the equation requires $(\pi\nabla)^s H^+$ to have modular weight $(0,-2s)$, however, this does not require $H^+$ itself to be modular invariant. If one were to require modular invariance of $H^+$, the only option would be $H^+\propto \EE_s$ but this would be too restrictive, in particular there would be no odd analogue.

Acting with $\pi\overline{\nabla}$ on the ansatz and using~\eqref{eq:Fmk} and \eqref{eq:CRcom} leads to the relations
\begin{align}
c_1^+ = \frac1{m(m{-}1)}\,,\hspace{20mm}
c_{s-1}^+ &= \frac1{k(k{-}1)}\,,\\
\big[m(m{-}1)-i(i{+}1)\big] c^+_{i+1} + \big[k(k{-}1)-(s{-}i)(s{-}i{-}1)\big] c^+_i &= \binom{s{-}1}{i} \quad\text{for $1\leq i\leq s{-}2$}\nn
\end{align}
for the coefficients $c^+_i$ that can be solved for recursively.
The complex conjugate of the Cauchy--Riemann equation~\eqref{eq:CRFp} 
can be obtained straightforwardly.

As examples of such Cauchy--Riemann equations we have
\begin{align}
(\pi\nabla)^2 \FFp{2}{2}{2} &= \frac12 (\pi \nabla \EE_2)^2\,,\\
(\pi\nabla)^3 \FFp{3}{2}{3} &= \frac12 (\pi\nabla)\EE_2 (\pi\nabla)^2 \EE_3 + (\Im\tau)^4 \GG_4 (\pi\nabla) \EE_3\,, \notag
\end{align}
which are equivalent to~\cite[Eqs.~(4.29), (4.35)]{Broedel:2018izr}. In the second one we have used~\eqref{eq:Ediff} to write $(\pi \nabla)^2 {\rm E}_2$ in terms of the holomorphic Eisenstein series $\GG_4$. In general, the term $(\pi\nabla)^i \EE_m$ in~\eqref{eq:CRFp} can also contain derivatives of $(\Im\tau)^{2m}\GG_{2m}$ since $s{-}1$ may be larger than $m$. 

In the two examples above one can check from the $q$-expansion or modular invariance that there are no homogeneous solutions $H^+$ present. Cases when $H^+\neq 0$ do occur and will be explored in detail in Part~II.

Note that the $\beta^{\rm sv}$ representations of a variety of Cauchy--Riemann 
derivatives $(\pi \nabla)^p \FFpm{s}{m}{k}$ and $(\pi \overline \nabla)^p \FFpm{s}{m}{k}$ are discussed in section~\ref{sec:odd.bsvrep}, examples are gathered in appendix \ref{app:CRderiv} and all derivatives for $m+k\leq 14$ are collected in the ancillary file.

%%%%%%%%%%%%%%%%%%%%%%%%%%%%%%%%%%%%%%%%%%%%%%%%%%%%%%%%%%%
\subsection{Reinstating depth-one terms}
\label{sec:4.2}
%%%%%%%%%%%%%%%%%%%%%%%%%%%%%%%%%%%%%%%%%%%%%%%%%%%%%%%%%%%

The Cauchy--Riemann equation~\eqref{eq:CRFp} can be used to obtain (candidate) expressions for $\FFp{s}{m}{k}$ in terms of the $\bsv$ of various depths. The depth-two part was fixed by the diagonalisation procedure in section~\ref{sec:3.2.2}. However, acting with the Cauchy--Riemann derivative on the depth-two terms does not generate only terms of the correct type.

The differential equation (\ref{rev.20}) that we reproduce here in rewritten form for convenience
\begin{align}
-4\pi \nabla
\bsvBR{j_1 &j_2}{k_1 &k_2}{\tau} &= (k_1{-}j_1{-}2) \bsvBR{j_1+1 &j_2}{k_1 &k_2}{\tau} 
+ (k_2{-}j_2{-}2) \bsvBR{j_1 &j_2+1}{k_1 &k_2}{\tau}  \notag \\
&\ \ \ \
-  \delta_{j_2,k_2-2 }(\tau{-}\bar \tau)^{k_2} {\rm G}_{k_2}(\tau) \bsvBR{j_1 }{k_1 }{\tau} 
\label{oct24.1}
\end{align}
shows that we obtain the holomorphic Eisenstein series $\GG_{k_2}$ as instances of $(\pi \nabla)^{k_2} \EE_{k_2}$ according to~\eqref{eq:Ediff} but they are multiplied by $\bsv$ of depth one that are not of the  $(\pi\nabla)^{\bullet} \EE_{\bullet}$ type as is required by~\eqref{eq:CRFp}. (A related statement is that the $\bsv$ are not lattice sums.)

The connection between the depth-one $\betasv{j\\k}$ and $(\pi\nabla)^\bullet \EE_\bullet$ was given in~\eqref{oct24.2} and involves additional depth-zero terms of the form $\zeta_{2s-1} y^{j+2-k}$. Looking back at this equation, we can therefore improve the Cauchy--Riemann derivative of the depth-two $\bsv$ by considering
\beq
\betasv{j_1 &j_2\\k_1 &k_2}  \rightarrow \hatbetasv{j_1 &j_2\\k_1 &k_2}  =\betasv{j_1 &j_2\\k_1 &k_2} 
 - \frac{ 2 \zeta_{k_1-1} }{(k_1{-}1) (4y)^{k_1-j_1-2}} \betasv{j_2\\ k_2}  \, .
\label{bsvd1.5}
\eeq
These combinations obey the differential equation
\begin{align}
-4\pi \nabla
\hatbsvBR{j_1 &j_2}{k_1 &k_2}{\tau} &= (k_1{-}j_1{-}2) \hatbsvBR{j_1+1 &j_2}{k_1 &k_2}{\tau} 
+ (k_2{-}j_2{-}2) \hatbsvBR{j_1 &j_2+1}{k_1 &k_2}{\tau}  \label{oct24.4} \\
&\ \ \ \
-  \delta_{j_2,k_2-2 }(\tau{-}\bar \tau)^{k_2} {\rm G}_{k_2}(\tau)  \Big( \bsvBR{j_1 }{k_1 }{\tau} 
-   \frac{ 2 \zeta_{k_1-1} }{(k_1{-}1) (4y)^{k_1-j_1-2}} \Big) \, , \notag
\end{align}
where the ${\rm G}_{k_2}(\tau)$ on the right-hand side is accompanied by one of the modular graph
forms (\ref{oct24.2}) of depth one. For the depth-two combinations $\betalagpm{j} $ in (\ref{eq:betalag}) 
relevant to the modular invariants $\FFpm{s}{m}{k}$, the analogue of (\ref{bsvd1.5}) is
\beq
\betalagpm{j}  \rightarrow \hatbetalagpm{j}  = \betalagpm{j} -2 \bigg(
\frac{ \zeta_{2m-1} }{(2m{-}1) (4y)^{j}} \bsvBR{k+j-m }{2k}{\tau}
\pm  \frac{ \zeta_{2k-1} }{(2k{-}1) (4y)^{k+j-m}} \bsvBR{j }{2m}{\tau}
\bigg) \, .
\label{oct24.5}
\eeq
Performing this substitution for the depth-two $\bsv$ that arise in the diagonalisation~\eqref{eq:tildeb} of the Laplacian therefore leads to an object whose Cauchy--Riemann derivatives are completely expressed in terms of depth-two $\hatbetasv{j_1&j_2\\k_1&k_2}$ and products of $(\pi \nabla)^\bullet \EE_\bullet$. 

The products on the right-hand sides of the repeated Cauchy--Riemann derivatives (\ref{eq:CRFp})
lead to shuffle combinations of depth-two $\bsv$ in the schematic form
\begin{align}
\hatbetasv{j_1 & j_2 \\ 2k & 2m} +\hatbetasv{j_2 & j_1 \\ 2m & 2k}  = \# (\pi\nabla)^\bullet \EE_m (\pi \nabla)^\bullet \EE_k + \frac{4 \zeta_{2m-1} \zeta_{2k-1}}{(2m{-}1)(2k{-}1) (4y)^{2m+2k-4-j_1-j_2}}\, ,
\end{align}
where the substitution rule (\ref{bsvd1.5}) captures the depth-one terms. However, the depth-zero terms $\sim \zeta_{2k-1} \zeta_{2m-1}$ (i.e.\ pure powers of $y$) need to be modified to obtain the correct 
Laurent polynomial of $\FFp{s}{m}{k}$ that we shall describe next. 

%%%%%%%%%%%%%%%%%%%%%%%%%%%%%%%%%%%%%%%%%%%%%%%%%%%%%%%%%%%
\subsection{Reinstating Laurent polynomials}
\label{sec:4.1}
%%%%%%%%%%%%%%%%%%%%%%%%%%%%%%%%%%%%%%%%%%%%%%%%%%%%%%%%%%%

From the modular properties of Cauchy--Riemann derivatives, we have derived the substitution 
rule (\ref{oct24.5}) that allows us to reconstruct the depth-one 
additions to the defining depth-two $\bsv$ terms in $\FFp{s}{m}{k}$. In this way, we have obtained a combination of depth-two and depth-one $\bsv$ that has the correct Cauchy--Riemann and Laplace equations up to terms of depth zero given by pure powers of $y$. The missing pure $y$-power terms can be conveniently inferred from
the Poincar\'e seeds $\seedp{s}{m}{k}$ we have constructed as a solution to the 
Laplace equation with our choice of boundary conditions: The Laurent polynomials of $\FFp{s}{m}{k}$ can be computed from the Poincar\'e seeds in (\ref{eq:genseed}),
using the conversion rules (\ref{conrul.1}) and (\ref{conrul.2}). 

In order to determine the additional $y$-powers that need to be added to the depth-two and depth-one $\bsv$ in order to match those of $\FFp{s}{m}{k}$, the only information we need are the Laurent-polynomial contributions of the $\bsv$ and the Laurent polynomial of $\FFp{s}{m}{k}$. The former are given by \cite{Gerken:2020yii}
\begin{align}
\bsvBR{j_1 }{k_1 }{\tau}  &=  \frac{ {\rm B}_{k_1}  j_1! (k_1{-}2{-}j_1)! (-4y)^{j_1+1} }{k_1! \, (k_1{-}1)!} 
+ O(q,\bar q)\,,
\notag \\
\bsvBR{j_1 &j_2}{k_1 &k_2}{\tau}   &= \frac{ {\rm B}_{k_1} {\rm B}_{k_2}  (j_1{+}j_2{+}1)!   (k_2{-}2{-}j_2)!  (-4y)^{j_1+j_2+2}  }{ (j_1{+}1) k_1! k_2!  (k_2{+}j_1)!} \label{more.bsvb} \\
&\ \ \ \ \times
 \, _3F_2\Big[
\begin{smallmatrix}
1 {+} j_1,\ 2 {+} j_1 {+} j_2, \  2 {+} j_1 {-} k_1 \\
2 {+} j_1,\ 1 {+} j_1 {+} k_2 
\end{smallmatrix} ; 1
\Big]  + O(q,\bar q) \notag
\, ,
\end{align}
where $_3F_2$ denotes a generalised hypergeometric function which, in the present case, always evaluates to a finite sum, yielding a rational number.

The Laurent polynomial of $\FFp{s}{m}{k}$ is determined from the explicit solution~\eqref{eq:genseed} by using (\ref{conrul.1}) and (\ref{conrul.2}) together with  the general formula~\eqref{eq:Iabr} for the Laurent polynomial of Poincar\'e seeds. As a result we obtain the Laurent polynomial for $2\leq m\leq k$,
\begin{align}
\FFp{s}{m}{k} &=
\frac{(-4)^{k{+}m} {\rm B}_{2m} {\rm B}_{2k} }{(k{+}m{-}s)(k{+}m{+}s{-}1) (2m)! (2k)!} y^{k+m}
-   \frac{2(-1)^{m} 4^{1{+}m{-}k}{\rm B}_{2m}  \Gamma(2k{-}1) \zeta_{2k{-}1}}{\Gamma(k) \Gamma(k) (m{-}k{+}s)(m{-}k{-}s{+}1)(2m)!} y^{1+m-k}
\nn\\
&\quad 
-   \frac{2(-1)^{k} 4^{1{+}k{-}m}{\rm B}_{2k}  \Gamma(2m{-}1) \zeta_{2m{-}1}}{\Gamma(m) \Gamma(m) (k{-}m{+}s)(k{-}m{-}s{+}1)(2k)!} y^{1+k-m}
\label{eq:FLP}\\
&\quad +  \frac{4^{3-m-k}\Gamma(2m{-}1)\Gamma(2k{-}1)\zeta_{2m{-}1}\zeta_{2k{-}1}}{[\Gamma(m)\Gamma(k)]^2 (k{+}m{-}s{-}1)(k{+}m{+}s{-}2)} y^{2-k-m}
+ c_{m,k}^{(s)} \zeta_{k+m+s-1} y^{1-s} + O(q,\bar{q})
\notag
\end{align}
with the rational coefficient
\begin{align}
 c_{m,k}^{(s)} = \frac{4^{2-s} (-1)^{m+s+1} {\rm B}_{s+m-k} {\rm B}_{k+m-s} {\rm B}_{k+s-m}(2s)! }{(s{+}m{-}k)\Gamma(m)\Gamma(s) {\rm B}_{2s} (k{+}m{-}s)! (k{+}s{-}m)!} \sum_{\ell=k-m+1}^{\min(k-1,s)} (-1)^\ell g^+_{m,k,\ell,s} \frac{\Gamma(\ell{+}s{-}1)}{\Gamma(\ell) (s{-}\ell)!}
\end{align}
in terms of the rational numbers $g^+_{m,k,\ell,s}$ defined in~\eqref{eq:fkm}.
For $m=k$ the terms $y^{1+m-k}$ and $y^{1+k-m}$ are both linear in $y$ and have the same coefficient that just doubles. All terms including the zeta values they contain, except for the one proportional to $y^{1-s}$, can be directly traced back to the Laurent polynomials of the source $\EE_m\EE_k$. We summarise the structure of the Laurent polynomial schematically as (recalling that $m <k$)
\begin{align}
\FFp{s}{m}{k}  \ \text{mod} \ O(q,\bar q)\ &\longleftrightarrow \  y^{m+k} \, , \  \zeta_{2m-1} y^{k-m+1} \, , \ \zeta_{2k-1} y^{m-k+1}
\, , \  \frac{ \zeta_{2m-1} \zeta_{2k-1} }{ y^{k+m-2} }  \, , \  \frac{ \zeta_{m+k+s-1} }{y^{s-1}} \, ,  \notag \\
 \FFp{s}{k}{k} \ \text{mod} \ O(q,\bar q) \  &\longleftrightarrow \ y^{2k} \, , \  \zeta_{2k-1} y 
\, , \  \frac{\zeta_{2k-1}^2 }{ y^{2k-2} }  \, , \  \frac{ \zeta_{2k+s-1} }{y^{s-1}} \, , \label{laurent}
\end{align}
where all terms have rational coefficients.  This means that all terms in the Laurent polynomial~\eqref{eq:FLP} have the same transcendental weight $m{+}k$, where both $\zeta_n$ and $y^n$ are assigned transcendental weight $n$.

We also note that the term $c_{m,k}^{(s)} \zeta_{m+k+s-1} $ along with $y^{1-s}$ in~\eqref{eq:FLP} 
is an instance of a multiple modular value associated
with the $\beta^{\rm sv}$ at depth two \cite{Brown:mmv, Brown:I, Brown:II, Saad:2020mzv} that we discuss in more detail in the companion Part~II. We shall always display these as the last terms in the examples below. Obtaining these correctly from the Poincar\'e-series approach is one of the central results of this paper.

The final step in constructing a combination of $\bsv$ that solves the Laplace equation~\eqref{eq:Fmk} is then adding terms to the Laurent polynomial obtained from depth two and depth one via~\eqref{more.bsvb} such that the correct Laurent polynomial~\eqref{eq:FLP} is obtained.
We summarise the steps by
\begin{enumerate}
\item 
For a given choice of $m\leq k$, and $s$ in the spectrum~\eqref{eq:Spectrum}, take the combination of $\betalagp{j}$ obtained from the diagonalisation~\eqref{eq:tildeb}. This solves the Laplace equation at depth two.
\item
Replace the $\betalagp{j}$ by $\hatbetalagp{j}$ as defined in~\eqref{oct24.5}. This solves the Laplace equation at depths two and one.
\item
Modify the pure $y$-power terms such that the correct Laurent polynomial~\eqref{eq:FLP} is obtained. After this step an exact solution to the Laplace equation is obtained.
\end{enumerate}
We denote the resulting combination of $\bsv$ of depths two, one and zero by $\cFFp{s}{m}{k}$.

Note that the second and third terms $\sim  \zeta_{2m-1} y^{k-m+1} , \ \zeta_{2k-1} y^{m-k+1}$ in 
(\ref{laurent}) do not solve the homogeneous Laplace equation $(\Delta - s(s{-}1)) F=0$ for the
values of $s$ in the spectrum (\ref{eq:Fmk}) of even $\FFp{s}{m}{k}$. Hence, their coefficients
are always determined by the source term in the Laplace equations (\ref{eq:Fmk}). The situation changes in the odd case, and we will see that Laurent monomials $\zeta_{2m-1} y^{k-m+1}$ signal additional solutions of
the homogeneous Laplace equation, i.e.\ the eigenvalue coincides with $k{-}m{+}1$, in the $\beta^{\rm sv}$ representations of $\FFm{s}{m}{k}$ to be constructed in section \ref{sec:odd.dpt1}.

%%%%%%%
\subsection{\texorpdfstring{Combinations $\cFFp{s}{m}{k}$ of $\beta^{\rm sv}$ versus modular invariants $\FFp{s}{m}{k}$}{Combinations checkFplus(s,m,k) of betasv versus modular invariants Fplus(s,m,k)}}
\label{sec:FcF}
%%%%%%%

The notation  $\cFFp{s}{m}{k}$ introduced above indicates that the combination of $\bsv$ does not need to be identical to $\FFp{s}{m}{k}$ but could be a `downgraded' version. Both $\FFp{s}{m}{k}$ and $\cFFp{s}{m}{k}$ are solutions to the same Laplace equation. However, while $\FFp{s}{m}{k}$ is modular invariant by construction as a Poincar\'e series, this is not necessarily 
true for $\cFFp{s}{m}{k}$ as it is built out of $\bsv$ that can have complicated modular S-transformation properties~\cite{Gerken:2020yii}. By the lower-depth terms in the modular transformations
(\ref{eq:bsv2prop}), it is not guaranteed that the combinations $\cFFp{s}{m}{k}$ obtained in this way are exactly modular. 

Another way of understanding this is to reconsider the Cauchy--Riemann equation~\eqref{eq:CRFp} that is satisfied by $\FFp{s}{m}{k}$ and that is compatible with the Laplace equation. From the way that $\cFFp{s}{m}{k}$ was constructed, it satisfies
\begin{align}
(\pi \nabla)^s \cFFp{s}{m}{k} = \sum_{i=1}^{s-1} c^+_i (\pi\nabla)^i \EE_m (\pi \nabla)^{s-i} \EE_k \,,
\end{align}
which has vanishing homogeneous term $H^+$. Even though the right-hand side of this equation has good modular properties and transforms with weight $(0,-2s)$ under ${\rm SL}_2(\ZZ)$, there is no guarantee that there is a modular invariant primitive to this equation. In fact, the Eichler--Shimura~theorem \cite{Eichler:1957,Shimura:1959} shows that in general the homogeneous term is needed, see for instance the discussion in the work of Brown~\cite{Brown:mmv, Brown:I}.

After multiplication by $y^{-2s}$, the homogeneous term $(\pi\nabla)^s H^+$ in (\ref{eq:CRFp}) must be a modular form of weight $(2s,0)$ and it must be holomorphic in order to be annihilated by $\pi\overline{\nabla}$, see~\eqref{eq:CRcom}. Thus we arrive at the strong requirement that 
\begin{align}
\label{eq:CRHp}
(\pi\nabla)^s H^+ = y^{2s} f
\end{align}
with $f$ a holomorphic modular form of weight $(2s,0)$. The space of holomorphic modular forms is very well studied and decomposes into holomorphic Eisenstein series $\GG_{2s}$ and cusp forms that arise for $2s\in\{12,16,18,\ldots\}$, see e.g.~\cite{ApostolTomM1976MfaD}. 

However, since $\cFFp{s}{m}{k}$ is already engineered to match the Laurent polynomial (\ref{eq:FLP})
of $\FFp{s}{m}{k}$, both $H^+$ and therefore $f$ has to vanish at the cusp. Hence, there is no room for 
the zero mode of $\GG_{2s} = 2 \zeta_{2s} + O(q)$, and $f$ cannot be a
holomorphic Eisenstein series.\footnote{Choosing $f$ to be a holomorphic Eisenstein series
would make the relation~\eqref{eq:CRHp} identical to~\eqref{eq:Ediff} and $H^+$ would become an
iterated integral of $\GG_{2s}$. Since we do not require $H^+$ to be modular invariant, it could differ from $\EE_s$. However, $H^+$ would necessarily contain a term of the form $y^s$ from the differential equation which is incompatible with the fact that the Laurent polynomial of $\FFp{s}{m}{k}$ is already accounted for by $\cFFp{s}{m}{k}$.} Therefore, the case $f\propto \GG_{2s}$ cannot arise for even functions $\FFp{s}{m}{k}$. 

We therefore conclude that the only cases when $\FFp{s}{m}{k}\neq \cFFp{s}{m}{k}$ is possible are associated with holomorphic cusp forms of weight $(2s,0)$, and in those cases the function $\cFFp{s}{m}{k}$ will not necessarily be modular invariant. This can also be understood as follows. The generating series of MGFs introduced in~\cite{Gerken:2020yii} implements conjectural matrix representations of Tsunogai's derivations~\cite{Tsunogai} and therefore does not contain certain depth-two combinations of $\beta^{\rm sv}$. The relations in the derivation algebra are known to be associated with holomorphic cusp forms as well~\cite{Pollack}. This discrepancy between MGFs and the $\FFp{s}{m}{k}$ is also hinted at by the different counting of two-loop modular graph functions $C_{a,b,c}$ discussed in section~\ref{sec:3.2.5} at the relevant weights.

The non-modular invariance of some $\cFFp{s}{m}{k}$ can moreover be traced back to the multiple modular values from the
$\beta^{\rm sv}$ of depth two that can go beyond MZVs and involve L-values of holomorphic 
cusp forms \cite{Brown2019}. 

In this paper we shall restrict to the modular invariant cases 
where $\cFFp{s}{m}{k}=\FFp{s}{m}{k}$ can be identified with MGFs. Cases with $\cFFp{s}{m}{k}\neq \FFp{s}{m}{k}$ 
in turn will be the subject of Part~II where iterated integrals of holomorphic cusp forms will play a key role.

\subsection{Examples}

We now give a few exemplary instances of the construction of the lower-depth terms and how the $\FFp{s}{m}{k}$ can be expressed in terms of the $\bsv$.

\subsubsection{\texorpdfstring{Examples at $m=k$}{Examples at m=k}}
\label{sec:4.2.1}

By applying the procedure outlined around (\ref{oct24.5}) and (\ref{more.bsvb}) to the simplest examples 
of $\FFp{s}{k}{k}$, we find the following completion of (\ref{oct20.3}),
\begin{align}
 \FFp{2}{2}{2} &=  18     \betasv{ 2 & 0\\4& 4} 
 -12  \zeta_{3}  \betasv{0\\4}+ \frac{ \zeta_{3}^2}{4 y^2} - \frac{ 5 \zeta_{5}}{12 y}\,,
\notag \\
 \FFp{2}{3}{3} &=   100 ( 2 \betasv{3& 1\\6& 6} - \betasv{4& 0\\6& 6}) +40 \zeta_{5}   \betasv{0\\6} - \frac{ 20 \zeta_{5}}{y}  \betasv{1\\6}  + \frac{\zeta_{5}^2}{32 y^4} - \frac{ 5 \zeta_{7}}{288 y}\,,
\label{oct24.11} \\
 \FFp{4}{3}{3} &=   25  (8 \betasv{3& 1\\6& 6} + \betasv{4& 0\\6& 6}) -10  \zeta_{5} \betasv{0\\6} - \frac{ 20\zeta_{5}}{y}  \betasv{1\\6}  + \frac{ 9 \zeta_{5}^2}{128 y^4 } - \frac{ 35 \zeta_{9}}{1152 y^3 } \, ,
 \notag
\end{align}
in agreement with the results for ${\rm E}_{2,2},{\rm E}_{3,3},{\rm E}_{3,3}'$ in \cite{Gerken:2020yii}.
As an example of our new results for $\FFp{s}{k}{k}$, 
the Poincar\'e sums over the seed functions (\ref{oct20.10}) yield 
\begin{align}
 \FFp{2}{4}{4} &=   490  (5 \betasv{4& 2\\8& 8} - 4 \betasv{5& 1\\8& 8} + 
   \betasv{6& 0\\8& 8}) -140  \zeta_{7} \betasv{0\\8} \notag \\
 &\ \ \ \ + \frac{140  \zeta_{7}}{y} \betasv{1\\8} - \frac{ 
 175  \zeta_{7}}{4 y^2}  \betasv{2\\8} + \frac{ 5 \zeta_{7}^2}{512 y^6} - \frac{ 5 \zeta_{9}}{3888 y}\,,
\label{oct24.12}  \\
 \FFp{4}{4}{4} &=   \frac{  490}{3}  (15 \betasv{4& 2\\8& 8} - 6 \betasv{5& 1\\8& 8} - 
   \betasv{6& 0\\8& 8}) +\frac{140}{3} \zeta_{7} \betasv{0\\8}  \notag \\
 &\ \ \ \ + \frac{  70  \zeta_{7}}{y }  \betasv{1\\8} - \frac{ 175 \zeta_{7}}{
 4 y^2} \betasv{2\\8}  + \frac{ 5 \zeta_{7}^2}{384 y^6 } - \frac{ 7 \zeta_{11}}{6912 y^3}\,,
\notag
\end{align}
where the terms $\sim \zeta_9/y$ and $\zeta_{11}/y^3$ in the kernel
of $( \Delta - s(s{-}1))  \FFp{s}{4}{4}$ are related to multiple modular values at depth two. 

Reinstating the lower-depth terms for the final eigenfunction with $s=6$ leads to
\begin{align}
 \cFFp{6}{4}{4} &=    \frac{98}{3}  (75 \betasv{4& 2\\8& 8} + 24 \betasv{5& 1\\8& 8} + 
   \betasv{6& 0\\8& 8}) -\frac{28}{3}  \zeta_{7} \betasv{0\\8} \notag \\
 &\ \ \ \  - \frac{  56  \zeta_{7}}{y } \betasv{1\\8} - \frac{ 175  \zeta_{7}}{ 4 y^2 } \betasv{2\\8} + \frac{ 25 \zeta_{7}^2}{768 y^6 } - \frac{ 5005 \zeta_{13}}{530688 y^5 } \,,
 \label{cF644}
\end{align}
with $ \zeta_{13}/y^5$ in the kernel of $( \Delta - s(s{-}1))$ at $s=6$.
This function can be checked to be non-invariant under modular transformations
and is one of the simplest examples outside the realm of MGFs where iterated integrals of cusp forms need to be added. The relation of the expression~\eqref{cF644} to Tsunogai's derivation algebra will be explored in Part~II.

The examples in~\eqref{oct24.12} are modular invariant which is why we write 
them as $\FFp{s}{m}{k}$ rather than $\cFFp{s}{m}{k}$. The Laurent polynomials of (\ref{oct24.11}) 
to (\ref{cF644}) resulting from (\ref{more.bsvb}) are gathered in appendix \ref{app:lau.1}.

\subsubsection{\texorpdfstring{Examples at $m<k$}{Examples at m<k}}
\label{sec:4.2.2}

Similarly, the above procedure to reinstate lower-depth terms 
leads to the completion of (\ref{oct20.5})~by
\begin{align}   
 \FFp{3}{2}{ 3}&= 
  30 (\betasv{2& 1\\4& 6} + \betasv{3& 0\\6& 4})
 -20  \zeta_{3} \betasv{1\\6} - \frac{  3  \zeta_{5}}{y}  \betasv{0\\4} + \frac{  \zeta_{5}}{360 } + \frac{  \zeta_{3} \zeta_{5}}{8 y^3} - \frac{  7 \zeta_{7}}{64 y^2}\,, \label{oct24.13} \\
 \FFp{4}{ 2}{ 4}&= 
  105 (\betasv{2& 2\\4& 8} + \betasv{4& 0\\8& 4}) -70  \zeta_{3} \betasv{2\\8} - \frac{ 15  \zeta_{7}}{8 y^2} \betasv{0\\4} + \frac{  \zeta_{7}}{480 y} + \frac{  5 \zeta_{3} \zeta_{7}}{64 y^4} - \frac{ 25 \zeta_{9}}{
 432 y^3} \, ,\notag 
\end{align}
in agreement with the results for ${\rm E}_{2,3},{\rm E}_{2,4}$ in \cite{Gerken:2020yii}.
The simplest cases that go beyond the state of the art include the
Poincar\'e sums over the seeds (\ref{oct20.12})
\begin{align} 
 \FFp{5}{ 2}{ 5}&= 
  378 (\betasv{2& 3\\4& 10} {+} \betasv{5& 0\\10& 4}) -252 \zeta_{3}  \betasv{3\\10} - \frac{  21 \zeta_{9}}{ 16 y^3 }  \betasv{0\\4}  + \frac{  \zeta_{9}}{640 y^2 }+ \frac{  7 \zeta_{3} \zeta_{9}}{128 y^5 } -\frac{  77 \zeta_{11}}{
 2048 y^4}\,,
 \label{oct24.14} 
\end{align}
and those over the seeds (\ref{oct20.13}),
\begin{align}
\FFp{3}{ 3}{ 4} &= 
  175 (2 \betasv{3& 2\\6& 8} - \betasv{4& 1\\6& 8} + 
     2 \betasv{4& 1\\8& 6} - \betasv{5& 0\\8& 6})  +70  \zeta_{5} \betasv{1\\8} - \frac{  35  \zeta_{5}}{y} \betasv{2\\8} \notag \\
 &\ \ \ \  + \frac{  25  \zeta_{7}}{2 y }  \betasv{0\\6} - \frac{   25 \zeta_{7}}{4 y^2 }  \betasv{1\\6}
     + \frac{  5 \zeta_{7}}{18144 } + \frac{  5 \zeta_{5} \zeta_{7}}{256 y^5  } - \frac{  49 \zeta_{9}}{
 11520 y^2 }\,, \notag \\
\FFp{5}{ 3}{ 4} &=  
  70 (5 \betasv{3& 2\\6& 8} + \betasv{4& 1\\6& 8} + 
     5 \betasv{4& 1\\8& 6} + \betasv{5& 0\\8& 6})   -28 \zeta_{5}  \betasv{1\\8} - \frac{  35 \zeta_{5}}{y}  \betasv{2\\8} \notag \\
 &\ \ \ \  - \frac{   5  \zeta_{7}}{y } \betasv{0\\6} - \frac{  25  \zeta_{7}}{ 4 y^2 } \betasv{1\\6} - \frac{ \zeta_{7}}{30240  }+  \frac{ 3 \zeta_{5} \zeta_{7}}{64 y^5 } - \frac{  77 \zeta_{11}}{ 4608 y^4}\,,
\label{oct24.15}\\
\FFp{4}{ 3}{ 5} &= 
  630 (2 \betasv{3& 3\\6& 10} - \betasv{4& 2\\6& 10} + 
     2 \betasv{5& 1\\10& 6} - \betasv{6& 0\\10& 6}) +252  \zeta_{5} \betasv{2\\10} - \frac{  
 126  \zeta_{5}}{y } \betasv{3\\10} \notag \\
 &\ \ \ \  + \frac{  35  \zeta_{9}}{ 4 y^2 } \betasv{0\\6} - \frac{  35  \zeta_{9}}{8 y^3 } \betasv{1\\6} + \frac{  \zeta_{9}}{4320 y } + \frac{  7 \zeta_{5} \zeta_{9}}{512 y^6 } - \frac{  5 \zeta_{11}}{ 2304 y^3 }\,.
\notag 
 \end{align}
Moreover, the above procedure yields two further examples of $\bsv$ combinations that are not modular invariant at eigenvalue $s=6$ and weight $m{+}k=8$,
\begin{align}
\cFFp{6}{ 2}{ 6}&=    1386 (\betasv{2& 4\\4& 12} {+} \betasv{6& 0\\12& 4}) -924  \zeta_{3} \betasv{4\\12} - \frac{  63  \zeta_{11}}{ 64 y^4 } \betasv{0\\4}  \notag \\
&\ \ \ \ + \frac{  7 \zeta_{11}}{5760 y^3} + \frac{  21 \zeta_{3} \zeta_{11}}{512 y^6 } 
- \frac{   9555 \zeta_{13}}{353792 y^5}\,,
\label{cF635}\\
\cFFp{6}{ 3}{ 5} &= 
  315 (4 \betasv{3& 3\\6& 10} + \betasv{4& 2\\6& 10} + 
     4 \betasv{5& 1\\10& 6} + \betasv{6& 0\\10& 6}) -126  \zeta_{5} \betasv{2\\10} - \frac{  
 126  \zeta_{5}}{y } \betasv{3\\10} \notag \\
 &\ \ \ \  - \frac{  35  \zeta_{9}}{ 8 y^2 } \betasv{0\\6} - \frac{  35  \zeta_{9}}{8 y^3 } \betasv{1\\6} - \frac{ \zeta_{9}}{24192 y  }+ \frac{  35 \zeta_{5} \zeta_{9}}{1024 y^6  } - \frac{   63063 \zeta_{13}}{5660672 y^5 }  \, .
\notag 
\end{align}
Still, the counting of $C_{a,b,c}$ in section \ref{sec:3.2.5} with two Laplace eigenfunctions at $s=6$ and
weight $a{+}b{+}c=8$ implies that two linear combinations of $\cFFp{6}{ 4}{ 4},\cFFp{6}{ 3}{ 5},\cFFp{6}{ 2}{ 6}$
must be modular graph functions. Indeed, the linear combinations
\begin{align}
\cFFp{6}{ 2}{ 6} - \frac{6}{35} \cFFp{6}{4}{4} &= \FFp{6}{ 2}{ 6} - \frac{6}{35} \FFp{6}{4}{4}\,,
\label{cF6xy} \\
\cFFp{6}{ 3}{ 5} + \frac{75}{112} \cFFp{6}{4}{4}  &= \FFp{6}{ 3}{ 5} + \frac{75}{112} \FFp{6}{4}{4} \,,
\notag
\end{align}
seen in the expressions (\ref{cabc.99}) for $C_{a,b,c}$ at weight 8 can be recovered from
the generating series \cite{Gerken:2020yii} after taking the relations among Tsunogai's derivations
into account. We have confirmed the modular invariance of (\ref{cF6xy}) both numerically from the
$q$-expansions of the iterated Eisenstein integrals in (\ref{cF644}), (\ref{cF635}) 
and analytically from the multiple modular values in the S-transformation of the $\beta^{\rm sv}$. We are 
indebted to Francis Brown for providing us with extensive data relevant for demonstrating this~\cite{FBpriv}.
 
The complete list of $\beta^{\rm sv}$ representations of 
$\FFp{s}{ m}{ k}$ and $\cFFp{s}{ m}{ k}$ with $m\leq k$ and
$m{+}k\leq 14$ can be found in an ancillary file in the arXiv submission and journal publication 
of this paper. The Laurent polynomials of (\ref{oct24.13}) to (\ref{cF635}) resulting from
(\ref{more.bsvb}) are gathered in appendix \ref{app:lau.2}.

%%%%%%%%%%%%%%%%%%%%%%%%%%%%%%%%%%%%%%%%%%%%%%%%%%%%%%%%%%%
\subsection{\texorpdfstring{The integration constants $\alpha$}{The integration constants alpha}}
\label{sec:4.3}
%%%%%%%%%%%%%%%%%%%%%%%%%%%%%%%%%%%%%%%%%%%%%%%%%%%%%%%%%%%

On the one hand, a major motivation to derive the above $\beta^{\rm sv}$ representations of $\FFpm{s}{ m}{ k}$
is to infer their Fourier expansion from those of the contributing iterated Eisenstein integrals. On the other hand,
the expression~\eqref{eq:bsv2} for the $\beta^{\rm sv}$ at depth two still involves antiholomorphic integration constants 
$\overline{\alpha[\begin{smallmatrix}  j_1& j_2 \\ k_1& k_2 \end{smallmatrix}]}$ that are only known up to $k_1{+}k_2 =12$ in the earlier literature \cite{Gerken:2020xfv}.

However, the Poincar\'e-series representations of $\FFpm{s}{ m}{ k}$ with seeds of the form
$y^r, y^r \Re {\cal E}_0$ (or $y^r \Im {\cal E}_0$ in the odd case studied later) imply that 
the iterated Eisenstein integrals at all depths add up to even or odd combinations, respectively. Since the holomorphic iterated Eisenstein integrals entering the $\beta^{\rm sv}[\begin{smallmatrix}  j_1& j_2 \\ k_1& k_2 \end{smallmatrix}]$ are completely
explicit from (\ref{eq:bsv2}), the $\overline{\alpha[\begin{smallmatrix}  j_1& j_2 \\ k_1& k_2 \end{smallmatrix}]}$ contributing to the $\FFpm{s}{ m}{ k}$ are determined by their reality properties. 
Hence, the $\beta^{\rm sv}$ representations
of $\FFp{s}{ m}{ k}$ such as (\ref{oct24.12}), (\ref{oct24.14}) and (\ref{oct24.15}) yield new examples of 
$\overline{\alpha[\begin{smallmatrix}  j_1& j_2 \\ k_1& k_2 \end{smallmatrix}]}$, and additional integration
constants will be inferred from the $\FFm{s}{ m}{ k}$ in section \ref{sec:odd.bsvrep}. The examples of
$\FFp{s}{ m}{ k}$ we gathered at $m{+}k\leq 14$ point towards the conjectural
closed formula in appendix \ref{appalp.B} for all even integration constants at arbitrary weight. 

\subsubsection{\texorpdfstring{Examples at $m=k$}{Examples at m=k}}
\label{sec:4.3.1}

Reality of $\FFp{2}{ 4}{ 4},\FFp{4}{ 4}{ 4}$ in (\ref{oct24.12})
and $\cFFp{6}{ 4}{ 4}$ in (\ref{cF644}) determines the integration constants
\begin{align}
\alpha[\begin{smallmatrix} 4& 2 \\ 8& 8 \end{smallmatrix}] &=0 \, ,&
\alpha[\begin{smallmatrix} 6& 0 \\ 8& 8 \end{smallmatrix}] &= \frac{2}{7} \zeta_7 \,{\cal E}_0(8)\,,  \notag \\ 
\alpha[\begin{smallmatrix} 4& 3 \\ 8& 8 \end{smallmatrix}] &=0  \, ,&
\alpha[\begin{smallmatrix} 6& 1 \\ 8& 8\end{smallmatrix}] &=  \frac{2}{7} \zeta_7\, {\cal E}_0(8, 0)\,,  \notag \\ 
\alpha[\begin{smallmatrix} 5& 1 \\ 8& 8 \end{smallmatrix}] &=0 \, , &
\alpha[\begin{smallmatrix} 6& 2 \\ 8& 8\end{smallmatrix}] &=  \frac{4}{7} \zeta_7\, {\cal E}_0(8, 0^2)\,,  \label{oct24.16}\\ 
\alpha[\begin{smallmatrix} 5& 2 \\ 8& 8 \end{smallmatrix}] &=0 \, , &
\alpha[\begin{smallmatrix} 6& 3 \\ 8& 8\end{smallmatrix}] &=  \frac{12}{7}  \zeta_7\, {\cal E}_0(8, 0^3)\,,  \notag \\ 
\alpha[\begin{smallmatrix} 5& 3 \\ 8& 8 \end{smallmatrix}] &=0 \, , &
\alpha[\begin{smallmatrix} 6& 4 \\ 8& 8\end{smallmatrix}] &=  \frac{48}{7}  \zeta_7\, {\cal E}_0(8, 0^4)\,,  \notag \\ 
\alpha[\begin{smallmatrix} 5& 4 \\ 8& 8 \end{smallmatrix}] &=0 \, , &
\alpha[\begin{smallmatrix} 6& 5 \\ 8& 8\end{smallmatrix}] &=  \frac{240}{7}  \zeta_7\, {\cal E}_0(8, 0^5) \,, \notag 
\end{align}
with $\alpha[\begin{smallmatrix}  j_1& j_2 \\ 2k& 2k \end{smallmatrix}] =  - \alpha[\begin{smallmatrix}  j_2& j_1 \\ 2k&2k \end{smallmatrix}]$ and no information on cases with $j_1{+}j_2< 6$. Similarly, the non-vanishing
$\alpha[\begin{smallmatrix}  j_1& j_2 \\ 10& 10 \end{smallmatrix}]$ with $8 \leq j_1{+}j_2 \leq 16$ and
$\alpha[\begin{smallmatrix}  j_1& j_2 \\ 12& 12 \end{smallmatrix}]$ with $10 \leq j_1{+}j_2 \leq 20$ identified
from the reality of $\cFFp{s}{ 5}{ 5}$ with $s=2,4,6,8$ and $\cFFp{s}{ 6}{ 6}$ with $s=2,4,6,8,10$ are
\begin{align}
\alpha[\begin{smallmatrix}  8& j \\ 10& 10 \end{smallmatrix}] &= \frac{ 2 }{9} \,j!\, \zeta_9 \,{\cal E}_0(10,0^{j})  \, , \ \ \ \ \ \ 0 \leq j\leq 7\,,
 \label{oct24.17} \\
\alpha[\begin{smallmatrix}  10& j \\ 12& 12 \end{smallmatrix}] &= \frac{ 2 }{11}  \,j!\, \zeta_{11}\, {\cal E}_0(12,0^{j}) 
 \, , \ \ \ \ \ \ 0 \leq j\leq 9 \, . \notag 
 \end{align}
Both the known cases of $\alpha[\begin{smallmatrix}  j_1& j_2 \\ 2k& 2k \end{smallmatrix}]$ in the ancillary
file of \cite{Gerken:2020xfv} and the new results (\ref{oct24.16}), (\ref{oct24.17}) point towards the
closed formula
\begin{align}
\alpha[\begin{smallmatrix}  2k-2& j \\ 2k& 2k \end{smallmatrix}] &= \frac{ 2 \zeta_{2k-1}}{2k{-}1}  j! {\cal E}_0(2k,0^{j})
= - \alpha[\begin{smallmatrix}  j&2k-2 \\ 2k& 2k \end{smallmatrix}]  \, , \ \ \ \ \ \ 0 \leq j\leq 2k{-}3\,,
 \label{oct24.18} \\
\alpha[\begin{smallmatrix}  j_1& j_2 \\ 2k&2k \end{smallmatrix}] &= 0 
 \, , \ \ \ \ \ \ 2k{-}2 \leq j_1{+}j_2\leq 4k{-}4 \ \ \text{with} \ \ j_1,j_2 \leq 2k{-}3  \, . \notag 
 \end{align}
On the one hand, the above strategy does not yield any constraints on the 
$\alpha[\begin{smallmatrix}  j_1& j_2 \\ 2k& 2k \end{smallmatrix}]$ with
$j_1{+}j_2 < 2k{-}2$. On the other hand, we expect the so far undetermined
$\alpha[\begin{smallmatrix}  j_1& j_2 \\ 2k& 2k \end{smallmatrix}]$ with $j_1{+}j_2 < 2k{-}2$ to
vanish based on transcendentality arguments: Given that
$\beta^{\rm sv}[\begin{smallmatrix}  j_1& j_2 \\ 2k& 2k \end{smallmatrix}]$ and $y$ have transcendental
weight $j_1{+}j_2{+}2$ and $1$, respectively, the weight of $\alpha[\begin{smallmatrix}  j_1& j_2 \\ 2k& 2k \end{smallmatrix}]$ is fixed to be $j_1{+}j_2{+}2$ by (\ref{eq:bsv2}). Hence, the transcendental weight
of the undetermined $\alpha[\begin{smallmatrix}  j_1& j_2 \\ 2k& 2k \end{smallmatrix}]$ is
$\leq 2k{-}1$, but reality of the underlying $\FFp{s}{ k}{ k}$ requires a factor of $\zeta_{2k-1}$ 
multiplying an antiholomorphic function that vanishes at the cusp. Since ${\cal E}_0(2k,0^p)$
carry transcendental weight $p{+}1$, there are no such functions of weight $\leq 0$ compatible
with the differential equations of $\FFp{s}{ k}{ k}$ to assemble the 
$\alpha[\begin{smallmatrix}  j_1& j_2 \\ 2k& 2k \end{smallmatrix}]$
with $j_1{+}j_2 < 2k{-}2$, that is why they are expected to vanish.

\subsubsection{\texorpdfstring{Examples at $m<k$}{Examples at m<k}}
\label{sec:4.3.2}

For the integration constants that become accessible from the reality of $\FFp{s}{ m}{ k}$ with $m<k$, it
will be convenient to employ the shorthands
\beq
\alpha^{0,j}_{m,k}
= \alpha[\begin{smallmatrix} 2m-2-j& k-m+j\\ 2m& 2k \end{smallmatrix}] 
+\alpha[\begin{smallmatrix}k+m-2-j &j\\ 2k& 2m \end{smallmatrix}] 
 \label{oct24.21}
 \eeq
for the combinations that mimic the $\betalagp{j} $ in (\ref{eq:betalag}).
Moreover, by analogy with repeated Cauchy--Riemann derivatives (\ref{oct24.1})
of the $\beta^{\rm sv}$, we furthermore introduce
\begin{align}
\alpha^{N,j}_{m,k}
= \sum_{a+b=N} \frac{ N! }{a! b!} \bigg\{ 
&\frac{ j! (  k {+} m {-}2 {-} j)!}{(  j {-} a)! (  k {+} m {-}2  {-} j {-} b)!}
   \alpha[\begin{smallmatrix}2 m - 2 - j + a &k - m + j + b \\ 2 m&    2 k  \end{smallmatrix}]   \label{oct24.22} \\
&\quad + \frac{(  k {-} m {+}j)! ( 2 m {-}2 {-} j)!}{(  k {-} m {+} j  {-}a)! ( 2 m {-}2  {-} j {-} b)!}
   \alpha[\begin{smallmatrix}k + m - 2 - j + a & j + b \\ 2 k & 2 m  \end{smallmatrix}] 
\bigg\}  \, ,\notag
\end{align}
subject to shuffle relations $\alpha^{N,j}_{m,k}= - \alpha^{N,2m-2-j}_{m,k}$.
As will be exemplified below and in appendix~\ref{appalp}, reality 
of $\cFFp{s}{ m}{ k}$ will fix all instances of $\alpha^{N,j}_{m,k}$ in (\ref{oct24.22}) with $N\geq 0$. 
In particular, the examples
{\allowdisplaybreaks \begin{align}
\alpha^{0,0}_{2, 3} &=
\frac{2}{3} \zeta_3\, {\cal E}_0(6, 0) \, ,
&\alpha^{0,0}_{2, 4}&=
\frac{4}{3} \zeta_3\, {\cal E}_0(8, 0^2)  \,,
\notag \\
\alpha^{1,0}_{2, 3}&=
4  \zeta_3\,{\cal E}_0(6, 0^2)  + \frac{2}{5} \zeta_5\, {\cal E}_0(4) \, ,
&\alpha^{1,0}_{2, 4}&=
16\zeta_3\, {\cal E}_0(8, 0^3) \,,
 \label{oct24.25} \\
\alpha^{2,0}_{2, 3}&=
16\zeta_3 \,{\cal E}_0(6, 0^3) + \frac{8}{5} \zeta_5\,{\cal E}_0(4, 0)  \, ,
&\alpha^{2,0}_{2, 4}&=
160  \zeta_3\,{\cal E}_0(8, 0^4) + \frac{4}{7}\zeta_7\, {\cal E}_0(4) \,,
\notag \\
&&\alpha^{3,0}_{2, 4}&=
960 \zeta_3 \,{\cal E}_0(8, 0^5) + \frac{24}{7}\zeta_7\, {\cal E}_0(4, 0) \,,
\notag
\end{align}}%
and $\alpha^{3,0}_{2, 3}=\alpha^{4,0}_{2, 4}=0$ resulting from reality of (\ref{oct24.13}) are consistent with the known 
expressions for $\alpha[\begin{smallmatrix}j_1 &j_2 \\ 4 & 6  \end{smallmatrix}] $
and $\alpha[\begin{smallmatrix}j_1 &j_2 \\ 4 & 8  \end{smallmatrix}]$ \cite{Gerken:2020yii, Gerken:2020xfv}.
Beyond this, reality of (\ref{oct24.14}) to (\ref{cF635}) yields new results
that are spelt out in appendix \ref{appalp.A} and line up with the conjectural closed formulae
\begin{align}
\alpha_{2,k}^{N,0} &=  \frac{2 \zeta_{2k-1} }{2k{-}1} \Big\{
\delta_{N,k-2} (k{-}2)! {\cal E}_0(4)
 + \delta_{N,k-1}  2 (k{-}1)! {\cal E}_0(4,0) 
\Big\}
\notag \\
& \ \ +  \frac{ 2 \zeta_3}{3} \frac{(k{-}2)!(k{-}1{+}N)!}{(k{-}1{-}N)!}  {\cal E}_0(2k,0^{k+N-2}) \, , \ \ \ \ \ \ k \geq 3\, , \ \ 0 \leq N \leq k{-}1 \notag \\
%%%%%
\alpha_{3,k}^{N,0} - 2 \alpha_{3,k}^{N,1} &=
 \frac{2 \zeta_{2k-1} }{2k{-}1} \Big\{
(k{-}3)! \delta_{N,k-3}  {\cal E}_0(6)
 +   2 (k{-}2)! \delta_{N,k-2} {\cal E}_0(6,0) 
\Big\} \label{oct24.26} \\
& \ \ + \frac{2 \zeta_5}{5}  \frac{ (k {-} 3)!  (k {-} 2 {+} N)! }{(k {-} N {-} 2)!} {\cal E}_0(2k,0^{k+N-3}) \, , \ \ \ \ \ \ k \geq 4\, , \ \ 0 \leq N \leq k{-}2 \notag \\
%%%%
\alpha_{3,k}^{N,1} &=  \frac{2 \zeta_{2k-1} }{2k{-}1} \Big\{
 (k{-}2)!  \delta_{N,k-2}{\cal E}_0(6,0)
 +  6 (k{-}1)!   \delta_{N,k-1}{\cal E}_0(6,0^2) 
  +  12 k! \delta_{N,k} {\cal E}_0(6,0^3) 
\Big\}\notag \\
& \ \ + \frac{2 \zeta_5}{5}  \frac{ N (k {-} 2)! ( k {-} 2 {+} N)! }{ (k {-} N)!}  {\cal E}_0(2k,0^{k+N-3}) \, , \ \ \ \ \ \ k \geq 4\, , \  \ 0 \leq N \leq k \, ,
\notag
\end{align}
where $\alpha_{2,k}^{k,0}, \alpha_{3,k}^{k+1,1}$ and $ \alpha_{3,k}^{k-1,0}- 2 \alpha_{3,k}^{k-1,1}$ vanish by
shuffle relations. We have tested (\ref{oct24.26}), and the later generalisation~\eqref{allalph.1}, to hold for all cases of $\alpha_{m,k}^{N,j}$ with $m{+}k\leq 14$.

However, the combinations (\ref{oct24.22}) only span a subspace of the $\alpha[\begin{smallmatrix}j_1 &j_2 \\ 2m & 2k  \end{smallmatrix}] $ since we did not yet investigate cases with $j_1{+}j_2<m{+}k{-}2$ or imaginary cusp forms at depth two. As will be detailed in section \ref{sec:odd.bsvrep}, the imaginary cusp forms associated with double integrals over $({\rm G}_4,{\rm G}_6)$ introduce terms of the form
$\zeta_3 \,{\cal E}_0(4,0^p)$ into some of the $\alpha[\begin{smallmatrix}j_1 &j_2 \\ 4 & 6  \end{smallmatrix}] $ \cite{Gerken:2020yii} which are absent in (\ref{oct24.25}). The $\alpha_{m,k}^{N,j}$ determined by
the reality of $\cFFp{s}{m}{k}$ 
are exclusively built from $  \zeta_{2m-1}\,{\cal E}_0(2k,0^p)$ and $\zeta_{2k-1}\,{\cal E}_0(2m,0^p)  $ whereas more general $\alpha[\begin{smallmatrix}j_1 &j_2 \\ 2m & 2k  \end{smallmatrix}] $
may also involve different combinations of zeta values and iterated Eisenstein integrals at depth one.
 
The expression for $\alpha[\begin{smallmatrix} 2&0 \\ 6&4  \end{smallmatrix}] $ in (\ref{exsalpha}) 
exemplifies that $\alpha[\begin{smallmatrix}j_1 &j_2 \\ 2m & 2k  \end{smallmatrix}] $ with
 $j_1{+}j_2 < m{+} k{-}2$ may be non-zero. Indeed, $m \neq k$ does not admit any transcendentality 
 argument for the vanishing of $\alpha[\begin{smallmatrix}j_1 &j_2 \\ 2m & 2k  \end{smallmatrix}] $
that do not occur in $\cFFpm{s}{ m}{ k}$. Already in the even case, the lowest-weight integration constant
$\alpha^{0,0}_{2, 3}  = \frac{2}{3}\zeta_3\,{\cal E}_0(6, 0)  $ contributing to $\FFp{3}{ 2}{ 3}$ still has
a non-trivial analogue $ \zeta_3\,{\cal E}_0(6) $ of lower transcendental weight which does
occur in $\alpha[\begin{smallmatrix}0 &2\\ 6 &4  \end{smallmatrix}]$ \cite{Gerken:2020yii}.

%%%%%%%%%%%%%%%%%%%%%%%%%%%%%%%%%%%%%%%%%%%%%%%%%%%%%%%%%%%
\section{\texorpdfstring{Laplacian, seed functions and lower-depth terms for $\FFm{s}{m}{k}$}{Laplacian, seed functions and lower-depth terms for Fminus(s,m,k)}}
\label{oddsection}
%%%%%%%%%%%%%%%%%%%%%%%%%%%%%%%%%%%%%%%%%%%%%%%%%%%%%%%%%%%

We shall now extend the analysis of the even modular invariants $\FFp{s}{ m}{ k}$ in
the previous sections to their odd counterparts $\FFm{s}{ m}{ k}$. As we will see, the
absence of Laurent polynomials in the expansion of $\FFm{s}{ m}{ k}$ and their Poincar\'e seeds 
around the cusp will simplify certain steps. At the same time, the iterated-integral representations
of odd modular invariants will turn out to pose additional challenges and introduce more diverse 
integration constants $\overline{\alpha[\ldots]}$ as compared to the even case.

%%%%%%%%%%%%%%%%%%%%%%%%%%%%%%%%%%%%%%%%%%%%%%%%%%%%%%%%%%%
\subsection{\texorpdfstring{Laplacian of odd combinations of $\bsv$}{Laplacian of odd combinations of betasv}}
\label{sec:Lapodd}
%%%%%%%%%%%%%%%%%%%%%%%%%%%%%%%%%%%%%%%%%%%%%%%%%%%%%%%%%%%

In the previous sections, we have solved the Laplace system~\eqref{eq:LapLag} for the even combinations $\betalagp{j}$ in terms of Poincar\'e seeds $\seedp{s}{m}{k}$. We now turn to the odd combinations $\betalagm{j}$ in \eqref{eq:betalag} with $m< k$ and $0\leq j \leq 2m{-}2$, since for $m=k$ all $\betalagm{j}$ vanish. One difference to the even case is that while $\betalagp{m-1}$ was a pure shuffle according to~\eqref{eq:shuffle}, this is no longer true for the odd combination. Instead we now find that
\begin{align}
\label{eq:oddshuffle}
\betalagm{m}-\betalagm{m-2} 
&=\betasv{m-2\\2m}\betasv{k\\2k} -\betasv{m\\2m}\betasv{k-2\\k}\nn\\
&= \frac{ (k{-}1)!(k{-}2)!(m{-}1)!(m{-}2)!}{(2k{-}1)! (2m{-}1)!} \frac{(\pi\nabla)\EE_k (\pi\overline{\nabla})\EE_m - (\pi\nabla)\EE_m (\pi\overline{\nabla})\EE_k}{y^2}  \\
&\ \ \ \ \ \ {\rm mod} \ {\rm lower} \ {\rm depth} \notag
\end{align}
is a combination of shuffles. In the second step we have used~\eqref{oct24.2} to express this in terms of Cauchy--Riemann derivatives of $\EE_m$ and $\EE_k$.

More generally, we define in analogy with~\eqref{prodd1} for $\ell \geq 1$ 
\begin{align}
\label{eq:Jmdef}
\Jm{\ell}{m}{k} = \frac{ (\pi\nabla)^\ell\EE_m (\pi\overline{\nabla})^\ell \EE_k
- (\pi\nabla)^\ell\EE_k (\pi\overline{\nabla})^\ell \EE_m}{2y^{2\ell}}\,,
\end{align}
which is the odd combination of the gradients of the Eisenstein series. It is purely imaginary and we have from~\eqref{eq:oddshuffle}
\begin{align}
\label{eq:oddshuffle2}
\betalagm{m}-\betalagm{m-2}  =  - 2 \frac{ (k{-}1)!(k{-}2)!(m{-}1)!(m{-}2)!}{(2k{-}1)! (2m{-}1)!} \Jm{1}{m}{k}
\ {\rm mod} \ {\rm lower} \ {\rm depth}\,,
\end{align}
which serves as the substitute for source term $\Jp{0}{m}{k}=\EE_m\EE_k$ appearing for the even combinations $\betalagp{m-1}$ in (\ref{eq:shuffle}). More generally, the odd counterpart of the dictionary (\ref{bsvjplus}) between $\beta^{{\rm sv}+}$ and $\Jp{\ell}{m}{k}$ is (for $1\leq \ell\leq m{-}1$)
\beq
 \Jm{\ell}{m}{k} = \frac{ (2m{-}1)! (2k{-}1)! }{2 \big[ (m{-}1)! (k{-}1)! \big]^2} \Big[ \prod_{j=1}^\ell (m{-}j) (k{-}j) \Big]
\Big( \betalagm{m-\ell-1}-\betalagm{m+\ell-1}  \Big)
\ {\rm mod} \ {\rm lower} \ {\rm depth}
\label{bsvminus}
\eeq
and the recurrence (\ref{altrec.10}) immediately carries over to the odd case upon replacing
$\Jp{\ell}{m}{k} \rightarrow \Jm{\ell}{m}{k}$ on both sides of the equation for $\ell \geq 1$ (noting
that $\Jm{0}{m}{k}=0$).

We take $\betalagm{j}$ in the range $0\leq j\leq m{-}1$ as 
shuffle-independent representatives which is one value more than in the even case studied in section~\ref{sec:3.1}.
This range for the superscripts $j$ is again such that the holomorphic Eisenstein series in~\eqref{eq:LapLag} 
never contribute to the Laplacians of $\betalagm{j}$. As a consequence of~\eqref{eq:LapLag}, the first time the source term $\Jm{1}{m}{k}$ is introduced by the Laplacian is $\Delta \betalagm{m-1}$. 
The Laplace system is now given in terms of an $m\times m$ matrix $M_{ji}^-$ as
\begin{align}
\label{eq:oddLap}
\Delta \betalagm{j} = \sum_{i=0}^{m-1} M_{ji}^- \betalagm{i} 
- 2\frac{[(m{-}1)!(k{-}1)!]^2}{(2m{-}1)!(2k{-}1)!}\delta_{j,m-1} \Jm{1}{m}{k}
\ {\rm mod} \ {\rm lower} \ {\rm depth}
\end{align}
with $0\leq j \leq m{-}1$ and
\begin{equation}\label{eq:oddM}
M^-_{ji} = 
\left\lbrace \begin{array}{ll} j \big(2m-j-1\big)+\big(m+k -2-j\big) \big(k-m+j+1\big) & \text{for $i=j$}\,, \\[1mm]
j\big(k-m+j\big) + \delta_{j,m-1} (m-1)(k-1) &\text{for $i=j-1$}\,,\\[1mm]
\big(2m-2-j\big)\big( m+k-j-2\big) & \text{for $i=j+1$}\,,\\[1mm]
0 & \text{otherwise}\,.
\end{array}\right.
\end{equation}
The contribution $\delta_{j,m-1}$ in the second line of (\ref{eq:oddM}) 
is due to having to form the combination~\eqref{eq:oddshuffle2} for the source. 

Examples of~\eqref{eq:oddLap} for small $m< k$ are
\begin{align}
\Delta \begin{pmatrix} \beta^{{\rm sv}-,0}_{2,3} \\ \beta^{{\rm sv}-,1}_{2,3} \end{pmatrix} 
&= \begin{pmatrix} 6 & 6 \\ 4 & 8 \end{pmatrix}  
\begin{pmatrix} \beta^{{\rm sv}-,0}_{2,3} \\ \beta^{{\rm sv}-,1}_{2,3} \end{pmatrix} 
- \frac1{90} \begin{pmatrix} 0 \\ \Jm{1}{2}{3} \end{pmatrix} \ {\rm mod} \ {\rm lower} \ {\rm depth}\,,\nn\\
\Delta \begin{pmatrix} \beta^{{\rm sv}-,0}_{2,4} \\ \beta^{{\rm sv}-,1}_{2,4} \end{pmatrix} 
&= \begin{pmatrix} 12 & 8 \\ 6 & 14 \end{pmatrix}  
\begin{pmatrix} \beta^{{\rm sv}-,0}_{2,4} \\ \beta^{{\rm sv}-,1}_{2,4} \end{pmatrix} 
- \frac1{420} \begin{pmatrix} 0 \\ \Jm{1}{2}{4} \end{pmatrix} \ {\rm mod} \ {\rm lower} \ {\rm depth}\,, \label{sampleqs}\\
\Delta \begin{pmatrix} \beta^{{\rm sv}-,0}_{3,4} \\ \beta^{{\rm sv}-,1}_{3,4} \\\beta^{{\rm sv}-,2}_{3,4} \end{pmatrix} 
&= \begin{pmatrix} 10 & 20&0 \\ 2& 16 & 12 \\0 &12&18 \end{pmatrix}  
\begin{pmatrix} \beta^{{\rm sv}-,0}_{3,4} \\ \beta^{{\rm sv}-,1}_{3,4} \\\beta^{{\rm sv}-,2}_{3,4} \end{pmatrix} 
- \frac1{2100} \begin{pmatrix} 0 \\0\\ \Jm{1}{3}{4} \end{pmatrix} \ {\rm mod} \ {\rm lower} \ {\rm depth} \,.
\notag
\end{align}
The matrix $M^-_{ji}$ can be diagonalised and a large number of examples suggests that the spectrum in the case of odd modular invariant combinations is given by
\begin{align}
\label{eq:Oddspec}
s(s{-}1) \,\,\mbox{with}\,\,s \in \left\lbrace k{-}m{+}1,k{-}m{+}3,\ldots,k{+}m{-}3,k{+}m{-}1 \right\rbrace\,.
\end{align}
As we have $m<k$ the value $s=1$ never occurs. 
Comparing with the spectrum~\eqref{eq:Spectrum} in the case of even modular invariants $\FFp{s}{m}{k}$, 
we see that odd modular invariants $\FFm{s}{m}{k}$ have the opposite correlation between $s$ 
and the transcendental weight $w=k{+}m$: The $\FFp{s}{m}{k}$ of even (odd) weight $w=k{+}m$ have even (odd)
$s$ whereas $\FFm{s}{m}{k}$ of even (odd) weight $w=k{+}m$ have odd (even) $s$. At odd weight $m{+}k=5$, for
instance, the eigenvalues are characterised by odd $s=3$ in case of $\FFp{3}{2}{3}$ but by even
$s=2,4$ in case of $\FFm{2}{2}{3},\FFm{4}{2}{3}$. Conversely, even weight $m{+}k=6$ gives rise to
even $s=4$ for $\FFp{4}{2}{4}$ and odd $s=3,5$ for $\FFm{3}{2}{4},\FFm{5}{2}{4}$.

We can proceed similarly to (\ref{eq:tildeb}) and study the diagonalisation of the matrix $M^-$ in (\ref{eq:oddM}) by writing the linear combination
\begin{equation}
\label{eq:tildebm}
\tilde{\beta}_{(s)}^- = \sum_{i=0}^{m-2} v^i_{(s)} \betalagm{i}\,,
\end{equation}
expressed in terms of the eigenvector $\mathbf{v}_{(s)}=\big(v_{(s)}^0,\ldots, v_{(s)}^{m-2}\big)^T$ of $(M^-)^T$.
With~\eqref{eq:tildebm} the Laplace equation (\ref{eq:oddLap}) reduces to
\begin{equation}\label{eq:LaplacianDiagm}
\Big(\Delta -s(s{-}1) \Big) \tilde{\beta}_{(s)}^-= \alpha_{(s)}\, \frac{  (\pi\nabla)\EE_m (\pi\overline{\nabla})\EE_k
- (\pi\nabla)\EE_k (\pi\overline{\nabla})\EE_m}{y^2} \ {\rm mod} \ {\rm lower} \ {\rm depth}
\end{equation}
for some rational coefficients $\alpha_{(s)}$.

Again, since $M^-$ is tridiagonal, the eigenvalue equation
$(M^-)^T \mathbf{v}_{(s)}  = s(s{-}1)\mathbf{v}_{(s)}$ 
translates into a three-term recurrence given by
\begin{align}
M_{i-1,i}^-\,v^{i-1}_{(s)} + \big(M_{ii}^- -s(s{-}1) \big)\,v^i_{(s)} + M_{i+1,i}^- \,v^{i+1}_{(s)} = 0\,,
\label{oct20.1minus}
\end{align}
with the boundary conditions $v^{-1}_{(s)} = v^{m}_{(s)}= 0$.
In order to have a non-zero solution for $\mathbf{v}_{(s)}$ we see that both $v_{(s)}^0$ and $v^{m-1}_{(s)}$ must be non-zero so in particular the constants $\alpha_{(s)}$ in equation (\ref{eq:LaplacianDiagm}) in front of the source term will never vanish.

As it happened in the even case, also in the odd sector and for generic $m\leq k$ and $s$ in (\ref{eq:Oddspec}) we do not have a closed-form solution. However, for $m< k$ and $s=k{-}m{+}1$, corresponding to the lowest possible eigenvalue in~\eqref{eq:Oddspec}, one can prove that
\begin{equation}
v^i_{(s)} =  \frac{ (-1)^i  \Gamma(2m{-}1)}{i!\, \Gamma(2m{-}i{-}1) (1{+}\delta_{i,m-1})}\,,
\label{oct20.2m}
\end{equation}
where we normalised $v^0_{(s)}=1$ obtaining an expression extremely similar to the even eigenvector \eqref{oct20.2} corresponding to the lowest eigenvalue.

For other configurations of $m$, $k$ and $s$ one can find closed expressions in a few instances and we have used them and large scans over matrices $M^-_{ji}$ given in~\eqref{eq:oddM} to test the claimed spectrum~\eqref{eq:Oddspec} in numerous cases.  

In conclusion, similarly to~\eqref{eq:Fmk}, we now study the solutions to the Laplace problem
\begin{align}
\label{eq:Fmkm}
\big(\Delta-s(s{-}1) \big)\FFm{s}{m}{k} &= \frac{  (\pi \nabla) \EE_m (\pi\overline{\nabla}) \EE_k
- (\pi \nabla) \EE_k (\pi\overline{\nabla}) \EE_m }{2y^2} \\
s &\in  \left\lbrace k{-}m{+}1,k{-}m{+}3,\ldots,k{+}m{-}3,k{+}m{-}1 \right\rbrace\,,\nn
\end{align}
where we wrote out the source $\Jm{1}{m}{k}$ for definiteness.  As suggested by the notation we are looking for modular invariant solutions $\FFm{s}{m}{k}$ that are odd under $\tau\to-\bar\tau$.
The source term on the right-hand side has transcendental weight $m{+}k$ which is also the transcendental weight of~$\FFm{s}{m}{k}$.

Similar to the even sector, equation~\eqref{eq:Fmkm} is the most general one to consider in the odd case when restricting to sources built out of $\Jm{\ell}{m}{k}$, since a source with $\ell>1$ can always be reduced to $\Jm{1}{m}{k}$ using  equation~\eqref{altrec.10} adapted to the odd case and after redefining the function $\FFm{s}{m}{k}$.

\subsection{Solution to odd Laplace equations via Poincar\'e series}
\label{sec:oddseed}

The strategy for solving~\eqref{eq:Fmkm} will be the same as in section~\ref{sec:3.2new}, i.e.\ we shall construct a Poincar\'e seed $\seedm{s}{m}{k}$ for $\FFm{s}{m}{k}$ and recall our assumption $m<k$. After having folded $ \nabla \EE_k$ and its complex conjugate (see section \ref{sec:fold.9} for the alternative folding of $ \nabla \EE_m$) a solution to~\eqref{eq:Fmkm} can be obtained in terms of an absolutely convergent Poincar\'e sum by solving
\begin{align}
\label{eq:seedmeq}
\big(\Delta -s(s{-}1) \big) \seedm{s}{m}{k} &= \Jseedm{1}{m}{k}\,,
\end{align}
where one has to insert the $\ell=1$ instance of the Poincar\'e seed
\begin{align}
\Jseedm{\ell}{m}{k} &= 
%-
 \frac{(-4)^k (k)_\ell {\rm B}_{2k}}{2 (2k)!}  y^{k-\ell } \big( (\pi\overline{\nabla})^{\ell}  - (\pi \nabla)^{\ell}   \big) 
 \EE_m\nn\\
&=
 (-1)^{k+\ell +1} \frac{  {\rm B}_{2k} (k)_\ell }{(2 k)!   \Gamma(m)} (4 y)^k
\sum_{n=1}^{\infty}  n^{m - 1} \sigma_{1-2m}(n) (q^n-\bar{q}^n)
\sum_{a=0}^{m-1}  (4 n y)^{-a}  \frac{ \Gamma(   m {+} a)}{a! \Gamma(m {-} a) }\notag\\
& \hspace{20mm}  \times \Big[-(a{+}1{-}\ell)_\ell +\sum_{s=0}^\ell \binom{\ell}{s} (a{+}1{-}\ell)_{\ell-s} (4 n y)^s\Big]  \label{eq:seedm.A}  \\
&=i (-1)^{k+\ell}  \frac{  {\rm B}_{2k}(k)_\ell \Gamma(2m)}{ (2k)! \Gamma(m)} (4y)^{k}\sum_{a=0}^{m-1}  \frac{\Gamma(m{+}a)}{a! \Gamma(m{-}a)} \notag \\
&\hspace{20mm} \times \sum_{s=1}^\ell \binom{\ell}{s} (a{+}1{-}\ell)_{\ell-s} \,(4 y)^{s-a} \Im[{\cal E}_0(2m, 0^{m+a-s-1}) ] \notag
\end{align}
of $\Jm{\ell}{m}{k}$ with $\ell \leq m{-}1$ obtained in analogy with section~\ref{sec:3.3}. Note that
there is no Laurent-polynomial part as this would be incompatible with being odd 
under $\tau\to -\bar\tau$ (that exchanges $q \leftrightarrow \bar{q}$ and keeps $y$ invariant).

The solution of~\eqref{eq:seedmeq} proceeds as in section~\ref{sec:3.2new} except for that there is no zero mode $c_0(y)$. We find
\begin{align}
\label{eq:seedm}
\seedm{s}{m}{k} = 
%-
 i (-1)^{k} \frac{  {\rm B}_{2k} (2m{-}1)! }{2\Gamma(2k) \Gamma(m)} \sum_{\ell=k-m+1}^{k} g^{-}_{m,k,\ell,s} (4y)^\ell \Im \mathcal{E}_0 (2m,0^{k+m-\ell-1})
\end{align}
with 
\begin{align}
g^{-}_{m,k,\ell,s} =  \frac{\Gamma(\ell)}{\Gamma(\ell{+}s)} \sum_{i=\ell}^{k}  \frac{ (\ell{+}1{-}s)_{i-\ell}\Gamma(s{+}i)\Gamma(m{+}k{-}i)}{\Gamma(k{-}i{+}1)\Gamma(i{+}1)\Gamma(m{-}k{+}i)}\, ,
\label{eq:seedm.B}
\end{align}
see (\ref{eq:fkm}) for the analogous coefficients $g^{+}_{m,k,\ell,s}$ in the even seed functions.

\subsubsection{\texorpdfstring{Examples of source terms at $m+k \leq 7$}{Examples of source terms at m+k<= 7}}

The simplest examples of the seeds $\Jseedm{\ell}{m}{k} $ in (\ref{eq:seedm.A}) with
$\ell \leq m{-}1$ are given by
\begin{align}
\Jseedm{1}{2}{3} &= \frac{ 16 i y^4}{105}  \Im \mathcal{E}_0 (4) +\frac{ 8 i y^3}{105}  \Im \mathcal{E}_0 (4, 0)\,,
\notag \\
\Jseedm{1}{2}{4} &= \frac{32 i y^5}{1575 } \Im \mathcal{E}_0 (4) + \frac{16 i y^4 }{1575 } \Im \mathcal{E}_0 (4, 0)\,,
\label{exminus.1} \\
\Jseedm{1}{2}{5} &= \frac{16 i y^6 }{6237 } \Im \mathcal{E}_0 (4) + \frac{8 i y^5 }{6237 }  \Im \mathcal{E}_0 (4, 0)\,,
\notag
\end{align}
as well as
\begin{align}
\Jseedm{1}{3}{4} &= \frac{ 64 i y^5}{315 }  \Im \mathcal{E}_0 (6, 0) + \frac{ 32 i y^4}{105 } \Im \mathcal{E}_0 (6, 0^2) + 
\frac{ 16 i y^3}{105 } \Im \mathcal{E}_0 (6, 0^3)\,,
\label{exminus.2}\\
\Jseedm{2}{3}{4} &= - \frac{ 256 i y^6}{63}  \Im \mathcal{E}_0 (6) - \frac{ 256 i y^5}{63 } \Im \mathcal{E}_0 (6, 0) - 
\frac{ 64 i y^4}{21 } \Im \mathcal{E}_0 (6, 0^2) - \frac{ 32 i y^3}{21}  \Im \mathcal{E}_0 (6, 0^3)\, .
\notag
\end{align}

\subsubsection{\texorpdfstring{Examples of Poincar\'e seeds at $m+k \leq 7$}{Examples of Poincar\'e seeds at m+k <= 7}}

The simplest examples of the seed functions (\ref{eq:seedm}) of the $\FFm{s}{m}{k} $ 
resulting from the expressions for $\Jseedm{1}{m}{k} $ in the previous section are given by
\begin{align}
\seedm{2}{2}{3}&= -\frac{4 i y^3}{315}  \Im \mathcal{E}_0 (4, 0) - \frac{ i y^2 }{63} \Im \mathcal{E}_0 (4, 0^2)\, , \ \
&\seedm{4}{2}{3}&= -\frac{4 i y^3}{315}  \Im \mathcal{E}_0 (4, 0)\,,
\notag\\
\seedm{3}{2}{4}&= - \frac{2 i y^4 }{1575} \Im \mathcal{E}_0 (4, 0) - \frac{ i y^3}{675}  \Im \mathcal{E}_0 (4, 0^2)\, , \ \
&\seedm{5}{2}{4}&= -\frac{2 i y^4 }{1575} \Im \mathcal{E}_0 (4, 0)\,,
\label{exminus.3}  \\
\seedm{4}{2}{5}&= - \frac{ 4 i y^5 }{31185} \Im \mathcal{E}_0 (4, 0) - \frac{i y^4 }{6930} \Im \mathcal{E}_0 (4, 0^2)\, , \ \ 
&\seedm{6}{2}{5}&= -\frac{4 i y^5}{31185}  \Im \mathcal{E}_0 (4, 0)\,,
\notag
\end{align}
as well as
\begin{align}
\seedm{2}{3}{4}&= -\frac{4 i y^4}{315}  \Im \mathcal{E}_0 (6, 0^2) - \frac{ 34 i y^3}{945}  \Im \mathcal{E}_0 (6, 0^3) - 
\frac{ i y^2}{27 } \Im \mathcal{E}_0 (6, 0^4)\,,
 \notag \\
\seedm{4}{3}{4}&= -\frac{4 i y^4}{315}  \Im \mathcal{E}_0 (6, 0^2) - \frac{ 8 i y^3}{315 }  \Im \mathcal{E}_0 (6, 0^3)\,,
\label{exminus.4} \\
\seedm{6}{3}{4}&= -\frac{4 i y^4}{315}  \Im \mathcal{E}_0 (6, 0^2) - \frac{ 2 i y^3}{315 } \Im \mathcal{E}_0 (6, 0^3)\, .
\notag
\end{align}

\subsection{Step form for odd Laplace system}
\label{sec:step.odd}

We shall now extend the step form of the Laplace equations in section \ref{sec:3.5} to odd
seed functions. The simplest examples for the step form in the odd case are obtained by
regrouping the results for seed functions in (\ref{exminus.1}) to (\ref{exminus.4}),
\begin{align}
y^2 \Im [ \mathcal{E}_0 (4, 0^2)] &= 63 i ( \seedm{2}{ 2}{ 3} -  \seedm{4}{ 2}{ 3} )\,,
\notag \\
y^3 \Im [ \mathcal{E}_0 (4, 0)] &=  \frac{315 i}{4}  \seedm{4}{ 2}{ 3}\,,
\label{exminus.5} \\
y^4 \Im [ \mathcal{E}_0 (4)] &= - \frac{ 315 i}{8} \seedm{4}{ 2}{ 3} - \frac{105 i }{16 } \Jseedm{1}{ 2}{ 3}\,,
\notag 
\end{align}
as well as
\begin{align}
y^3 \Im [ \mathcal{E}_0 (4, 0^2)] &= 675 i  ( \seedm{3}{ 2}{ 4} -   \seedm{5}{ 2}{ 4} )\,,
\notag \\
y^4 \Im [ \mathcal{E}_0 (4, 0)] &=  \frac{1575 i}{2} \seedm{5}{ 2}{ 4}\,,
\label{exminus.6} \\
y^5 \Im [ \mathcal{E}_0 (4)] &= -\frac{1575 i}{4}  \seedm{5}{ 2}{ 4} -  \frac{1575 i}{32} \Jseedm{1}{ 2}{ 4}\, ,
\notag 
\end{align}
and the analogous examples with $m{+}k=7,8$ can be found in appendix \ref{app:ex.9}.

The analogous step form involving  $\seedm{s}{ m}{ k}$ and $\Jseedm{\ell}{ m}{ k}$ at general
$m<k$ follows the logic of the even case in section \ref{sec:3.5}: The starting point is the $j=0$ case
of $\betalagm{j} $ in (\ref{eq:betalag}) which requires the maximum number of Laplace actions until
a holomorphic Eisenstein series can be factored out. As the odd counterpart of (\ref{oct21.1}), we have
\beq
\sum_{\gamma \in B(\ZZ)\backslash {\rm SL}(2,\ZZ)} \Big[
 y^{k-m+1}  \Im[{\cal E}_0({2m},0^{2m-2}) ]  \, \Big]_\gamma =i \rho_{m,k}\,
\betalagm{0} \ \text{mod lower depth}
\label{oct21.1odd}
\eeq
with the same rational prefactor $\rho_{m,k}$ given by (\ref{rhoexpr}) as in the even case.
Starting from (\ref{oct21.1odd}), repeated action of the shifted Laplace operator ${\cal O}_\ell = - \frac{1}{4\ell}(\Delta - \ell(\ell{-}1))$ introduced in (\ref{oct21.3}) yields $y^{k-m+1+r}  \Im[{\cal E}_0({2m},0^{2m-2-r}) ]$ on the left-hand side.
This follows from
\begin{align}
\Delta \Big( y^n   \Im[{\cal E}_0({2m}) ]  \Big) &= \frac{ 4n  y^{n+1} }{(2\pi i)^{2m}} \Im[{\rm G}_{2m}]+ n(n{-}1) y^n  \Im[{\cal E}_0({2m}) ]\,,
\label{d1lemma.odd}\\
\Delta \Big( y^n   \Im[{\cal E}_0({2m,0^p}) ]  \Big) &= -  4n   y^{n+1}  \Im[{\cal E}_0({2m,0^{p-1}}) ] + n(n{-}1) y^n  \Im[{\cal E}_0({2m,0^p}) ] \, , \ \ \ \ \ \ p \neq 0\,,\nn 
\end{align}
with $\Im[{\rm G}_{2m} ]= \Im[{\rm G}_{2m}^0 ]$, see (\ref{d1lemma}) for the even counterpart. On the right-hand side,
we obtain a sequence of ${\cal O}_\ell$ acting on $\betalagm{0}$ which can be simplified via 
(\ref{eq:LapLag}),
($r=1,2,\ldots,2m{-}2$)
\begin{align}
&\sum_{\gamma \in B(\ZZ)\backslash {\rm SL}(2,\ZZ)}
\Big[   y^{k-m+1+r}   \Im[{\cal E}_0({2m},0^{2m-2-r}) ]  \Big]_\gamma  \label{oct21.5.odd}\\
& \ \ \ \ \ \
\sim  {\cal O}_{k-m+r} {\cal O}_{k-m+r-1} \ldots {\cal O}_{k-m+2} {\cal O}_{k-m+1} \betalagm{0}  \ \text{mod lower depth} 
\notag
\end{align}
with $r=1,2,\ldots,2m{-}2$, see (\ref{oct21.5}) for the even counterpart.
After $2m{-}1$ Laplace actions on (\ref{oct21.1odd}), one arrives at the
following analogue of (\ref{notthis}) and (\ref{notthis2}), 
\begin{align}
\frac{1}{(2\pi i)^{2m}} &\PS \left[y^{k+m} \Im(\GG_{2m}(\tau))  \right]_\gamma  \notag \\
& \! \! \! \! \!  \! \! \! \! \! =   \frac{i (2k)! \Gamma(k) }{2 (-4)^k {\rm B}_{2k} \Gamma(k{+}m)} \bigg\{
\frac{ {\rm G}_{2m} }{(2\pi i)^{2m}} (\pi \overline \nabla)^m {\rm E}_k
- \frac{\overline{ {\rm G}_{2m} } }{(2\pi i)^{2m}} (\pi  \nabla)^m {\rm E}_k
\bigg\}  \label{notthis2.odd} \\
& \! \! \! \! \!  \! \! \! \! \! = - \frac{i (2k)!(2k{-}1)!(k{-}m)}{2 (-4)^{k+m}{\rm B}_{2k} \Gamma(k{+}m) (k{-}m)!} \left\{
(\tau{-}\bar\tau)^{2m}  \GG_{2m} \betasv{ k-m-1\\2k}
- \frac{ \overline{\GG_{2m} }}{(2\pi i)^{2m}}\betasv{k+m-1\\2k}  \right\} \notag \\
&\ \ \ \ \   \MLD \, ,\notag
\end{align}
where we used~\eqref{oct24.2} in the last step.

\subsection{Comparison with cuspidal MGFs}
\label{sec:comp.odd}

In the same way as the $\FFp{s}{m}{k}$ were compared with the modular graph functions $C_{a,b,c}$ in section \ref{sec:3.2.5}, we shall now relate the simplest $\FFm{s}{m}{k}$ to imaginary MGFs. A variety of imaginary 
and thereby cuspidal MGFs have been identified in \cite{DHoker:2019txf, Gerken:2020yii}. Due to the absence
of $\FFm{s}{m}{k}$ at weight $m{+}k = 4$ (the source vanishes) we will
spell out the seed functions for their bases at $m{+}k=5,6$.

\subsubsection{Weight $m{+}k=5$}

At weight $m{+}k=5$, the shuffle irreducible imaginary MGFs are spanned by the quantities
${\rm B}_{2,3}$ in section 5 of \cite{Gerken:2020yii} and 
$\mathcal{A}_{1,2;5}$ in section 6 of \cite{DHoker:2019txf} subject to the Laplace equations
\begin{align}
(\Delta-2) (2{\rm B}_{2,3} +3 \mathcal{A}_{1,2;5}) & = 42 \mathcal{A}_{1,2;5}\,, \notag \\
(\Delta - 6) \mathcal{A}_{1,2;5} &= 2(3 \mathcal{A}_{1,2;5}-  \Jm{1}{2}{3} )
\,,
\label{cspfm.6} \\
(\Delta - 12) (3 \mathcal{A}_{1,2;5}-\Jm{1}{2}{3} )
 &= - \frac{3}{\pi^4}
 \Big( {\rm G}_4 (\pi \overline \nabla)^2 {\rm E}_3 
- \overline{ {\rm G}_4 }(\pi  \nabla)^2 {\rm E}_3  \Big) \, ,
\notag
\end{align}
see in particular equations (5.19a) and (5.19b) of \cite{Gerken:2020yii}.
This can be lined up with the Laplace equations (\ref{sampleqs}) of the $\betalagm{j}$ at $(m,k)=(2,3)$,
consistent with the $\beta^{\rm sv}$ representations (see \cite{Gerken:2020yii} for
the lower-depth terms)
\begin{align}
2{\rm B}_{2,3} +3 \mathcal{A}_{1,2;5} & =1260\Big( \betasv{2 & 1 \\ 4 &6 } - \betasv{3 & 0 \\ 6 &4 }\Big)\, \text{mod lower depth}\,, \notag \\
 \mathcal{A}_{1,2;5} &\notag = 180\Big( \betasv{1 & 2 \\ 4 &6 } - \betasv{2 & 1 \\ 6 &4 }\Big)+120\Big( \betasv{2 & 1 \\ 4 &6 } - \betasv{3 & 0 \\ 6 &4 }\Big)\,\text{mod lower depth}\,,\\
 3 \mathcal{A}_{1,2;5}-\Jm{1}{2}{3} 
 & = 180\Big( \betasv{0 & 3 \\ 4 &6 } - \betasv{1 & 2 \\ 6 &4 }\Big)+ 540\Big( \betasv{1 & 2 \\ 4 &6 } - \betasv{2 & 1 \\ 6 &4 }\Big)\\
 &\notag\phantom{=}+180\Big( \betasv{2 & 1 \\ 4 &6 } - \betasv{3 & 0 \\ 6 &4 }\Big)\,\text{mod lower depth}\,.
\end{align}
By translating the right-hand sides into the leading-depth terms of the $\FFm{s}{m}{k}$, we identify
\begin{align}
{\cal A}_{1,2;5} = - 2 \FFm{4}{2}{3}  \, , \ \ \ \ \ \
 {\rm B}_{2,3} = \frac{ 21}{5}  \FFm{2}{2}{3}
 - \frac{ 6}{5}  \FFm{4}{2}{3}  \, .
 \label{Ftolatt}
\end{align}
Moreover, (\ref{cspfm.6}) matches the Laplace system of the seed functions in (\ref{exminus.5}),
which suggests the Poincar\'e-series representations
\begin{align}
 \sum_{\gamma \in B(\ZZ)\backslash {\rm SL}(2,\ZZ)} \Big[y^2  \Im[{\cal E}_0({4, 0^2}) ]  \Big]_\gamma &= \frac{ 15 i}{2 } (2{\rm B}_{2,3} +3 {\cal A}_{1,2;5})\,,
 \notag \\
 \sum_{\gamma \in B(\ZZ)\backslash {\rm SL}(2,\ZZ)} \Big[y^3 \Im[{\cal E}_0({4, 0}) ]  \Big]_\gamma &= - \frac{ 315 i }{8} {\cal A}_{1,2;5} \,,
 \label{oct20.17Cusp}\\
 \sum_{\gamma \in B(\ZZ)\backslash {\rm SL}(2,\ZZ)} \Big[y^4 \Im[{\cal E}_0({4}) ]  \Big]_\gamma &=  \frac{ 105 i}{16}(3 {\cal A}_{1,2;5} - \Jm{1}{2}{3} )
 \,.
\notag 
\end{align}
Note that (\ref{Ftolatt}) can be solved to express the Laplace eigenfunctions via 
imaginary combinations of lattice sums (see (\ref{rev.03}) for our notation for lattice momenta)
\begin{align}
\FFm{4}{2}{3}  &= - \frac{1}{2}{\cal A}_{1,2;5} = - \frac{ i}{3} \bigg( \frac{ \Im \tau }{\pi } \bigg)^5 \Im
\sum_{p_1,p_2,p_3 \in \Lambda'} \frac{ \delta(p_1{+}p_2{+}p_3) }{p_2^2 p_3^3 \bar p_1^3 \bar p_3^2}\,,
 \notag \\
\FFm{2}{2}{3}  &=- \frac{1}{7}{\cal A}_{1,2;5} + \frac{ 5}{21} {\rm B}_{2,3} \,,
\label{Ftolatt.2} \\
&= \frac{ 5}{63} \Jm{1}{2}{3} + \frac{ 2i}{21}\bigg( \frac{ \Im \tau }{\pi } \bigg)^5 \Im\bigg\{
5\sum_{p_1,p_2,p_3,p_4 \in \Lambda'} \frac{ \delta(p_1{+}p_2{+}p_3{+}p_4) }{p_2 p_3^2 p_4^2 
\bar p_1 \bar p_2 \bar p_4^3}
- \sum_{p_1,p_2,p_3 \in \Lambda'} \frac{ \delta(p_1{+}p_2{+}p_3) }{p_2^2 p_3^3 \bar p_1^3 \bar p_3^2}
\bigg\} \, .\notag
\end{align}
By the exhaustive scan of weight-five MGFs in \cite{Gerken:2020aju}, the quadruple sum over four
lattice momenta $p_1,p_2,p_3,p_4$ in $\FFm{2}{2}{3}$ cannot be reduced to simpler lattice sums
over $\leq 3$ momenta.

\subsubsection{Weight $m{+}k=6$}

The same kind of discussion applies to the imaginary cusp forms in the $(m,k)=(2,4)$ sector:
The imaginary parts of the complex modular graph forms ${\rm B}_{2,4}$ and
${\rm B}_{2,4}'$ introduced in section 9.2 of \cite{Gerken:2020aju} obey a Laplace system\footnote{We are grateful to Jan Gerken for providing the Laplace equations of ${\rm B}_{2,4}$ and ${\rm B}_{2,4}'$.}
\begin{align}
\Delta( {\rm B}_{2,4} - \overline{{\rm B}}_{2,4}) &= 20 ( {\rm B}_{2,4} - \overline{{\rm B}}_{2,4}) + 4 \,\Jm{1}{2}{4} \,,
\label{cspfm.11} \\
\Delta ( {\rm B}_{2,4}' - \overline{{\rm B}}_{2,4}') &= 6 ( {\rm B}_{2,4} ' - \overline{{\rm B}}_{2,4}')
-180 ( {\rm B}_{2,4} - \overline{{\rm B}}_{2,4})   \,, \notag
\end{align}
consistent with the Poincar\'e-series representations
\begin{align}
 \sum_{\gamma \in B(\ZZ)\backslash {\rm SL}(2,\ZZ)} \Big[y^3  \Im[{\cal E}_0({4, 0^2}) ]  \Big]_\gamma &=
 \frac{ 105i }{8}  ({\rm B}_{2,4}' - \overline{ {\rm B}}_{2,4}' )\,,
 \notag \\
 \sum_{\gamma \in B(\ZZ)\backslash {\rm SL}(2,\ZZ)} \Big[y^4 \Im[{\cal E}_0({4, 0}) ]  \Big]_\gamma &= 
  \frac{1575 i}{8}  ({\rm B}_{2,4}  -  \overline{{\rm B}}_{2,4}) \,,
 \label{may20.17Cusp}\\
 \sum_{\gamma \in B(\ZZ)\backslash {\rm SL}(2,\ZZ)} \Big[y^5 \Im[{\cal E}_0({4}) ]  \Big]_\gamma &= 
 - \frac{1575 i}{32} (2{\rm B}_{2,4}  -  2\overline{{\rm B}}_{2,4} + \Jm{1}{2}{4} ) \, .
\notag 
\end{align}
Note that the Laplace action on the last lines of (\ref{oct20.17Cusp}) and (\ref{may20.17Cusp}) yields 
\begin{align}
  \sum_{\gamma \in B(\ZZ)\backslash {\rm SL}(2,\ZZ)}\Big[  (-y^5)  \, \Im {\rm G}_4\Big]_\gamma
&= \frac{ 315 i }{ 16} \Big( {\rm G}_4 (\pi \overline \nabla)^2 {\rm E}_3 
- \overline{ {\rm G}_4 } (\pi  \nabla)^2 {\rm E}_3  \Big)\,,
\label{cspfm.5}
 \\
 \sum_{\gamma \in B(\ZZ)\backslash {\rm SL}(2,\ZZ)} \Big[(- y^6)  \, \Im {\rm G}_4\Big]_\gamma
&= \frac{ 945 i }{ 8} \Big( {\rm G}_4 (\pi \overline \nabla)^2 {\rm E}_4 
- \overline{ {\rm G}_4 } (\pi  \nabla)^2 {\rm E}_4  \Big)\, ,
\notag 
\end{align}
consistent with (\ref{notthis2.odd}).
Unfortunately, it currently appears challenging to confirm
(\ref{oct20.17Cusp}) and (\ref{may20.17Cusp}) by direct computation:
From~\cite{Iwaniec:2002,Fleig:2015vky} we know that the Poincar\'e sums over the seeds $y^a \Im[{\cal E}_0({2m, 0^b}) ] $ must involve more complicated Kloosterman sums and, in particular, the results of \cite{Dorigoni:2019yoq} cannot be applied to directly determine these imaginary cusp forms. For the moment, we leave them as conjectures supported by their consistency with the Laplacian which commutes with the convergent Poincar\'e sums over $y^a \Im[{\cal E}_0({4,0^b}) ] $.

\subsection{Reinstating lower depth for odd modular invariants}
\label{sec:odd.dpt1}

Our next step is to reinstate the $\beta^{\rm sv}$ of depth $\leq 1$ into the iterated-integral representation
of the odd functions $\FFm{s}{m}{k}$, following an extension of the strategy for the even case in section \ref{sec:4}.
At leading depth two, the $\FFm{s}{m}{k}$ are expressed in terms of the $\betalagm{j}$ as we showed in section~\ref{sec:Lapodd} by diagonalising the corresponding Laplace system~\eqref{eq:Fmkm}. The $\FFm{s}{m}{k}$ must have vanishing Laurent polynomial (and in fact vanishing Fourier zero mode, see also section \ref{sec:q-series}) since they are by definition odd under the transformation $\tau\to-\bar\tau$ that sends $y\to y$ and $q\leftrightarrow \bar q$. 

In order to determine the lower-depth $\bsv$ in compliance with a vanishing Laurent polynomial we again resort to the Cauchy--Riemann equation that is compatible with the Laplace system~\eqref{eq:Fmkm}. The generalisation of~\eqref{eq:CRFp} to the odd case is
\begin{align}
\label{eq:CRFm}
(\pi \nabla)^s \FFm{s}{m}{k} = \sum_{i=0}^s c_i^- (\pi \nabla)^i \EE_m (\pi\nabla)^{s-i} \EE_k + (\pi \nabla)^s H^-\,,
\end{align}
where $H^-$ is a homogeneous solutions to the Laplace equation $(\Delta-s(s{-}1))H^-=0$ that is not required to be modular invariant on its own, only its $s^{\rm th}$ Cauchy--Riemann derivative has to be modular. 

The coefficients $c_i^-$ appearing in~\eqref{eq:CRFm} now have to satisfy
\begin{align}
c_0^- =-\frac12\,,\hspace{20mm} c_s^-&=\frac12\,,\\
\big[m(m{-}1)-i(i{+}1) \big]c_{i+1}^- + \big[k(k{-}1)-(s{-}i)(s{-}i{-}1) \big] c_i^- &= \binom{k}{2}\binom{s-1}{i-1}-\binom{m}{2}\binom{s-1}{i+1}\nn
\end{align}
for $0\leq i\leq s{-}1$. These equations can be solved by iteration. 

In order to write a solution to the Laplace equation that is expressed in terms of the $\bsv$ at all depths, we proceed as in section~\ref{sec:4}. Starting from the depth-two terms $\betalagm{j}$ that come from the diagonalisation of the Laplacian, we first perform the substitution $\betalagm{j}\to\hatbetalagm{j}$ from~\eqref{oct24.5}. This substitution generates a specific set of depth-one terms that are constructed so that the Cauchy--Riemann derivative gives modular expressions $(\pi\nabla)^\bullet \EE_\bullet$, compatible with~\eqref{eq:CRFm}. This produces a solution to the Laplace equation at depths two and one.

We then compute the Laurent polynomial of this combination of depth-two and depth-one terms using the degeneration limits~\eqref{more.bsvb} of the $\bsv$. We know that $\FFm{s}{m}{k}$ has a vanishing Laurent polynomial since it is odd and therefore all depth-zero $y$-powers coming from the combinations of the $\hatbetalagm{j}$ must be cancelled. 

Non-positive powers of $y$ can simply be cancelled by adding their negatives to the $\hatbetalagm{j}$ combinations since they are in the kernel of $(\pi\nabla)^s$. This is also consistent with the fact that non-positive powers of $y$ are ubiquitous in the generating series of MGFs (see~\cite{Gerken:2019cxz,Gerken:2020yii}). 

If the Laurent polynomial of the $\hatbetalagm{j}$ features positive powers of $y$, the only possibility 
compatible with the Laplace and Cauchy--Riemann equations is $y^s$. 
Since the generating series of MGF in~\cite{Gerken:2020yii} does not introduce positive powers of $y$ in isolation\footnote{The coefficients of $\beta^{\rm sv}$ in the expansion of the generating series $Y^\tau$ in~\cite{Gerken:2020yii} involve non-positive powers of $y$ from the operator $\exp( - \frac{ \ep_0 }{4y})$ acting on a suitable initial value at $\tau \rightarrow i \infty$.}, the appropriate way of removing $y^s$ from the Laurent polynomial is to add a
suitable multiple of $\betasv{s-1\\2s}$. Indeed, the depth-one contributions in the substitution rule (\ref{oct24.5})
applied to a single $\hatbetalagm{j}$ in $\FFm{s}{m}{k}$ introduces the
positive power $\zeta_{2m-1} y^{k-m+1}$ at the cusp. This power $s=k{-}m{+}1$ is 
part of the spectrum (\ref{eq:Fmkm}) in the odd case but does not occur in
the even spectrum (\ref{eq:Fmk}). That is why only the odd functions
$\FFm{k-m+1}{m}{k}$ require corrections of the Laurent polynomial via
$\betasv{k-m\\2k-2m+2}$, and we did not encounter such terms
for the even $\FFp{s}{m}{k}$.

As we know from~\eqref{oct24.2}, $\betasv{s-1\\2s}$ is proportional to $\EE_s$ up to a term involving $\zeta_{2s-1}y^{1-s}$. This power of $y$ is in the kernel of both $(\Delta - s(s{-}1))$ and $(\pi \nabla)^s$ such that an
extra term proportional to $\betasv{s-1\\2s}$ will add a contribution with $(\pi\nabla)^sH^-\propto y^{2s} \GG_{2s}$ to the Cauchy--Riemann equation (\ref{eq:CRFm}). As already anticipated, apart from holomorphic Eisenstein series, we will
also find iterated integrals of holomorphic cusp forms in the homogeneous solutions $H^-$ relevant to
$\FFm{s}{m}{k}$ with $s \geq 6$ to be discussed in Part~II.

Since we are seeking an odd function with vanishing Laurent polynomial, the addition of the $\betasv{s-1\\2s}$ needs to be combined with antiholomorphic corrections $\overline{ {\cal E}_0(2s,0^p)}$ from the integration constants $\overline{\alpha[\begin{smallmatrix} j_1& j_2 \\ k_1&k_2 \end{smallmatrix}] }$ at depth two. As will be detailed
in section \ref{sec:odd.bsvrep}, the $\overline{\alpha[\ldots]}$ contributing to $\FFm{k{-}m{+}1}{m}{k}$
are tailored to effectively flip the sign of the $\overline{ {\cal E}_0(2s,0^p)}$ in (\ref{eq:EII}) for $\EE_s$.
In this way, the $q,\bar q$-terms of $\betasv{s-1\\2s}$ are promoted to the odd analogue $\EEodd_s$ of the non-holomorphic Eisenstein series $\EE_s$ \eqref{eq:EII} which we define as
\beq
\EEodd_s (\tau) = 
- 2i \frac{\Gamma(2s)}{\Gamma(s)} \sum_{a=0}^{s-1}  (4y)^{-a} \frac{\Gamma(s{+}a)}{a!\Gamma(s{-}a)} \Im \mathcal{E}_0\big(2s, 0^{s-1+a};\tau\big)\,.
\label{oddsols}
\eeq
These odd functions solve the desired eigenvalue equation $(\Delta - s(s{-}1))\EEodd_s =0$ but 
are not modular invariant. In summary, for all values of $m<k$ considered, the subtraction
of $\betasv{s-1\\2s}$ needed for a vanishing Laurent polynomial of $\FFm{s}{m}{k}$ is 
associated with $\zeta_{m+k-s} \EEodd_s $ and occurs only for the eigenvalue $s=k{-}m{+}1$.

In case of the even $\FFp{s}{m}{k}$, the matching of the Laurent polynomial
with the results from the Poincar\'e series in section \ref{sec:4.1} could be achieved
solely in terms of non-positive powers of $y$. For the MGFs among the $\FFp{s}{m}{k}$,
one can give a heuristic explanation of why the even case did not involve any analogue 
of the above $\betasv{s-1\\2s}$ beyond the substitution rule (\ref{bsvd1.5}): We have checked up to $m{+}k\leq 14$ that all the $\FFp{s}{m}{k}$ with an MGF representation are expressible in terms of $C_{a,b,c}$, ${\rm E}_s$
and $\zeta_{2n-1}$, i.e.\ in terms of sums over no more than three lattice momenta. 
The space of $\FFp{s}{m}{k}$ with an MGF representation differs from the space of {\it all} $\FFp{s}{m}{k}$ by
iterated integrals of holomorphic cusp forms (cf.\ Part II), and the counting in (\ref{dimsmatch}) therefore suggests
that at arbitrary weight, all $\FFp{s}{m}{k}$ in the MGF subspace are expressible in terms of $C_{a,b,c}$. 
While all even $\FFp{s}{m}{k}$ that enjoy lattice-sum expressions must therefore admit a representation
in terms of three lattice momenta, the lattice-sum representations of the odd $\FFm{s}{m}{k}$ in turn may necessitate four or more lattice momenta as in the weight-five example (\ref{Ftolatt.2}). 
These extra momenta give room for the additional 
complexity of having $\betasv{s-1\\2s}$, e.g.\ via independent appearances of 
${\rm G}_{2s}$ multiplying odd zeta values 
in the Cauchy--Riemann derivatives of certain $\FFm{s}{m}{k}$.

In conclusion, performing the above steps we arrive at a combination of $\bsv$ that we call $\cFFm{s}{m}{k}$ that is odd under $\tau\to-\bar\tau$ and solves the correct Laplace equation. As in the even case, we are not guaranteed that the function is invariant under modular transformations and so it may differ from the modular invariant $\FFm{s}{m}{k}$ that was constructed from the Poincar\'e seed. In the present work we shall focus on cases where $\cFFm{s}{m}{k}=\FFm{s}{m}{k}$ and relegate the other cases to the companion Part~II.

\subsubsection{\texorpdfstring{Examples at weight $m{+}k=5$}{Examples at weight m+k=5}}

We shall now apply the prescription above to the simplest odd functions
with Poincar\'e-series representations in section \ref{sec:comp.odd}.
At weight five, one arrives at
{\allowdisplaybreaks 
\begin{align}
\FFm{2}{2}{3} &= -90 \left( \betasv{1& 2\\ 4 & 6} - \betasv{2& 1\\4& 6} - \betasv{2& 1\\ 6 &4} + \betasv{3 & 0\\ 6 & 4} \right)\nn\\*
&\quad  - 60 \betasv{1\\6} \zeta_3 + \frac{15\zeta_3}{y}\betasv{2\\ 6} 
+ \frac{9\zeta_5}{y}\betasv{0\\4} - \frac{9\zeta_5}{4y^2} \betasv{1\\ 4}  -\frac{5\zeta_3 }{7} \betasv{1 \\ 4} - \frac{\zeta_5}{40} \,,\notag\\
\FFm{4}{2}{3}&= -30 \left(3\betasv{1& 2\\4& 6} +2 \betasv{2&  1\\4 & 6} -3 \betasv{2&  1\\6 & 4} -2\betasv{3&  0\\6 & 4} \right)  \label{odd1st}\\*
&\quad + 
 40 \zeta_3 \betasv{1\\6}  +  \frac{15\zeta_3}{y} \betasv{2\\6} - \frac{6\zeta_5}{y} \betasv{0\\4} - \frac{9\zeta_5}{4y^2}\betasv{1\\4} 
 +   \frac{\zeta_5}{360} \,,
\notag
\end{align}}%
where the terms beyond the reach of the substitution rule (\ref{bsvd1.5}) are
$-\frac{5\zeta_3 }{7} \betasv{1 \\ 4} - \frac{\zeta_5}{40} $ for $\FFm{2}{2}{3}$
and $ \frac{\zeta_5}{360} $ for $\FFm{4}{2}{3}$. The multiples of $\zeta_5$ are
examples of the non-positive powers of $y$ which are added by hand to the
$\beta^{\rm sv}$ representations of $\FFm{s}{m}{k}$. The other term
$\zeta_3\betasv{1 \\ 4}$ in the second line of (\ref{odd1st}) together with the 
antiholomorphic integration constants $\overline{\alpha[\ldots]}$ to be detailed 
in section \ref{sec:odd.bsvrep} conspire to the desired multiple of $\zeta_3 y^2$ 
and the odd quantity $\zeta_3 \EEodd_2$ 
in (\ref{oddsols}) to get a vanishing Laurent polynomial for $\FFm{2}{2}{3}$.
In the resulting Cauchy--Riemann equation
\beq
 (\pi \nabla)^2 \FFm{2}{2}{3}=
 \frac{1}{2} (\pi \nabla {\rm E}_2) (\pi \nabla {\rm E}_3) 
 - \frac{1}{2} {\rm E}_2 (\pi \nabla)^2 {\rm E}_3
 + (\Im \tau)^4 {\rm G}_4 \bigg( 3 {\rm E}_3 + \frac{5}{7} \zeta_3 \bigg)  \, ,
 \label{nabf223}
\eeq
the last term $\sim \zeta_3 ( \Im \tau)^4 {\rm G}_4$ corresponds to
$(\pi \nabla)^2 H^-$ in the notation of (\ref{eq:CRFm}).
The appearance of such extra terms is consistent with the quadruple lattice sum for $\FFm{2}{2}{3}$
in (\ref{Ftolatt.2}) while the simpler $\bsv$ representation of $\FFm{4}{2}{3}$ without
any analogue of $\EEodd_2$ lines up with the sum over three lattice momenta in (\ref{Ftolatt.2}).
Note that (\ref{odd1st}) reproduces the $\beta^{\rm sv}$ representations of the odd
MGFs ${\rm B}_{2,3}$ and $\mathcal{A}_{1,2;5}$ in \cite{Gerken:2020yii} through the dictionary (\ref{Ftolatt}).

\subsubsection{\texorpdfstring{Examples at weight $m{+}k=6$}{Examples at weight m+k=6}}

The next examples are
\begin{align}
\FFm{3}{2}{4} &= -420 \left(\betasv{1& 3\\ 4& 8} -\betasv{2 & 2 \\ 4 & 8} - \betasv{3& 1\\ 8& 4} + \betasv{4 &  0 \\8 & 4} \right)\nn\\
&\quad 
 + \frac{70\zeta_3}{y}\betasv{3\\ 8} 
 + \frac{15\zeta_7}{2y^2}\betasv{0\\ 4} 
 - \frac{15\zeta_7}{8y^3} \betasv{1\\4}  - 280 \zeta_3 \betasv{2\\ 8} 
  -\frac{7\zeta_3}{2}\betasv{2\\6}  - \frac{\zeta_7}{48 y}\,, \notag \\
  %%%
 \FFm{5}{2}{4} &= -420 \betasv{1 & 3\\4 & 8} - 315 \betasv{2 & 2\\4 & 8} + 
 420 \betasv{3&  1\\8 & 4} + 315 \betasv{4 & 0\\ 8 & 4} \label{odd4th}\\
 &\quad  + 
 210\zeta_3 \betasv{2\\8}  + \frac{70\zeta_3}{y} \betasv{3\\8}  - \frac{45\zeta_7}{8y^2}\betasv{0\\4}  -\frac{15\zeta_7}{8 y^3}\betasv{1\\4} + 
 \frac{\zeta_7}{288 y} \,.
\notag
\end{align}
The Laurent polynomials are adjusted to vanish by means of the last two terms $-\frac{7\zeta_3}{2}\betasv{2\\6}  - \frac{\zeta_7}{48 y}$ in $\FFm{3}{2}{4}$ and the last term $ \frac{\zeta_7}{288 y} $ in $\FFm{5}{2}{4}$. As detailed in section \ref{sec:comp.odd} all these functions given here are modular invariant and expressible in terms of 
the MGFs ${\rm B}_{2,4}$ and ${\rm B}'_{2,4}$
introduced in section 9.2 of \cite{Gerken:2020aju}
\beq
 \FFm{3}{2}{4}  =  \frac{1}{4}({\rm B}_{2,4} -\overline{{\rm B}}_{2,4} ) + \frac{ 7}{360}({\rm B}'_{2,4} -\overline{{\rm B}}'_{2,4} )
 \, , \ \ \ \ \ \ 
 \FFm{5}{2}{4}  = \frac{1}{4}({\rm B}_{2,4} -\overline{{\rm B}}_{2,4} )\, .
 \label{quadsum}
 \eeq
Upon reinstating their real parts $\mbox{Re} \, {\rm B}_{2,4} =-6\FFp{4}{2}{4}- \EE_2 \EE_4 $ and $\mbox{Re} \, {\rm B}_{2,4}' =180\FFp{4}{2}{4}-3 \zeta_3 \EE_3 $ \cite{Gerken:2020aju}, we obtain new $\beta^{\rm sv}$ representations
\begin{align}
{\rm B}_{2,4}' &=
37800 \betasv{2 & 2 \\ 4& 8} - 25200 \zeta_3 \betasv{2 \\ 8}
 -  \frac{ 9 \zeta_3 \zeta_5}{4 y^2} - \frac{ \zeta_7}{4 y} 
 + \frac{ 225 \zeta_3 \zeta_7}{16 y^4} - \frac{ 125  \zeta_9}{12 y^3}\,,
 \label{b24beta} \\
{\rm B}_{2,4} &= 
 -1680 \betasv{1& 3 \\ 4& 8} - 
 1260 \betasv{2& 2 \\ 4& 8} + 840 \zeta_3 \betasv{2 \\ 8}
  +  \frac{ 280 \zeta_3}{y} \betasv{3 \\8}  -  \frac{ \zeta_7}{180 y} - 
 \frac{35 \zeta_3 \zeta_7}{32 y^4} + \frac{ 25 \zeta_9}{72 y^3} \, .
 \notag
\end{align}
The lattice-sum representations of ${\rm B}_{2,4}$ and ${\rm B}_{2,4}'$ given in
section 9.2 of \cite{Gerken:2020aju} involve three and four lattice momenta, respectively.
The quadruple sum in ${\rm B}_{2,4}'$ only enters $ \FFm{3}{2}{4} $ (but not $ \FFm{5}{2}{4} $)
via (\ref{quadsum}) and can be viewed as triggering the term $\sim \zeta_3 \betasv{2\\6} $ in the second
line of (\ref{odd4th}) associated with the odd non-modular Laplace eigenfunction $\EEodd_3$.

\subsubsection{Higher weight}

Examples at weight $m{+}k=7,8$ are spelt out in appendix \ref{app:fminus}.
Starting from weight $m{+}k=7$, the odd Laplace eigenfunctions may involve
eigenvalues $s\geq 6$ associated with holomorphic cusp forms in the 
Cauchy--Riemann equation (\ref{eq:CRFm}). Indeed, the Poincar\'e series
$\FFm{6}{2}{5}$ and $\FFm{6}{3}{4}$ cannot be individually
identified with MGFs. The combinations of $\beta^{\rm sv}$ in
appendix \ref{app:fminus.1} instead refer to $\cFFm{6}{2}{5}$ 
and $\cFFm{6}{3}{4}$ and only the particular combination
\beq
 3\cFFm{6}{2}{5}+\cFFm{6}{3}{4} = 3\FFm{6}{2}{5}+\FFm{6}{3}{4}
\eeq
is modular invariant and expressible via MGFs. The individual modular
invariant completions $\FFm{6}{2}{5}$ and $\FFm{6}{3}{4}$ via primitives of the holomorphic
cusp form at weight 12 are discussed in Part~II.

All the $\FFm{s}{m}{k}$ at $m{+}k=8$ with $\beta^{\rm sv}$ representations in appendix \ref{app:fminus.2}
are MGFs whereas each weight $m{+}k\geq 9$ features at least one Laplace eigenvalue $s$ with
iterated integrals of holomorphic cusp forms in the modular invariant completion discussed in Part~II.
The $\beta^{\rm sv}$ representations of all the $\cFFm{s}{m}{k} $ up to and including
weight $m{+}k=14$ are given in an ancillary file in the arXiv submission and journal publication of this work.
Note that higher-weight examples starting from $\FFm{6}{2}{7}$ may have contributions 
from both $\EEodd_s$ and holomorphic cusp forms.

%%%%%%%%%%%%%%%%%%%%%%%%%%%%%%%%
\subsection{\texorpdfstring{Completing integration constants $\alpha$ at depth two}{Completing integration constants alpha at depth two}}
\label{sec:odd.bsvrep}
%%%%%%%%%%%%%%%%%%%%%%%%%%%%%%%%

The combined spectra of $\FFp{s}{ m}{ k}$ and $\FFm{s}{ m}{ k}$ in (\ref{eq:Fmk}) and (\ref{eq:Fmkm})
with given $m< k$ involve all eigenvalues $s\in\{k{-}m{+}1,k{-}m{+}2,\ldots,k{+}m{-}1\}$ with multiplicity one,
leading to the total number of $2m{-}1$ even or odd modular invariants. By imposing the reality properties 
$\overline{\FFpm{s}{ m}{ k}}=\pm \FFpm{s}{ m}{ k}$ on their $\beta^{\rm sv}$ sector $\cFFpm{s}{ m}{ k}$, 
we can solve for all the antiholomorphic integration constants 
$\overline{\alpha[\begin{smallmatrix}  j_1& j_2 \\ 2m& 2k \end{smallmatrix}]}=-
\overline{\alpha[\begin{smallmatrix}  j_2& j_1 \\ 2k& 2m \end{smallmatrix}]}$ at depth two
with $j_1{+}j_2 \geq m{+}k{-}2$, thereby filling some of the gaps in section \ref{sec:4.3}.

In the previous subsection, some of the odd functions $\FFm{s}{ m}{ k}$ were seen to 
feature odd Laplace eigenfunctions $\zeta_{m+k-s}\EEodd_s$, signalled by the need to 
cancel Laurent monomials $y^s$ via $\betasv{s-1\\2s}$. The antiholomorphic $\overline{ {\cal E}_0(2s,0^b)}$
in the expression (\ref{oddsols}) for these $\EEodd_s$ receive essential contributions from the $\overline{\alpha[\begin{smallmatrix}  j_1& j_2 \\ 2m& 2k \end{smallmatrix}]}$ which violate the pattern of the $\overline{\alpha_{m,k}^{N,j}}$ (\ref{oct24.22}) in the even case:
Every term in the expressions (\ref{oct24.26}) or (\ref{allalph.1}) for $\overline{\alpha_{m,k}^{N,j}}$ is of the form $ \zeta_{2m-1}\overline{ {\cal E}_0(2k,0^p)}$ or 
$ \zeta_{2k-1} \overline{{\cal E}_0(2m,0^p)}$ which clearly differ from the additional terms 
$\zeta_{m+k-s} \overline{{\cal E}_0(2s,0^p)}$ related to $\zeta_{m+k-s}\EEodd_s$ in the odd case. 
These additional terms occur for Laplace eigenvalue $s=k{-}m{+}1$ and thereby introduce
iterated Eisenstein integrals $\zeta_{m+k-s} \overline{{\cal E}_0(2s,0^p)}=
\zeta_{2m-1} \overline{{\cal E}_0(2k{-}2m{+}2,0^p)}$. The simplest examples
are the terms $ \zeta_3  \overline{ {\cal E}_0(4,0^p) }$ in
\beq
\overline{ \alpha[\begin{smallmatrix} 3&0 \\ 6&4 \end{smallmatrix}]  } 
= - \frac{ \zeta_3}{210} \overline{ {\cal E}_0(4,0) }
\, , \ \ \ \ \ \ 
\overline{ \alpha[\begin{smallmatrix} 4&0 \\ 6&4 \end{smallmatrix}]  }
= - \frac{ 2 \zeta_3}{105}  \overline{ {\cal E}_0(4,0^2) } + \frac{ 2 \zeta_5}{5} \overline{ {\cal E}_0(4) }
\label{exexotic}
\eeq
due to $\zeta_3\EEodd_2$ in $\FFm{2}{2}{ 3}$ since they deviate from the ingredients $ \zeta_3  \overline{ {\cal E}_0(6,0^p) }$ and $ \zeta_5  \overline{ {\cal E}_0(4,0^p) }$ of the
even cases $\overline{ \alpha^{N,0}_{2,3}}$ seen in (\ref{oct24.25}).

\subsubsection{\texorpdfstring{The missing cases with $j_1{+}j_2 < m{+}k{-}2$}{The missing cases with j1+j2 < m+k-2}}

As a next step, it remains to determine the
$\overline{\alpha[\begin{smallmatrix}  j_1& j_2 \\ 2m& 2k \end{smallmatrix}]}$ 
with $j_1{+}j_2 < m{+}k{-}2$. These only occur in the antiholomorphic
derivatives $(\pi \overline{\nabla} )^p \FFpm{s}{ m}{ k}$ with $0<p<s$. Since there is no
simple differential equation for $\overline{\nabla} $ action on the $\bsv$, the
antiholomorphic derivatives of $\FFpm{s}{ m}{ k}$ have to be determined
on the basis of the Laplace equations: For this purpose, 
we use a variant of~\eqref{eq:CRcom} that can also be viewed as the depth-two extension of (\ref{altrec.6}),
\begin{align}
\pi \nabla \bigg( \frac{ ( \pi \overline{\nabla} )^p \FFp{s}{ m}{ k}}{y^{2p}} \bigg) &= (s{-}p)(s{+}p{-}1) 
\frac{ ( \pi \overline{\nabla} )^{p-1} \FFp{s}{ m}{ k}}{y^{2p-2}} +  \frac{ ( \pi \overline{\nabla} )^{p-1}\Jp{0}{m}{k}}{y^{2p-2}} \,,
\notag\\
\pi \nabla \bigg( \frac{ ( \pi \overline{\nabla} )^p \FFm{s}{ m}{ k}}{y^{2p}} \bigg) &= (s{-}p)(s{+}p{-}1) 
\frac{ ( \pi \overline{\nabla} )^{p-1} \FFm{s}{ m}{ k}}{y^{2p-2}} +  \frac{ ( \pi \overline{\nabla} )^{p-1}\Jm{1}{m}{k}}{y^{2p-2}} \, .
\label{d2exten}
\end{align}
These equations serve as a recursion to determine higher antiholomorphic derivatives $\overline{\nabla} ^p\FFpm{s}{ m}{ k}$ from lower ones $\overline{\nabla}^{p-1}\FFpm{s}{ m}{ k}$ and 
the known $\beta^{\rm sv}$ representations of the derivatives of the sources.

Since we are here interested only in the $\bsv$ part that contains the integration constants, the following discussion is solely based on the $\bsv$-solutions $\cFFpm{s}{m}{k}$ to the differential equation and the fact that sometimes $\cFFpm{s}{m}{k}\neq \FFpm{s}{m}{k}$ does not affect the conclusions.
 We shall write $\FFpm{s}{m}{k}$ for simplicity even though the whole argument only relies on $\cFFpm{s}{m}{k}$.

In the first place,~\eqref{d2exten} only gives the $\nabla$-derivative of the initially
unknown $\big(\overline{\nabla}^p \FFpm{s}{ m}{ k}\big)/y^{2p}$ in terms of the $(p{-}1)$-th derivative. 
By making an ansatz for the depth-one and depth-two terms in $\big(\overline{\nabla}^p \FFpm{s}{ m}{ k}\big)/y^{2p}$ constructed out of $\bsv$ and products of derivatives of Eisenstein series, we can fix a solution at depth one and depth two if the right-hand side is known.
For $p=1$, we have construced $\bsv$-representatives of the right-hand sides of~\eqref{d2exten} in sections~\ref{sec:4} and~\ref{sec:odd.dpt1}, respectively, and in general we shall use~\eqref{d2exten} to determine the antiholomorphic derivatives iteratively.

The remaining information required for fixing a unique $\bsv$-representative of $\big(\overline{\nabla}^p \FFpm{s}{ m}{ k}\big)/y^{2p}$ at any step are the depth-zero terms that are fixed by the known Laurent polynomial and
\beq
( \pi \overline{\nabla} )^p \FFp{s}{ m}{ k} = 
( \pi  \nabla)^p \FFp{s}{ m}{ k}+O(q,\bar q)\, , \ \ \ \ \ \
( \pi \overline{\nabla} )^p \FFm{s}{ m}{ k} = O(q,\bar q) \, ,
\label{misslaurent}
\eeq
see (\ref{eq:FLP}) for the Laurent polynomials of $\FFp{s}{ m}{ k} $ while those of $\FFm{s}{m}{k}$ vanish.

By equating the full $\bsv$ representations of
\beq
\label{eq:ccF}
\overline{\nabla}^p\FFpm{s}{ m}{ k} = \pm \overline{ \nabla^p\FFpm{s}{ m}{ k} }\, ,
\eeq
one can determine the complex-conjugation properties of the depth-two $\bsv$ occurring in this relation. Considering this equation for fixed $m\leq k$ but all possible values of $s$ and $0\leq p<s$ shows that all $\betasv{j_1&j_2\\2m&2k}$ with $j_1{+}j_2\leq m{+}k{-}2$ occur and their exact complex-conjugation properties are then fixed by~\eqref{eq:ccF}. Since complex conjugation reflects the range of $(j_1,j_2)$, see~\eqref{eq:bsv2prop}, and squares to one, we can use this to determine the exact complex-conjugation properties of all $\bsv$ for fixed $m$ and $k$. The resulting formul\ae{} for the complex conjugation of the $\bsv$ are presented up to $k_1{+}k_2\leq 28$ in the ancillary file.

By comparing these complex-conjugation properties with the abstract~\eqref{eq:ccbsv2}, we can then fix  
the antiholomorphic $\overline{\alpha[\ldots]}$ in (\ref{eq:bsv2}). 
A variety of representative examples of $\nabla^p\FFpm{s}{ m}{ k}$ and $\overline \nabla^p\FFpm{s}{ m}{ k}$
can be found in appendix \ref{app:CRderiv}, and all Cauchy--Riemann derivatives for $m{+}k\leq 14$ are given in the ancillary file. We note that for $p\geq m$, antiholomorphic Eisenstein series and their derivatives can appear explicitly in (\ref{eq:ccF}).

This strategy gives access to all the $\overline{\alpha[\begin{smallmatrix}  j_1& j_2 \\ 2m& 2k \end{smallmatrix}]}$
at depth two, for any $2\leq m\leq k$ as well as $0\leq j_1\leq 2m{-}2$ and $0\leq j_2\leq 2k{-}2$. The case with
$m=k$ is considerably simpler than the generic one with $m<k$, and the associated integration constants have already been determined in section \ref{sec:4.3.1}. The complete set of such $\overline{\alpha[\begin{smallmatrix}  j_1& j_2 \\ 2m& 2k \end{smallmatrix}]}$ with $m{+}k\leq 14$ and $m\leq k$ can be found as an ancillary file which also repeats the cases with $m{+}k\leq 6$ from the arXiv submission of \cite{Gerken:2020xfv} for completeness.

\subsubsection{\texorpdfstring{A conjectural pattern among the $\alpha$}{A conjectural pattern among the alpha}}

Investigating the outcome of the above algorithm leads to the following conjectural identity
\beq
-\frac{\pi \overline{ \nabla } }{4y^2} \overline{ \alpha[\begin{smallmatrix} j_1& j_2 \\ k_1& k_2 \end{smallmatrix}]  }
= j_1 \, \overline{ \alpha[\begin{smallmatrix} j_1{-}1& j_2 \\ k_1& k_2 \end{smallmatrix}]  }
+ j_2\, \overline{ \alpha[\begin{smallmatrix} j_1& j_2{-}1 \\ k_1& k_2 \end{smallmatrix}]  }
\ {\rm mod} \ \overline{{\rm G}^0_k}
\label{modG0k}
\eeq
that we have checked for $k_1{+}k_2\leq 28$ and all admissible values of $j_1,j_2$. The differential 
operator on the left-hand side simply removes the terminal zero of
\beq
-\frac{\pi \overline{ \nabla } }{4y^2} \,  \overline{ {\cal E}_0(2k,0^p) } = \overline{ {\cal E}_0(2k,0^{p-1}) } \, , \ \ \ \ \ \ 
p \geq 1 
\eeq
and acts for $p=0$ as
\begin{align}
-\frac{\pi \overline{ \nabla } }{4y^2} \,  \overline{ {\cal E}_0(2k) } = -\frac{\overline{\GG_{2k}^0}}{(2\pi i)^{2k}} \,.
\end{align}
In order not to keep track of the antiholomorphic Eisenstein series in $\overline{\GG_{2k}^0}$ we have added 
the disclaimer ${\rm mod} \ \overline{{\rm G}^0_k}$ in (\ref{modG0k}). For instance, this amounts to
dropping the last term $\sim \overline{ {\rm G}_4^0}$ in
\beq
-\frac{\pi \overline{ \nabla } }{4y^2} \overline{ \alpha[\begin{smallmatrix} 4&0 \\ 6&4 \end{smallmatrix}]  }
= 4 \overline{ \alpha[\begin{smallmatrix} 3&0 \\ 6&4 \end{smallmatrix}]  } - \frac{ 2 \zeta_5}{5} \frac{\overline{{\rm G}_4^0}}{(2\pi i)^4}\, ,
\eeq
see (\ref{exexotic}) for the $\overline{\alpha[\ldots]}$ on both sides.
We expect a simple explanation of the observation (\ref{modG0k}) once the $\bsv$ and
their integration constants are related to Brown's single-valued iterated Eisenstein 
integrals \cite{Brown:mmv, Brown:I, Brown:II}.

In fact, the conjectural identity~\eqref{modG0k} can be exploited to generate the missing cases of $\overline{\alpha[\begin{smallmatrix}  j_1& j_2 \\ 2m& 2k \end{smallmatrix}]}$ 
with $j_1{+}j_2 < m{+}k{-}2$. As the differential operator lowers the $j$-labels on  $\overline{\alpha[\ldots]}$ we can determine lower labels from higher labels ${\rm mod} \ \overline{{\rm G}^0_k}$. Since all of their instances with higher labels $j_1{+}j_2 \geq m{+}k{-}2$ are fixed by the reality properties of $\FFpm{s}{m}{k}$, one can take advantage of (\ref{modG0k}) to bypass the investigation of the reality property~\eqref{eq:ccF} for $p>0$ which requires constructing the Cauchy--Riemann derivatives of $\FFpm{s}{m}{k}$ first. Note that the $\overline{\alpha[\ldots]}$ never contain unintegrated $\overline{{\rm G}^0_k}$, so the conjectural identity~\eqref{modG0k} fixes them completely in terms of $\overline{ {\cal E}_0(2k,0^p)}$.

%%%%%%%%%%%%%%%%%%%%%%%%%%%%%%%%%%%%%%%%%%%%%%%%%%%%%%%%%%%
\section{Exhausting the seed functions}
\label{sec:4alt}
%%%%%%%%%%%%%%%%%%%%%%%%%%%%%%%%%%%%%%%%%%%%%%%%%%%%%%%%%%%

In section \ref{sec:3}, we have constructed Poincar\'e seed functions for the
$\FFpm{s}{m}{k}$ by folding the Eisenstein series, i.e.\ replacing it by its Poincar\'e series~\eqref{eq:PSEk}, of higher weight $k\geq m$ in
the inhomogeneous term $\Jp{0}{m}{k}$ and $\Jm{1}{m}{k}$ of the Laplace equation (\ref{eq:Fmk}) and~\eqref{eq:Fmkm}, respectively. 
This choice of folding leads to convergent Poincar\'e sums, and we have given
similar seeds for the other $\Jpm{\ell}{m}{k}$ at higher values of $\ell$.
When consistently folding the Eisenstein series
of higher weight in the even case, the resulting seed functions 
(\ref{eq:genseed}) and (\ref{eqseedR}) turn out to exhaust the $y^a \Re[{\cal E}_0(2m,0^b)]$
with $a{+}b \geq 2m{-}1$ and $a\geq1,\ b\geq 0$, see (\ref{oct21.5}) and (\ref{oct21.6}) for the accompanying term $\sim y^{k+m}$. A similar statement holds for the odd case and seeds of the form $y^a \Im[{\cal E}_0(2m,0^b)]$, see~(\ref{eq:seedm.A}) and \eqref{eq:seedm}.

In this section, we will discuss the role of certain $y^a \Re[{\cal E}_0(2m,0^b)]$ and $y^a \Im[{\cal E}_0(2m,0^b)]$
with $a\geq1,\ b\geq 0$ but $ a{+}b < 2m{-}1$ as alternative seed functions for 
$\FFpm{s}{m}{k}$ and $\Jpm{\ell}{m}{k}$ with $m\neq k$. These cases arise from
folding the Eisenstein series of {\it lower} weight $m<k$ in the source term 
of the Laplace equations (\ref{eq:Fmk}) and~\eqref{eq:Fmkm} for $\FFpm{s}{m}{k}$.

%%%%%%%%%%%%%%%%%%%%%%%%%%%%%%%%%%%%%%%%%%%%%%%%%%%%%%%%%%%
\subsection{Overview of seed functions with convergent Poincar\'e sums}
\label{sec:fold.1}
%%%%%%%%%%%%%%%%%%%%%%%%%%%%%%%%%%%%%%%%%%%%%%%%%%%%%%%%%%%

Tables \ref{chess.1} to \ref{chess.3} below give samples of the leading-depth terms $\betalagpm{j}$ in (\ref{eq:betalag}) that were found to arise from seed functions $y^a \Re[{\cal E}_0(2m,0^b)]$ and $y^a \Im[{\cal E}_0(2m,0^b)]$. The (red) crosses in the tables refer to cases that cannot be covered by $\betalagpm{j}$ and we shall comment on what kind of modular objects these are as well as on their more general seeds in section~\ref{sec:5.3}.

\begin{table}[h]
\begin{center}
\tikzpicture
\draw(-2.5,-2.1)node{\underline{$m=2$:}};
\draw(-0.04,1) -- (-0.04,-4.9);
\draw(0.04,1) -- (0.04,-4.9);
\draw(-1,0.04) -- (5.8,0.04);
\draw(-1,-0.04) -- (5.8,-0.04);
\draw(-0.04,0.04) -- (-1,1);
\draw(-0.75,0.3) node{$a$};
\draw(-0.3,0.75) node{$b$};
\draw(1.6,1.1) -- (1.6,-4.9);
\draw(3.2,1.1) -- (3.2,-4.9);
\draw(4.8,1.1) -- (4.8,-4.9);
\draw(-1,-0.7) -- (5.8,-0.7);
\draw(-1,-1.4) -- (5.8,-1.4);
\draw(-1,-2.1) -- (5.8,-2.1);
\draw(-1,-2.8) -- (5.8,-2.8);
\draw(-1,-3.5) -- (5.8,-3.5);
\draw(-1,-4.2) -- (5.8,-4.2);
\draw(0.8,0.5)node{0};
\draw(2.4,0.5)node{1};
\draw(4.0,0.5)node{2};
\draw(5.3,0.5)node{3};
\draw(-0.5,-0.35)node{1};
\draw(-0.5,-1.05)node{2};
\draw(-0.5,-1.75)node{3};
\draw(-0.5,-2.45)node{4};
\draw(-0.5,-3.15)node{5};
\draw(-0.5,-3.85)node{6};
\draw(-0.5,-4.55)node{7};
% 1st beta^svs %%
\draw(0.8,-0.35)node{$\times$};
\draw(0.8,-1.05)node{$\rcross$};
\draw(0.8,-1.75)node{$\Big.\beta^{{\rm sv},\, j\leq 2}_{2,2} \Big.$};
\draw(0.8,-2.45)node{$\beta^{{\rm sv},\, j\leq 2}_{2,3} \Big.$};
\draw(0.8,-3.15)node{$\beta^{{\rm sv},\, j\leq 2}_{2,4} \Big.$};
\draw(0.8,-3.85)node{$\beta^{{\rm sv},\, j\leq 2}_{2,5} \Big.$};
\draw(0.8,-4.55)node{$\beta^{{\rm sv},\, j\leq 2}_{2,6} \Big.$};
% 2nd beta^svs
\draw(2.4,-0.35)node{$\rcross$};
\draw(2.4,-1.05)node{$\beta^{{\rm sv},\, j\leq 1}_{2,2} \Big.$};
\draw(2.4,-1.75)node{$\beta^{{\rm sv},\, j\leq 1}_{2,3}$};
\draw(2.4,-2.45)node{$\beta^{{\rm sv},\, j\leq 1}_{2,4}$};
\draw(2.4,-3.15)node{$\beta^{{\rm sv},\, j\leq 1}_{2,5}$};
\draw(2.4,-3.85)node{$\beta^{{\rm sv},\, j\leq 1}_{2,6}$};
\draw(2.4,-4.55)node{$\beta^{{\rm sv},\, j\leq 1}_{2,7}$};
% 3rd beta^svs
\draw(4.0,-0.35)node{$\beta^{{\rm sv},\, j=0}_{2,2} \Big.$};
\draw(4.0,-1.05)node{$\beta^{{\rm sv},\, j=0}_{2,3}$};
\draw(4.0,-1.75)node{$\beta^{{\rm sv},\, j=0}_{2,4}$};
\draw(4.0,-2.45)node{$\beta^{{\rm sv},\, j=0}_{2,5}$};
\draw(4.0,-3.15)node{$\beta^{{\rm sv},\, j=0}_{2,6}$};
\draw(4.0,-3.85)node{$\beta^{{\rm sv},\, j=0}_{2,7}$};
\draw(4.0,-4.55)node{$\beta^{{\rm sv},\, j=0}_{2,8}$};
\draw(5.3,-0.35)node{$\times$};
\draw(5.3,-1.05)node{$\times$};
\draw(5.3,-1.75)node{$\times$};
\draw(5.3,-2.45)node{$\times$};
\draw(5.3,-3.15)node{$\times$};
\draw(5.3,-3.85)node{$\times$};
\draw(5.3,-4.55)node{$\times$};
\draw[blue](3.2,-0.7)node{$\swarrow$}node[above]{\footnotesize${\cal O}_1 \ \ \ \ $};
\draw[blue](1.6,-1.4)node{$\swarrow$}node[above]{\footnotesize${\cal O}_2 \ \ \ \ $};
\draw[blue](3.2,-1.4)node{$\swarrow$}node[above]{\footnotesize${\cal O}_2 \ \ \ \ $};
\draw[blue](1.6,-2.1)node{$\swarrow$}node[above]{\footnotesize${\cal O}_3 \ \ \ \ $};
\draw[blue](3.2,-2.1)node{$\swarrow$}node[above]{\footnotesize${\cal O}_3 \ \ \ \ $};
\draw[blue](1.6,-2.8)node{$\swarrow$}node[above]{\footnotesize${\cal O}_4 \ \ \ \ $};
\draw[blue](3.2,-2.8)node{$\swarrow$}node[above]{\footnotesize$\ldots \ \ \ \ $};
\draw[blue](1.6,-3.5)node{$\swarrow$}node[above]{\footnotesize$\ldots \ \ \ \ $};
\endtikzpicture
\end{center}
\caption{\textit{Leading-depth terms $\beta^{{\rm sv},\, j}_{2,k} $ obtained from Poincar\'e sums
over $y^a  {\cal E}_0(4,0^b)$, together with the action of the operators $\mathcal{O}_\ell$ defined in~\eqref{oct21.3}.}}
\label{chess.1}
\end{table}

The $(m,k)$ sectors of iterated Eisenstein integrals over ${\rm G}_{2m}$ and
 ${\rm G}_{2k}$ are spread out across the diagonals $a{+}b=m{+}k{-}1 \geq 2m{-}1$ of 
 the tables. For instance, the $\beta^{{\rm sv}}_{2,3} $
 referring to double integrals over ${\rm G}_4$ and ${\rm G}_6$ as in (\ref{eq:betalag})
 cover the diagonal with $a{+}b=4$ in table \ref{chess.1}, bounded by $0\leq b \leq 2$.
The inequalities in the superscripts of $\beta^{{\rm sv},\, j\leq 1}_{2,2} $ or
$\beta^{{\rm sv},\, j\leq 2}_{2,3}$ in table \ref{chess.1} are a shorthand for the specific linear combinations
of $\FFpm{s}{m}{k}$ in (\ref{oct21.5}) that are generated by the Poincar\'e sums over the $y^a  {\cal E}_0(4,0^b)$
in question. From the step forms for even Poincar\'e seeds in (\ref{oct20.15}) and (\ref{oct20.17}), for instance, we can read off
\begin{align}
\beta^{{\rm sv},\, j\leq 1}_{2,2}&\rightarrow - \frac{15}{4} (2 \FFp{2}{2}{2} + \Jp{0}{2}{2}) = - \frac{135}{2} 
(\beta^{{\rm sv}+,\, 0}_{2,2}  + 2 \beta^{{\rm sv}+,\,   1}_{2,2} ) \MLD \,, \\
\beta^{{\rm sv},\, j\leq 2}_{2,3}&\rightarrow \frac{105}{16} (3\Jp{0}{2}{3}+ \Jp{1}{2}{3}) =  \frac{4725}{4} 
(\beta^{{\rm sv}+,\, 0}_{2,3}  + 3 \beta^{{\rm sv}+,\,   1}_{2,3} + \beta^{{\rm sv}+,\,   2}_{2,3} ) \MLD 
\notag
\end{align}
and their counterparts in the odd case follow from step forms as in (\ref{exminus.5}) and (\ref{exminus.6}).

The same notation applies to tables \ref{chess.2} and \ref{chess.3} to indicate the
schematic form of the Poincar\'e sums over $y^a  {\cal E}_0(6,0^b)$
and $y^a {\cal E}_0(8,0^b)$, leading to $\beta^{\rm sv}_{3,k}$ and $\beta^{\rm sv}_{4,k}$, 
respectively. In the following subsections, we will
discuss the fields marked by $?$ in the diagonal $a{+}b=4$ of table \ref{chess.2} and 
the two diagonals $a{+}b=5,6$ of table \ref{chess.3} for which we have not yet spelt out 
a Poincar\'e series. The black and red cross signposts will be discussed in sections~\ref{sec:5.2} and~\ref{sec:5.3}, respectively.

\begin{table}[h]
\begin{center}
\tikzpicture
\draw(-2.5,-2.1)node{\underline{$m=3$:}};
\draw(-0.04,1) -- (-0.04,-4.9);
\draw(0.04,1) -- (0.04,-4.9);
\draw(-1,0.04) -- (9,0.04);
\draw(-1,-0.04) -- (9,-0.04);
\draw(-0.04,0.04) -- (-1,1);
\draw(-0.75,0.3) node{$a$};
\draw(-0.3,0.75) node{$b$};
\draw(1.6,1.1) -- (1.6,-4.9);
\draw(3.2,1.1) -- (3.2,-4.9);
\draw(4.8,1.1) -- (4.8,-4.9);
\draw(6.4,1.1) -- (6.4,-4.9);
\draw(8,1.1) -- (8,-4.9);
\draw(-1,-0.7) -- (9.0,-0.7);
\draw(-1,-1.4) -- (9.0,-1.4);
\draw(-1,-2.1) -- (9.0,-2.1);
\draw(-1,-2.8) -- (9.0,-2.8);
\draw(-1,-3.5) -- (9.0,-3.5);
\draw(-1,-4.2) -- (9.0,-4.2);
\draw(0.8,0.5)node{0};
\draw(2.4,0.5)node{1};
\draw(4.0,0.5)node{2};
\draw(5.6,0.5)node{3};
\draw(7.2,0.5)node{4};
\draw(8.5,0.5)node{5};
\draw(-0.5,-0.35)node{1};
\draw(-0.5,-1.05)node{2};
\draw(-0.5,-1.75)node{3};
\draw(-0.5,-2.45)node{4};
\draw(-0.5,-3.15)node{5};
\draw(-0.5,-3.85)node{6};
\draw(-0.5,-4.55)node{7};
% 1st beta^svs %% INSERT STUFF FROM HERE
\draw(0.8,-0.35)node{$\times$};
\draw(0.8,-1.05)node{$\times$};
\draw(0.8,-1.75)node{$\rcross$};
\draw(0.8,-2.45)node{?};
\draw(0.8,-3.15)node{$\beta^{{\rm sv},\, j\leq 4}_{3,3} \Big.$};
\draw(0.8,-3.85)node{$\beta^{{\rm sv},\, j\leq 4}_{3,4} \Big.$};
\draw(0.8,-4.55)node{$\beta^{{\rm sv},\, j\leq 4}_{3,5} \Big.$};
% 2nd beta^svs
\draw(2.4,-0.35)node{$\times$};
\draw(2.4,-1.05)node{$\rcross$};
\draw(2.4,-1.75)node{?};
\draw(2.4,-2.45)node{$\beta^{{\rm sv},\, j\leq 3}_{3,3} \Big.$};
\draw(2.4,-3.15)node{$\beta^{{\rm sv},\, j\leq 3}_{3,4} \Big.$};
\draw(2.4,-3.85)node{$\beta^{{\rm sv},\, j\leq 3}_{3,5} \Big.$};
\draw(2.4,-4.55)node{$\beta^{{\rm sv},\, j\leq 3}_{3,6} \Big.$};
% 3rd beta^svs
\draw(4.0,-0.35)node{$\rcross$};
\draw(4.0,-1.05)node{?};
\draw(4.0,-1.75)node{$\beta^{{\rm sv},\, j\leq 2}_{3,3} \Big.$};
\draw(4.0,-2.45)node{$\beta^{{\rm sv},\, j\leq 2}_{3,4} \Big.$};
\draw(4.0,-3.15)node{$\beta^{{\rm sv},\, j\leq 2}_{3,5} \Big.$};
\draw(4.0,-3.85)node{$\beta^{{\rm sv},\, j\leq 2}_{3,6} \Big.$};
\draw(4.0,-4.55)node{$\beta^{{\rm sv},\, j\leq 2}_{3,7} \Big.$};
\draw(5.6,-0.35)node{?};
\draw(5.6,-1.05)node{$\beta^{{\rm sv},\, j\leq 1}_{3,3} \Big.$};
\draw(5.6,-1.75)node{$\beta^{{\rm sv},\, j\leq 1}_{3,4} \Big.$};
\draw(5.6,-2.45)node{$\beta^{{\rm sv},\, j\leq 1}_{3,5} \Big.$};
\draw(5.6,-3.15)node{$\beta^{{\rm sv},\, j\leq 1}_{3,6} \Big.$};
\draw(5.6,-3.85)node{$\beta^{{\rm sv},\, j\leq 1}_{3,7} \Big.$};
\draw(5.6,-4.55)node{$\beta^{{\rm sv},\, j\leq 1}_{3,8} \Big.$};
\draw(7.2,-0.35)node{$\beta^{{\rm sv},\, j=0}_{3,3} \Big.$};
\draw(7.2,-1.05)node{$\beta^{{\rm sv},\, j=0}_{3,4} \Big.$};
\draw(7.2,-1.75)node{$\beta^{{\rm sv},\, j=0}_{3,5} \Big.$};
\draw(7.2,-2.45)node{$\beta^{{\rm sv},\, j=0}_{3,6} \Big.$};
\draw(7.2,-3.15)node{$\beta^{{\rm sv},\, j=0}_{3,7} \Big.$};
\draw(7.2,-3.85)node{$\beta^{{\rm sv},\, j=0}_{3,8} \Big.$};
\draw(7.2,-4.55)node{$\beta^{{\rm sv},\, j=0}_{3,9} \Big.$};
% END OF INSERT STUFF
\draw(8.5,-0.35)node{$\times$};
\draw(8.5,-1.05)node{$\times$};
\draw(8.5,-1.75)node{$\times$};
\draw(8.5,-2.45)node{$\times$};
\draw(8.5,-3.15)node{$\times$};
\draw(8.5,-3.85)node{$\times$};
\draw(8.5,-4.55)node{$\times$};
\draw[blue](6.4,-0.7)node{$\swarrow$}node[above]{\footnotesize${\cal O}_1 \ \ \ \ $};
\draw[blue](4.8,-1.4)node{$\swarrow$}node[above]{\footnotesize${\cal O}_2 \ \ \ \ $};
\draw[blue](3.2,-2.1)node{$\swarrow$}node[above]{\footnotesize${\cal O}_3 \ \ \ \ $};
\draw[blue](1.6,-2.8)node{$\swarrow$}node[above]{\footnotesize${\cal O}_4 \ \ \ \ $};
\draw[blue](6.4,-1.4)node{$\swarrow$}node[above]{\footnotesize${\cal O}_2 \ \ \ \ $};
\draw[blue](4.8,-2.1)node{$\swarrow$}node[above]{\footnotesize${\cal O}_3 \ \ \ \ $};
\draw[blue](3.2,-2.8)node{$\swarrow$}node[above]{\footnotesize${\cal O}_4 \ \ \ \ $};
\draw[blue](1.6,-3.5)node{$\swarrow$}node[above]{\footnotesize${\cal O}_5 \ \ \ \ $};
\draw[blue](6.4,-2.1)node{$\swarrow$}node[above]{\footnotesize${\cal O}_3 \ \ \ \ $};
\draw[blue](4.8,-2.8)node{$\swarrow$}node[above]{\footnotesize${\cal O}_4 \ \ \ \ $};
\draw[blue](3.2,-3.5)node{$\swarrow$}node[above]{\footnotesize${\cal O}_5 \ \ \ \ $};
\draw[blue](1.6,-4.2)node{$\swarrow$}node[above]{\footnotesize${\cal O}_6 \ \ \ \ $};
\draw[blue](6.4,-2.8)node{$\swarrow$}node[above]{\footnotesize$\ldots \ \ \ \ $};
\draw[blue](4.8,-3.5)node{$\swarrow$}node[above]{\footnotesize$\ldots \ \ \ \ $};
\draw[blue](3.2,-4.2)node{$\swarrow$}node[above]{\footnotesize$\ldots \ \ \ \ $};
\endtikzpicture
\end{center}
\caption{\textit{Leading-depth terms $\beta^{{\rm sv},\, j}_{3,k} $ obtained from Poincar\'e sums
over $y^a {\cal E}_0(6,0^b)$, together with the action of the operators $\mathcal{O}_\ell$ defined in~\eqref{oct21.3}.}}
\label{chess.2}
\end{table}
\begin{table}[h]
\begin{center}
\tikzpicture
\draw(-2.5,-2.1)node{\underline{$m=4$:}};
\draw(-0.04,1) -- (-0.04,-5.6);
\draw(0.04,1) -- (0.04,-5.6);
\draw(-1,0.04) -- (12.2,0.04);
\draw(-1,-0.04) -- (12.2,-0.04);
\draw(-0.04,0.04) -- (-1,1);
\draw(-0.75,0.3) node{$a$};
\draw(-0.3,0.75) node{$b$};
\draw(1.6,1.1) -- (1.6,-5.6);
\draw(3.2,1.1) -- (3.2,-5.6);
\draw(4.8,1.1) -- (4.8,-5.6);
\draw(6.4,1.1) -- (6.4,-5.6);
\draw(8,1.1) -- (8,-5.6);
\draw(9.6,1.1) -- (9.6,-5.6);
\draw(11.2,1.1) -- (11.2,-5.6);
\draw(-1,-0.7) -- (12.2,-0.7);
\draw(-1,-1.4) -- (12.2,-1.4);
\draw(-1,-2.1) -- (12.2,-2.1);
\draw(-1,-2.8) -- (12.2,-2.8);
\draw(-1,-3.5) -- (12.2,-3.5);
\draw(-1,-4.2) -- (12.2,-4.2);
\draw(-1,-4.9) -- (12.2,-4.9);
\draw(0.8,0.5)node{0};
\draw(2.4,0.5)node{1};
\draw(4.0,0.5)node{2};
\draw(5.6,0.5)node{3};
\draw(7.2,0.5)node{4};
\draw(8.8,0.5)node{5};
\draw(10.4,0.5)node{6};
\draw(11.7,0.5)node{7};
\draw(-0.5,-0.35)node{1};
\draw(-0.5,-1.05)node{2};
\draw(-0.5,-1.75)node{3};
\draw(-0.5,-2.45)node{4};
\draw(-0.5,-3.15)node{5};
\draw(-0.5,-3.85)node{6};
\draw(-0.5,-4.55)node{7};
\draw(-0.5,-5.25)node{8};
% 1st beta^svs %% INSERT STUFF FROM HERE
\draw(0.8,-0.35)node{$\times$};
\draw(0.8,-1.05)node{$\times$};
\draw(0.8,-1.75)node{$\times$};
\draw(0.8,-2.45)node{$\rcross$};
\draw(0.8,-3.15)node{?};
\draw(0.8,-3.85)node{?};
\draw(0.8,-4.55)node{$\beta^{{\rm sv},\, j\leq 6}_{4,4} \Big.$};
\draw(0.8,-5.25)node{$\beta^{{\rm sv},\, j\leq 6}_{4,5} \Big.$};
% 2nd beta^svs
\draw(2.4,-0.35)node{$\times$};
\draw(2.4,-1.05)node{$\times$};
\draw(2.4,-1.75)node{$\rcross$};
\draw(2.4,-2.45)node{?};
\draw(2.4,-3.15)node{?};
\draw(2.4,-3.85)node{$\beta^{{\rm sv},\, j\leq 5}_{4,4} \Big.$};
\draw(2.4,-4.55)node{$\beta^{{\rm sv},\, j\leq 5}_{4,5} \Big.$};
\draw(2.4,-5.25)node{$\beta^{{\rm sv},\, j\leq 5}_{4,6} \Big.$};
% 3rd beta^svs
\draw(4.0,-0.35)node{$\times$};
\draw(4.0,-1.05)node{$\rcross$};
\draw(4.0,-1.75)node{?};
\draw(4.0,-2.45)node{?};
\draw(4.0,-3.15)node{$\beta^{{\rm sv},\, j\leq 4}_{4,4} \Big.$};
\draw(4.0,-3.85)node{$\beta^{{\rm sv},\, j\leq 4}_{4,5} \Big.$};
\draw(4.0,-4.55)node{$\beta^{{\rm sv},\, j\leq 4}_{4,6} \Big.$};
\draw(4.0,-5.25)node{$\beta^{{\rm sv},\, j\leq 4}_{4,7} \Big.$};
% 4th beta^svs
\draw(5.6,-0.35)node{$\rcross$};
\draw(5.6,-1.05)node{?};
\draw(5.6,-1.75)node{?};
\draw(5.6,-2.45)node{$\beta^{{\rm sv},\, j\leq 3}_{4,4} \Big.$};
\draw(5.6,-3.15)node{$\beta^{{\rm sv},\, j\leq 3}_{4,5} \Big.$};
\draw(5.6,-3.85)node{$\beta^{{\rm sv},\, j\leq 3}_{4,6} \Big.$};
\draw(5.6,-4.55)node{$\beta^{{\rm sv},\, j\leq 3}_{4,7} \Big.$};
\draw(5.6,-5.25)node{$\beta^{{\rm sv},\, j\leq 3}_{4,8} \Big.$};
% 5th beta^svs
\draw(7.2,-0.35)node{?};
\draw(7.2,-1.05)node{?};
\draw(7.2,-1.75)node{$\beta^{{\rm sv},\, j\leq 2}_{4,4} \Big.$};
\draw(7.2,-2.45)node{$\beta^{{\rm sv},\, j\leq 2}_{4,5} \Big.$};
\draw(7.2,-3.15)node{$\beta^{{\rm sv},\, j\leq 2}_{4,6} \Big.$};
\draw(7.2,-3.85)node{$\beta^{{\rm sv},\, j\leq 2}_{4,7} \Big.$};
\draw(7.2,-4.55)node{$\beta^{{\rm sv},\, j\leq 2}_{4,8} \Big.$};
\draw(7.2,-5.25)node{$\beta^{{\rm sv},\, j\leq 2}_{4,9} \Big.$};
% 6th beta^svs
\draw(8.8,-0.35)node{?};
\draw(8.8,-1.05)node{$\beta^{{\rm sv},\, j\leq 1}_{4,4} \Big.$};
\draw(8.8,-1.75)node{$\beta^{{\rm sv},\, j\leq 1}_{4,5} \Big.$};
\draw(8.8,-2.45)node{$\beta^{{\rm sv},\, j\leq 1}_{4,6} \Big.$};
\draw(8.8,-3.15)node{$\beta^{{\rm sv},\, j\leq 1}_{4,7} \Big.$};
\draw(8.8,-3.85)node{$\beta^{{\rm sv},\, j\leq 1}_{4,8} \Big.$};
\draw(8.8,-4.55)node{$\beta^{{\rm sv},\, j\leq 1}_{4,9} \Big.$};
\draw(8.8,-5.25)node{$\beta^{{\rm sv},\, j\leq 1}_{4,10} \Big.$};
% 7th beta^svs
\draw(10.4,-0.35)node{$\beta^{{\rm sv},\, j=0}_{4,4} \Big.$};
\draw(10.4,-1.05)node{$\beta^{{\rm sv},\, j=0}_{4,5} \Big.$};
\draw(10.4,-1.75)node{$\beta^{{\rm sv},\, j=0}_{4,6} \Big.$};
\draw(10.4,-2.45)node{$\beta^{{\rm sv},\, j=0}_{4,7} \Big.$};
\draw(10.4,-3.15)node{$\beta^{{\rm sv},\, j=0}_{4,8} \Big.$};
\draw(10.4,-3.85)node{$\beta^{{\rm sv},\, j=0}_{4,9} \Big.$};
\draw(10.4,-4.55)node{$\beta^{{\rm sv},\, j=0}_{4,10} \Big.$};
\draw(10.4,-5.25)node{$\beta^{{\rm sv},\, j=0}_{4,11} \Big.$};
% END OF INSERT STUFF
\draw(11.7,-0.35)node{$\times$};
\draw(11.7,-1.05)node{$\times$};
\draw(11.7,-1.75)node{$\times$};
\draw(11.7,-2.45)node{$\times$};
\draw(11.7,-3.15)node{$\times$};
\draw(11.7,-3.85)node{$\times$};
\draw(11.7,-4.55)node{$\times$};
\draw(11.7,-5.25)node{$\times$};
\draw[blue](9.6,-0.7)node{$\swarrow$}node[above]{\footnotesize${\cal O}_1 \ \ \ \ $};
\draw[blue](8.0,-1.4)node{$\swarrow$}node[above]{\footnotesize${\cal O}_2 \ \ \ \ $};
\draw[blue](6.4,-2.1)node{$\swarrow$}node[above]{\footnotesize${\cal O}_3 \ \ \ \ $};
\draw[blue](4.8,-2.8)node{$\swarrow$}node[above]{\footnotesize${\cal O}_4 \ \ \ \ $};
\draw[blue](3.2,-3.5)node{$\swarrow$}node[above]{\footnotesize${\cal O}_5 \ \ \ \ $};
\draw[blue](1.6,-4.2)node{$\swarrow$}node[above]{\footnotesize${\cal O}_6 \ \ \ \ $};
\draw[blue](9.6,-1.4)node{$\swarrow$}node[above]{\footnotesize${\cal O}_2 \ \ \ \ $};
\draw[blue](8.0,-2.1)node{$\swarrow$}node[above]{\footnotesize${\cal O}_3 \ \ \ \ $};
\draw[blue](6.4,-2.8)node{$\swarrow$}node[above]{\footnotesize${\cal O}_4 \ \ \ \ $};
\draw[blue](4.8,-3.5)node{$\swarrow$}node[above]{\footnotesize${\cal O}_5 \ \ \ \ $};
\draw[blue](3.2,-4.2)node{$\swarrow$}node[above]{\footnotesize${\cal O}_6 \ \ \ \ $};
\draw[blue](1.6,-4.9)node{$\swarrow$}node[above]{\footnotesize${\cal O}_7 \ \ \ \ $};
\draw[blue](9.6,-2.1)node{$\swarrow$}node[above]{\footnotesize${\cal O}_3 \ \ \ \ $};
\draw[blue](8.0,-2.8)node{$\swarrow$}node[above]{\footnotesize${\cal O}_4 \ \ \ \ $};
\draw[blue](6.4,-3.5)node{$\swarrow$}node[above]{\footnotesize${\cal O}_5 \ \ \ \ $};
\draw[blue](4.8,-4.2)node{$\swarrow$}node[above]{\footnotesize${\cal O}_6 \ \ \ \ $};
\draw[blue](3.2,-4.9)node{$\swarrow$}node[above]{\footnotesize${\cal O}_7 \ \ \ \ $};
\draw[blue](9.6,-2.8)node{$\swarrow$}node[above]{\footnotesize$\ldots \ \ \ \ $};
\draw[blue](8.0,-3.5)node{$\swarrow$}node[above]{\footnotesize$\ldots  \ \ \ \ $};
\draw[blue](6.4,-4.2)node{$\swarrow$}node[above]{\footnotesize$\ldots  \ \ \ \ $};
\draw[blue](4.8,-4.9)node{$\swarrow$}node[above]{\footnotesize$\ldots  \ \ \ \ $};
\endtikzpicture
\end{center}
\caption{\textit{Leading-depth terms $\beta^{{\rm sv},\, j}_{4,k} $ obtained from Poincar\'e sums
over $y^a {\cal E}_0(8,0^b)$, together with the action of the operators $\mathcal{O}_\ell$ defined in~\eqref{oct21.3}.}}
\label{chess.3}
\end{table}
%

%%%%%%%%%%%%%%%%%%%%%%%%%%%%%%%%%%%%%%%%%%%%%%%%%%%%%%%%%%%
\subsection{Alternative folding}
\label{sec:fold.2}
%%%%%%%%%%%%%%%%%%%%%%%%%%%%%%%%%%%%%%%%%%%%%%%%%%%%%%%%%%%

In order to propose a Poincar\'e series for the $?$-fields in tables
\ref{chess.2} and \ref{chess.3} (corresponding to seeds of the form $y^a  {\cal E}_0(2m,0^b)$
with $a\geq1,\ b\geq 0$ but in particular $ a{+}b < 2m{-}1$), we will generalise the step form of Laplace
equations in sections \ref{sec:3.5} and \ref{sec:step.odd}. By repeated action of the
Laplace operator, even seed functions $y^a \Re[{\cal E}_0(2m,0^b)]$
with such $a\geq1,\ b\geq 0$ but $ a{+}b < 2m{-}1$ can be eventually mapped to
\begin{align}
 \label{endcasc} 
&\quad 
 -\frac{1}{(2\pi i)^{2m}}\PS \left[y^{k+m} \Re \GG_{2m}(\tau)  \right]_\gamma \nn\\
& = \frac{(2k)! (k{-}1)! (m{-}1)! }{2 (-4)^{k+m} {\rm B}_{2k} (k{+}m{-}1)! (2m{-}1)! y^{2m}} \Big[
(\pi \overline \nabla)^m {\rm E}_m (\pi \nabla)^m {\rm E}_k + (\pi \overline \nabla)^m {\rm E}_k (\pi \nabla)^m {\rm E}_m 
\Big] \notag \\
&= \frac{(2k)! (k{-}1)! (m{-}1)! }{ (-4)^{k+m} {\rm B}_{2k} (k{+}m{-}1)! (2m{-}1)! } \Jp{m}{k}{m}\,.
\end{align}
The Poincar\'e sum converges for $k{+}m>1$. 
Here, $k{+}m=a{+}b{+}1$ by~\eqref{d1lemma} which, thanks to the bound $a{+}b <2m{-}1 $, implies that $k=a{+}b{+}1{-}m<m$. 
The resulting functions $\Jp{m}{k}{m}$ defined in~\eqref{prodd1} with $m>k$ crucially depart from
the earlier cases  throughout sections~\ref{sec:3} to~\ref{sec:4}. Given that $(\pi \nabla)^m {\rm E}_k$ in (\ref{endcasc}) with $m>k$ involves derivatives of ${\rm G}_{2k}$, a rewriting in terms of $\bsv$ is no longer possible as opposed to~\eqref{notthis2}. 

The seed function on the left-hand side of~\eqref{endcasc} stems from $\mathcal{O}_{k+m-1}\big(y^{k+m-1} \Re \mathcal{E}_0(2m) -\frac{2 {\rm B}_{2m}}{(2m)!} y^{k+m}\big)$ according to (\ref{oct21.4}), where the shifted
Laplace operators ${\cal O}_\ell \sim \Delta -\ell(\ell{-}1)$ were defined in (\ref{oct21.3}). The seed function $-\frac{y^{k+m} }{(2\pi i)^{2m}} \Re \GG_{2m}(\tau)$ is the endpoint of the cascade of ${\cal O}_\ell$ operators
with adjacent values of $\ell$ described around (\ref{oct21.5}). Even though $\Jp{m}{k}{m}$ is not expressible in terms of $\betalagp{j}$, the number $s$ of Cauchy--Riemann derivatives
of $\Jp{s}{k}{m}$ grows by one with each ${\cal O}_\ell$, see (\ref{altrec.10}), so we
can attain the term $\Jp{m}{k}{m}$ in $m$ steps from $\Jp{0}{k}{m}$ which corresponds to
$\beta^{{\rm sv}+, \, k-1}_{k,m}$ modulo lower depth. The latter in turn is generated in
$k{-}1$ steps of applying suitable ${\cal O}_\ell$ to $\beta^{{\rm sv}+, \, 0}_{k,m}$, so we know
that the cascade of Laplace equations for the associated seed functions that terminates with (\ref{endcasc}) has length $k{+}m{-}1$,
\beq
{\cal O}_{k+m-1} {\cal O}_{k+m-2} \ldots {\cal O}_2{\cal O}_1 \Big( b_{m,k}y^{m+k}+ y \Re \! \big[  {\cal E}_0(2m,0^{k+m-2})\big] \Big) = 
-\frac{y^{k+m} }{(2\pi i)^{2m}} \Re \GG_{2m}(\tau) \, .
\label{casc.1}
\eeq
In this expression, it is important that $m>k$ for $\Re \! \big[  {\cal E}_0(2m,0^{k+m-2})\big]$ to be a standard iterated integral since we would otherwise violate the bound $p \leq 2m{-}2$ of ${\cal E}_0(2m,0^{p})$ of Brown's iterated integrals (see section \ref{sec:5.2} for a brief discussion of such cases with $p > 2m{-}2$). The subscripts of the ${\cal O}_\ell$ are engineered such that the terms $\sim n (n{-}1)y^n$ in (\ref{d1lemma}) 
drop out. In the first $0\leq r\leq 2k{-}1$ steps, the Poincar\'e series
\begin{align}
&\sum_{\gamma \in B(\ZZ)\backslash {\rm SL}(2,\ZZ)}
\Big[  b_{m,k}^r y^{k+m} + y^{1+r}   \Re[{\cal E}_0({2m},0^{k+m-2-r}) ]  \Big]_\gamma  \label{casc.2}\\
& \ \ \ \ \ \
\sim  {\cal O}_{r} {\cal O}_{r-1} \ldots {\cal O}_{2} {\cal O}_{1} \beta^{{\rm sv}+, \, 0}_{m,k}  \ \text{mod lower depth} 
\notag
\end{align}
is expressible in terms of the real modular invariants $\FFp{s}{k}{m}$ and $\Jp{\ell}{k}{m}$ with $\ell \leq k$ at depth two. The rational prefactors in (\ref{casc.1}) and (\ref{casc.2}) are given by 
\begin{align}
b_{m,k} &=b_{m,k}^0=  -\frac{2 (-4)^{  k + m-2}
   {\rm B}_{2m}\, ( k {+} m{-} 2)!}{ (2 m)! ( 2 k {+} 2 m{-}3)!}\,,
\label{casc.3} \\
b_{m,k}^r &= -\frac{2(-4)^{ k +  m - r-2}
 {\rm B}_{2m} \,( k {+} m{-}2)! ( k {+} m {+} r{-}1)!}{ r! (2 m)!  ( k {+} m {-} r {-} 1)! (2 k {+} 
    2 m{-}3)!}\,, \notag
\end{align}
see (\ref{oct21.5}) and (\ref{oct21.6}) for the counterparts in the earlier choice of folding.
The depth-one part of the seed functions in (\ref{casc.2}) is always of the form $y^a \Re[{\cal E}_0(2m,0^b)]$
with $a{+}b=k{+}m{-}1$ which obeys indeed $ a{+}b < 2m{-}1$ as advertised above since $k<m$. For the same
reason, the iterated Eisenstein integral $\Re[{\cal E}_0({2m},0^{2m-2})]$ with the maximal number of zeros
which is combined with $\zeta_{2m-1}$ in (\ref{oct21.1}) does not occur in (\ref{casc.1}) and (\ref{casc.2}).

The same strategy applies in the odd case, based on a variant of (\ref{notthis2.odd}) with $m>k$.
The cascade of shifted Laplacians then reads
\begin{align}
{\cal O}_{k+m-1} {\cal O}_{k+m-2} \ldots {\cal O}_2{\cal O}_1\Big(  y \Im \! \big[  {\cal E}_0(2m,0^{k+m-2})\big] \Big)&= 
-\frac{y^{k+m} }{(2\pi i)^{2m}} \Im \GG_{2m}(\tau) \, ,
\label{oddcasc}
\\
\sum_{\gamma \in B(\ZZ)\backslash {\rm SL}(2,\ZZ)}
\Big[  \Im \! \big[{\cal E}_0({2m},0^{k+m-2-r}) \big]  \Big]_\gamma  
&\sim  {\cal O}_{r} {\cal O}_{r-1} \ldots {\cal O}_{2} {\cal O}_{1} \beta^{{\rm sv}-, \, 0}_{m,k}  \ \text{mod lower depth} 
\notag
\end{align}
instead of (\ref{casc.1}) and (\ref{casc.2}), without the term $\sim y^{m+k}$ in the seed.

\subsubsection{Even examples}

Based on (\ref{endcasc}), (\ref{casc.1}) and (\ref{casc.2}), there
is a unique proposal for the Poincar\'e sums over even seeds $b_{m,k}^r y^{k+m} + y^{1+r}   \Re[{\cal E}_0({2m},0^{k+m-2-r}) ] $
with $k<m$ which is consistent with the Laplace equations.
Although from their defining relations~\eqref{eq:Fmk}, \eqref{eq:Fmkm}, \eqref{prodd1} and~\eqref{eq:Jmdef} we have that  $\FFpm{s}{m}{k}$ and $\Jpm{\ell}{m}{k}$ are symmetric under the interchange of $m$ and $k$, we shall adopt the convention that we write the smaller value first in the subscript. 
According to this convention, the notation  $\FFpm{s}{k}{m}$  then signals  that  $m>k$.

The Poincar\'e sums over the iterated integrals discussed then can be seen to provide alternative seed functions
$\seedpalt{s}{ k}{ m}$ and $\Jseedpalt{\ell}{ k}{ m}$ of
\begin{align}
\FFpm{s}{k}{m}(\tau)  &= \PS \seedpmalt{s}{k}{m}(\gamma\cdot \tau)\,,
\label{defaltseed} \\
\Jpm{\ell}{k}{m}(\tau)  &= \PS \Jseedpmalt{\ell}{k}{m} (\gamma\cdot \tau)\,,
\notag
\end{align}
where $2\leq k<m$. 
In the simplest cases within this range, the above reasoning leads to
\begin{align}
 \frac{ y^5}{198450}+ y \Re[{\cal E}_0(6, 0^{3}) ]  
&\cong \frac{3}{2} \seedpalt{3}{ 2}{ 3} \,,\notag
 \\
- \frac{ y^5}{39690}+ y^2 \Re[{\cal E}_0(6, 0^{2}) ]  
&\cong -\frac{3}{8} (6 \seedpalt{3}{ 2}{ 3} + \Jseedpalt{0}{ 2}{ 3})\,,
\label{casc.4}\\
\frac{  y^5}{17640 }+ y^3 \Re[{\cal E}_0(6, 0) ] 
&\cong\frac{3}{32 }(12 \seedpalt{3}{ 2}{ 3} + 6 \Jseedpalt{0}{ 2}{ 3} + \Jseedpalt{1}{ 2}{ 3})\,,
   \notag
\\
- \frac{ y^5}{15120}+ y^4 \Re[{\cal E}_0(6) ]  
&\cong -\frac{1}{128} (36 \Jseedpalt{0}{ 2}{ 3} + 12 \Jseedpalt{1}{ 2}{ 3} + \Jseedpalt{2}{ 2}{ 3})\,,
\notag
\end{align}
where we have indicated through the equivalence relation $\cong$ that the two sides may differ
by terms that sum to zero under the Poincar\'e sum in (\ref{defaltseed}). This will become important for
comparison with the seed functions to be presented in sections \ref{sec:fold.3} and \ref{sec:fold.9} below.
Examples at higher weight $m{+}k=6$ and $7$ include
\begin{align}
 \frac{ y^6}{35721000}+ y \Re[{\cal E}_0(8, 0^{4}) ]  
&\cong \frac{3}{28}  \seedpalt{4}{ 2}{ 4} \,, \notag
\\
 - \frac{ y^6}{4762800 }+y^2 \Re[{\cal E}_0(8, 0^{3}) ]
&\cong -\frac{3}{112} (12 \seedpalt{4}{ 2}{ 4} + \Jseedpalt{0}{ 2}{ 4})\,,
\notag \\
\frac{  y^6}{1360800 }+y^3 \Re[{\cal E}_0(8, 0^{2}) ] 
&\cong\frac{3}{448} (60 \seedpalt{4}{ 2}{ 4} + 12 \Jseedpalt{0}{ 2}{ 4} + \Jseedpalt{1}{ 2}{ 4})\,,
\label{casc.5}
\\
- \frac{ y^6}{680400 }+y^4 \Re[{\cal E}_0(8, 0) ] &\cong -\frac{1}{1792}
(360 \seedpalt{4}{ 2}{ 4} + 180 \Jseedpalt{0}{ 2}{ 4} + 30 \Jseedpalt{1}{ 2}{ 4} +  \Jseedpalt{2}{ 2}{ 4})\,,
   \notag
\\
 \frac{  y^6}{604800 }+y^5 \Re[{\cal E}_0(8) ]  &\cong
\frac{1}{28672} (1440 \Jseedpalt{0}{ 2}{ 4} + 360 \Jseedpalt{1}{ 2}{ 4} + 24 \Jseedpalt{2}{ 2}{ 4} + 
 \Jseedpalt{3}{ 2}{ 4}) \,,
 \notag
 \end{align}
as well as
\begin{align}
-  \frac{y^7}{196465500} +y  \Re[{\cal E}_0(8, 0^{5}) ] 
&\cong\frac{9}{14} (\seedpalt{3}{ 3}{ 4} - \seedpalt{5}{ 3}{ 4})\,,
\notag \\
  \frac{y^7}{18711000 }+y^2 \Re[{\cal E}_0(8, 0^{4}) ] 
&\cong -\frac{9}{28} (3 \seedpalt{3}{ 3}{ 4} - 10 \seedpalt{5}{ 3}{ 4})\,,
\notag \\
-  \frac{y^7}{3742200 }+y^3 \Re[{\cal E}_0(8, 0^{3}) ] 
&\cong\frac{9}{224} (12 \seedpalt{3}{ 3}{ 4} - 180 \seedpalt{5}{ 3}{ 4} - 7 \Jseedpalt{0}{ 3}{ 4})\,,
\label{casc.6} \\
 \frac{y^7}{1247400 }+y^4 \Re[{\cal E}_0(8, 0^{2}) ] 
&\cong\frac{3}{64} (180 \seedpalt{5}{ 3}{ 4} + 18 \Jseedpalt{0}{ 3}{ 4} + \Jseedpalt{1}{ 3}{ 4})\,,
\notag \\
-  \frac{y^7}{665280 }+y^5 \Re[{\cal E}_0(8, 0) ]  &\cong
-\frac{3 }{1024} (1440 \seedpalt{5}{ 3}{ 4} + 360 \Jseedpalt{0}{ 3}{ 4} + 40 \Jseedpalt{1}{ 3}{ 4} + 
    \Jseedpalt{2}{ 3}{ 4})\,,
   \notag  
\\
 \frac{ y^7}{604800} + y^6 \Re[{\cal E}_0(8) ] &\cong 
\frac{3 }{20480}  (3600 \Jseedpalt{0}{ 3}{ 4} + 600 \Jseedpalt{1}{ 3}{ 4} + 30 \Jseedpalt{2}{ 3}{ 4} + 
   \Jseedpalt{3}{ 3}{ 4}) \, .
   \notag
\end{align}
Note that the Poincar\'e sums (\ref{defaltseed}) of $\Jseedpalt{2}{ 2}{ 3} ,\Jseedpalt{2}{ 2}{ 4} ,
\Jseedpalt{3}{ 2}{ 4} ,\Jseedpalt{3}{ 3}{ 4}$ in last line of (\ref{casc.4}) and (\ref{casc.6}) as well as 
the last two lines of (\ref{casc.5}) involve explicit appearances of holomorphic Eisenstein
series, e.g.
\beq
\Jp{2}{2}{3}  = \frac{(\pi \nabla)^2 {\rm E}_2 (\pi \overline \nabla)^2 {\rm E}_3 +
(\pi \nabla)^2 {\rm E}_3 (\pi \overline \nabla)^2 {\rm E}_2   }{2y^4}
= \frac{3}{\pi^4} \big( {\rm G}_4 (\pi \overline \nabla)^2 {\rm E}_3 + \overline{ {\rm G}_4} (\pi  \nabla)^2 {\rm E}_3 \big)
\label{holEex}
\eeq
as a consequence of (\ref{eq:Ediff}).
The examples of this section exhaust the fields marked by $?$ in tables \ref{chess.2} and \ref{chess.3} as we can see in tables \ref{chess.2ALT} and \ref{chess.3ALT}. The quotation marks of ``$\beta^{{\rm sv},\, j\leq 3}_{2,3} \Big. $'' in tables \ref{chess.2ALT} and further cells of table \ref{chess.3ALT} are a reminder that the definition (\ref{eq:betalag}) leads to $\beta^{\rm sv}[\begin{smallmatrix} j_1 &j_2\\ k_1 &k_2 \end{smallmatrix}]$
with $j_i$ outside the admissible range $0 \leq j_i \leq k_i{-}2$ and that we get (derivatives of) holomorphic Eisenstein series as in (\ref{holEex}).

\begin{table}[h]
\begin{center}
\tikzpicture
\draw(-0.04,1) -- (-0.04,-4.9);
\draw(0.04,1) -- (0.04,-4.9);
\draw(-1,0.04) -- (9,0.04);
\draw(-1,-0.04) -- (9,-0.04);
\draw(-0.04,0.04) -- (-1,1);
\draw(-0.75,0.3) node{$a$};
\draw(-0.3,0.75) node{$b$};
\draw(1.6,1.1) -- (1.6,-4.9);
\draw(3.2,1.1) -- (3.2,-4.9);
\draw(4.8,1.1) -- (4.8,-4.9);
\draw(6.4,1.1) -- (6.4,-4.9);
\draw(8,1.1) -- (8,-4.9);
\draw(-1,-0.7) -- (9.0,-0.7);
\draw(-1,-1.4) -- (9.0,-1.4);
\draw(-1,-2.1) -- (9.0,-2.1);
\draw(-1,-2.8) -- (9.0,-2.8);
\draw(-1,-3.5) -- (9.0,-3.5);
\draw(-1,-4.2) -- (9.0,-4.2);
\draw(0.8,0.5)node{0};
\draw(2.4,0.5)node{1};
\draw(4.0,0.5)node{2};
\draw(5.6,0.5)node{3};
\draw(7.2,0.5)node{4};
\draw(8.5,0.5)node{5};
\draw(-0.5,-0.35)node{1};
\draw(-0.5,-1.05)node{2};
\draw(-0.5,-1.75)node{3};
\draw(-0.5,-2.45)node{4};
\draw(-0.5,-3.15)node{5};
\draw(-0.5,-3.85)node{6};
\draw(-0.5,-4.55)node{7};
% 1st beta^svs %% INSERT STUFF FROM HERE
\draw(0.8,-0.35)node{$\times$};
\draw(0.8,-1.05)node{$\times$};
\draw(0.8,-1.75)node{$\rcross$};
\draw(0.8,-2.45)node{``$\beta^{{\rm sv},\, j\leq 3}_{2,3} \Big. $''};
\draw(0.8,-3.15)node{$\beta^{{\rm sv},\, j\leq 4}_{3,3} \Big.$};
\draw(0.8,-3.85)node{$\beta^{{\rm sv},\, j\leq 4}_{3,4} \Big.$};
\draw(0.8,-4.55)node{$\beta^{{\rm sv},\, j\leq 4}_{3,5} \Big.$};
% 2nd beta^svs
\draw(2.4,-0.35)node{$\times$};
\draw(2.4,-1.05)node{$\rcross$};
\draw(2.4,-1.75)node{$\beta^{{\rm sv},\, j\leq 2}_{2,3} \Big.$};
\draw(2.4,-2.45)node{$\beta^{{\rm sv},\, j\leq 3}_{3,3} \Big.$};
\draw(2.4,-3.15)node{$\beta^{{\rm sv},\, j\leq 3}_{3,4} \Big.$};
\draw(2.4,-3.85)node{$\beta^{{\rm sv},\, j\leq 3}_{3,5} \Big.$};
\draw(2.4,-4.55)node{$\beta^{{\rm sv},\, j\leq 3}_{3,6} \Big.$};
% 3rd beta^svs
\draw(4.0,-0.35)node{$\rcross$};
\draw(4.0,-1.05)node{$\beta^{{\rm sv},\, j\leq 1}_{2,3} \Big.$};
\draw(4.0,-1.75)node{$\beta^{{\rm sv},\, j\leq 2}_{3,3} \Big.$};
\draw(4.0,-2.45)node{$\beta^{{\rm sv},\, j\leq 2}_{3,4} \Big.$};
\draw(4.0,-3.15)node{$\beta^{{\rm sv},\, j\leq 2}_{3,5} \Big.$};
\draw(4.0,-3.85)node{$\beta^{{\rm sv},\, j\leq 2}_{3,6} \Big.$};
\draw(4.0,-4.55)node{$\beta^{{\rm sv},\, j\leq 2}_{3,7} \Big.$};
% 4th beta^svs
\draw(5.6,-0.35)node{$\beta^{{\rm sv},\, j=0}_{2,3} \Big.$};
\draw(5.6,-1.05)node{$\beta^{{\rm sv},\, j\leq 1}_{3,3} \Big.$};
\draw(5.6,-1.75)node{$\beta^{{\rm sv},\, j\leq 1}_{3,4} \Big.$};
\draw(5.6,-2.45)node{$\beta^{{\rm sv},\, j\leq 1}_{3,5} \Big.$};
\draw(5.6,-3.15)node{$\beta^{{\rm sv},\, j\leq 1}_{3,6} \Big.$};
\draw(5.6,-3.85)node{$\beta^{{\rm sv},\, j\leq 1}_{3,7} \Big.$};
\draw(5.6,-4.55)node{$\beta^{{\rm sv},\, j\leq 1}_{3,8} \Big.$};
% 5nd beta^svs
\draw(7.2,-0.35)node{$\beta^{{\rm sv},\, j=0}_{3,3} \Big.$};
\draw(7.2,-1.05)node{$\beta^{{\rm sv},\, j=0}_{3,4} \Big.$};
\draw(7.2,-1.75)node{$\beta^{{\rm sv},\, j=0}_{3,5} \Big.$};
\draw(7.2,-2.45)node{$\beta^{{\rm sv},\, j=0}_{3,6} \Big.$};
\draw(7.2,-3.15)node{$\beta^{{\rm sv},\, j=0}_{3,7} \Big.$};
\draw(7.2,-3.85)node{$\beta^{{\rm sv},\, j=0}_{3,8} \Big.$};
\draw(7.2,-4.55)node{$\beta^{{\rm sv},\, j=0}_{3,9} \Big.$};
% END OF INSERT STUFF
\draw(8.5,-0.35)node{$\times$};
\draw(8.5,-1.05)node{$\times$};
\draw(8.5,-1.75)node{$\times$};
\draw(8.5,-2.45)node{$\times$};
\draw(8.5,-3.15)node{$\times$};
\draw(8.5,-3.85)node{$\times$};
\draw(8.5,-4.55)node{$\times$};
\draw[blue](4.8,-0.7)node{$\swarrow$}node[above]{\footnotesize${\cal O}_1 \ \ \ \ $};
\draw[blue](3.2,-1.4)node{$\swarrow$}node[above]{\footnotesize${\cal O}_2 \ \ \ \ $};
\draw[blue](1.6,-2.1)node{$\swarrow$}node[above]{\footnotesize${\cal O}_3 \ \ \ \ $};
\draw[blue](6.4,-0.7)node{$\swarrow$}node[above]{\footnotesize${\cal O}_1 \ \ \ \ $};
\draw[blue](4.8,-1.4)node{$\swarrow$}node[above]{\footnotesize${\cal O}_2 \ \ \ \ $};
\draw[blue](3.2,-2.1)node{$\swarrow$}node[above]{\footnotesize${\cal O}_3 \ \ \ \ $};
\draw[blue](1.6,-2.8)node{$\swarrow$}node[above]{\footnotesize${\cal O}_4 \ \ \ \ $};
\draw[blue](6.4,-1.4)node{$\swarrow$}node[above]{\footnotesize${\cal O}_2 \ \ \ \ $};
\draw[blue](4.8,-2.1)node{$\swarrow$}node[above]{\footnotesize${\cal O}_3 \ \ \ \ $};
\draw[blue](3.2,-2.8)node{$\swarrow$}node[above]{\footnotesize${\cal O}_4 \ \ \ \ $};
\draw[blue](1.6,-3.5)node{$\swarrow$}node[above]{\footnotesize${\cal O}_5 \ \ \ \ $};
\draw[blue](6.4,-2.1)node{$\swarrow$}node[above]{\footnotesize${\cal O}_3 \ \ \ \ $};
\draw[blue](4.8,-2.8)node{$\swarrow$}node[above]{\footnotesize${\cal O}_4 \ \ \ \ $};
\draw[blue](3.2,-3.5)node{$\swarrow$}node[above]{\footnotesize${\cal O}_5 \ \ \ \ $};
\draw[blue](1.6,-4.2)node{$\swarrow$}node[above]{\footnotesize${\cal O}_6 \ \ \ \ $};
\draw[blue](6.4,-2.8)node{$\swarrow$}node[above]{\footnotesize$\ldots \ \ \ \ $};
\draw[blue](4.8,-3.5)node{$\swarrow$}node[above]{\footnotesize$\ldots \ \ \ \ $};
\draw[blue](3.2,-4.2)node{$\swarrow$}node[above]{\footnotesize$\ldots \ \ \ \ $};
\endtikzpicture
\end{center}
\caption{\textit{Leading-depth terms $\beta^{{\rm sv},\, j}_{3,k} $, or alternatively $\beta^{{\rm sv},\, j}_{k,3} $, obtained from Poincar\'e sums
over $y^a  {\cal E}_0(6,0^b)$ including the alternative folding seeds.}}
\label{chess.2ALT}
\end{table}
\begin{table}[h]
\begin{center}
\tikzpicture
\draw(-0.04,1) -- (-0.04,-5.6);
\draw(0.04,1) -- (0.04,-5.6);
\draw(-1,0.04) -- (12.2,0.04);
\draw(-1,-0.04) -- (12.2,-0.04);
\draw(-0.04,0.04) -- (-1,1);
\draw(-0.75,0.3) node{$a$};
\draw(-0.3,0.75) node{$b$};
\draw(1.6,1.1) -- (1.6,-5.6);
\draw(3.2,1.1) -- (3.2,-5.6);
\draw(4.8,1.1) -- (4.8,-5.6);
\draw(6.4,1.1) -- (6.4,-5.6);
\draw(8,1.1) -- (8,-5.6);
\draw(9.6,1.1) -- (9.6,-5.6);
\draw(11.2,1.1) -- (11.2,-5.6);
\draw(-1,-0.7) -- (12.2,-0.7);
\draw(-1,-1.4) -- (12.2,-1.4);
\draw(-1,-2.1) -- (12.2,-2.1);
\draw(-1,-2.8) -- (12.2,-2.8);
\draw(-1,-3.5) -- (12.2,-3.5);
\draw(-1,-4.2) -- (12.2,-4.2);
\draw(-1,-4.9) -- (12.2,-4.9);
\draw(0.8,0.5)node{0};
\draw(2.4,0.5)node{1};
\draw(4.0,0.5)node{2};
\draw(5.6,0.5)node{3};
\draw(7.2,0.5)node{4};
\draw(8.8,0.5)node{5};
\draw(10.4,0.5)node{6};
\draw(11.7,0.5)node{7};
\draw(-0.5,-0.35)node{1};
\draw(-0.5,-1.05)node{2};
\draw(-0.5,-1.75)node{3};
\draw(-0.5,-2.45)node{4};
\draw(-0.5,-3.15)node{5};
\draw(-0.5,-3.85)node{6};
\draw(-0.5,-4.55)node{7};
\draw(-0.5,-5.25)node{8};
% 1st beta^svs %% INSERT STUFF FROM HERE
\draw(0.8,-0.35)node{$\times$};
\draw(0.8,-1.05)node{$\times$};
\draw(0.8,-1.75)node{$\times$};
\draw(0.8,-2.45)node{$\rcross$};
\draw(0.8,-3.15)node{``$\beta^{{\rm sv},\, j\leq 4}_{2,4} \Big.$''};
\draw(0.8,-3.85)node{``$\beta^{{\rm sv},\, j\leq 5}_{3,4} \Big.$''};
\draw(0.8,-4.55)node{$\beta^{{\rm sv},\, j\leq 6}_{4,4} \Big.$};
\draw(0.8,-5.25)node{$\beta^{{\rm sv},\, j\leq 6}_{4,5} \Big.$};
% 2nd beta^svs
\draw(2.4,-0.35)node{$\times$};
\draw(2.4,-1.05)node{$\times$};
\draw(2.4,-1.75)node{$\rcross$};
\draw(2.4,-2.45)node{``$\beta^{{\rm sv},\, j\leq 3}_{2,4} \Big.$''};
\draw(2.4,-3.15)node{$\beta^{{\rm sv},\, j\leq 4}_{3,4} \Big.$};
\draw(2.4,-3.85)node{$\beta^{{\rm sv},\, j\leq 5}_{4,4} \Big.$};
\draw(2.4,-4.55)node{$\beta^{{\rm sv},\, j\leq 5}_{4,5} \Big.$};
\draw(2.4,-5.25)node{$\beta^{{\rm sv},\, j\leq 5}_{4,6} \Big.$};
% 3rd beta^svs
\draw(4.0,-0.35)node{$\times$};
\draw(4.0,-1.05)node{$\rcross$};
\draw(4.0,-1.75)node{$\beta^{{\rm sv},\, j\leq 2}_{2,4} \Big.$};
\draw(4.0,-2.45)node{$\beta^{{\rm sv},\, j\leq 3}_{3,4} \Big.$};
\draw(4.0,-3.15)node{$\beta^{{\rm sv},\, j\leq 4}_{4,4} \Big.$};
\draw(4.0,-3.85)node{$\beta^{{\rm sv},\, j\leq 4}_{4,5} \Big.$};
\draw(4.0,-4.55)node{$\beta^{{\rm sv},\, j\leq 4}_{4,6} \Big.$};
\draw(4.0,-5.25)node{$\beta^{{\rm sv},\, j\leq 4}_{4,7} \Big.$};
% 4th beta^svs
\draw(5.6,-0.35)node{$\rcross$};
\draw(5.6,-1.05)node{$\beta^{{\rm sv},\, j\leq 1}_{2,4} \Big.$};
\draw(5.6,-1.75)node{$\beta^{{\rm sv},\, j\leq 2}_{3,4} \Big.$};
\draw(5.6,-2.45)node{$\beta^{{\rm sv},\, j\leq 3}_{4,4} \Big.$};
\draw(5.6,-3.15)node{$\beta^{{\rm sv},\, j\leq 3}_{4,5} \Big.$};
\draw(5.6,-3.85)node{$\beta^{{\rm sv},\, j\leq 3}_{4,6} \Big.$};
\draw(5.6,-4.55)node{$\beta^{{\rm sv},\, j\leq 3}_{4,7} \Big.$};
\draw(5.6,-5.25)node{$\beta^{{\rm sv},\, j\leq 3}_{4,8} \Big.$};
% 5th beta^svs
\draw(7.2,-0.35)node{$\beta^{{\rm sv},\, j=0}_{2,4} \Big.$};
\draw(7.2,-1.05)node{$\beta^{{\rm sv},\, j\leq 1}_{3,4} \Big.$};
\draw(7.2,-1.75)node{$\beta^{{\rm sv},\, j\leq 2}_{4,4} \Big.$};
\draw(7.2,-2.45)node{$\beta^{{\rm sv},\, j\leq 2}_{4,5} \Big.$};
\draw(7.2,-3.15)node{$\beta^{{\rm sv},\, j\leq 2}_{4,6} \Big.$};
\draw(7.2,-3.85)node{$\beta^{{\rm sv},\, j\leq 2}_{4,7} \Big.$};
\draw(7.2,-4.55)node{$\beta^{{\rm sv},\, j\leq 2}_{4,8} \Big.$};
\draw(7.2,-5.25)node{$\beta^{{\rm sv},\, j\leq 2}_{4,9} \Big.$};
% 6th beta^svs
\draw(8.8,-0.35)node{$\beta^{{\rm sv},\, j=0}_{3,4} \Big.$};
\draw(8.8,-1.05)node{$\beta^{{\rm sv},\, j\leq 1}_{4,4} \Big.$};
\draw(8.8,-1.75)node{$\beta^{{\rm sv},\, j\leq 1}_{4,5} \Big.$};
\draw(8.8,-2.45)node{$\beta^{{\rm sv},\, j\leq 1}_{4,6} \Big.$};
\draw(8.8,-3.15)node{$\beta^{{\rm sv},\, j\leq 1}_{4,7} \Big.$};
\draw(8.8,-3.85)node{$\beta^{{\rm sv},\, j\leq 1}_{4,8} \Big.$};
\draw(8.8,-4.55)node{$\beta^{{\rm sv},\, j\leq 1}_{4,9} \Big.$};
\draw(8.8,-5.25)node{$\beta^{{\rm sv},\, j\leq 1}_{4,10} \Big.$};
% 7th beta^svs
\draw(10.4,-0.35)node{$\beta^{{\rm sv},\, j=0}_{4,4} \Big.$};
\draw(10.4,-1.05)node{$\beta^{{\rm sv},\, j=0}_{4,5} \Big.$};
\draw(10.4,-1.75)node{$\beta^{{\rm sv},\, j=0}_{4,6} \Big.$};
\draw(10.4,-2.45)node{$\beta^{{\rm sv},\, j=0}_{4,7} \Big.$};
\draw(10.4,-3.15)node{$\beta^{{\rm sv},\, j=0}_{4,8} \Big.$};
\draw(10.4,-3.85)node{$\beta^{{\rm sv},\, j=0}_{4,9} \Big.$};
\draw(10.4,-4.55)node{$\beta^{{\rm sv},\, j=0}_{4,10} \Big.$};
\draw(10.4,-5.25)node{$\beta^{{\rm sv},\, j=0}_{4,11} \Big.$};
% END OF INSERT STUFF
\draw(11.7,-0.35)node{$\times$};
\draw(11.7,-1.05)node{$\times$};
\draw(11.7,-1.75)node{$\times$};
\draw(11.7,-2.45)node{$\times$};
\draw(11.7,-3.15)node{$\times$};
\draw(11.7,-3.85)node{$\times$};
\draw(11.7,-4.55)node{$\times$};
\draw(11.7,-5.25)node{$\times$};
\draw[blue](6.4,-0.7)node{$\swarrow$}node[above]{\footnotesize${\cal O}_1 \ \ \ \ $};
\draw[blue](4.8,-1.4)node{$\swarrow$}node[above]{\footnotesize${\cal O}_2 \ \ \ \ $};
\draw[blue](3.2,-2.1)node{$\swarrow$}node[above]{\footnotesize${\cal O}_3 \ \ \ \ $};
\draw[blue](1.6,-2.8)node{$\swarrow$}node[above]{\footnotesize${\cal O}_4 \ \ \ \ $};
\draw[blue](8.0,-0.7)node{$\swarrow$}node[above]{\footnotesize${\cal O}_1 \ \ \ \ $};
\draw[blue](6.4,-1.4)node{$\swarrow$}node[above]{\footnotesize${\cal O}_2 \ \ \ \ $};
\draw[blue](4.8,-2.1)node{$\swarrow$}node[above]{\footnotesize${\cal O}_3 \ \ \ \ $};
\draw[blue](3.2,-2.8)node{$\swarrow$}node[above]{\footnotesize${\cal O}_4 \ \ \ \ $};
\draw[blue](1.6,-3.5)node{$\swarrow$}node[above]{\footnotesize${\cal O}_5 \ \ \ \ $};
\draw[blue](9.6,-0.7)node{$\swarrow$}node[above]{\footnotesize${\cal O}_1 \ \ \ \ $};
\draw[blue](8.0,-1.4)node{$\swarrow$}node[above]{\footnotesize${\cal O}_2 \ \ \ \ $};
\draw[blue](6.4,-2.1)node{$\swarrow$}node[above]{\footnotesize${\cal O}_3 \ \ \ \ $};
\draw[blue](4.8,-2.8)node{$\swarrow$}node[above]{\footnotesize${\cal O}_4 \ \ \ \ $};
\draw[blue](3.2,-3.5)node{$\swarrow$}node[above]{\footnotesize${\cal O}_5 \ \ \ \ $};
\draw[blue](1.6,-4.2)node{$\swarrow$}node[above]{\footnotesize${\cal O}_6 \ \ \ \ $};
\draw[blue](9.6,-1.4)node{$\swarrow$}node[above]{\footnotesize${\cal O}_2 \ \ \ \ $};
\draw[blue](8.0,-2.1)node{$\swarrow$}node[above]{\footnotesize${\cal O}_3 \ \ \ \ $};
\draw[blue](6.4,-2.8)node{$\swarrow$}node[above]{\footnotesize${\cal O}_4 \ \ \ \ $};
\draw[blue](4.8,-3.5)node{$\swarrow$}node[above]{\footnotesize${\cal O}_5 \ \ \ \ $};
\draw[blue](3.2,-4.2)node{$\swarrow$}node[above]{\footnotesize${\cal O}_6 \ \ \ \ $};
\draw[blue](1.6,-4.9)node{$\swarrow$}node[above]{\footnotesize${\cal O}_7 \ \ \ \ $};
\draw[blue](9.6,-2.1)node{$\swarrow$}node[above]{\footnotesize$\ldots \ \ \ \ $};
\draw[blue](8.0,-2.8)node{$\swarrow$}node[above]{\footnotesize$\ldots \ \ \ \ $};
\draw[blue](6.4,-3.5)node{$\swarrow$}node[above]{\footnotesize$\ldots \ \ \ \ $};
\draw[blue](4.8,-4.2)node{$\swarrow$}node[above]{\footnotesize$\ldots \ \ \ \ $};
\draw[blue](3.2,-4.9)node{$\swarrow$}node[above]{\footnotesize$\ldots \ \ \ \ $};
\endtikzpicture

\end{center}
\caption{\textit{Leading-depth terms $\beta^{{\rm sv},\, j}_{4,k} $, or alternatively $\beta^{{\rm sv},\, j}_{k,4} $, obtained 
from Poincar\'e sums over $y^a {\cal E}_0(8,0^b)$ including the alternative seeds.}}
\label{chess.3ALT}
\end{table}

\subsubsection{Odd examples}

As similar step-form strategy applies to the odd case (\ref{oddcasc}) as well for deriving a 
unique proposal to the odd Poincar\'e sums over $y^{1+r}  \Im[{\cal E}_0({2m},0^{k+m-2-r}) ] $
with $k<m$ that preserves the Laplace equations of $\FFm{s}{k}{m}$ and $\Jm{\ell}{k}{m}$.
For the simplest cases $(k,m) =(2,3), (2,4), (3,4)$, the seed functions $\seedmalt{s}{ k}{ m}$ 
and $\Jseedmalt{\ell}{ k}{ m}$ defined in (\ref{defaltseed}) are determined by
\begin{align}
y \Im[{\cal E}_0(6, 0^3)] &\cong
-\frac{3 i}{10} \seedmalt{2}{ 2}{ 3} + \frac{3i}{10 }\seedmalt{4}{ 2}{ 3}\,,
\notag \\
y^2 \Im[{\cal E}_0(6, 0^2)] &\cong
\frac{3i}{20} \seedmalt{2}{ 2}{ 3} - \frac{9i}{10} \seedmalt{4}{2}{ 3}  \,,
\label{oddstp.1}  \\
y^3 \Im[{\cal E}_0(6, 0)] &\cong
\frac{9i}{8 } \seedmalt{4}{2}{ 3} + \frac{3i}{32 } \Jseedmalt{1}{2}{3} \,,
\notag \\
y^4 \Im[{\cal E}_0(6)]&\cong
-\frac{i}{128} (72 \seedmalt{4}{2}{ 3} + 12 \Jseedmalt{1}{2}{ 3} + \Jseedmalt{2}{2}{3})\,,
\notag
\end{align}
as well as
\begin{align}
 y  \Im[{\cal E}_0(8, 0^4)] &\cong
-\frac{3i}{196} \seedmalt{3}{2}{ 4} + \frac{3i}{196} \seedmalt{5}{ 2}{ 4} \,,
\notag \\
y^2  \Im[{\cal E}_0(8, 0^3)] &\cong
\frac{9i}{392 } \seedmalt{3}{2}{ 4} - \frac{15i}{196 } \seedmalt{5}{2}{ 4} \,,
\notag \\
 y^3 \Im[{\cal E}_0(8, 0^2)]&\cong
-\frac{9i}{784} \seedmalt{3}{2}{ 4} +  \frac{135i}{784 } \seedmalt{5}{2}{ 4} + \frac{3i}{448 } \Jseedmalt{1}{2}{ 4}  \,,
\label{oddstp.2}  \\
y^4 \Im[{\cal E}_0(8, 0)] &\cong
-\frac{45i}{224} \seedmalt{5}{2}{ 4} - \frac{15i}{896 } \Jseedmalt{1}{2}{ 4} - \frac{i}{1792} \Jseedmalt{2}{2}{ 4}\,,
\notag \\
y^5 \Im[{\cal E}_0(8)]&\cong
\frac{ i}{ 28672}  (2880 \seedmalt{5}{2}{ 4} + 360 \Jseedmalt{1}{2}{ 4}
 + 24 \Jseedmalt{2}{2}{ 4} +  \Jseedmalt{3}{2}{ 4})
\notag
\end{align}
and
\begin{align}
y \Im[{\cal E}_0(8, 0^5)] &\cong
-\frac{9i}{140} \seedmalt{2}{3}{ 4} + \frac{i}{10} \seedmalt{4}{3}{ 4} - \frac{i}{28} \seedmalt{6}{3}{ 4} \,,
\notag \\
y^2 \Im[{\cal E}_0(8, 0^4)] &\cong
\frac{9i}{280 } \seedmalt{2}{3}{ 4} - \frac{3i}{10 } \seedmalt{4}{3}{ 4} +  \frac{15i}{56 } \seedmalt{6}{3}{ 4} \,,
\notag \\
y^3  \Im[{\cal E}_0(8, 0^3)]&\cong
\frac{3i}{8 } \seedmalt{4}{3}{ 4} - \frac{15i}{16 } \seedmalt{6}{3}{ 4}  \,,
\label{oddstp.3} \\
y^4  \Im[{\cal E}_0(8, 0^2)] &\cong
-\frac{3i}{16} \seedmalt{4}{3}{ 4} +  \frac{15i}{8 } \seedmalt{6}{3}{ 4} + \frac{3i}{64} \Jseedmalt{1}{3}{ 4} \,,
\notag \\
y^5 \Im[{\cal E}_0(8, 0)]&\cong
-\frac{135i}{64} \seedmalt{6}{3}{ 4} -  \frac{15i}{128 } \Jseedmalt{1}{3}{ 4} - \frac{3 i }{1024 } \Jseedmalt{2}{3}{ 4}\,,
\notag \\
y^6 \Im[{\cal E}_0(8)]&\cong
\frac{ 3 i}{20480 }  (7200 \seedmalt{6}{3}{ 4} + 600 \Jseedmalt{1}{3}{ 4} + 30 \Jseedmalt{2}{3}{ 4} +  \Jseedmalt{3}{3}{ 4}) \, .
\notag 
\end{align}
The equivalence relation $\cong$ again indicates that the statements hold up to 
terms that sum to zero under the Poincar\'e sums (\ref{defaltseed}), we will presently give more details on this issue.

%%%%%%%%%%%%%%%%%%%%%%%%%%%%%%%%%%%%%%%%%%%%%%%%%%%%%%%%%%%
\subsection{Relations to earlier even seed functions}
\label{sec:fold.3}
%%%%%%%%%%%%%%%%%%%%%%%%%%%%%%%%%%%%%%%%%%%%%%%%%%%%%%%%%%%

The arguments above yield seed functions whose Poincar\'e sums agree with those of the seeds found in section \ref{sec:3.3.1}. The non-uniqueness of seed functions is well-known. Since a Poincar\'e series is a sum over images under ${\rm SL}(2,\mathbb{Z})$, any of the images is in principle equally well suited since
\begin{align}
\sum_{\gamma\in B(\ZZ)  \backslash {\rm SL}(2,\ZZ)} \seeed(\gamma\tau) = \sum_{\gamma\in  (\gamma_0^{-1} B(\ZZ)\gamma_0) \backslash  {\rm SL}(2,\ZZ)} \seeed(\gamma_0\gamma\tau)
\end{align}
for any $\gamma_0\in {\rm SL}(2,\ZZ)$. However, as shown this will in general change the stabiliser in the Poincar\'e sum to a conjugate Borel subgroup. Other seeds with the same Borel stabiliser can be constructed if one allows for divergent seeds that are to be interpreted via analytic continuation. According to~\eqref{eq:FR}, the seeds $y^s$ and $y^{1-s}$ both yield Poincar\'e series that are proportional to $\EE_s$, although only one of the Poincar\'e sums is convergent. Nevertheless, one could formally write down the seed
\begin{align}
\label{eq:rh0}
\seeed(\tau) = y^s - \frac{\pi^{2s-1/2}\Gamma(s-1/2)\zeta_{2s-1} }{ \Gamma(s)\zeta_{2s}} y^{1-s}\,,
\end{align}
which has vanishing Poincar\'e sum, if we sum the two terms individually and combine them after analytic continuation using~\eqref{eq:FR}. Here, $y^{1-s}$ is not a single ${\rm SL}(2,\ZZ)$ image of $y^s$ but an infinite sum of images.

The seed functions $\seedpalt{s}{m}{k}$ deduced from the step form of the Laplace system are of the same nature as $y^{1-s}$ in the example when compared to $\seedp{s}{m}{k}$ in (\ref{eq:genseed}). 
In the case of $\FFp{s}{m}{k}$, we can also obtain an alternative, non-convergent seed by folding $\EE_m$ instead $\EE_k$ as we did in section~\ref{sec:3.3.1} where now $k\geq m$. The alternative seed obtained in this way differs from the one obtained from the step form. However, as we shall show, the difference between the two has a vanishing Poincar\'e sum. By performing the same steps as in section~\ref{sec:3.2new} but folding $\EE_m$ to $y^m$ one can derive the following form for $\seedpalt{s}{m}{k}$:
\begin{align}
\seedpalt{s}{m}{k} &= (-1)^{k+m} \frac{ {\rm B}_{2k} {\rm B}_{2m}  (4y)^{k+m} }{(2k)! (2m)! (\mu_{k+m}-\mu_s) } 
-(-1)^{m} \frac{4 {\rm B}_{2m} (2k{-}3)! \zeta_{2k-1}  (4y)^{m+1-k} }{(2m)!(k{-}2)!(k{-}1)! (\mu_{m-k+1}-\mu_s)} 
\nn\\
&\quad -(-1)^m \frac{2{\rm B}_{2m} \Gamma(2k)}{(2m)!\Gamma(k)} \sum_{\ell=1}^{m-1} g^+_{k,m,\ell,s} (4y)^\ell \Re[\mathcal{E}_0(2k,0^{k+m-1-\ell})]  \label{eq:genseedpalt} \\
&\quad + (-1)^m  \frac{2{\rm B}_{2m} \Gamma(2k)}{(2m)!\Gamma(k)}  \sum_{\ell=m+1-k}^0 \tilde{g}^+_{k,m,\ell,s} (4y)^\ell \Re[\mathcal{E}_0(2k,0^{k+m-1-\ell})]\,.
\notag
\end{align}
 We recall $\mu_s=s(s{-}1)$ and the coefficients $g^+_{k,m,\ell,s}$ were defined in~\eqref{eq:fkm}, but importantly, the order of $m$ and $k$ in the alternative seed is swapped. The new coefficients appearing in the last line are given by
\begin{align}
\tilde{g}^+_{k,m,\ell,s} = \sum_{i=m+1-k}^\ell \frac{(i)_{\ell-i} \Gamma(k{+}m{-}i)}{(1{-}s{+}i)_{\ell-i} (s{+}i)_{\ell-i} \Gamma(m{+}1{-}i)\Gamma(k{+}i{-}m)(\mu_i{-}\mu_s)}\,.
\end{align}
In comparison to~\eqref{eq:genseed}, the first two lines of~\eqref{eq:genseedpalt} are simply obtained by the interchange $m \leftrightarrow k$, but the third line is new and contains non-positive powers of $y$ accompanying the iterated integrals. We stress that the Poincar\'e sum over this seed function is not absolutely convergent, but has to be interpreted with care as we shall explain below.

Applying the formula~\eqref{eq:genseedpalt} together with (\ref{eqseedR}) at $m\leftrightarrow k$
we obtain the following form of the examples in (\ref{casc.4}) to (\ref{casc.6}):
\begin{align}
&y \Re[{\cal E}_0(6, 0^{3}) ] + \frac{ 1}{2}  \big( \Re[{\cal E}_0(6, 0^4) ]  - \tfrac{ \zeta_5}{5!} \big) + \frac{ y^5}{198450}
=
\frac{3}{2} \seedpalt{3}{ 2}{ 3}\,, \notag
 \\
&y^2 \Re[{\cal E}_0(6, 0^{2}) ] - \frac{ y^5}{39690} 
=
-\frac{3}{8} (6 \seedpalt{3}{ 2}{ 3} + \Jseedpalt{0}{ 2}{ 3})\,,
\label{casc.14}\\
&y^3 \Re[{\cal E}_0(6, 0) ] +\frac{  y^5}{17640 }
=
\frac{3}{32 }(12 \seedpalt{3}{ 2}{ 3} + 6 \Jseedpalt{0}{ 2}{ 3} + \Jseedpalt{1}{ 2}{ 3})\,,
   \notag
\\
&y^4 \Re[{\cal E}_0(6) ] - \frac{ y^5}{15120} 
= -\frac{1}{128} (36 \Jseedpalt{0}{ 2}{ 3} + 12 \Jseedpalt{1}{ 2}{ 3} + \Jseedpalt{2}{ 2}{ 3})\,,
\notag
\end{align}
as well as
% 24
\begin{align}
&y \Re[{\cal E}_0(8, 0^{4}) ] + \frac{ 3}{2} \Re[{\cal E}_0(8, 0^{5}) ] + 
\frac{  3}{4 y} \big( \Re[{\cal E}_0(8, 0^6) ]  - \tfrac{ \zeta_7}{7!} \big) + \frac{ y^6}{35721000}
=
\frac{3}{28}  \seedpalt{4}{ 2}{ 4}\,,
 \notag
\\
&
y^2 \Re[{\cal E}_0(8, 0^{3}) ] - \frac{ 3}{4} \Re[{\cal E}_0(8, 0^{5}) ] - 
\frac{  3}{8 y} \big( \Re[{\cal E}_0(8, 0^6) ]  - \tfrac{ \zeta_7}{7!} \big) - \frac{ y^6}{4762800 }
 =
-\frac{3}{112} (12 \seedpalt{4}{ 2}{ 4} + \Jseedpalt{0}{ 2}{ 4})\,,
\notag \\
&
y^3 \Re[{\cal E}_0(8, 0^{2}) ] +\frac{  y^6}{1360800 }
=
\frac{3}{448} (60 \seedpalt{4}{ 2}{ 4} + 12 \Jseedpalt{0}{ 2}{ 4} + \Jseedpalt{1}{ 2}{ 4})\,,
\label{casc.15}
\\
&y^4 \Re[{\cal E}_0(8, 0) ] - \frac{ y^6}{680400 }= -\frac{1}{1792}
(360 \seedpalt{4}{ 2}{ 4} + 180 \Jseedpalt{0}{ 2}{ 4} + 30 \Jseedpalt{1}{ 2}{ 4} +  \Jseedpalt{2}{ 2}{ 4})\,,
   \notag
\\
&y^5 \Re[{\cal E}_0(8) ] + \frac{  y^6}{604800 } =
\frac{1}{28672} (1440 \Jseedpalt{0}{ 2}{ 4} + 360 \Jseedpalt{1}{ 2}{ 4} + 24 \Jseedpalt{2}{ 2}{ 4} + 
 \Jseedpalt{3}{ 2}{ 4}) 
 \notag
 \end{align}
and
\begin{align}
&y  \Re[{\cal E}_0(8, 0^{5}) ] -  \frac{y^7}{196465500} + 
  \frac{1}{2} \big( \Re[{\cal E}_0(8, 0^6) ]  - \tfrac{ \zeta_7}{7!} \big)
=\frac{9}{14} (\seedpalt{3}{ 3}{ 4} - \seedpalt{5}{ 3}{ 4}) \,, 
\notag
 \\
&
y^2 \Re[{\cal E}_0(8, 0^{4}) ] +  \frac{y^7}{18711000 }
=-\frac{9}{28} (3 \seedpalt{3}{ 3}{ 4} - 10 \seedpalt{5}{ 3}{ 4})\,,
\notag \\
&
y^3 \Re[{\cal E}_0(8, 0^{3}) ] -  \frac{y^7}{3742200 }
=
\frac{9}{224} (12 \seedpalt{3}{ 3}{ 4} - 180 \seedpalt{5}{ 3}{ 4} - 7 \Jseedpalt{0}{ 3}{ 4})\,,
\label{casc.16} \\
&
y^4 \Re[{\cal E}_0(8, 0^{2}) ] +  \frac{y^7}{1247400 }
=
\frac{3}{64} (180 \seedpalt{5}{ 3}{ 4} + 18 \Jseedpalt{0}{ 3}{ 4} + \Jseedpalt{1}{ 3}{ 4})\,,
\notag \\
&
y^5 \Re[{\cal E}_0(8, 0) ] -  \frac{y^7}{665280 }
 =
-\frac{3 }{1024} (1440 \seedpalt{5}{ 3}{ 4} + 360 \Jseedpalt{0}{ 3}{ 4} + 40 \Jseedpalt{1}{ 3}{ 4} + 
    \Jseedpalt{2}{ 3}{ 4})\,,
   \notag  
\\
&y^6 \Re[{\cal E}_0(8) ] + \frac{ y^7}{604800} = 
\frac{3 }{20480}  (3600 \Jseedpalt{0}{ 3}{ 4} + 600 \Jseedpalt{1}{ 3}{ 4} + 30 \Jseedpalt{2}{ 3}{ 4} + 
   \Jseedpalt{3}{ 3}{ 4})\,.
   \notag
\end{align}
These towers of equations are identical to many of the corresponding equations in (\ref{casc.4}), (\ref{casc.5}), and (\ref{casc.6}), but some of them are augmented by extra terms involving non-positive powers of $y$.

\subsubsection{\texorpdfstring{Equivalence of the seed functions in the sectors  $(2,3)$ and $(3,4)$}{Equivalence of the seed functions in the sectors  (2,3) and (3,4)}}

We now compare these different towers and show that they differ merely by `red-herrings', namely non-vanishing seed functions which, upon Poincar\'e summation, interpreted via analytic continuation, give rise to a vanishing modular function, similar to~\eqref{eq:rh0}.

Let us start by comparing (\ref{casc.4}) with (\ref{casc.14}). 
The difference  appears on the first line of~\eqref{casc.14} and is given by the red-herring
\beq
{\rm rh}_1 = \frac{ 1}{2}  \big( \Re[{\cal E}_0(6, 0^4) ]  - \tfrac{ \zeta_5}{5!} \big)\,.
\label{redher}
\eeq
We know that, in both (\ref{casc.4}) and (\ref{casc.14}), we must apply $\mathcal{O}_1 = -\frac{1}{4} \Delta$ to go from the first line to the second. Consistency of both expressions hinges on the fact that ${\rm rh}_1$ and its Poincar\'e sum lie in the kernel of $\mathcal{O}_1$ just like ${\rm E}_1$, or rather its regulated version ${\rm E}_0$.

Let us first notice that using (\ref{eq:E0sigma}) we can rewrite the iterated integral as
\beq
\frac{1}{2}\Re[{\cal E}_0(6, 0^4) ]  = -\frac{1}{5!} \sum_{n=1}^{\infty} \sigma_{-5}( n  )\Big[\sqrt{n  \Im \tau} K_{0-1/2}(2\pi n \Im \tau) \Big(e^{2\pi i n \Re \tau}+e^{-2\pi i n  \Re\tau}\Big) \Big]\,,
\eeq
by using the asymptotic expansion of the Bessel function~\eqref{eq:Besselexp}.
Furthermore, we notice the triviality $\tfrac{ \zeta_5}{5!}  =\tfrac{ \zeta_5}{5!}  y^{0}$.
From the usual expansion for the Eisenstein series ${\rm E}_k$ given in (\ref{eq:FEk}) and the above statements, we realise that the red-herring is comprised of two different terms but both appearing as points (or rather infinite sums thereof) on the Poincar\'e orbit (\ref{eq:PSEk}) for ${\rm E}_0$.

It is not surprising then that, by suitable analytic continuation of equations (\ref{eq:PSEk}) and (\ref{Besselsum}), we arrive at
\beq
\sum_{\gamma \in B(\ZZ)\backslash {\rm SL}(2,\ZZ)} \bigg[ \frac{ 1}{2}  \big( \Re[{\cal E}_0(6, 0^4) ]  - \tfrac{ \zeta_5}{5!} \big) 
\bigg]_{\gamma}= - \frac{1}{2}\frac{\zeta_5}{ 5!}( {\rm E}_0 - {\rm E}_0) = 0\,.
\eeq
Hence, the two seed systems (\ref{casc.4}) and (\ref{casc.14}) are completely equivalent modulo the very convoluted vanishing Poincar\'e sum over the red-herring (\ref{redher}).

The situation is identical when comparing the next towers (\ref{casc.6}) with (\ref{casc.16}) where the only red-herring appears on the first line and is given by
\beq
 {\rm rh}_2 = \frac{ 1}{2}  \big( \Re[{\cal E}_0(8, 0^6) ]  - \tfrac{ \zeta_7}{7!} \big)\,,
 \eeq
which again belongs to the kernel of $\mathcal{O}_1$ and has a vanishing Poincar\'e sum thanks to (\ref{eq:PSEk}) and (\ref{Besselsum}). Hence, the seed systems (\ref{casc.6}) and (\ref{casc.16}) are indeed identical.

\subsubsection{\texorpdfstring{Equivalence of the seed functions in the sector $(2,4)$}{Equivalence of the seed functions in the sector (2,4)}}

Finally when we compare (\ref{casc.5}) with (\ref{casc.15}) we see that two red-herrings appear
\begin{align}
{\rm rh}_3 & = \frac{ 3}{2} \Re[{\cal E}_0(8, 0^{5}) ] + 
\frac{  3}{4 y} \big( \Re[{\cal E}_0(8, 0^6) ]  - \tfrac{ \zeta_7}{7!} \big)\,,\\
 {\rm rh}_4 & =-\frac{ 3}{4} \Re[{\cal E}_0(8, 0^{5}) ] -
\frac{  3}{8 y} \big( \Re[{\cal E}_0(8, 0^6) ]  - \tfrac{ \zeta_7}{7!} \big)\,, \notag
\end{align}
related by
\begin{align}
\mathcal{O}_1 {\rm rh}_3 &= {\rm rh}_4\,, \notag \\
\mathcal{O}_2 {\rm rh}_4 & = 0\,.
\end{align}
From the second equation, and the previous discussion, we anticipate that ${\rm rh}_4$ is related
to ${\rm E}_2$ (or equivalently ${\rm E}_{-1}$), and in fact this turns out to be the case.
We notice that the sum of iterated integrals can be rewritten as
\begin{align}
-\frac{ 3}{4} &\Re[{\cal E}_0(8, 0^{5}) ] -
\frac{  3}{8 y} \Re[{\cal E}_0(8, 0^6) ]  = \frac{2}{7!} \frac{3}{4}  \sum_{n=1}^{\infty} n \,\sigma_{-7}( n ) \\
&\times \Big[\,\sqrt{ n \Im \tau} K_{2-1/2}(2\pi n \Im \tau) \Big(e^{2\pi i n \Re \tau}+e^{-2\pi i n  \Re \tau}\Big) \Big]\,,
\notag
\end{align}
while the remaining term in ${\rm rh}_4$ can be rewritten trivially as $\frac{3}{8}\tfrac{ \zeta_7}{7!}  y^{1-2}$.

As before we realise that the second red-herring is comprised of two different terms but both appearing as points (or rather infinite sums thereof) on the Poincar\'e orbit (\ref{eq:PSEk}) for ${\rm E}_2$.

We can use (\ref{Besselsum}) to perform the Poincar\'e sum for the Bessel function above and, after that, analytically continue the Dirichlet sum over $n$ to arrive at
\beq
\sum_{\gamma \in B(\ZZ)\backslash {\rm SL}(2,\ZZ)} \bigg[{-}\frac{ 3}{4} \Re[{\cal E}_0(8, 0^{5}) ] -
\frac{  3}{8 y} \big( \Re[{\cal E}_0(8, 0^6) ]  - \tfrac{ \zeta_7}{7!} \big) \bigg]_\gamma = \frac{3}{8}\frac{\zeta_7}{7! }\frac{1}{\zeta_3}({\rm E}_2 -{\rm E}_2) = 0\,.
\eeq
In conclusion, the red-herring ${\rm rh}_4$ yields a vanishing Poincar\'e sum, only written in a very convoluted way, and we can then safely omit it from the seed functions and from the Laplace system above. One can similarly get rid of ${\rm rh}_3$ which is in the kernel of ${\cal O}_1$ after discarding ${\rm rh}_4$ such that the systems (\ref{casc.5}) and (\ref{casc.15}) become completely identical.

\subsubsection{General even seed functions without red-herrings}

Note that the general situation can quickly appear more complicated, with multiple linearly independent red-herrings appearing when we compare the alternative folding for the Laplace system in step form (\ref{casc.2}) and the inhomogenous Laplace system (\ref{eq:seedeq}). We present such an example in appendix \ref{app:RH}.
However, it is always possible to isolate these red-herring seeds with non-positive powers of $y$ and show that their Poincar\'e sums vanish identically as explained above. 

Hence, our construction of the seed functions by following the step form is completely equivalent to 
(\ref{eq:genseedpalt}). In fact, the minimal seeds in a step form
can be given in closed form by truncating (\ref{eq:genseedpalt}) to the terms with positive powers of $y$,
\begin{align}
\seedpalt{s}{m}{k} &\cong (-1)^{k+m} \frac{ {\rm B}_{2k} {\rm B}_{2m}  (4y)^{k+m} }{(2k)! (2m)! (\mu_{k+m}-\mu_s) } 
 \label{trunc.seed} \\
&\quad -(-1)^m \frac{2{\rm B}_{2m} \Gamma(2k)}{(2m)!\Gamma(k)} \sum_{\ell=1}^{m-1} g^+_{k,m,\ell,s} (4y)^\ell \Re[\mathcal{E}_0(2k,0^{k+m-1-\ell})] \notag
\end{align}
and similarly dropping non-positive powers of $y$ in the prescription
(\ref{eqseedR}) for $\Jseedp{\ell}{m}{k}$ with $m\leftrightarrow k$.

\subsection{Relations to earlier odd seed functions}
\label{sec:fold.9}

The discussion of the odd case is very similar to that of the even case: On top of the step forms
in (\ref{oddstp.1}) to (\ref{oddstp.3}), one can construct alternative seeds for $\FFm{s}{m}{k}$ by 
analysing the Laplace system as in section~\ref{sec:oddseed} and performing the alternative folding
of $\nabla {\rm E}_m$ and $\overline \nabla {\rm E}_m$ with $m<k$. The seed one obtains in this way is 
\begin{align}
\seedmalt{s}{m}{k} &=  -i (-1)^m \frac{2{\rm B}_{2m} \Gamma(2k)}{(2m)!\Gamma(k)} \sum_{\ell=1}^{m} g^-_{k,m,\ell,s} (4y)^\ell \Im[\mathcal{E}_0(2k,0^{k+m-1-\ell})] \label{eq:genseedmalt}\\
&\quad -i (-1)^m  \frac{2{\rm B}_{2m} \Gamma(2k)}{(2m)!\Gamma(k)}  \sum_{\ell=m+2-k}^0 \tilde{g}^-_{k,m,\ell,s} (4y)^\ell \Im[\mathcal{E}_0(2k,0^{k+m-1-\ell})]\,,\notag
\end{align}
where the coefficients $g^-_{k,m,\ell,s}$ were defined in~\eqref{eq:seedm.B} and the new coefficients appearing in the second line are given by
\begin{align}
\tilde{g}^-_{k,m,\ell,s} = \sum_{i=m+1-k}^\ell \frac{(i)_{\ell-i} \Gamma(k{+}m{+}1{-}i)}{(1{-}s{+}i)_{\ell-i} (s{+}i)_{\ell-i} \Gamma(m{-}i)\Gamma(k{-}1{+}i{-}m)(\mu_i{-}\mu_s)}\,.
\end{align}
Again, the terms in the last line of (\ref{eq:genseedmalt}) with non-positive powers of $y$ are new in the alternative folding while the first line is just given by applying $m \leftrightarrow k$ to~\eqref{eq:seedm}.
With the general formula (\ref{eq:genseedmalt}) and (\ref{eq:seedm.A}) for $\Jseedm{\ell}{m}{k} $ 
at $m\leftrightarrow k$, 
one reproduces the results (\ref{oddstp.1}) and (\ref{oddstp.3}) of the step form in the sectors $(m,k)=(2,3)$ 
and $(3,4)$. In comparison to (\ref{oddstp.2}), however, one
additionally finds the red-herring
\beq
{\rm rh}_{\rm odd} = \Im[{\cal E}_0(8, 0^5)]
\label{oddredherr}
\eeq
in both $\Jseedmalt{\ell}{2}{ 4}$ and
\begin{align}
\seedmalt{3}{2}{4}  &= \frac{56i}{3} y^2 \Im[\mathcal{E}_0(8, 0^3)] + \frac{280i}{3} y \Im[\mathcal{E}_0(8, 0^4)] -  280 i \Im[\mathcal{E}_0(8, 0^5)]\,,\nn\\
\seedmalt{5}{2}{4}  &= \frac{56i}{3} y^2 \Im[\mathcal{E}_0(8, 0^3)] + 28 i y \Im[\mathcal{E}_0(8, 0^4)] - 84 i \Im[\mathcal{E}_0(8, 0^5)]\,.
\end{align}
The red-herring (\ref{oddredherr}) has a vanishing Poincar\'e sum because
the sum in~\eqref{Besselsum} is insensitive to the sign of the Fourier mode $n$. Therefore any non-positive powers of $y$ times an imaginary part of an $\mathcal{E}_0(2k,0^p)$ can always be arranged into a sum of Bessel functions multiplied by $e^{2\pi i n \Re\tau} - e^{-2\pi i n \Re \tau}$ as in
\begin{align}
i \Im[\mathcal{E}_0(8, 0^5)] &= -\frac{1}{7!} \sum_{n=1}^\infty \sigma_{7}(n) n^{-6} (q^n-\bar{q}^n)\\
&=  -\frac{2}{7!} \sum_{n=1}^\infty \sigma_{7}(n) n^{-6} \sqrt{n \Im \tau}  K_{-1/2}(2\pi n \Im \tau) (e^{2\pi i n \Re\tau} - e^{-2\pi i n \Re \tau})\,.
\nn
\end{align}
The Poincar\'e sum using~\eqref{Besselsum} formally just produces the difference of two identical expressions.
Performing the  Poincar\'e sum of these terms for each $n>0$ separately using~\eqref{Besselsum} leads to
\begin{align}
\PS i \Im[\mathcal{E}_0(8, 0^5)] \Big|_\gamma &= -\frac{2}{7!} \sum_{n=1}^\infty \sigma_{7}(n) n^{-6}  \left(\frac{6 \sigma_{-1}(n) \EE_0}{ n^{-1}} - \frac{6 \sigma_{-1}(n) \EE_0}{ n^{-1}} \right) = 0 \,,
\end{align}
where we have used analytic continuation. Doing the calculation this way gives that the Poincar\'e sum vanishes Fourier mode by Fourier mode. This is a formal argument since the alternative Poincar\'e seed is divergent to start with and only defined by analytic continuation.

More generally, the simplest representatives of odd seeds in a step form
can be given in closed form by truncating (\ref{eq:genseedmalt}) to the terms with positive powers of $y$,
\begin{align}
\seedmalt{s}{m}{k} &\cong  -i (-1)^m \frac{2{\rm B}_{2m} \Gamma(2k)}{(2m)!\Gamma(k)} \sum_{\ell=1}^{m} g^-_{k,m,\ell,s} (4y)^\ell \Im[\mathcal{E}_0(2k,0^{k+m-1-\ell})] \label{mineq:genseedmalt}
\end{align}
and similarly dropping non-positive powers of $y$ in the prescription
(\ref{eq:seedm.A}) for $-\Jseedm{\ell}{m}{k}$ with $m\leftrightarrow k$.

%%%%%%%%%%%%%%%%%%%%%%%%%%%%%%%%%%%%%%%%%%%%%%%%%%%%%%%%%%%
\section{Further directions}
\label{sec:5}
%%%%%%%%%%%%%%%%%%%%%%%%%%%%%%%%%%%%%%%%%%%%%%%%%%%%%%%%%%%

In this section we want to comment on some interesting future directions opened up by our analysis.

%%%%%%%%%%%%%%%%%%%%%%%%%%%%%%%%%%%%%%%%%%%%%%%%%%%%%%%%%%%
\subsection{Exponentially suppressed terms}
\label{sec:q-series}
%%%%%%%%%%%%%%%%%%%%%%%%%%%%%%%%%%%%%%%%%%%%%%%%%%%%%%%%%%%

Let us remind the reader that all the modular objects studied in this work can be written in an expansion around the cusp of the form
\begin{align}
\label{eq:MGFq}
\summ(\tau) = \sum_{a,b\geq 0} d_{a,b}(\Im\tau) q^a \bar{q}^b\,,
\end{align}
where $d_{a,b}(\Im\tau)$ are Laurent polynomials in $\Im\tau$. This can also be arranged into a Fourier expansion with respect to the periodic variable $\Re\tau$ according to
\begin{equation}
\summ(\tau) = \sum_{\ell\in\mathbb{Z}} a_\ell(\Im \tau) e^{2\pi i \ell \Re\tau}\,,
\end{equation}
where the $\ell^{\rm th}$-Fourier coefficient $a_\ell(\Im \tau)$ takes the form
\begin{equation}
a_\ell(\Im\tau) = e^{- 2 \pi | \ell| \Im \tau} \sum_{n=\min\{0,-\ell\}}^\infty (q\bar{q})^n d_{\ell+n,n}(\Im\tau)
\,,\label{eq:Fourier}
\end{equation} 
where $(q\bar{q})^n = e^{-4\pi n \Im\tau}$ and we refer to such terms 
with $n\geq 1$ as \textit{non-perturbative} (at the cusp $\Im\tau\to \infty$) by slight 
abuse of terminology. For most of the present work we have focused on the Laurent polynomial in 
the zeroth Fourier mode, i.e.\ the $\ell=0,n=0$ sector.

We now want to turn our attention to two different type of exponentially suppressed corrections: firstly analysing the $(q\bar{q})^n$ terms in the zeroth Fourier mode, i.e.\ the $\ell=0,n\neq0$ sector, and secondly the perturbative coefficients in the non-zero mode, i.e.\ the $\ell\neq0,n=0$ sectors in~\eqref{eq:Fourier}.

\subsubsection{Non-perturbative terms in the Fourier zero mode}

Let us start with the $(q\bar{q})^n$ terms in the zero-mode sector $a_0(\Im\tau) =  \sum_{n=0}^\infty (q\bar{q})^n d_{n,n}(\Im\tau)$. Firstly, from the integral representation \eqref{eq:bsv2}, we can easily see that the only term possibly 
containing powers of both $q$ and $\bar{q}$ is given by
\begin{align}
\label{eq:bsv2qqb}
& \betasv{j_1 &j_2\\k_1 &k_2} \Big\vert_{q^{>0} \bar{q}^{>0}}\subset - \frac{(2\pi i)^{-2}}{(4y)^{k_1+k_2-j_1-j_2-4}} \\
&\ \ \  \times \int\limits^{i\infty}_\tau \dd\tau_2(\tau{-}\tau_2)^{k_2-j_2-2} (\bar\tau{-}\tau_2)^{j_2}  \GG_{k_2}(\tau_2) \int\limits^{-i\infty}_{\bar\tau} \dd\bar\tau_1  (\tau{-}\bar\tau_1)^{k_1-j_1-2}(\bar\tau{-}\bar\tau_1)^{j_1}\overline{\GG_{k_1}(\tau_1)}\,.\nn
\end{align}
With the notation $q^{>0} \bar{q}^{>0}$ we mean all terms in the expansion that have positive $q$ \textit{and} $\bar{q}$ powers, although not necessarily the same.

After some algebra we can rewrite these integrals and make use of \eqref{eq:E0depth1} to isolate all the terms containing both $q$ and $\bar{q}$, arriving at
\begin{align}\label{eq:beta2qqb}
\betasv{j_1 &j_2\\k_1 &k_2} \Big\vert_{q^{>0}\bar{q}^{>0}}= &\sum_{A=0}^{j_2} \sum_{B=0}^{k_1-j_1-2} (k_2{-}j_2{-}2{+}A)!(j_1{+}B)! \binom{j_2}{A} \binom{k_1{-}j_1{-}2}{B} \\
&\nn\ \times (4y)^{2+2j_2-k_2-A-B}   {\cal E}_0(k_2,0^{k_2-j_2-2+A})\,\overline{ {\cal E}_0(k_1,0^{j_1+B})}\,.
\end{align}
Using our definition \eqref{eq:betalag} we can write
\begin{align}
\betalagpm{j}\Big\vert_{q^{>0}\bar{q}^{>0}} &\nn=\betasv{2m-2-j& k-m+j\\ 2m& 2k}\Big\vert_{q^{>0}\bar{q}^{>0}}   \pm  \betasv{k+m-2-j &j\\ 2k& 2m}  \Big\vert_{q^{>0}\bar{q}^{>0}} \\
& = \sum_{A=0}^{j+k-m} \sum_{B=0}^{j} (4y)^{2+2j-2m-A-B} (k{+}m{+}A{-}j{-}2)!(2m{+}B{-}j{-}2)! \label{eq:betalagqqb} \\
& \qquad\qquad \times \binom{ j {+}k{-}m}{A}\binom{ j}{ B} \Big({\cal E}_0(2k,0^{k+m+A-j-2}) \overline{ {\cal E}_0(2m,0^{2m+B-j-2})} \pm\cc\Big) \, . \nn
\end{align}
Following the discussion in sections \ref{sec:4.2} and \ref{sec:odd.dpt1}, it is clear that lower-depth terms
cannot possibly contain both $q$ and $\bar{q}$ in $\FFpm{s}{m}{k}$. Hence, we have that $\FFpm{s}{m}{k}\vert_{q^{>0}\bar{q}^{>0}}$ can only come from its depth-two part, given by a suitable rational linear combination of $\betalagpm{j}$.\footnote{The possible homogeneous contributions $H^+$ and $H^-$ in the Cauchy--Riemann equation never contribute to $q^{n_1} \bar{q}^{n_2}$ with $n_1,n_2>0$ either as will become clearer in Part~II, i.e.\ the results of this subsection apply equally to $\cFFpm{s}{m}{k}$ and $\FFpm{s}{m}{k}$.}

In particular, from \eqref{eq:betalagqqb} we notice that for all Fourier modes, the $(q\bar{q})^n$ terms contain only rational coefficients times powers of $y$.
For example we have
\begin{align}
\FFp{2}{2}{2}\Big\vert_{q^{>0}\bar{q}^{>0}} &= \frac{9}{2y^2} {\cal E}_0(4,0^2) \,\overline{ {\cal E}_0(4,0^2) }\,, \notag\\
\FFp{3}{2}{3}\Big\vert_{q^{>0}\bar{q}^{>0}} &= \frac{45}{2y^3} \Big(y \,{\cal E}_0(6,0^3) \,\overline{ {\cal E}_0(4,0^2) }+
{\cal E}_0(6,0^4)\,\overline{ {\cal E}_0(4,0^2) }+\cc \Big)\,,\\
\FFm{2}{2}{3} \Big\vert_{q^{>0}\bar{q}^{>0}} &\nn = -\frac{45}{2y^2}\Big(8y^2\, {\cal E}_0(6,0^2)\,\overline{{\cal E}_0(4,0)}+4y\, {\cal E}_0(6,0^2)\,\overline{{\cal E}_0(4,0^2)}+12y \,{\cal E}_0(6,0^3)\,\overline{{\cal E}_0(4,0)}\\
&\qquad\qquad +3 {\cal E}_0(6,0^3)\,\overline{{\cal E}_0(4,0^2)}+6 {\cal E}_0(6,0^4)\,\overline{{\cal E}_0(4,0)}-\cc\Big)\,.
\notag
\end{align}
Furthermore, from the $q$-series representation \eqref{eq:E0sigma}, it is very simple to isolate the purely $(q\bar{q})^n$ terms in the zeroth Fourier mode, given by
\begin{align}
\betasv{j_1 &j_2\\k_1 &k_2} \Big\vert_{(q\bar{q})^{>0}}&=\sum_{n=1}^\infty \frac{4j_1! (k_2{-}j_2{-}2)!}{(k_1{-}1)!(k_2{-}1)!} \sigma_{k_1-1}(n)\sigma_{k_2-1}(n) (q\bar{q})^n  \label{fzmodes} \\
&\nn \quad \times  (4y)^{j_1+j_2+2}   U(j_1{+}1,k_1;4ny) U(k_2{-}j_2{-}1,k_2;4ny) \,,
\end{align}
with $U(j,k;z)$ Kummer's confluent hypergeometric function, which reduces to a polynomial of degree $k{-}1$ in $1/z$ for $j,k\in\mathbb{N}$ such that $0\leq j \leq k{-}2$.
The notation $|_{(q\bar{q})^{>0}}$ on the left-hand side of (\ref{fzmodes}) and below refers to terms that contain $(q\bar{q})^n$ with $n>0$ but no other separate powers of $q$ or $\bar{q}$ such that this is the non-perturbative part of the Fourier zero mode.

From the previous expression \eqref{eq:betalagqqb}, we can then obtain the $(q\bar{q})^n$ terms appearing in the zeroth Fourier mode of $\betalagp{j}$
\begin{subequations}
\label{botheq:betapqqb}
\begin{align}
\betalagm{j}\Big\vert_{(q\bar{q})^{>0}} &\label{eq:betamqqb0}=\betasv{2m-2-j& k-m+j\\ 2m& 2k}\Big\vert_{(q\bar{q})^{>0}}    -  \betasv{k+m-2-j &j\\ 2k& 2m}  \Big\vert_{(q\bar{q})^{>0}}  = 0 \,,\\
\betalagp{j}\Big\vert_{(q\bar{q})^{>0}} &\nn=\betasv{2m-2-j& k-m+j\\ 2m& 2k} \Big\vert_{(q\bar{q})^{>0}}   +  \betasv{k+m-2-j &j\\ 2k& 2m}  \Big\vert_{(q\bar{q})^{>0}} \\
&\label{eq:betapqqb0} = \frac{8 (2m{-}j{-}2)! (k{+}m{-}j{-}2)!}{(2m{-}1)!(2k{-}1)!} \sum_{n=1}^\infty \sigma_{2m-1}(n)\sigma_{2k-1}(n) (q\bar{q})^n\\
&\nn\qquad\quad \times  (4y)^{k+m}  U(2m{-}j{-}1,2m;4ny) U(k{+}m{-}j{-}1,2k;4ny)  \,.
\end{align}
\end{subequations}
Thanks to these equations we then deduce that the $(q\bar{q})^n$ Fourier zero-mode sector of any $\FFpm{s}{m}{k}$ can be written as a finite polynomial in $1/y$ with rational coefficients involving the product of two divisors sums. This also follows from inspection of~\eqref{eq:beta2qqb}.

In particular we have
\begin{equation}
\FFm{s}{m}{k}\Big\vert_{(q\bar{q})^{>0}}  = 0\,,\label{eq:FFmqqb}
\end{equation}
so that the full Fourier zero mode, perturbative and non-perturbative, of $\FFm{s}{m}{k}$ vanishes identically, 
which is of course expected from the fact that $\FFm{s}{m}{k}$ should be odd under the involution $\tau\to -\bar{\tau}$.
For the even modular invariant functions, by contrast, we have for example
\begin{align}
\FFp{2}{2}{2}\Big\vert_{(q\bar{q})^{>0}} &= -C_{2,1,1}\Big\vert_{(q\bar{q})^{>0}} = \sum_{n=1}^\infty \frac{\sigma_{-3}(n)^2}{2y^2} (q\bar{q})^n  \,,\\
\FFp{3}{2}{3}\Big\vert_{(q\bar{q})^{>0}} &=-4C_{3,1,1}\Big\vert_{(q\bar{q})^{>0}} = \sum_{n=1}^\infty \frac{\sigma_{-3}(n)\sigma_{-5}(n)}{4y^3}(1+ny) (q\bar{q})^n\,, \notag
\end{align}
reproducing precisely the results of \cite{Broedel:2018izr, DHoker:2019txf}, as well as \cite{Dorigoni:2019yoq} where resurgent analysis was used to reconstruct the $(q\bar{q})^n$ sector from a suitable analytic continuation of the asymptotic perturbative Laurent expansion. It would be very interesting to extend the discussion of \cite{Dorigoni:2019yoq} to the general seeds \eqref{eq:genseed} presented in this work.

\subsubsection{Non-perturbative terms in the Fourier non-zero mode}

In a similar fashion we can also derive the non-zero mode perturbative coefficients, i.e.\ the $\ell\neq0,n=0$ sectors following the Fourier decomposition \eqref{eq:Fourier}.
To proceed we can take the integral representation for the depth-two part \eqref{eq:bsv2} and isolate all the terms containing only $q$ (or alternatively only $\bar{q}$), similarly to \eqref{eq:beta2qqb} just discussed. 
For the $\overline{\alpha[\cdots]}$ we can use the explicit representation in terms of $\overline{{\cal E}_0}$, with $q$-series given by \eqref{eq:E0sigma}, as discussed in sections \ref{sec:4.3} and \ref{sec:odd.bsvrep}. 

Finally, for the depth-one part, we can use \eqref{eq:betasvE0} to write directly
\begin{align}
\betasv{j \\k} \Big\vert_{q^{>0}\bar{q}^0} &\nn=  \sum_{A=0}^{j} (k{-}j{-}2{+}A)! \binom{ j}{ A} (4y)^{2{+}2j{-}k{-}A} {\cal E}_0(k,0^{k-j-2+A})\,,\\
\betasv{j \\k} \Big\vert_{\bar{q}^{>0}\bar{q}^0} & = \sum_{B=0}^{k-j-2} (j{+}B)! \binom{k{-} j{-}2}{B} (4y)^{-B}
\overline{ {\cal E}_0(k,0^{j+B})} \,.
\end{align}
Here, the notation means that only terms with positive powers of $q$ but no powers of $\bar{q}$ are considered in the general expansion. A term with $q^\ell \bar{q}^0$ is in the $\ell$-th Fourier mode and its coefficient is perturbative with respect to $y$, i.e.\ it does not include the non-perturbative terms with $(q\bar{q})^n=e^{-4n y}$ for $n>0$.

Putting all these pieces together we can present examples such as
\begin{align}
&\FFp{2}{2}{2}\Big\vert_{q^{>0}\bar{q}^0}  = q\Big({-} \frac{y}{45} -\frac{1}{30} - \frac{1}{60y}  + \frac{\zeta_3}{2 y^2}\Big) +q^2\Big({-} \frac{y}{40} +\frac{77}{160}+ \frac{397}{640 y}  + \frac{4+9\zeta_3}{
 16 y^2} \Big) +O(q^3)\,,\nn\\
& \FFp{3}{2}{3}\Big\vert_{q^{>0}\bar{q}^0}  = q\Big({-} \frac{y^2}{1890} -\frac{y}{180} - \frac{37}{2520} -\frac{1}{56 y} -
 \frac{1+ 14\zeta_3}{112 y^2} +\frac{\zeta_3+\zeta_5}{8 y^3}\Big) \nn\\
 &\quad+q^2\Big({-}\frac{y^2}{1680}-\frac{11y}{960}+\frac{4033}{26880}+\frac{1307}{3584 y}+\frac{4891+1848\zeta_3}{14336 y^2}+\frac{32+33\zeta_3+36\zeta_5}{256 y^3} \Big) +O(q^3)\,,\nn\\
&\FFm{2}{2}{3}\Big\vert_{q^{>0}\bar{q}^0} = q\Big(\frac{2 y^3}{945}-\frac{8y^2}{945}-\frac{y}{30} -\frac{11-60\zeta_3}{252}-\frac{11+192\zeta_3}{504y} - \frac{3\zeta_3-6\zeta_5}{8y^{2}}\Big) \\
&\quad + q^2 \Big(\frac{y^3}{210}-\frac{3y^2}{70}-\frac{11 y}{160} +\frac{5(37+96\zeta_3)}{896}+\frac{1977 -6912 \zeta_3}{3584 y}+\frac{3(16-33\zeta_3+72\zeta_5)}{128 y^2}\Big) +O(q^3)\,,\nn
\end{align}
where obviously we have $\FFp{s}{m}{k} \vert_{ q^0\bar{q}^{>0}} = + \overline{\FFp{s}{m}{k} \vert_{q^{>0}\bar{q}^0}}$ while $\FFm{s}{m}{k} \vert_{q^0\bar{q}^{>0}} = - \overline{\FFm{s}{m}{k}\vert_{q^{>0}\bar{q}^0}}$.

\subsubsection{Preview to Part~II: L-values in the Fourier expansion}

Unlike for the $(q\bar{q})^n$ sector, we notice now that due to the presence of the $\overline{\alpha[\cdots]}$ integration constants, the perturbative coefficients in the non-zero Fourier mode sectors are not purely rational any longer and contain also single-valued zetas.
Furthermore, as already anticipated in section \ref{sec:3.2.5} and presented in full detail in Part~II, whenever we consider a Poincar\'e sum such that $\FFpm{s}{m}{k} \neq \cFFpm{s}{m}{k}$ the difference $\FFpm{s}{m}{k} - \cFFpm{s}{m}{k}$ will involve some special iterated integral of holomorphic (and antiholomorphic) cusp forms.
This is already evident from the perturbative coefficients in the non-zero Fourier mode sectors of these modular objects $\FFpm{s}{m}{k} \neq \cFFpm{s}{m}{k}$ which will now involve special completed L-values.
As an appetiser for Part II we have for example:
\begin{align}
\FFp{6}{4}{4} \Big\vert_{q^{>0}\bar{q}^0} &\nn= q\Big[{-}\frac{y^3}{85050}-\frac{y^2}{56700}-\frac{y}{113400}-\frac{11}{10800}+ \frac{7\lambda}{20730}+\Big({-}\frac{11}{1440} +\frac{7\lambda}{2764}\Big) y^{-1} \\
&\nn\phantom{=}+\Big({-}\frac{77}{2880}+\frac{49\lambda}{5528}\Big)y^{-2} +\Big({-}\frac{77}{1440}+\frac{49 \lambda}{2764}\Big)y^{-3} +\Big({-}\frac{77}{1280}+\frac{441 \lambda}{22112} +\frac{5\zeta_7}{192}\Big) y^{-4} \\
&\phantom{=}+\Big({-}\frac{77}{2560}+\frac{441\lambda}{44224}+\frac{29\zeta_7}{384}\Big)y^{-5} +\frac{25\zeta_7}{384}y^{-6}\Big]+O(q^2)\,,
\end{align}
where for compactness we use the shorthand $\lambda = \Lambda(\Delta_{12}, 13) / \Lambda(\Delta_{12},11)$ for a ratio of completed L-values\footnote{The completed L-function of a holomorphic cusp form $\Delta(\tau)=\sum_{n=1}^{\infty} a(n) q^n$ of modular weight $2s$ is defined~by
\begin{align}
\Lambda(\Delta,t) = (2\pi)^{-t} \Gamma(t) \sum_{n=1}^{\infty} a(n) n^{-t}\,,\nn
\end{align}
where the sum converges absolutely for $\Re(t)>s{+}\frac{1}{2}$ \cite{Hecke:1937, Deligne:1974} and can be extended to a meromorphic function.} for $\Delta_{12}(\tau)$, the Ramanujan cusp form of modular weight $12$.

As a final comment for this section we want to stress that, given the general seeds \eqref{eq:genseed} and \eqref{eq:seedm}, it is in principle possible to reconstruct the whole Fourier mode decomposition (\ref{eq:MGFq}) for the corresponding $\FFpm{s}{m}{k}$ and not just its Laurent polynomial. The mapping from seed to generic Fourier mode is given in \eqref{eq:nonzeromode}. However, the careful reader will notice the presence of some challenging Kloosterman sums \eqref{eq:Kloos} in this integral transform thus considerably complicating the story.  At the present time no general result is known for the non-zero Fourier modes, unlike our general formula for Laurent polynomial \eqref{eq:Iabr}. 
Our previous discussion implies that a careful analysis of these Kloosterman sums should produce zeta values as well as more complicated ratios of completed L-values amongst the perturbative coefficients in the non-zero Fourier mode sectors. Needless to say it would be extremely interesting to explore this direction.

%%%%%%%%%%%%%%%%%%%%%%%%%%%%%%%%%%%%%%%%%%%%%%%%%%%%%%%%%%%
\subsection{Even cusp forms at depth two}
\label{sec:5.1}
%%%%%%%%%%%%%%%%%%%%%%%%%%%%%%%%%%%%%%%%%%%%%%%%%%%%%%%%%%%

Given the results for the Laurent polynomials of $\FFp{s}{m}{k}$ in section \ref{sec:4.1}, we can
construct infinite families of real cusp forms, i.e.\ modular invariants that are even under $\tau\to-\bar\tau$. These complement the imaginary or odd cusp forms $\FFm{s}{m}{k}$, $\Jm{\ell}{m}{k}$ (along with their antecedents in \cite{DHoker:2019txf, Gerken:2020yii}) and the first even cusp forms identified in \cite{Gerken:2020aju}. The construction is based on a simple counting argument for the dimension of the vector space given by the Laurent polynomials appearing for the even modular invariants discussed so far at a given weight.  

From our analysis (\ref{laurent}) we know that each $\FFp{s}{m}{k}$ with $m<k$ contains five different Laurent monomials, and four for $m=k$. The products $\Jp{\ell}{m}{k}$ of depth-one objects defined in~\eqref{prodd1} for $\ell\geq 0$ contribute
the same types of Laurent monomials (with $m<k$)
\begin{align}
\Jp{\ell}{m}{k}  \ \text{mod} \ O(q,\bar q)\ &\longleftrightarrow \  y^{m+k} \, , \  \zeta_{2m-1} y^{k-m+1} \, , \ \zeta_{2k-1} y^{m-k+1}\, , \  \frac{ \zeta_{2m-1} \zeta_{2k-1} }{ y^{k+m-2} }  \,,
 \notag \\
 \Jp{\ell}{k}{k} \ \text{mod} \ O(q,\bar q) \  &\longleftrightarrow \ y^{2k} \, , \  \zeta_{2k-1} y 
\, , \  \frac{\zeta_{2k-1}^2 }{ y^{2k-2} }   \, , \label{laurentH}
\end{align}
though their relative coefficients will differ from those of $\FFp{s}{m}{k}$, and the terms $\zeta_{m+k+s-1} y^{1-s}$ of the latter are absent. The exact general formula for $m\leq k$ similar to~\eqref{eq:FLP} is here
\begin{align}
\Jp{\ell}{m}{k}  &= \frac{(-4)^{k{+}m} {\rm B}_{2m}{\rm B}_{2k} (k)_\ell (m)_\ell}{(2m)!(2k)!} y^{k+m} 
- \frac{(-1)^k 4^{2{-}m{+}k} {\rm B}_{2k} (2m{-}3)! (1{-}m)_\ell (k)_\ell \zeta_{2m{-}1}}{(m{-}1)!(m{-}2)!(2k)!} y^{1-m+k}\nn\\
&\quad - \frac{(-1)^m 4^{2{-}k{+}m} {\rm B}_{2m} (2k{-}3)! (1{-}k)_\ell (m)_\ell \zeta_{2k{-}1}}{(k{-}1)!(k{-}2)!(2m)!} y^{1-k+m}  \label{eq:HLP} \\
&\quad +\frac{4^{4{-}m{-}k} (2m{-}3)!(2k{-}3)! (1{-}m)_\ell (1{-}k)_\ell \zeta_{2m{-}1}\zeta_{2k{-}1}}{(m{-}1)!(m{-}2)!(k{-}1)!(k{-}2)!} y^{2-k-m} +O(q,\bar{q}) \notag \,,
\end{align}
where the middle two terms coalesce for $m=k$.

From the admissible values of $s$ in (\ref{eq:Fmk}), the collection of all Laurent polynomials (\ref{laurent}) of $\FFp{s}{m}{k}$ appearing at a fixed weight $w=k{+}m$ involves degrees $(2{-}w,3{-}w,5{-}w,\ldots,w{-}3,w)$. Most of the coefficients were found to be rational multiples of odd zeta values 
\beq
\mathcal{V}_{{\rm F}^+ \oplus {\rm J}^+}(w) {=} \mbox{span}_\mathbb{Q}\left\{\FFp{s}{m}{k} , \, \Jp{\ell}{m}{k}   \,\middle\vert\,w=k+m\right\}\ \underset{\text{LP}}{\rightarrow} \ \zeta_3 y^{w-3}, \ \zeta_5 y^{w-5} , \ldots , \zeta_{2w-5} y^{5-w}, \ \zeta_{2w-3} y^{3-w}
\label{laurentH.1}
\eeq
except for the purely rational coefficient of $y^w$ and products $\zeta_{2m-1}\zeta_{2k-1}$ multiplying $y^{2-w}$ in
any $\FFp{s}{m}{k} $ or $ \Jp{\ell}{m}{k} $ with $k{+}m=w$. More precisely, there are $\lfloor \frac{w}{2}\rfloor -1$ distinct products $\zeta_{2m-1}\zeta_{2k-1}$ compatible with $2\leq m\leq k$ and fixed weight $k{+}m=w$.
Together with $y^w$ and the $w{-}2$ Laurent monomials in (\ref{laurentH.1}), we arrive at a total of
\beq
n^\mathbb{Q}_{{\rm F}^+\oplus {\rm J}^+} (w)=  w+\left\lfloor\frac{w}{2}\right\rfloor-2
\label{laurentH.2}
\eeq
$\mathbb Q$-independent monomials in the Laurent polynomials of all even modular invariants
$\FFp{s}{m}{k} $ and $ \Jp{\ell}{m}{k} $ of weight $k{+}m=w$, see table \ref{TLcount.1} for examples. This counting is based on the standard transcendentality conjectures for MZVs 
\cite{GilFresan}, i.e.\ we have assumed that all odd zeta values and their bilinears are
linearly independent over $\mathbb{Q}$.

\begin{table}[h]
\begin{center}
\begin{tabular}{c||c|c|c}
weight $w$ & zeta values of depth $\leq 1$ & bilinears in  $\zeta_{2k-1}$ & $n^\mathbb{Q}_{{\rm F}^+\oplus {\rm J}^+} (w)$ 
\\[2mm]\hline\hline
4 &$1 , \ \zeta_3, \ \zeta_5$ &$\zeta_3^2$ &4  \\\hline
5 &$1 , \ \zeta_3, \ \zeta_5, \ \zeta_7$ &$\zeta_3 \zeta_5$ &5  \\\hline
6 &$1 , \ \zeta_3, \ \zeta_5, \ \zeta_7, \ \zeta_9$ &$\zeta_3 \zeta_7, \ \zeta_5^2$ &7  \\\hline
7 &$1 , \ \zeta_3, \ \zeta_5, \ \zeta_7, \ \zeta_9, \ \zeta_{11}$ &$\zeta_3 \zeta_9, \ \zeta_5 \zeta_7$ &8  \\\hline
8 &$1 , \ \zeta_3, \ \zeta_5, \ \zeta_7, \ \zeta_9, \ \zeta_{11}, \ \zeta_{13}$ &$\zeta_3 \zeta_{11}, \ \zeta_5 \zeta_9, \ \zeta_7^2$ &10  \\\hline
9 &$1 , \ \zeta_3, \ \zeta_5, \ \zeta_7, \ \zeta_9, \ \zeta_{11}, \ \zeta_{13}, \ \zeta_{15}$ &$\zeta_3 \zeta_{13}, \ \zeta_5 \zeta_{11}, \ \zeta_7 \zeta_9$ &11  \\\hline
10 &$1 , \ \zeta_3, \ \zeta_5, \ \zeta_7, \ \zeta_9, \ \zeta_{11}, \ \zeta_{13}, \ \zeta_{15}, \ \zeta_{17}$ &$\zeta_3 \zeta_{15}, \ \zeta_5 \zeta_{13}, \ \zeta_7 \zeta_{11} , \ \zeta_9^2$ &13
\end{tabular}
\end{center}
\caption{\textit{Zeta values occurring in the Laurent polynomials of even modular invariants of depth two.}}
\label{TLcount.1}
\end{table}

However, similar to the discussion in section \ref{sec:3.2.5},
the total number of $\FFp{s}{m}{k} $ and $ \Jp{\ell}{m}{k} $ appearing at a fixed weight $k{+}m=w$
follows the very simple counting
\begin{align}
\dim \mathcal{V}_{{\rm F}^+}(w) &=  \sum_{s=2}^{w-2} \dim \mathcal{V}_{{\rm F}^+}(w,s) = \frac{1}{2}\Big( \left\lfloor\frac{w}{2} \right\rfloor^2-\left\lfloor\frac{w}{2} \right\rfloor \Big)\,, \notag \\
\dim \mathcal{V}_{{\rm J}^+}(w) & =\sum_{m=2}^{\lfloor \frac{w}{2}\rfloor} m=   \frac{1}{2}\Big( \left\lfloor\frac{w}{2} \right\rfloor^2+\left\lfloor\frac{w}{2} \right\rfloor-2 \Big)\,.
\label{laurentH.3}
\end{align}
We have used equation (\ref{eq:dimF}) for  $\dim \mathcal{V}_{{\rm F}^+}(w,s)$, while we simply counted the number of depth-two $\Jp{\ell}{m}{k}$ with $0\leq \ell < \mbox{min}(m,k)$ at fixed weight. By comparing with the counting of
$\mathbb Q$-independent Laurent monomials in (\ref{laurentH.2}), we conclude that
\begin{align}
\dim \mathcal{V}_{{\rm F}^+\oplus{\rm J}^+}(w) &=
\dim \mathcal{V}_{{\rm F}^+}(w)+ \dim \mathcal{V}_{{\rm J}^+}(w) = \left\lfloor\frac{w}{2} \right\rfloor^2 - 1 
\label{laurentH.4}\\
&\geq n^\mathbb{Q}_{{\rm F}^+\oplus {\rm J}^+}(w) =  w+\left\lfloor\frac{w}{2}\right\rfloor-2\,,\qquad w= 6 \ \ {\rm or} \ \ w\geq 8\,,
\notag
\end{align}
so for weight $w=6$ and $w\geq 8$, we have more depth-two modular invariant objects than possible Laurent polynomials. This means that for sufficiently large weights, the space of even modular invariants spanned by the linearly independent functions  $\FFp{s}{m}{k}$ and $\Jp{\ell}{m}{k}$ must contain some even cusp forms, see table \ref{TLcount.2}. 

For odd $w$, there is the additional possibility of adding the constant  $\zeta_w$ to the space of (single-valued) modular invariants which corresponds to the vanishing eigenvalue of the Laplacian. Instances of the appearance of $\zeta_w$ can be seen in~\eqref{cabc.1} and \eqref{z7rel}. To take this into account, we introduce the following notation for the combined bookkeeping
\beq
\dim \mathcal{V}_{{{\rm F}^+}\oplus{\rm J}^+\oplus\zeta}(w) = \dim \mathcal{V}_{{{\rm F}^+}\oplus{\rm J}^+}(w) + w-2 \left\lfloor\frac{w}{2} \right\rfloor =   \dim \mathcal{V}_{{{\rm F}^+}\oplus{\rm J}^+}(w) + \left\{
\begin{array}{cl}
  1  &w \ \text{odd} \, , \\[1mm]
0     & w \ \text{even}\, .
\end{array}
\right.
\eeq
In fact, the difference
\beq
\dim \mathcal{V}_{{{\rm F}^+}\oplus{\rm J}^+\oplus\zeta}(w)  - n^\mathbb{Q}_{{\rm F}^+\oplus {\rm J}^+}(w)
= \left\lfloor\frac{w}{2} \right\rfloor^2 -3\left\lfloor\frac{w}{2} \right\rfloor  + 1
\label{laurentH.5}
\eeq
is a lower bound on the number of cusp forms at fixed $w=k{+}m$. Larger numbers are conceivable
since it is not a priori clear if the Laurent polynomials of the relevant
$\FFp{s}{m}{k}$ and $\Jp{\ell}{m}{k}$ are linearly independent over $\mathbb Q$. We have tested
up to and including weight $w=28$ that the bound is saturated, i.e.\ that (\ref{laurentH.5}) is the actual number of even cusp forms. At odd weight $w \geq 7$, an additional cusp can be formed by adding a rational multiple of $\zeta_w$ to the general $\mathbb Q$-linear combination of $\FFp{s}{m}{k}$ and $\Jp{\ell}{m}{k}$.

\begin{table}[h]
\begin{center}
\begin{tabular}{c||c|c|c|c|c}
weight $w$ &$\dim \mathcal{V}_{{\rm F}^+}(w)$ &$ \dim \mathcal{V}_{{\rm J}^+}(w) $ 
&$\dim \mathcal{V}_{{{\rm F}^+}\oplus{\rm J}^+\oplus\zeta}(w)$
&$n^\mathbb{Q}_{{\rm F}^+\oplus {\rm J}^+} (w)$
&\#(cusp forms) 
\\[1mm]\hline\hline
4 &1 &2 &3 &4 &0
 \\\hline
5 &1 &2 &4 &5 &0
 \\\hline
6 &3 &5 &8 &7&1
 \\\hline
7 &3 &5 &9 &8 &1
 \\\hline
8 &6 &9 &15 &10 &5
 \\\hline
9 &6 &9 &16 &11 &5
 \\\hline
10 &10 &14 &24 &13 &11 
\end{tabular}
\end{center}
\caption{\textit{Extracting the minimum number of even cusp forms at weight $w$ by comparing the counting of Laurent monomials in table \ref{TLcount.1} with the total number of $\FFp{s}{m}{k} $ and $ \Jp{\ell}{m}{k} $. At each odd weight $w\geq 7$, an additional even cusp form can be formed by adding multiples of $\zeta_w$ to the combinations of modular invariants at depth 2.}}
\label{TLcount.2}
\end{table}
\begin{itemize}
\item For example at weight $w=6$ we have $\dim \mathcal{V}_{{\rm F}^+}(6)+ \dim \mathcal{V}_{{\rm J}^+}(6)  = 8$ while the dimension over the rationals of the possible Laurent polynomials is $n^\mathbb{Q}_{{\rm F}^+\oplus {\rm J}^+}(6)  = 7$. Since the latter are linearly independent over $\mathbb Q$, we obtain a single cusp form which is fairly simple to construct by cancelling the Laurent polynomials of $\FFp{s}{m}{k}$ and $\Jp{\ell}{m}{k}$ in the combination
\begin{align}
{\rm S}_2^{(6)} &\label{eq:cuspS2} = -54\FFp{4}{2}{4}  - \frac{162}{25} \FFp{2}{3}{3} + \frac{720}{7}\FFp{4}{3}{3} -\frac{269}{50} \Jp{0}{3}{3} - \frac{3739}{2100}\Jp{1}{3}{3} + \frac{1}{840}\Jp{2}{3}{3}+ \frac{9}{4}\Jp{1}{2}{4} \,.
\end{align}
This is one of the even cusp forms of weight 6 presented in equation (9.8b) of \cite{Gerken:2020aju}\footnote{The identification of (\ref{eq:cuspS2}) with (9.8b) of \cite{Gerken:2020aju} relies on the change of basis (cf.\ (\ref{oct20.4}) and (\ref{oct20.6}))
\[
{\rm E}_{2,4} = -54\FFp{4}{2}{4}\,,\qquad {\rm E}_{3,3} = -\frac{18}{5} (\FFp{2}{3}{3} - \FFp{4}{3}{3})\,,\qquad
 {\rm E}_{3,3}' = -\frac{27}{25} \FFp{2}{3}{3} + \frac{12}{25}\FFp{4}{3}{3}\,.
\]}, where additional cusp forms at that weight have been constructed from admixtures of modular graph forms of depth three.

The counting in table \ref{TLcount.2} leads to the following further examples based on the
fact that $\mathbb Q$-linear combinations of $\FFp{s}{m}{k}$ and $\Jp{\ell}{m}{k}$ span the whole
set of admissible Laurent polynomials. This is checked on a case-by-case basis at weight $w\leq 28$:

\item At weight $w=7$ we have $\dim \mathcal{V}_{{\rm F}^+}(7)+ \dim \mathcal{V}_{{\rm J}^+}(7)  = 8 = n^\mathbb{Q}_{{\rm F}^+\oplus {\rm J}^+}(7)$ so we would not expect any even cusp at depth two if it was not for the possibility of combining $\zeta_7$ with $\FFp{s}{m}{k}$ and $\Jp{\ell}{m}{k}$. Hence, a single even cusp form can be found 
\begin{align}
{\rm S}_1^{(7)} &=
23 \zeta_7 -37635840 \FFp{3}{3}{4} +  230519520 \FFp{5}{3}{4} - 102453120 \FFp{5}{2}{5}  \label{laurentH.8} \\
& \ \ \ \
 - 10869120 \Jp{0} {3}{ 4}  - 1773180 \Jp{1}{3}{4} + 
 2034 \Jp{2}{3}{4} +  2561328 \Jp{1}{2}{5}  \,.\notag
\end{align}
\item At weight $w=8$, the space of even cusp forms at depth two is five-dimensional by
$\dim \mathcal{V}_{{\rm F}^+}(8)+ \dim \mathcal{V}_{{\rm J}^+}(8)  = 15$ and $n^\mathbb{Q}_{{\rm F}^+\oplus {\rm J}^+}(8)=10$, and one can pick the following basis: {\allowdisplaybreaks
\begin{align}
{\rm S}^{(8)}_{1} &=  \FFp{4}{4}{4}  -\frac{\FFp{2}{4}{4}}{7}  - \frac{9501862 \FFp{4}{3}{5}}{63035} 
- \frac{142086685 \FFp{6}{2}{6}}{529494} + 
\frac{ 2436903800 \FFp{6}{3}{5} }{3744279} - \frac{5 \FFp{6}{4}{4}}{14} \notag \\*
& \ \  - 
\frac{ 17440506049 \Jp{0}{3}{5}}{655248825} + 
\frac{ 28417337 \Jp{1}{2}{6}}{6353928} - 
\frac{ 27880409152 \Jp{1}{3}{5}}{9828732375 }+ 
\frac{ 482042723 \Jp{2}{3}{5}}{117944788500} \,,\notag \\
{\rm S}^{(8)}_{2} &= \frac{ 20}{7} \FFp{2}{4}{4} + \frac{2046167552 \FFp{4}{3}{5}}{567315 }+ 
\frac{ 5115418880 \FFp{6}{2}{6}}{794241 }- \frac{ 5311016000 \FFp{6}{3}{5}}{340389} - \frac{ 90}{7} \FFp{6}{4}{4}  
\notag  \\*
& \ \ {+} 
\frac{ 37987131796 \Jp{0}{3}{5}}{59568075} {+} \Jp{0}{4}{4} {-} 
\frac{ 255770944 \Jp{1}{2}{6}}{2382723 }{+} 
\frac{ 60832177033 \Jp{1}{3}{5}}{893521125} {-}
\frac{ 525859921 \Jp{2}{3}{5}}{5361126750} \,,\notag \\
{\rm S}^{(8)}_{3} &= -\frac{120}{7} \FFp{2}{4}{4} - 
\frac{ 4778052136 \FFp{4}{3}{5}}{189105} - 
\frac{ 11945130340 \FFp{6}{2}{6}}{264747} + 
\frac{ 12420404800 \FFp{6}{3}{5}}{113463} \label{laurentH.9}  \\*
& \ \ -\frac{ 720}{7} \FFp{6}{4}{4} - 
\frac{ 88794207428 \Jp{0}{3}{5}}{19856025} + 
\frac{ 597256517 \Jp{1}{2}{6}}{794241} - 
\frac{ 142388765144 \Jp{1}{3}{5}}{297840375} 
\notag \\*
& \ \ + \Jp{1}{4}{4}  + \frac{ 613427464 \Jp{2}{3}{5}}{893521125 }\,,
\notag \\
{\rm S}^{(8)}_{4} &= \frac{ 1920}{7} \FFp{2}{4}{4} + \frac{17886117392 \FFp{4}{3}{5}}{37821} 
+ \frac{ 223576467400 \FFp{6}{2}{6}}{264747} - 
\frac{ 232237840000 \FFp{6}{3}{5}}{113463} \notag \\*
& \ \  - 
\frac{ 3600}{7} \FFp{6}{4}{4} + 
\frac{ 332347746616 \Jp{0}{3}{5}}{3971205} - 
\frac{ 11178823370 \Jp{1}{2}{6}}{794241} + 
\frac{ 532024449268 \Jp{1}{3}{5}}{59568075}  \notag \\*
& \ \ - \frac{ 2503432058 \Jp{2}{3}{5}}{178704225 }+ \Jp{2}{4}{4}\,, \notag \\
{\rm S}^{(8)}_{5} &= 
2880 \Jp{0}{4}{4} - 180 \Jp{1}{4}{4} - 
 36 \Jp{2}{4}{4} + \Jp{3}{4}{4} \,.\notag 
\end{align}}
\end{itemize}
Note that only four linear combinations spanned by
${\rm S}^{(8)}_{1},\ldots, {\rm S}^{(8)}_{5}$ are expressible
in terms of MGFs since $\FFp{6}{4}{4},\FFp{6}{3}{5}$ and $\FFp{6}{2}{6}$ carry admixtures of
holomorphic cusp forms detailed in Part~II that cancel from the combinations (\ref{cF6xy}).

By the linear independence of the $\betalagp{j}$ of depth two entering via $\FFp{s}{m}{k},\Jp{\ell}{m}{k}$, none of the even cusp forms in the counting of this section vanishes. Moreover, they do not satisfy any obvious Laplace-type equations. This is due to the fact that we are using $\Jp{\ell}{m}{k}$ with $\ell < \mbox{min}(m,k)$ which do not close
under $(\Delta - \lambda)$ for any eigenvalue $\lambda$, see section \ref{sec:3.2.1}.
Despite all of these cusp forms having vanishing perturbative expansion, i.e.\ vanishing Laurent polynomials, we believe it still should be possibly to use a similar resurgent analysis as described in \cite{Dorigoni:2019yoq,Dorigoni:2020oon}, to suitably deform the perturbative expansion and use it to retrieve all of the non-perturbative, i.e.\ $( q \bar{q})^n$ terms. 
In particular we note that, unlike what happens in \eqref{eq:FFmqqb} for odd cusp forms $\FFm{s}{m}{k}$, these new even cusp forms will generically have non-vanishing $(q\bar{q})^n$ terms in the zeroth Fourier mode.

Note that $C_{a,b,c}$ and ${\rm E}_w$ do not add any even cusp forms to the
$\mathbb Q$-span of $\zeta_w, \FFp{s}{m}{k}$ and $\Jp{\ell}{m}{k}$: First, $C_{a,b,c}$ can be rewritten in terms of
$\zeta_w, \FFp{s}{m}{k}$ and ${\rm E}_w$ with $w= m{+}k=a{+}b{+}c$ according to (\ref{eq:decomp}). Second, the
Laurent monomial $ \zeta_{2w-1} y^{1-w}$ of ${\rm E}_w$ in (\ref{eq:FEk}) cannot be compensated
by any of $\zeta_w, \FFp{s}{m}{k},\Jp{\ell}{m}{k}$, so the coefficients of ${\rm E}_w$ in the cusp forms
of interest must be zero. Third, products $\zeta_a {\rm E}_{w-a}$ with odd $a\geq 3$ 
also introduce additional bilinears $\zeta_a \zeta_{2w-2a-1}$ into the Laurent polynomials which do not arise from 
$\FFp{s}{m}{k}$ and $\Jp{\ell}{m}{k}$ such that their $\mathbb Q$-coefficients have to vanish separately
in cusp forms.\footnote{This can be seen as follows: The combined weight of $\zeta_a \zeta_{2w-2a-1}$ 
due to $\zeta_a {\rm E}_{w-a}$ is $2w{-}a{-}1 \leq 2w{-}4$ (since $a\geq 3$) 
which is strictly lower than the weight $2w{-}2$ of the $\zeta_{2k-1}\zeta_{2m-1}$ from $\FFp{s}{m}{k}$ 
and $\Jp{\ell}{m}{k}$.}

As a final comment, we notice that additional even cusp forms can be formed from
$\mathbb Q[{\rm MZV}]$-linear rather than $\mathbb Q$-linear combinations of 
$\zeta_w, \FFp{s}{m}{k}$ and $\Jp{\ell}{m}{k}$. For instance, an additional even cusp
form built from $\FFp{s}{m}{k},\Jp{\ell}{m}{k}$ at $m{+}k=6$ beyond (\ref{eq:cuspS2}) 
and those in \cite{Gerken:2020aju} is given by
\begin{align}
\tilde {\rm S}_2 &= \zeta_5^2\Big( \frac{9}{2}\FFp{2}{3}{3}  - 
\frac{3069}{1400}\Jp{0}{3}{3} + \frac{673}{2800} \Jp{1}{3}{3} + \frac{1373}{5600}\Jp{2}{3}{3} -\frac{45}{4} \Jp{0}{2}{4}  - \frac{45}{16}\Jp{1}{2}{4} \Big) \notag \\
&\phantom{=}+ 
  \frac{7 \zeta_3\zeta_7}{32} \Big(72 \Jp{0}{3}{3} + 8 \Jp{1}{3}{3} - \Jp{2}{3}{3}\Big) \,.
\label{cforms}
\end{align}
Although the $\tau$-dependence is carried by modular graph forms of depth two and weight six,
the cusp form (\ref{cforms}) should be understood as having depth four and weight sixteen
by the coefficients in $\mathbb{Q}[\zeta_5^2]$ and $\mathbb{Q}[\zeta_3\zeta_7]$.

%%%%%%%%%%%%%%%%%%%%%%%%%%%%%%%%%%%%%%%%%%%%%%%%%%%%%%%%%%%
\subsection{Overly integrated seed functions}
\label{sec:5.2}
%%%%%%%%%%%%%%%%%%%%%%%%%%%%%%%%%%%%%%%%%%%%%%%%%%%%%%%%%%%

From the comprehensive collection of seed functions and their associated Poincar\'e series in tables \ref{chess.1} to \ref{chess.3ALT}, one might wonder about the cross signposts $\times$ in the remaining cells:
What type of seed functions and corresponding modular objects correspond to the black crosses filling the 
right side, and the red crosses (and beyond) on the top left corner?

Let us first discuss the infinite class of modular objects associated with seed functions that
exceed the maximum number of zeros ${\cal E}_0(2m,0^{2m-2})$ encountered in earlier sections,
\beq
y^a {\cal E}_0(2m,0^b)
\,, \ \ \ \ \ \ a\geq1 \, , \ \ \ \ \ \ b>2m{-}2\, ,
\label{overseed}
\eeq
i.e.\ the black crosses left as signposts for the infinitely extended right side of the tables. We shall address the red crosses and beyond in the next section.
The class of seeds~\eqref{overseed} will be informally referred to as ``overly integrated'' iterated Eisenstein integrals:
The term ``overly'' is understood in comparison with the iterated Eisenstein integrals over kernels $\tau^j {\rm G}_k(\tau)$
with $0 \leq j\leq k{-}2$ \cite{Brown:mmv, Brown:I, Brown:II}, where the restriction on $j$
ensures nice modular properties in terms of iterated integrals of the form $ {\cal E}_0(2m,0^b)$.
There is nothing wrong \textit{per se} in considering overly integrated iterated integrals, however, their modular S-transformation involves an infinite series of ${\cal E}_0(2m,0^c)$ with no upper bound on $c$ \cite{Dorigoni:2020oon}.

The modular invariant functions associated with the overly integrated seeds (\ref{overseed})
can be understood from our discussion in section \ref{sec:3.5} of the Laplace system in steps form.
We focus on the case of $\Re[{\cal E}_0]$ for definiteness and can go from the rightmost columns in tables~\ref{chess.1} to \ref{chess.3ALT} to the neighbouring columns on their left by considering 
\beq
\mathcal{O}_{a} \Big( y^a \Re[{\cal E}_0(2m,0^{2m-2})] \Big) =y^{a+1}\Re[{\cal E}_0(2m,0^{2m-3})] \,,
\eeq
see (\ref{oct21.3}) for the shifted Laplacians ${\cal O}_a$ and as depicted in table \ref{chess.1Over} for the $m=2$ case.

\begin{table}[h]
\begin{center}
\tikzpicture
\draw(-2.5,-2.1)node{\underline{$m=2$:}};
\draw(-0.04,1) -- (-0.04,-4.9);
\draw(0.04,1) -- (0.04,-4.9);
\draw(-1,0.04) -- (5.8,0.04);
\draw(-1,-0.04) -- (5.8,-0.04);
\draw(-0.04,0.04) -- (-1,1);
\draw(-0.75,0.3) node{$a$};
\draw(-0.3,0.75) node{$b$};
\draw(1.6,1.1) -- (1.6,-4.9);
\draw(3.2,1.1) -- (3.2,-4.9);
\draw(4.8,1.1) -- (4.8,-4.9);
\draw(-1,-0.7) -- (5.8,-0.7);
\draw(-1,-1.4) -- (5.8,-1.4);
\draw(-1,-2.1) -- (5.8,-2.1);
\draw(-1,-2.8) -- (5.8,-2.8);
\draw(-1,-3.5) -- (5.8,-3.5);
\draw(-1,-4.2) -- (5.8,-4.2);
\draw(0.8,0.5)node{0};
\draw(2.4,0.5)node{1};
\draw(4.0,0.5)node{2};
\draw(5.3,0.5)node{3};
\draw(-0.5,-0.35)node{1};
\draw(-0.5,-1.05)node{2};
\draw(-0.5,-1.75)node{3};
\draw(-0.5,-2.45)node{4};
\draw(-0.5,-3.15)node{5};
\draw(-0.5,-3.85)node{6};
\draw(-0.5,-4.55)node{7};
% 1st beta^svs %% INSERT STUFF FROM HERE
\draw(0.8,-0.35)node{$\times$};
\draw(0.8,-1.05)node{$\rcross$};
\draw(0.8,-1.75)node{$\Big.\beta^{{\rm sv},\, j\leq 2}_{2,2} \Big.$};
\draw(0.8,-2.45)node{$\beta^{{\rm sv},\, j\leq 2}_{2,3} \Big.$};
\draw(0.8,-3.15)node{$\beta^{{\rm sv},\, j\leq 2}_{2,4} \Big.$};
\draw(0.8,-3.85)node{$\beta^{{\rm sv},\, j\leq 2}_{2,5} \Big.$};
\draw(0.8,-4.55)node{$\beta^{{\rm sv},\, j\leq 2}_{2,6} \Big.$};
% 2nd beta^svs
\draw(2.4,-0.35)node{$\rcross$};
\draw(2.4,-1.05)node{$\beta^{{\rm sv},\, j\leq 1}_{2,2} \Big.$};
\draw(2.4,-1.75)node{$\beta^{{\rm sv},\, j\leq 1}_{2,3}$};
\draw(2.4,-2.45)node{$\beta^{{\rm sv},\, j\leq 1}_{2,4}$};
\draw(2.4,-3.15)node{$\beta^{{\rm sv},\, j\leq 1}_{2,5}$};
\draw(2.4,-3.85)node{$\beta^{{\rm sv},\, j\leq 1}_{2,6}$};
\draw(2.4,-4.55)node{$\beta^{{\rm sv},\, j\leq 1}_{2,7}$};
% 3rd beta^svs
\draw(4.0,-0.35)node{$\beta^{{\rm sv},\, j=0}_{2,2} \Big.$};
\draw(4.0,-1.05)node{$\beta^{{\rm sv},\, j=0}_{2,3}$};
\draw(4.0,-1.75)node{$\beta^{{\rm sv},\, j=0}_{2,4}$};
\draw(4.0,-2.45)node{$\beta^{{\rm sv},\, j=0}_{2,5}$};
\draw(4.0,-3.15)node{$\beta^{{\rm sv},\, j=0}_{2,6}$};
\draw(4.0,-3.85)node{$\beta^{{\rm sv},\, j=0}_{2,7}$};
\draw(4.0,-4.55)node{$\beta^{{\rm sv},\, j=0}_{2,8}$};
\draw(5.3,-0.35)node{$\times$};
\draw(5.3,-1.05)node{$\times$};
\draw(5.3,-1.75)node{$\times$};
\draw(5.3,-2.45)node{$\times$};
\draw(5.3,-3.15)node{$\times$};
\draw(5.3,-3.85)node{$\times$};
\draw(5.3,-4.55)node{$\times$};
\draw[blue](3.2,-0.7)node{$\swarrow$}node[above]{\footnotesize${\cal O}_1 \ \ \ \ $};
\draw[blue](1.6,-1.4)node{$\swarrow$}node[above]{\footnotesize${\cal O}_2 \ \ \ \ $};
\draw[red](4.8,-0.7)node{$\swarrow$}node[above]{\footnotesize${\cal O}_1 \ \ \ \ $};
\draw[blue](3.2,-1.4)node{$\swarrow$}node[above]{\footnotesize${\cal O}_2 \ \ \ \ $};
\draw[blue](1.6,-2.1)node{$\swarrow$}node[above]{\footnotesize${\cal O}_3 \ \ \ \ $};
\draw[red](6.0,-0.7)node{$\swarrow$}node[above]{\footnotesize${\cal O}_1 \ \ \ \ $};
\draw[red](4.8,-1.4)node{$\swarrow$}node[above]{\footnotesize${\cal O}_2 \ \ \ \ $};
\draw[blue](3.2,-2.1)node{$\swarrow$}node[above]{\footnotesize${\cal O}_3 \ \ \ \ $};
\draw[blue](1.6,-2.8)node{$\swarrow$}node[above]{\footnotesize${\cal O}_4 \ \ \ \ $};
\draw[red](6.0,-1.4)node{$\swarrow$}node[above]{\footnotesize$\ldots \ \ \ \ $};
\draw[red](4.8,-2.1)node{$\swarrow$}node[above]{\footnotesize${\cal O}_3 \ \ \ \ $};
\draw[blue](3.2,-2.8)node{$\swarrow$}node[above]{\footnotesize$\ldots \ \ \ \ $};
\draw[blue](1.6,-3.5)node{$\swarrow$}node[above]{\footnotesize$\ldots \ \ \ \ $};
\endtikzpicture
\end{center}
\caption{\textit{Filling in the chessboard with the action of the operators $\mathcal{O}_\ell$ defined in~\eqref{oct21.3}.}}
\label{chess.1Over}
\end{table}

\subsubsection{\texorpdfstring{Example in the $(2,3)$ sector}{Example in the (2,3) sector}}

For example using table \ref{chess.1}, together with the results of section \ref{sec:3.5}, we can construct the tower of equations relevant for $\beta^{\rm sv}_{2,3}$ 
\begin{align}
\mathcal{O}_2 \Big[y^2 \Big( \Re[{\cal E}_0(4,0^2)] - \frac{\zeta_3}{6}\Big) +\frac{y^5}{945} \Big] &=y^3\Re[{\cal E}_0(4,0)]-\frac{y^5}{420} \,, \notag \\
\mathcal{O}_3 \Big[y^3\Re[{\cal E}_0(4,0)]-\frac{y^5}{420} \Big] &= y^4\Re[{\cal E}_0(4)]+\frac{y^5}{360}\,,\label{oldtower1}\\
\mathcal{O}_4\Big[y^4\Re[{\cal E}_0(4)]+\frac{y^5}{360}\Big] &= -\frac{y^5}{(2\pi i)^4} \Re{\rm G}_4\,,
\notag
\end{align}
or equivalently after Poincar\'e summation:
\begin{align}
\mathcal{O}_2 \Big(315 \FFp{3}{2}{3}\Big) &= -\frac{315}{8} \Big(\FFp{3}{2}{3}+\Jp{0}{2}{3}\Big)\,, \notag \\
\mathcal{O}_3\Big[ {-}\frac{315}{8} \Big(\FFp{3}{2}{3}+\Jp{0}{2}{3}\Big)\Big] &= \frac{105}{16}\Big(3\Jp{0}{2}{3}+\Jp{1}{2}{3}\Big)\, , \label{oldtower2}\\
\mathcal{O}_4 \Big[ \frac{105}{16}\Big(3\Jp{0}{2}{3}+\Jp{1}{2}{3}\Big)\Big] &=-\frac{315}{256}\Big(\GG_{4} (\pi \overline{\nabla})^2 \EE_3+ \overline{\GG}_{4} (\pi \nabla)^2 \EE_3 \Big)\,.
\notag 
\end{align}
These towers can be seen as being generated by moving along a diagonal of table \ref{chess.1Over}, starting from the top right corner.

With a similar reasoning we can then start populating the crosses on the infinite right side of all these tables.
For example to determine the top right cross in table \ref{chess.1Over} we want to find a seed $f$ subject to
the Laplace equation
\beq
\mathcal{O}_1 f = y^2\Big(\Re[{\cal E}_0(4,0^2)] - \frac{\zeta_3}{6} \Big)+ \frac{y^5}{945}\,,
\label{inhomoequ}
\eeq
which is for instance solved by
\beq
f = y\Re[{\cal E}_0(4,0^3)] +\frac{\zeta_3}{3} y^2-\frac{y^5}{4725}\,,\label{eq:OverlySeed}
\eeq
and, as expected, we start generating overly integrated iterated integrals.\footnote{The 
modular invariant solutions to the homogeneous version $\mathcal{O}_1f=0$ of (\ref{inhomoequ}) include constants, so that we could in principle add zeta values to the solutions (\ref{eq:OverlySeed}). We do not include them here as we have no other means to determine them and our focus is on displaying the overly integrated integrals.}

Alternatively, the modular function $F$ defined by the Poincar\'e summation over this seed $f$ must obey
\beq
\mathcal{O}_1 F = -\frac{1}{4}\Delta F = 315 \FFp{3}{2}{3}\,.
\label{newlapeq}
\eeq
Interestingly enough, we can still use equations (\ref{conrul.1}) and (\ref{conrul.2}) applied to the seed (\ref{eq:OverlySeed}) to obtain the perturbative expansion for this new modular function
\begin{align} F &\notag= -\frac{y^5}{4725} + \frac{\zeta_3}{3} y^2   +\frac{735\, \zeta_7}{32 y^2}- \frac{105\, \zeta_3\zeta_5}{8 y^3}\\
&\label{eq:TLOI1}\quad - 
\frac{7\, \zeta_5}{2} \Big[  \log\Big(\frac{y}{\pi^2}\Big) +\frac{\zeta'_2}{\zeta_2} + \frac{\zeta'_4}{\zeta_4} - 
\frac{\zeta'_5}{\zeta_5}\Big] + O(q,\bar q) \,.
\end{align}
To obtain these results we have taken the limits in the order discussed in appendix~\ref{app:Poincare}.
Note that we can rewrite \eqref{newlapeq} making use of the known Laplace equation for $\FFp{3}{2}{3}$ arriving at
\begin{equation}
\Delta \FFp{1}{2}{3} = \EE_2 \EE_3 
\quad\text{with} \quad \FFp{1}{2}{3} = \frac1{210} F + \FFp{3}{2}{3}\,,
\end{equation}
thus effectively extending from below the spectrum \eqref{int:Fmk} all the way to $s = 1$.

\subsubsection{\texorpdfstring{Example in the $(2,4)$ sector}{Example in the (2,4) sector}}

As an extension of the previous example, we can continue filling in the rightmost column of table \ref{chess.1} diagonal by diagonal. For example for the next two crosses on the diagonal of $\beta^{\rm sv}_{2,4}$ just below the one considered, we must have two new seeds $f_1\,,f_2$ satisfying
\begin{align}
\mathcal{O}_1 f_1 &= f_2\,, \notag\\
\mathcal{O}_2 f_2 & = y^3 \Big( \Re[{\cal E}_0(4,0^2)] - \frac{\zeta_3}{6} \Big) +\frac{y^6}{810}\,,
\end{align}
with solutions
\begin{align}
f_1 & = y \Re[{\cal E}_0(4,0^4)] -\frac{2\zeta_3}{9} y^3 +\frac{2y^6}{42525}\,, \notag\\
f_2 &\label{eq:OverlySeed2} =  y^2 \Re[{\cal E}_0(4,0^3)]+\frac{\zeta_3}{3} y^3 - \frac{y^6}{2835}\,.
\end{align}
Alternatively, this diagonal can be understood from the modular functions $F_1$ and $F_2$ constructed from such seeds $f_1$ and $f_2$ via Poincar\'e summation:
\begin{align}
\mathcal{O}_1 F_1 &=-\frac{1}{4} \Delta F_1 = F_2\,,\\
\mathcal{O}_2 F_2 & = -\frac{1}{8} (\Delta - 2) F_2 = 4725 \,\FFp{4}{2}{4}\,.
\notag
\end{align}
These equations can be rewritten as an inhomogeneous Laplace system by making use of the known Laplace equation satisfied by $\FFp{4}{2}{4}$:
\begin{subequations}
\begin{align}
(\Delta-2)  \FFp{2}{2}{4} &=  {\rm E}_2 {\rm E}_4  &\text{with} \quad \FFp{2}{2}{4} &=  \frac{1}{3780}F_2+\FFp{4}{2}{4}\,,\\
\Delta \FFp{0}{2}{4} &=  {\rm E}_2 {\rm E}_4  &\text{with} \quad \FFp{0}{2}{4} &=  \frac{1}{7560}F_1 +\frac1{3780} F_2 +  \FFp{4}{2}{4}\,.
\end{align}
\end{subequations}
Just like in the previous example we see that these overly integrated seeds produce modular invariant objects which allow us to extend from below the 
spectrum \eqref{int:Fmk} with the new eigenvalues $s \in \{ (k{-}m)\ \mbox{mod}\ 2, \ldots , k{-}m{-}2, k{-}m \}$, in this case $s\in \{0,2\}$.
Note, however, that although these new modular objects are solving very similar inhomogenous Laplace equations, they are very different in nature from all the $\FFp{s}{m}{k}$ studied so far as one can easily anticipate from their expansions near the cusp.

Once again we can use equations (\ref{conrul.1}) and (\ref{conrul.2}) applied to the seeds in (\ref{eq:OverlySeed2}) to obtain the perturbative expansion for these new modular functions, and for the functions $F_1$ and $F_2$ just presented we obtain
\begin{subequations}
\label{eq:bothTLO}
\begin{align}
F_1 &\label{eq:TLOI2} = \frac{2 y^6}{42525} - \frac{2\,\zeta_3}{9} y^3 - \frac{10\, \zeta_3 \zeta_5}{3}- \frac{875\, \zeta_9}{12 y^3}+\frac{525 \,\zeta_3 \zeta_7}{16 y^4}\\
&\notag \phantom{=}+ \frac{35\,\zeta_7}{y}\Big[1+\log\Big(\frac{y}{\pi^2}\Big)  +
   2  \frac{\zeta_4'}{\zeta_4}-\frac{\zeta_7'}{\zeta_7}\Big] + O(q,\bar q)\,,\\
F_2 &\label{eq:TLOI3}= - \frac{y^6}{2835} + \frac{\zeta_3}{3} y^3  +\frac{875\, \zeta_9}{4 y^3}- \frac{2625\, \zeta_3\zeta_7}{16 y^4} \\
&\phantom{=}+\notag \frac{35 \,\zeta_7}{2y} \Big[ \frac{1}{2}- \log\Big(\frac{y}{\pi^2}\Big)   - 
   2  \frac{\zeta_4'}{\zeta_4}+\frac{\zeta_7'}{\zeta_7} \Big] + O(q,\bar q)\,.
\end{align}
\end{subequations}
Note that the expansions near the cusp $y\gg 1$ for these new modular objects (\ref{eq:TLOI1}) and (\ref{eq:bothTLO}) are not Laurent polynomials any longer due to the appearance of new logarithmic contributions. Furthermore, the coefficients feature new interesting combinations related to the derivatives of the Riemann zeta, e.g.\ $\zeta'_4/\zeta_4$ and $\zeta'_7/\zeta_7$ above. Such combinations $\zeta'_k/\zeta_k$ have appeared from the 
integration over modular parameters in genus-one amplitudes of closed strings \cite{Green:2008uj, DHoker:2015gmr, DHoker:2019blr} and open strings \cite{Hohenegger:2017kqy}. Moreover, the transcendentality properties of these terms and accompanying harmonic sums have been discussed in \cite{DHoker:2019blr}.

Both these novelties appeared in \cite{Dorigoni:2020oon} precisely in the context of the perturbative expansion for overly integrated iterated Eisenstein integrals.
It would be tempting to interpret these modular objects in terms of an extension of the 
$\beta^{{\rm sv},\, j}_{m,k}$ in (\ref{eq:betalag}) to $j\leq -1$. However, we should stress that the 
$\beta^{\rm sv}[\begin{smallmatrix} j_1 &j_2\\ k_1 &k_2 \end{smallmatrix}]$ in (\ref{eq:bsv2})  have not been defined for such extensions and in fact never appear in the configuration-space integrals of closed-string genus-one amplitudes.
Nonetheless, it would be extremely interesting to understand better their properties and whether they play any role in string theory.

%%%%%%%%%%%%%%%%%%%%%%%%%%%%%%%%%%%%%%%%%%%%%%%%%%%%%%%%%%%
\subsection{``The red crosses and beyond''}
\label{sec:5.3}
%%%%%%%%%%%%%%%%%%%%%%%%%%%%%%%%%%%%%%%%%%%%%%%%%%%%%%%%%%%
As a final comment on our discussion of seed functions, we want to give further details regarding the seeds and associated modular functions related to the top left diagonal with red crosses in tables \ref{chess.1}--\ref{chess.3ALT}
and what lies beyond that.
From our general discussion it follows that the seeds associated with the red crosses are all of the form
\beq 
y^a \Re[{\cal E}_0(2m,0^{m-a})]\, , \ \ \ \ \ \ 
a\geq1 \, .
\label{adepd}
\eeq
However, if we try to apply our formula (\ref{conrul.1}) to obtain the Laurent polynomials from these seeds we immediately face the obstacle that these diverge when $a$ tends to an integer.
The reason for this divergence lies in the fact that we are now starting to explore, as already discussed in section \ref{sec:fold.2}, the realm of non-convergent Poincar\'e sums which have to be interpreted as analytic continuations.

To better understand what is going on we can focus on the whole red-cross diagonal whose starting-point seed would be naively given by $y^0 \Re[{\cal E}_0(2m,0^{m})]$. This putative seed is not present in the tables
\ref{chess.1}--\ref{chess.3ALT} but would appear just one check-box higher along the red-cross diagonal.
From this seed, it would be tempting to apply $\mathcal{O}_0$ to generate $y^1 \Re[{\cal E}_0(2m,0^{m-1})]$ if it was not for $\mathcal{O}_0$ being ill-defined. However, the above discussion suggests an approach to fix this
by analytically continuing the power $a$ of $y$ away from integers, i.e.\ by studying
\begin{equation}
\label{eq:Oepsilon}
\mathcal{O}_\epsilon \Big( y^\epsilon \Re[{\cal E}_0(2m,0^{m})]\Big) =-\frac{1}{4\epsilon}[\Delta-\epsilon(\epsilon{-}1)]\,y^\epsilon \Re[{\cal E}_0(2m,0^{m})] =  y^{1+\epsilon} \Re[{\cal E}_0(2m,0^{m-1})]\,.
\end{equation}
As explained in appendix~\ref{app:Poincare}, we regulate such seeds by analytically continuing the overall power of $y$ only. The key point is that the Poincar\'e sum and the $\epsilon\to 0$ limit do not commute, in particular
\beq
 \lim_{\epsilon\to0} \Delta\sum_{\gamma \in B(\ZZ)\backslash {\rm SL}(2,\ZZ)}    \Big[ y^\epsilon \Re[{\cal E}_0(2m,0^{m})]\Big]_\gamma \neq \sum_{\gamma \in B(\ZZ)\backslash {\rm SL}(2,\ZZ)}\Delta \lim_{\epsilon\to0}    \Big[ y^\epsilon \Re[{\cal E}_0(2m,0^{m})]\Big]_\gamma \,.
\eeq
It is easy to see that 
\beq
\Delta \lim_{\epsilon\to0}    y^\epsilon \Re[{\cal E}_0(2m,0^{m})]  = 0\,,
\eeq
while with the use of (\ref{conrul.1}) we were able to deduce that
\begin{equation}\label{eq:Poinc0m}
\sum_{\gamma \in B(\ZZ)\backslash {\rm SL}(2,\ZZ)}  \Big[ \Re[{\cal E}_0(2m,0^{m})]  \Big]_\gamma = \frac{6(m{-}2)!}{m(2m{-}1)!}
{\rm E}_m\, , \ \ \ \ \ \ m>2 \, .
\end{equation}
We believe it should be possible to prove this identity by a similar analytic continuation that led to the proof~\cite{Bossard:2017kfv} of (\ref{Besselsum}), and we have checked that the Laurent polynomials match for $3\leq m \leq 20$. However, this does not exclude that the two expressions differ by some cusp form.

The case $m=2$ is particular. The reason is that with the use of (\ref{Besselsum}) we can prove that
\begin{equation}\label{eq:Poinc02m}
\sum_{\gamma \in B(\ZZ)\backslash {\rm SL}(2,\ZZ)} \Big[ \Re[{\cal E}_0(2m,0^{2m-2})]  \Big]_\gamma  = \frac{\zeta_{2m-1}}{(2m{-}1)!}{\rm E}_0 \, , \ \ \ \ \ \ 
m>2 \,.
\end{equation}
For $m=2$ we notice that $2m{-}2 = m$ such that equations (\ref{eq:Poinc0m}) and (\ref{eq:Poinc02m}) become degenerate, and the correct answer is the sum of the two expressions (which notably violates uniform transcendentality):
\begin{equation}
\sum_{\gamma \in B(\ZZ)\backslash {\rm SL}(2,\ZZ)}  \Big[\Re[{\cal E}_0(4,0^{2})] \Big]_\gamma = \frac{1}{2}{\rm E}_2 +\frac{\zeta_3}{3!}\,.
\end{equation}
We do not have a proof for the above statement, but we have checked that indeed the Laurent polynomial we produce is the correct one.

With these results at hand we can go back to our analytically continued diagonal action (\ref{eq:Oepsilon}) and we can understand why the red-crosses diagonal gives rise to divergent Laurent polynomials
\begin{align}
\sum_{\gamma \in B(\ZZ)\backslash {\rm SL}(2,\ZZ)}  \Big[ y^{1+\epsilon} \Re[{\cal E}_0(2m,0^{m})]  \Big]_\gamma&\notag= -\frac{1}{4 \epsilon} \Delta\sum_{\gamma \in B(\ZZ)\backslash {\rm SL}(2,\ZZ)}    \Big[ y^\epsilon \Re[{\cal E}_0(2m,0^{m})]\Big]_\gamma + O(\epsilon^0)\\
& = -\frac{3}{2 \epsilon}\frac{(m{-}1)!}{(2m{-}1)!} {\rm E}_m +O(\epsilon^0)\,,
\end{align}
which again we have checked at the level of Laurent polynomials using (\ref{conrul.1}) regularised via analytic continuation for the power of $y$ in the seed.

It would be tempting to interpret the red crosses as originating from $\beta^{{\rm sv},\, j}_{m,k}$ with $k\leq 1$
which one may relate to the depth-two versions of the iterated integrals (\ref{eq:E0depth1}) containing ${\rm G}_2$, and, as we move further up from the red crosses diagonal, ${\rm G}_0\,,\,{\rm G}_{-2}$ and so on, all intended as their associated $q$-series, similarly to the discussion in \cite{Dorigoni:2020oon}.
There is no evidence for these strange objects to appear in any direct perturbative string-theory computation.

%%%%%%%%%%%%%%%%%%%%%%%%%%%%%%%%%%%%%%%%%%%%%%%%%%%%%
\section{Conclusions}
\label{sec:last}
%%%%%%%%%%%%%%%%%%%%%%%%%%%%%%%%%%%%%%%%%%%%%%%%%%%%%%

In this work, we have systematically extended the representation of non-holomorphic Eisenstein series ${\rm E}_k$ 
as Poincar\'e sums over $(\Im \tau)^k$ to even and odd modular invariants of depth two. The notion of depth refers to the iterated-Eisenstein-integral representations, i.e.\ we exhaust the modular invariant functions built from double-integrals over holomorphic Eisenstein series. Our depth-two targets are spanned by modular invariant bilinears in ${\rm E}_k$ and their Cauchy--Riemann derivatives together with solutions $ \FFpm{s}{m}{k}$ to inhomogeneous Laplace eigenvalue equations of the same type that are known from two-loop modular graph functions \cite{DHoker:2015gmr}. We stress that the modular invariants constructed in this work extend beyond the realm of modular graph forms. As will be further explored in Part~II,
this is reflected in the depth-one integrals of holomorphic cusp forms
contributing to some of the $ \FFpm{s}{m}{k}$. While the results in this work apply to integer $m,k$, 
generalisations to half-odd integer values of the Laplace equations under discussion
play an important role for string dualities \cite{Green:2005ba,Green:2014yxa,Chester:2020vyz}.

These depth-two modular invariants $\FFpm{s}{m}{k}$ are obtained from Poincar\'e sums over
iterated Eisenstein integrals at depth one over a single kernel 
 $\tau^j {\rm G}_k$ with $0\leq j \leq k{-}2$ \cite{Brown:mmv}. In fact, these seed functions
are organised according to the real or imaginary parts of the convergent iterated Eisenstein integrals, 
where the cuspidal combination $\GG_k^0(\tau)= {\rm G}_k(\tau) - 2 \zeta_k$ is integrated between 1 and
$k{-}1$ times. In this way, we expose Fourier decompositions of the seeds with all non-zero modes in the
form of $(q^n \pm \bar q^n)$ with $q= e^{2\pi i \tau}$, bypassing the powers of $q\bar q$ in
earlier seed functions for two-loop modular graph forms \cite{DHoker:2019txf}.
Our results support the general expectation that Poincar\'e-series
representations of modular invariants at depth $\ell$ admit seed functions built from
iterated Eisenstein integrals at depth 
$\ell{-}1$ and below, which is here worked out for $\ell=2$.

Our work contributes to a structural understanding of the interplay between non-holomorphic modular forms and 
iterated integrals. At the same time, the new Poincar\'e-series representations of modular invariants are useful 
for practical calculations, to integrate over the modular parameter $\tau$ in the low-energy expansion of
closed-string genus-one amplitudes in flat spacetime. 
Poincar\'e series play a prominent role in the Rankin--Selberg--Zagier method for
such $\tau$-integrals of modular-invariant functions~\cite{MR656029,Bump:2005,Green:1999pv,Angelantonj:2011br,Angelantonj:2012gw}.
We emphasise that infinite families of the $ \FFpm{s}{m}{k}$ go beyond the two-loop modular graph functions
all of which have been integrated over $\tau$ in \cite{DHoker:2019mib}. The basis $\FFpm{s}{m}{k}$ of functions gives an alternative method for determining their $\tau$-integrals by exploiting their simple Laplace equations~\eqref{int:Fmk} and~\eqref{int:Fmkm}. We also anticipate that our approach based on Poincar\'e seeds will be useful for higher-depth generalisations.

This work suggests various directions of follow-up research. 
A first problem is to connect the combinations $\beta^{\rm sv}$ of iterated Eisenstein integrals \cite{Gerken:2020yii} used in this work to Brown's construction of non-holomorphic modular forms \cite{Brown:mmv, Brown:I, Brown:II}. In this way, the organisation of modular graph forms in Brown's work via
tensor products of ${\rm SL}(2)$ representations may have an echo at the level of the seed functions
in Poincar\'e-series representations at arbitrary depth.

Another important follow-up question concerns the generalisation of modular graph forms to 
single-valued functions of torus punctures $z_1,z_2,\ldots$, so-called elliptic modular graph functions \cite{DHoker:2018mys, DHoker:2020tcq, Basu:2020pey, Basu:2020iok,DHoker:2020aex}.
It would be interesting to investigate Poincar\'e-series representations of elliptic modular graph functions, where
the seed functions will depend on the co-moving coordinates $(u_j,v_j) \in \mathbb R^2$ of $z_j = u_j \tau + v_j$.
In particular, one may speculate about similar correlations between the depth of iterated-integral representations of both the elliptic modular graph functions and their Poincar\'e seed.

%%%%%%%%%%%%%%%%%%%%%%%%%%%%%%%%%%%%%%%
%%%%%%%%%%%%%%%%%%%%%%%%%%%%%%%%%%%%%%%
%%%%%%%%%%%%%%%%%%%%%%%%%%%%%%%%%%%%%%%

\medskip
\subsection*{Acknowledgements}
We wish to thank Francis Brown for his generous help in understanding his results on multiple 
modular values, providing us with explicit examples beyond the ones published in~\cite{Brown2019} 
and valuable comments on an earlier version of this work. 
We are grateful to Eric D'Hoker, Nikolaos Diamantis, Mehregan Doroudiani, Jan Gerken, Martijn Hidding, Kim Klinger-Logan, Nils Matthes, Stephen D.\ Miller, Bram Verbeek, and Federico Zerbini for inspiring discussions and/or collaboration on related topics.
DD and OS thank the Albert Einstein Institute Potsdam for kind hospitality and creating a stimulating atmosphere. 
OS is supported by the European Research Council under ERC-STG-804286 UNISCAMP.

\appendix

%%%%%%%%%%%%%%%%%%%%%%%%%%%%%%%%%%%%%%%%%%%%%%%%%%%%%%%%%%%
\section{From Poincar\'e seeds to Laurent polynomials}
%%%%%%%%%%%%%%%%%%%%%%%%%%%%%%%%%%%%%%%%%%%%%%%%%%%%%%%%%%%
\label{app:Poincare}

In this appendix, we review how to obtain the Fourier expansion of a modular function from that of its seed, with particular emphasis on the zero-mode sector.
Given the Fourier expansions
\begin{subequations}
\begin{align}
\summ(\tau) &= \sum_{\ell\in\mathbb{Z}} a_\ell(\tau_2) e^{2\pi i \ell \tau_1} = \sum_{\gamma \in B(\ZZ) \setminus  {\rm SL}(2,\ZZ)} \seeed(\gamma \tau)\,,\\
\seeed(\tau) &= \sum_{\ell\in\mathbb{Z}} c_\ell(\tau_2) e^{2\pi i \ell \tau_1}\,,
\end{align}
\end{subequations}
with $\tau_1=\Re \tau$ and $\tau_2 = \Im \tau $, the Fourier modes $a_\ell(\tau_2)$ can be reconstructed from 
those of the seed $c_\ell(\tau_2)$ using the well-known result~\cite{Iwaniec:2002,Fleig:2015vky}:
\begin{align}
a_\ell(\tau_2) &=\label{eq:nonzeromode} c_\ell(\tau_2) + \sum_{d=1}^\infty\sum_{n\in\mathbb{Z}} S(n,\ell;d) \int_{\mathbb{R}} e^{-2\pi i \ell \omega -2\pi i n \frac{\omega}{d^2 (\tau_2^2+\omega^2)}} c_n\Big(\frac{\tau_2}{d^2(\tau_2^2+\omega^2)}\Big)\dd \omega\,.
\end{align}
Here $S(n,\ell;d)$ denotes in general a Kloosterman sum
\begin{equation}
S(n,\ell;d) = \sum_{r\in (\mathbb{Z}/d\mathbb{Z})^\times} e^{2\pi i (n r + \ell r^{-1}) / d}\,,\label{eq:Kloos}
\end{equation}
which is a finite sum over all $0\leq r <d$ that are coprime to $d$, such that $r$ has a multiplicative inverse, denoted by $r^{-1}$, in $ (\mathbb{Z}/d\mathbb{Z})^\times$.

In particular we have that the zero-mode $a_0(\tau_2)$ can be expressed as
\begin{align}
a_0(\tau_2) &= c_0(\tau_2) + \sum_{d=1}^\infty \sum_{n\in\mathbb{Z}} \sum_{r\in (\mathbb{Z}/d\mathbb{Z})^\times} e^{2\pi i n r/d} \int_{\mathbb{R}} e^{-2\pi i n \frac{\omega}{d^2 (\tau_2^2+\omega^2)}} c_n\Big(\frac{\tau_2}{d^2(\tau_2^2+\omega^2)}\Big)\dd \omega\,.
\end{align}
We split this expression into $a_0 = I_0+ I$, where the first contribution $I_0$ entirely stems from the zero-mode $c_0$ of the seed function
\begin{equation}
I_0 = c_0(\tau_2) +\tau_2 \sum_{d=1}^\infty \sum_{r\in (\mathbb{Z}/d\mathbb{Z})^\times} \int_{\mathbb{R}} c_0\Big(\frac{1}{\tau_2 d^2(1+t^2)}\Big)\dd t \,,
\end{equation}
and the second contribution $I$ comes from all the non-zero modes $c_n$ with $n\neq 0$
\begin{equation}
\label{eq:Kloost1}
I = \tau_2  \sum_{d=1}^\infty \sum_{n\neq0} \sum_{r\in (\mathbb{Z}/d\mathbb{Z})^\times} e^{2\pi i n r/d} \int_{\mathbb{R}} e^{-2\pi n \frac{i t}{\tau_2 d^2 (1+t^2)}} c_n\Big(\frac{1}{\tau_2 d^2(1+t^2)}\Big)\dd t\,,
\end{equation}
where in both integrals we changed variables $\omega = \tau_2\,t$.

In all the cases we will discuss, the Fourier modes of the seed functions will  be of the form
\begin{subequations}
\begin{align}
c_0(y) &\label{eq:Seed0Gen}=  (\pi \tau_2)^r = y^r\,,\\
c_\ell (y) & \label{eq:SeedNZGen}=   \sigma_{a}(\vert \ell \vert) (4\pi \vert \ell \vert )^b \tau_2 ^r e^{-2\pi \vert \ell \vert \tau_2}= \sigma_{a}(\vert \ell \vert) (4\pi \vert \ell \vert )^b (y/\pi)^r e^{-2 \vert \ell \vert y}\,,
\end{align}
\end{subequations}
with $a,b,r\in \CC$ and $y =\pi \tau_2 $, or finite linear combinations of these seeds.
In order to compute the Laurent zero-mode for the associated modular form we need to use\footnote{In this equation and below, we write the Riemann zeta function as $\zeta(s)$ instead of $\zeta_s$ in order to make the various different arguments more legible.}
\begin{subequations}
\begin{align}
I_0(r) &\label{eq:I0}=  y^r + 
\frac{ (-16)^{1 - r} ( 2 r)! ( 2 r {-}3)!}{
{\rm B}_{2r}  (r {-} 2)! (r {-} 1)!} \zeta(2r{-}1) y^{1 - r}\,, \\
I(a,b,r) &=  \frac{2^{3-2r+2b}\pi }{\Gamma(r)} \Big(\frac{y}{\pi}\Big)^{1+b-r}\Bigg[ \frac{y}{\pi^2} \frac{\Gamma(b{+}1)\Gamma(2r{-}b{-}2)}{\Gamma(r{-}b{-}1)} \frac{\zeta(2r{-}a{-}2b{-}2)\zeta(1{-}a)}{\zeta(2r{-}a{-}2b{-}1)} \nn\\
&\hspace{2mm} + \left(\frac{y}{\pi^2}\right)^{a+1} \frac{\Gamma(a{+}b{+}1)\Gamma(2r{-}a{-}b{-}2)}{\Gamma(r{-}a{-}b{-}1)}\frac{\zeta(2r {-}a{-}2b{-}2)\zeta(a{+}1)}{\zeta(2r{-}a{-}2b{-}1)} \nn\\
&\hspace{2mm} +\left(\frac{\pi^2}{y}\right)^b 
\sum_{n \geq 0} \left(\frac{-\pi^2}{y}\right)^{n} \frac{\Gamma(2r{+}n{-}1)}{n! \cdot \Gamma(r{+}n)}  \label{eq:Iabr}\\
&\hspace{20mm}\times \frac{\zeta({-}b{-}n)\zeta({-}a{-}b{-}n)\zeta(2r{-}a{-}b{+}n{-}1)\zeta(2r{-}b{+}n{-}1)}{\zeta(2r{+}2n)\zeta(2r{-}a{-}2b{-}1)} \bigg]\,, \notag
\end{align}
\end{subequations}
where $I(a,b,r)$ was derived in \cite{Dorigoni:2019yoq} (note that the variable $y$ used in the reference corresponds to $\tau_2$ and not the current $y = \pi \tau_2$). The expression for $I_0$ is proportional to the usual Laurent polynomial of non-holomorphic Eisenstein series $\EE_r$, see~\eqref{eq:FEk}, and we shall refer to the term with $y^{1-r}$ as the Weyl reflected term in the following, due to its origin in the general theory of Eisenstein series.

A few comments are in order. First to derive both $I_0(r)$ and $I(a,b,r)$ we assumed that $\Re(r)>1$ for the integral and the Dirichlet series over $n$ in~\eqref{eq:Kloost1} to both be convergent. We will, however, encounter  cases for which this is not true.
For example when considering seeds for ``diagonal'' modular invariant functions, i.e.\ solutions to $(\Delta - \lambda) F = {\rm E}_k^2$, we will see that  $r=1$. In this case $I_0(r)$ diverges because the Dirichlet series over $n$ produces $\zeta(2r{-}1)\to \zeta(1)$. The correct way to proceed would be to introduce a regulator so that we fold
$(\Delta - \lambda(\epsilon)) F = {\rm E}_k{\rm E}_{k+\epsilon}$ effectively shifting $r=1{+}\epsilon$, thus regulating $I_0$, see~\cite{Dorigoni:2019yoq} for the details.

Similarly, we can produce divergent sums when we consider seeds for ``non-diagonal'' modular invariant functions, i.e.\ solutions to $(\Delta - \lambda) F = {\rm E}_m{\rm E}_k$ with $m< k$. We can obtain a seed by folding ${\rm E}_m$ instead of $\EE_k$ and this will effectively gives us $r=1{+}m{-}k\leq 0$ and hence the integral in $I_0(r)$ will be divergent. This divergence is due to the Weyl reflected term which for $r\in\mathbb{Z}$ with $r\leq 0$ contains a $\Gamma(r)$ and $\zeta(2r)$ in the numerator.
Again a regulator is needed so that $r = 1{+}m{-}k{+}\epsilon$, as we mentioned for the case $r=1$, so that the Weyl reflected term produces a finite result.
The Weyl reflected term $\propto y^{1-r}$ in $I_0(r)$ combines with the $n=0$ term in (\ref{eq:Iabr}). For the regulated zero mode these terms cancel. 

The contribution $I(a,b,r)$ from the non-zero modes contains a term linear in $y$ in the first line of (\ref{eq:Iabr}), and it corresponds to what Zagier calls the Riemann term \cite{ZagierApp}. Its importance was discussed in \cite{Dorigoni:2019yoq}, extending \cite{Ahlen:2018wng}.

We should also stress that  the expression (\ref{eq:Iabr}) for the contribution $I(a,b,r)$ coming from the non-zero modes is in general an infinite, asymptotic series. However, in all the cases which we will be discussing here it will actually truncate. The reason for this truncation comes from the fact that the parameters $a,b$ will always be integers for us, in particular since we are dealing with Eisenstein series and their iterated integrals, we will always encounter integers powers of $\ell$ and divisors functions with odd indices, i.e.\ $a,b\in\mathbb{Z}$ and $a$ odd in (\ref{eq:SeedNZGen}).  
By means of the functional identity $I(a,b,r)=I(-a,a{+}b,r)$ we can always arrange for the second argument to be odd provided that $a$ is odd. 
The infinite series in expression (\ref{eq:Iabr}) then contains the combination $\zeta(-b{-}n)\zeta(-a{-}b{-}n)$ which  identically vanishes for $n$ large enough  when $a$ and $b$ are integers of opposite parity.

When specialising (\ref{eq:Iabr}) to integer values of its parameters we can directly plug in the value for $a$, while $b$ should always be considered as a limit. At the very end, if necessary as discussed above, we can take the integer-value limit for the last parameter $r$.

Note that the asymptotic nature of this series is nonetheless of crucial importance for deriving the non-perturbative corrections to the zero-mode using resurgence methods, see \cite{Dorigoni:2019yoq,Dorigoni:2020oon}, we will not
however discuss this issue here.

%%%%%%%%%%%%%%%%%%%%%%%%%%%%%%%%%%%%%%%%%%%%%%%%%%%%%%%%%%%
\section{Two-loop modular graph functions at weight eight}
\label{app:cabc}
%%%%%%%%%%%%%%%%%%%%%%%%%%%%%%%%%%%%%%%%%%%%%%%%%%%%%%%%%%%

In this appendix, we gather higher-weight examples of the discussion of two-loop
modular graph functions $C_{a,b,c}$ in sections \ref{sec:3.2.5} and \ref{sec:4.1}.

%%%%%%%%%%%%%%%%%%%%%%%%%%%%%%%%%%%%%%%%%%%%%%%%%%%%%%%%%%%
\subsection{\texorpdfstring{Expansions in terms of $\FFp{s}{m}{k}$ at weight 8}{Expansions in terms of Fplus(s,m,k) at weight 8}}
\label{app:cabc.1}
%%%%%%%%%%%%%%%%%%%%%%%%%%%%%%%%%%%%%%%%%%%%%%%%%%%%%%%%%%%

By comparing the Laplace equations (\ref{lapsec.3}) and (\ref{eq:Fmk}) of the $C_{a,b,c}$
and $\FFp{s}{m}{k}$, we find the following relations at weight $a{+}b{+}c=8$ (see (\ref{cabc.1})
to (\ref{cabc.3}) for lower-weight analogues):
\begin{align}
C_{3,3,2} &= 
 - \frac{ 9}{5} \FFp{2}{ 4}{ 4} + \frac{54}{5} \FFp{4}{ 4}{ 4} -\frac{1 }{3} {\rm E}_8 \,,
\notag \\
C_{4,2,2} &= 
  \frac{18}{35} \FFp{2}{ 4}{ 4} - \frac{16}{3} \FFp{4}{ 3}{ 5} - 
 \frac{74}{5} \FFp{4}{ 4}{ 4} + \frac{64}{3} \FFp{6}{ 3}{ 5} + \frac{100}{7} \FFp{6}{ 4}{ 4}- \frac{115 }{66}{\rm E}_8 \,,
\notag \\
C_{4,3,1} &= 
 - \frac{9}{35} \FFp{2}{ 4}{ 4} + \frac{8}{3} \FFp{4}{ 3}{ 5} - 
 \frac{8}{5} \FFp{4}{ 4}{ 4} - \frac{32}{3} \FFp{6}{ 3}{ 5} - \frac{50}{7} \FFp{6}{ 4}{ 4} + \frac{107 }{66}{\rm E}_8 \,,
\label{cabc.99} \\
C_{5,2,1} &=
  \frac{27}{70} \FFp{2}{ 4}{ 4} - \frac{16}{3} \FFp{4}{ 3}{ 5} + 
 \frac{7}{5} \FFp{4}{ 4}{ 4} - \frac{8}{3} \FFp{6}{ 3}{ 5} - \frac{25}{14} \FFp{6}{ 4}{ 4} + \frac{37 }{44}{\rm E}_8 \,,
\notag \\
C_{6,1,1} &= 
 - \frac{9}{70} \FFp{2}{ 4}{ 4} + \frac{8}{3} \FFp{4}{ 3}{ 5} + 
 \frac{1}{5} \FFp{4}{ 4}{ 4} - 10 \FFp{6}{ 2}{ 6} - \frac{8}{3} \FFp{6}{ 3}{ 5} - \frac{1}{14} \FFp{6}{ 4}{ 4} + \frac{2573 }{1716}{\rm E}_8 \,.
\notag 
\end{align}
As elaborated around (\ref{cabc.0}), these relations do not suffice to express all the 
$\FFp{s}{m}{8-m}$ at weight eight in terms of $C_{a,b,c}$ and ${\rm E}_w$. 

%%%%%%%%%%%%%%%%%%%%%%%%%%%%%%%%%%%%%%%%%%%%%%%%%%%%%%%%%%%
\subsection{Laurent polynomials at weight 6 and 7}
\label{app:cabc.2}
%%%%%%%%%%%%%%%%%%%%%%%%%%%%%%%%%%%%%%%%%%%%%%%%%%%%%%%%%%%

In this appendix, we gather the Laurent polynomials of $C_{a,b,c}$ with $a{+}b{+}c=6$ and $7$
which are known from \cite{DHoker:2017zhq} and partially from \cite{Green:2008uj, DHoker:2016quv}. 
The decompositions in (\ref{cabc.2}) and (\ref{cabc.3}) along with the Laurent polynomials of the 
$\FFp{s}{m}{k}$ in sections \ref{sec:4.2.1} and \ref{sec:4.2.2} reproduce the weight-six expressions
\begin{align}
C_{2,2,2} &= \frac{ 38 y^6}{91216125}  + \frac{ \zeta_{7}}{24 y}
 - \frac{  7 \zeta_{9}}{16 y^3 } + \frac{ 15 \zeta_{5}^2}{16 y^4} - \frac{ 81 \zeta_{11}}{128 y^5 } +O(q,\bar q) \,,
\notag \\
C_{3,2,1} &= \frac{ 43 y^6}{58046625  }+ \frac{  y \zeta_{5}}{630 } + 
 \frac{\zeta_{7}}{144 y} + \frac{ 7 \zeta_{9}}{64 y^3 } - \frac{ 17 \zeta_{5}^2}{64 y^4 } + \frac{ 99 \zeta_{11}}{256 y^5} +O(q,\bar q) \,,
\label{app.b21} \\
C_{4,1,1} &=  \frac{808 y^6}{638512875} + \frac{ y^3 \zeta_{3}}{4725 } - \frac{ y \zeta_{5}}{1890 } 
+ \frac{ \zeta_{7}}{720 y } - \frac{ 15 \zeta_{3} \zeta_{7}}{
 32 y^4 } + \frac{ 23 \zeta_{9}}{64 y^3 } - \frac{ \zeta_{5}^2}{64 y^4  } + \frac{ 167 \zeta_{11}}{256 y^5 }  +O(q,\bar q) \, ,
\notag
\end{align}
and the following ones at weight seven:
\begin{align}
C_{3,2,2} &= 
\frac{4 y^7}{127702575 }+ \frac{ \zeta_{7}}{756 } + \frac{ 
 7 \zeta_{9}}{480 y^2 } - \frac{ 33 \zeta_{11}}{128 y^4  } + \frac{ 21 \zeta_{5} \zeta_{7}}{32 y^5 } - \frac{ 253 \zeta_{13}}{
 512 y^6} +O(q,\bar q) \,,
\notag \\
C_{3,3,1} &= 
\frac{8 y^7}{127702575  }+ \frac{ \zeta_{7}}{378 }   - \frac{ 
 7 \zeta_{9}}{480 y^2 } + \frac{ 33 \zeta_{11}}{128 y^4  } - \frac{ 21 \zeta_{5} \zeta_{7}}{32 y^5 }+ \frac{ 11 \zeta_{13}}{
 16 y^6} +O(q,\bar q) \,,
\notag \\
C_{4,2,1} &= 
\frac{46 y^7}{638512875 } + \frac{ y^2 \zeta_{5}}{6300 } -\frac{ \zeta_{7}}{1512 } + \frac{ 7 \zeta_{9}}{480 y^2 } + \frac{ 11 \zeta_{11}}{
 256 y^4 } - \frac{ 
 3 \zeta_{5} \zeta_{7}}{16 y^5 }  + \frac{ 11 \zeta_{13}}{32 y^6 } +O(q,\bar q) \,,
\label{app.b22} \\
C_{5,1,1} &= 
\frac{244 y^7}{1915538625 }+\frac{ 2 y^4 \zeta_{3}}{93555 } - \frac{ y^2 \zeta_{5}}{18900} + 
\frac{ \zeta_{7}}{7560 }- \frac{ \zeta_{9}}{2880 y^2 }  +\frac{ 253 \zeta_{11}}{768 y^4 }  \notag \\
&\ \ \ \ \ \ \ \ \ \  - \frac{ 
 7 \zeta_{3} \zeta_{9}}{16 y^5 } -\frac{ 3 \zeta_{5} \zeta_{7}}{64 y^5}  + \frac{ 
 661 \zeta_{13}}{1024 y^6 } +O(q,\bar q)
\, .
\notag 
\end{align}
%

%%%%%%%%%%%%%%%%%%%%%%%%%%%%%%%%%%%%%%%%%%%%%%%%%%%%%%%%%%%
\section{Examples of Poincar\'e seed functions up to weight 8}
\label{app:ex}
%%%%%%%%%%%%%%%%%%%%%%%%%%%%%%%%%%%%%%%%%%%%%%%%%%%%%%%%%%%

In this appendix, we complement the examples in the main text to gather 
seed functions of depth one for all the $\FFpm{s}{m}{k}$ and $\Jpm{\ell}{m}{k}$ 
with $m{+}k\leq 8$ and $\ell < m$.

%%%%%%%%%%%%%%%%%%%%%%%%%%%%%%%%%%%%%%%%%%%%%%%%%%%%%%%%%%%
\subsection{\texorpdfstring{Cases with $\FFp{s}{m}{k}$ at $m=k=4$}{Cases with Fplus(s,m,k) at m=k}}
\label{app:ex.1}
%%%%%%%%%%%%%%%%%%%%%%%%%%%%%%%%%%%%%%%%%%%%%%%%%%%%%%%%%%%

By combining the seed functions (\ref{oct20.10}) with those in (\ref{eqseedR}),
we can isolate iterated Eisenstein integrals over ${\rm G}_8$ as follows:
{\allowdisplaybreaks
\begin{align}
\frac{y^8}{1277025750}+y \Big( \Re[{\cal E}_0({8, 0^6}) ]  - \frac{ \zeta_7}{7!} \Big) &=  \frac{3}{7} (9 \seedp{2}{ 4}{ 4} - 14 \seedp{4}{ 4}{ 4} + 5 \seedp{6}{ 4}{ 4})\,,
\notag \\
-\frac{y^8}{91216125}+y^2 \Re[{\cal E}_0({8, 0^5}) ] &= -\frac{9}{14} (3 \seedp{2}{ 4}{ 4} - 28 \seedp{4}{ 4}{ 4} + 25 \seedp{6}{ 4}{ 4})   \,,
\notag \\
\frac{y^8}{13513500}+y^3 \Re[{\cal E}_0({8, 0^4}) ] &= -\frac{45}{4} (2 \seedp{4}{ 4}{ 4} - 5 \seedp{6}{ 4}{ 4})\,,
\notag \\
-\frac{y^8}{3243240}+y^4 \Re[{\cal E}_0({8, 0^3}) ] &= \frac{45}{16} (4 \seedp{4}{ 4}{ 4} - 40 \seedp{6}{ 4}{ 4} - \Jseedp{0}{ 4}{ 4})\,,
\label{appB.1} \\
\frac{y^8}{1179360}+ y^5 \Re[{\cal E}_0({8, 0^2}) ] &=   \frac{45}{128} (360 \seedp{6}{ 4}{ 4} + 24 \Jseedp{0}{ 4}{ 4} + \Jseedp{1}{ 4}{ 4})\,,
 \notag \\
-\frac{y^8}{655200}+y^6 \Re[{\cal E}_0({8, 0}) ] &=  -\frac{9}{512} (3600 \seedp{6}{ 4}{ 4} + 600 \Jseedp{0}{ 4}{ 4} + 50 \Jseedp{1}{ 4}{ 4} +   \Jseedp{2}{ 4}{ 4})\,,
 \notag\\
\frac{y^8}{604800}+ y^7 \Re[{\cal E}_0({8}) ] &= \frac{3 }{4096} (7200 \Jseedp{0}{ 4}{ 4} + 900 \Jseedp{1}{ 4}{ 4} + 36 \Jseedp{2}{ 4}{ 4} + 
   \Jseedp{3}{ 4}{ 4})\,.
  \notag
\end{align}}%
%
 
%%%%%%%%%%%%%%%%%%%%%%%%%%%%%%%%%%%%%%%%%%%%%%%%%%%%%%%%%%%
\subsection{\texorpdfstring{Cases with $\FFp{s}{m}{k}$ at $m<k$}{Cases with Fplus(s,m,k) at m<k}}
\label{app:ex.2}
%%%%%%%%%%%%%%%%%%%%%%%%%%%%%%%%%%%%%%%%%%%%%%%%%%%%%%%%%%%

Similarly, the seed functions (\ref{oct20.12}) and (\ref{oct20.13}) together with those in (\ref{eqseedR})
allow us to isolate various iterated Eisenstein integrals as follows:
\begin{align}
 \frac{2 y^7}{1485 }+y^4 \Big( \Re[{\cal E}_0({4, 0^2}) ]  - \frac{ \zeta_3}{6} \Big)  &=  62370 \seedp{5}{ 2}{ 5}\,,
\notag \\
- \frac{y^7}{396} +y^5 \Re[{\cal E}_0({4, 0}) ] &=   -\frac{31185}{8} (8 \seedp{5}{ 2}{ 5} + \Jseedp{0}{ 2}{ 5})\,,
\label{appB.2}\\
\frac{y^7}{360} +y^6 \Re[{\cal E}_0({4}) ]  &= 
\frac{6237}{16} (5 \Jseedp{0}{ 2}{ 5} + \Jseedp{1}{ 2}{ 5}) \, ,
\notag 
\end{align}
and
\begin{align}
\frac{y^8}{702}+y^5 \Big( \Re[{\cal E}_0({4, 0^2}) ]  - \frac{ \zeta_3}{6} \Big)   &=  \frac{1064188125 }{1382} \seedp{6}{ 2}{ 6}\,,
\notag \\
 - \frac{y^8}{390 }+y^6 \Re[{\cal E}_0({4, 0}) ] &=
-\frac{212837625 }{5528} (10 \seedp{6}{ 2}{ 6} + \Jseedp{0}{ 2}{ 6})\,,
\label{appB.3} \\
\frac{ y^8}{360 }+y^7 \Re[{\cal E}_0({4}) ]  &= 
\frac{70945875 }{22112} (6 \Jseedp{0}{ 2}{ 6} + \Jseedp{1}{ 2}{ 6}) \,,
\notag 
 \end{align}
as well as 
\begin{align}
- \frac{ y^7}{467775}+y^2 \big( \Re[{\cal E}_0(6, 0^4) ] - \tfrac{ \zeta_5}{5!} \big) 
&=
270 (\seedp{3}{ 3}{ 4} - \seedp{5}{ 3}{ 4})\,,
\notag \\
 \frac{ y^7}{93555}+y^3 \Re[{\cal E}_0(6, 0^{3}) ] 
&=
-\frac{135}{2} (2 \seedp{3}{ 3}{ 4} - 9 \seedp{5}{ 3}{ 4})\,,
\notag \\
- \frac{ y^7}{31185}+y^4 \Re[{\cal E}_0(6, 0^{2}) ] 
&= -\frac{315}{8} (18 \seedp{5}{ 3}{ 4} + \Jseedp{0}{ 3}{ 4})\,,
\label{appB.4} \\
\frac{ y^7}{16632}+y^5 \Re[{\cal E}_0(6, 0) ] 
&= \frac{315}{64} (72 \seedp{5}{ 3}{ 4} + 12 \Jseedp{0}{ 3}{ 4} + \Jseedp{1}{ 3}{ 4})\,,
\notag \\
-  \frac{ y^7}{15120}+y^6 \Re[{\cal E}_0(6) ] 
&=
-\frac{63}{256} (120 \Jseedp{0}{ 3}{ 4} + 20 \Jseedp{1}{ 3}{ 4} + \Jseedp{2}{ 3}{ 4}) \, ,\notag
\end{align}
and
% 35
{\allowdisplaybreaks
\begin{align}
- \frac{  2 y^8}{675675}+y^3 \big( \Re[{\cal E}_0(6, 0^4) ]  - \tfrac{ \zeta_5}{5!} \big) 
&=
4158 (\seedp{4}{ 3}{ 5} - \seedp{6}{ 3}{ 5})\,,
\notag \\
\frac{  y^8}{81081 }+y^4 \Re[{\cal E}_0(6, 0^{3}) ] 
&=
-2079 (\seedp{4}{ 3}{ 5} - 4 \seedp{6}{ 3}{ 5})\,,
\notag \\
- \frac{ y^8}{29484 }+y^5 \Re[{\cal E}_0(6, 0^{2}) ] 
&=
-\frac{6237}{16} (24 \seedp{6}{ 3}{ 5} + \Jseedp{0}{ 3}{ 5})\,,
\label{appB.5} \\
+ \frac{  y^8}{16380 }+y^6 \Re[{\cal E}_0(6, 0) ] 
&=
\frac{6237}{160} (120 \seedp{6}{ 3}{ 5} + 15 \Jseedp{0}{ 3}{ 5} + \Jseedp{1}{ 3}{ 5})\,,
\notag \\
 - \frac{  y^8}{15120 }+y^7 \Re[{\cal E}_0(6) ]
&=
-\frac{2079 }{1280} (180 \Jseedp{0}{ 3}{ 5} + 24 \Jseedp{1}{ 3}{ 5} + \Jseedp{2}{ 3}{ 5}) \, .
   \notag
\end{align}}%
%

%%%%%%%%%%%%%%%%%%%%%%%%%%%%%%%%%%%%%%%%%%%%%%%%%%%%%%%%%%%
\subsection{\texorpdfstring{Cases with $\FFm{s}{m}{k}$ at $m{+}k=7,8$}{Cases with Fplus(s,m,k) at m+k=7,8}}
\label{app:ex.9}
%%%%%%%%%%%%%%%%%%%%%%%%%%%%%%%%%%%%%%%%%%%%%%%%%%%%%%%%%%%

In this appendix, we provide additional examples for the
discussion in section \ref{sec:step.odd}. The step form
for odd seed functions in (\ref{exminus.5}) and (\ref{exminus.6})
generalises as follows to weight $m{+}k=7$,
\begin{align}
y^4 \Im \! \big[ \mathcal{E}_0 (4, 0^2)\big] &= 6930 i  ( \seedm{4}{ 2}{ 5} -   \seedm{6}{ 2}{ 5})\,,
\notag \\
y^5 \Im \! \big[ \mathcal{E}_0 (4, 0)\big] &= \frac{ 31185 i }{4} \seedm{6}{ 2}{ 5}\,,
\label{exminus.8} \\
y^6 \Im \! \big[ \mathcal{E}_0 (4)\big] &= - \frac{ 31185 i}{8}  \seedm{6}{ 2}{ 5} -  \frac{ 6237 i}{16 } \Jseedm{1}{ 2}{ 5}\, ,
\notag 
\end{align}
and
\begin{align}
y^2 \Im \! \big[ \mathcal{E}_0 (6,0^4)\big] &= 27 i \seedm{2}{ 3}{ 4} - 42 i \seedm{4}{ 3}{ 4} + 
  15 i \seedm{6}{ 3}{ 4}\,,
\notag \\
y^3 \Im \! \big[ \mathcal{E}_0 (6,0^3)\big] &=  \frac{105 i}{2} \seedm{4}{ 3}{ 4} -  \frac{105 i}{2 } \seedm{6}{ 3}{ 4}\,,
\notag  \\
y^4 \Im \! \big[ \mathcal{E}_0 (6, 0^2)\big] &= -\frac{105 i}{4} \seedm{4}{ 3}{ 4} + 105 i \seedm{6}{ 3}{ 4}\,,
\label{exminus.9} \\
y^5 \Im \! \big[ \mathcal{E}_0 (6, 0)\big] &= -\frac{945 i }{8} \seedm{6}{ 3}{ 4} - \frac{ 315 i }{64 } \Jseedm{1}{ 3}{ 4}\,,
\notag \\
y^6 \Im \! \big[ \mathcal{E}_0 (6)\big] &=  \frac{ 945 i }{16} \seedm{6}{ 3}{ 4} + \frac{315i }{64} \Jseedm{1}{ 3}{ 4} + 
\frac{ 63 i }{256 } \Jseedm{2}{ 3}{ 4}\, .
\notag  
\end{align}
The analogous expressions at weight $m{+}k=8$ are given by
\begin{align}
y^5 \Im \! \big[ \mathcal{E}_0 (4, 0^2)\big] &= \frac{ 96744375 i }{1382}  \seedm{5}{ 2}{ 6} 
-    \frac{96744375 i }{1382}  \seedm{7}{ 2}{ 6}\,,
\notag \\
y^6 \Im \! \big[ \mathcal{E}_0 (4, 0)\big] &=  \frac{212837625 i }{2764} \seedm{7}{ 2}{ 6}\,,
\label{exminus.8a} \\
y^7 \Im \! \big[ \mathcal{E}_0 (4)\big] &= 
 -  \frac{212837625 i }{5528}  \seedm{7}{ 2}{ 6} -  \frac{ 70945875 i }{22112 }\Jseedm{1}{ 2}{ 6}\, ,
\notag 
\end{align}
and
\begin{align}
y^3 \Im \! \big[ \mathcal{E}_0 (6,0^4)\big] &= 
297 i \seedm{3}{ 3}{ 5} - 486 i \seedm{5}{ 3}{ 5} + 
 189 i \seedm{7}{ 3}{ 5} \,,
\notag \\
y^4 \Im \! \big[ \mathcal{E}_0 (6,0^3)\big] &=  567 i ( \seedm{5}{ 3}{ 5} -   \seedm{7}{ 3}{ 5} )\,,
\notag  \\
y^5 \Im \! \big[ \mathcal{E}_0 (6, 0^2)\big] &= - \frac{567 i}{2}  \seedm{5}{ 3}{ 5} +  \frac{8505 i}{8}  \seedm{7}{ 3}{ 5}\,,
\label{exminus.9a} \\
y^6 \Im \! \big[ \mathcal{E}_0 (6, 0)\big] &=  - \frac{18711 i}{16}  \seedm{7}{ 3}{ 5} -  \frac{6237 i}{160}  \Jseedm{1}{ 3}{ 5}\,,
\notag \\
y^7 \Im \! \big[ \mathcal{E}_0 (6)\big] &=   \frac{18711 i}{32}  \seedm{7}{ 3}{ 5} +  \frac{6237 i}{160}  \Jseedm{1}{ 3}{ 5} 
+  \frac{ 2079 i }{1280} \Jseedm{2}{ 3}{ 5}
\, .
\notag  
\end{align}
%

%%%%%%%%%%%%%%%%%%%%%%%%%%%%%%%%%%%%%%%%%%%%%%%%%%%%%%%%%%%
\section{\texorpdfstring{Examples of Laurent polynomials of $\FFp{s}{m}{k}$}{Examples of Laurent polynomials of Fplus(s,m,k)}}
\label{app:lau}
%%%%%%%%%%%%%%%%%%%%%%%%%%%%%%%%%%%%%%%%%%%%%%%%%%%%%%%%%%%

In this appendix, we gather the Laurent polynomials of all modular invariant $\FFp{s}{m}{k}$ with $m{+}k\leq 8$ that can be obtained from the general formula~\eqref{eq:FLP}. 
For those $\FFp{s}{m}{k}$ with a representation in terms of $\zeta_w,{\rm E}_w$ and 
$C_{a,b,c}$, the subsequent Laurent polynomials are checked to
be consistent with the results of \cite{Green:2008uj, DHoker:2016quv, DHoker:2017zhq}.
For $\FFm{s}{m}{k}$ the Laurent polynomial vanishes identically.

The expressions below also represent the Laurent polynomials of $\cFFp{s}{m}{k}$ even when these functions are not modular invariant. This follows from our method for reinstating the lower-depth terms in section~\ref{sec:4.1}, engineered so that $\cFFp{s}{m}{k}$ yields the same Laurent polynomials as $\FFp{s}{m}{k}$. 
The only possible discrepancy between $\cFFp{s}{m}{k}$ and $\FFp{s}{m}{k}$ lies fully in the $O(q,\bar{q})$ sector. In other words, the expressions below can also be obtained by taking the degeneration limit (\ref{more.bsvb}) of the $\beta^{\rm sv}$.

%%%%%%%%%%%%%%%%%%%%%%%%%%%%%%%%%%%%%%%%%%%%%%%%%%%%%%%%%%%
\subsection{\texorpdfstring{Cases with $\FFp{s}{m}{k}$ at $m=k$}{Cases with Fplus(s,m,k) at m=k}}
\label{app:lau.1}
%%%%%%%%%%%%%%%%%%%%%%%%%%%%%%%%%%%%%%%%%%%%%%%%%%%%%%%%%%%

The expressions for $\FFp{s}{k}{k}$ with $k\leq 3$ in (\ref{oct24.11}) lead to the Laurent 
polynomials
\begin{align}
\FFp{2}{ 2 }{ 2} &= \frac{ y^4}{20250 } -     \frac{ y \zeta_{3}}{45} +  \frac{ \zeta_{3}^2}{4 y^2} 
-  \frac{ 5 \zeta_{5}}{12 y} + O(q,\bar q) \,,
\notag \\
\FFp{2}{ 3 }{ 3} &=  \frac{y^6}{6251175} -    \frac{y \zeta_{5}}{630} + \frac{ \zeta_{5}^2}{32 y^4}
 -  \frac{ 5 \zeta_{7}}{ 288 y}+ O(q,\bar q) \,,
\label{app:lau.3} \\
\FFp{4}{ 3 }{ 3} &=  \frac{ 2 y^6}{8037225 } -  \frac{ y \zeta_{5}}{3780  }+  \frac{9 \zeta_{5}^2}{128 y^4}
 - \frac{  35 \zeta_{9}}{1152 y^3 }+O(q,\bar q) \,,
 \notag
\end{align}
equivalent to those of ${\rm E}_{2,2},{\rm E}_{3,3},{\rm E}_{3,3}'$ the literature, and the examples (\ref{oct24.12}) at $k=4$ yield
\begin{align}
\FFp{2}{ 4 }{ 4} &=  \frac{ y^8}{1205583750 } -  \frac{ y \zeta_{7}}{7560 }
 +  \frac{ 5 \zeta_{7}^2}{512 y^6 } -  \frac{  5 \zeta_{9}}{3888 y }+O(q,\bar q) \,,
\notag \\
\FFp{4}{ 4 }{ 4} &=  \frac{ y^8}{982327500 } -  \frac{ y \zeta_{7}}{45360 } +  \frac{ 5 \zeta_{7}^2}{384 y^6 } -  \frac{ 
 7 \zeta_{11}}{6912 y^3 }+O(q,\bar q) \,,
\label{app:lau.4}  \\
\FFp{6}{ 4 }{ 4} &=  \frac{ y^8}{580466250 } -  \frac{ y \zeta_{7}}{113400 } +  \frac{ 25 \zeta_{7}^2}{768 y^6 }
- \frac{  5005 \zeta_{13}}{530688 y^5 }+ O(q,\bar q)\, ,
\notag
\end{align}
where the Laurent polynomial of the $\beta^{\rm sv}$-combination $\cFFp{6}{ 4 }{ 4}$
in (\ref{cF644}) is identical to that of the modular invariant $\FFp{6}{ 4 }{ 4}$.

%%%%%%%%%%%%%%%%%%%%%%%%%%%%%%%%%%%%%%%%%%%%%%%%%%%%%%%%%%%
\subsection{\texorpdfstring{Cases with $\FFp{s}{m}{k}$ at $m<k$}{Cases with Fplus(s,m,k) at m<k}}
\label{app:lau.2}
%%%%%%%%%%%%%%%%%%%%%%%%%%%%%%%%%%%%%%%%%%%%%%%%%%%%%%%%%%%

The expressions for $\FFp{s}{m}{k}$ with $m{+}k\leq 6$ in (\ref{oct24.13}) reproduce the known
Laurent polynomials
\begin{align}
\FFp{3}{ 2 }{ 3} &= \frac{ y^5}{297675 } - \frac{ y^2 \zeta_3}{1890} - \frac{ \zeta_5}{360 } 
+\frac{ \zeta_3 \zeta_5}{ 8 y^3 } - \frac{ 7 \zeta_7}{64 y^2 }  +O(q,\bar q)  \,,\label{app:lau.5} \\
\FFp{4}{ 2 }{ 4} &=  \frac{y^6}{3827250} - \frac{ y^3 \zeta_3}{28350} - \frac{ \zeta_7}{720 y } 
+\frac{  5 \zeta_3 \zeta_7}{64 y^4 } - \frac{ 25 \zeta_9}{432 y^3 }+O(q,\bar q)  \, ,\notag 
\end{align}
equivalent to those of ${\rm E}_{2,3},{\rm E}_{2,4}$. At weight seven,
(\ref{oct24.14}) and (\ref{oct24.15}) lead to the expressions
\begin{align}
\FFp{5}{ 2 }{ 5} &= \frac{y^7}{46309725} - \frac{ y^4 \zeta_3}{374220 } - \frac{ \zeta_9}{1152 y^2} 
+\frac{  7 \zeta_3 \zeta_9}{128 y^5} - \frac{ 77 \zeta_{11}}{2048 y^4 } +O(q,\bar q)  \,,\notag \\
 \FFp{3}{ 3 }{ 4} &= \frac{ y^7}{80372250 } - \frac{ y^2 \zeta_5}{25200 } - \frac{ \zeta_7}{4536 } + \frac{ 
 5 \zeta_5 \zeta_7}{256 y^5 } - \frac{ 49 \zeta_9}{11520 y^2 } +O(q,\bar q) \,, \label{app:lau.6a} \\
 \FFp{5}{ 3 }{ 4} &= \frac{ y^7}{49116375 } - \frac{ y^2 \zeta_5}{113400 } - \frac{ \zeta_7}{15120 } + \frac{ 
 3 \zeta_5 \zeta_7}{64 y^5 } - \frac{ 77 \zeta_{11}}{4608 y^4 } +O(q,\bar q) \,, \notag 
\end{align}
and the weight-eight combinations of $\beta^{\rm sv}$ in (\ref{oct24.15}) and (\ref{cF635}) yield
\begin{align}
\FFp{6}{ 2 }{ 6} &= \frac{ 691 y^8}{373530031875 } - \frac{ 691 y^5 \zeta_3}{3192564375 } - \frac{ 7 \zeta_{11}}{
 11520 y^3 } + \frac{ 21 \zeta_3 \zeta_{11}}{512 y^6 } - \frac{ 9555 \zeta_{13}}{
 353792 y^5 } +O(q,\bar q) \,,\notag \\
\FFp{4}{ 3 }{ 5} &= \frac{ y^8}{972504225 } - \frac{ y^3 \zeta_5}{374220 } - \frac{ \zeta_9}{8640 y } + \frac{ 
 7 \zeta_5 \zeta_9}{512 y^6 } - \frac{ 5 \zeta_{11}}{2304 y^3 } +O(q,\bar q)  \,, \label{app:lau.6b} \\
\FFp{6}{ 3 }{ 5} &= \frac{ 2 y^8}{1149323175 } - \frac{ y^3 \zeta_5}{1496880 } - \frac{ \zeta_9}{24192 y } + \frac{ 
 35 \zeta_5 \zeta_9}{1024 y^6 } - \frac{ 63063 \zeta_{13}}{5660672 y^5 } +O(q,\bar q) \, ,\notag 
\end{align}
where the Laurent polynomials of the $\beta^{\rm sv}$-combinations $\cFFp{s}{ m }{ k}$ are identical
to those of the modular invariants $\FFp{s}{ m }{ k}$.

%%%%%%%%%%%%%%%%%%%%%%%%%%%%%%%%%%%%%%%%%%%%%%%%%%%%%%%%%%%
\section{\texorpdfstring{Examples of $\beta^{\rm sv}$ representations of $\FFm{s}{m}{k}$}{Examples of betasv representations of Fminus(s,m,k)}}
\label{app:fminus}
%%%%%%%%%%%%%%%%%%%%%%%%%%%%%%%%%%%%%%%%%%%%%%%%%%%%%%%%%%%

In this appendix, we spell out the $\beta^{\rm sv}$ representation of further $\FFm{s}{m}{k}$
besides the examples (\ref{odd1st}) and (\ref{odd4th}) at weights $m{+}k\leq 6$. The complete list for weights $m{+}k\leq 14$ can be found in the ancillary file. When the combination of $\bsv$ is not modular invariant, we denote the corresponding function by $\cFFm{s}{m}{k}$ in accordance with the discussion in section~\ref{sec:odd.dpt1}.

%%%%%%%%%%%%%%%%%%%%%%%%%%%%%%%%%%%%%%%%%%%%%%%%%%%%%%%%
\subsection{\texorpdfstring{Examples at $m{+}k=7$}{Examples at m+k=7}}
\label{app:fminus.1}
%%%%%%%%%%%%%%%%%%%%%%%%%%%%%%%%%%%%%%%%%%%%%%%%%%%%%%%%

At weight $m{+}k=7$, the $\beta^{\rm sv}$ representations of
$\FFm{s}{m}{k}$ constructed from the prescription of section \ref{sec:odd.dpt1}
are given by
\begin{align}
\FFm{4}{2}{5} &= -1890 \betasv{1& 4\\4& 10} + 
   1890 \betasv{2& 3\\4& 10} + 1890 \betasv{4& 1\\10& 4} - 
   1890 \betasv{5& 0\\10& 4} \notag\\
   &\ \ \  - \frac{ 175}{11 }\betasv{3\\8} \zeta_{3} - 
   1260 \zeta_{3} \betasv{3\\10}  + \frac{315  \zeta_{3}}{ y} \betasv{4\\10} 
   + \frac{105  \zeta_{9}}{16 y^3} \betasv{0\\4}
    - \frac{105  \zeta_{9}}{64 y^4} \betasv{1\\4}  - \frac{7 \zeta_{9}}{384 y^2} \notag \,,
\\
\cFFm{6}{2}{5} &= -1890 \betasv{1& 4\\4& 10} - 
   1512 \betasv{2& 3\\4& 10} + 1890 \betasv{4& 1\\10& 4} + 
   1512 \betasv{5& 0\\10& 4}   \\
   &\ \ \ + 1008  \zeta_{3} \betasv{3\\10} + 
\frac{   315  \zeta_{3}}{y} \betasv{4\\10}  - 
\frac{   21  \zeta_{9}}{4 y^3} \betasv{0\\4} - 
\frac{   105  \zeta_{9}}{64 y^4} \betasv{1\\4} + \frac{7 \zeta_{9}}{1920 y^2}\notag\,,
\end{align}
as well as
\begin{align}
\FFm{2}{3}{4} &= -2100 \betasv{2& 3\\6& 8} + 
   2800 \betasv{3& 2\\6& 8} + 2100 \betasv{3& 2\\8& 6} - 
   700 \betasv{4& 1\\6& 8} - 2800 \betasv{4& 1\\8& 6}  \notag \\
   &\ \ \ \ + 
   700 \betasv{5& 0\\8& 6} -  \frac{\zeta_{5} }{12} \betasv{1\\4} + 
   280 \zeta_{5} \betasv{1\\8}  - \frac{280  \zeta_{5}}{y} \betasv{2\\8} + 
   \frac{105 \zeta_{5}}{2 y^2}  \betasv{3\\8} \notag   \\
   &\ \ \ \
  - \frac{   50 \zeta_{7}}{y} \betasv{0\\6}  + \frac{50  \zeta_{7}}{   y^2} \betasv{1\\6} - \frac{75  \zeta_{7}}{8 y^3} \betasv{2\\6} 
   - \frac{5 \zeta_{7}}{1512} \notag\,,
\\
\FFm{4}{3}{4} &= -2100 \betasv{2& 3\\6& 8} + 
   1050 \betasv{3& 2\\6& 8} + 2100 \betasv{3& 2\\8& 6} + 
   1050 \betasv{4& 1\\6& 8} - 1050 \betasv{4& 1\\8& 6} \notag \\
   &\ \ \ \  - 
   1050 \betasv{5& 0\\8& 6} - 420  \zeta_{5}\betasv{1\\8} - 
   \frac{105 \zeta_{5}}{y}  \betasv{2\\8} + \frac{105  \zeta_{5}}{   2 y^2}\betasv{3\\8}  \\
   &\ \ \ \
   + \frac{75  \zeta_{7}}{y} \betasv{0\\6}
    + \frac{75  \zeta_{7}}{4 y^2} \betasv{1\\6} - 
 \frac{  75  \zeta_{7}}{8 y^3} \betasv{2\\6} + \frac{5 \zeta_{7}}{6048 }\,,
\notag \\
\cFFm{6}{3}{4} &= -2100 \betasv{2& 3\\6& 8} - 
   2100 \betasv{3& 2\\6& 8} + 2100 \betasv{3& 2\\8& 6} - 
   210 \betasv{4& 1\\6& 8} + 2100 \betasv{4& 1\\8& 6}  \notag \\
   &\ \ \ \ + 
   210 \betasv{5& 0\\8& 6} + 84 \zeta_{5} \betasv{1\\8} + 
   \frac{210 \zeta_{5}}{y} \betasv{2\\8}  + \frac{105 \zeta_{5}}{   2 y^2}  \betasv{3\\8} \notag \\
   &\ \ \ \
    - \frac{15 \zeta_{7}}{y}  \betasv{0\\6} - 
  \frac{ 75  \zeta_{7}}{2 y^2} \betasv{1\\6} - 
   \frac{75  \zeta_{7}}{8 y^3} \betasv{2\\6} - \frac{ \zeta_{7}}{15120}\, . \notag
   \end{align}

%%%%%%%%%%%%%%%%%%%%%%%%%%%%%%%%%%%%%%%%%%%%%%%%%%%%%%%%
\subsection{\texorpdfstring{Examples at $m{+}k=8$}{Examples at m+k=8}}
\label{app:fminus.2}
%%%%%%%%%%%%%%%%%%%%%%%%%%%%%%%%%%%%%%%%%%%%%%%%%%%%%%%%

The weight-eight instances of $\FFm{s}{m}{k}$ in terms of $\beta^{\rm sv}$ are given by
{\allowdisplaybreaks 
\begin{align}
\FFm{5}{2}{6} &= -8316 \betasv{1& 5\\4& 12} + 
   8316 \betasv{2& 4\\4& 12} + 8316 \betasv{5& 1\\12& 4} - 
   8316 \betasv{6& 0\\12& 4} \notag \\*
   &\ \ \ \ - 
  \frac{ 22803 \zeta_{3}}{325} \betasv{4\\10}  - 
   5544 \zeta_{3} \betasv{4\\12}  + \frac{1386 \zeta_{3}}{y}   \betasv{5\\12}\notag \\*
   &\ \ \ \
    + \frac{189  \zeta_{11}}{32 y^4} \betasv{0\\4}
    - \frac{189  \zeta_{11}}{128 y^5} \betasv{1\\4} 
    - \frac{21 \zeta_{11}}{1280 y^3} \,,
\\
\FFm{7}{2}{6} &= -8316 \betasv{1& 5\\4& 12} - 
   6930 \betasv{2& 4\\4& 12} + 8316 \betasv{5& 1\\12& 4} + 
   6930 \betasv{6& 0\\12& 4} \notag \\*
   &\ \ \ \ + 4620  \zeta_{3}  \betasv{4\\12}+ 
   \frac{1386 \zeta_{3}}{y}  \betasv{5\\12} 
    - \frac{315  \zeta_{11}}{64 y^4} \betasv{0\\4}
    - \frac{189 \zeta_{11}}{128 y^5} \betasv{1\\4}
   + \frac{7 \zeta_{11}}{1920 y^3} \notag\,,
   \end{align}}%
as well as
{\allowdisplaybreaks
\begin{align}
\FFm{3}{3}{5} &= -9450 \betasv{2& 4\\6& 10} + 
   12600 \betasv{3& 3\\6& 10} - 3150 \betasv{4& 2\\6& 10} + 
   9450 \betasv{4& 2\\10& 6} - 12600 \betasv{5& 1\\10& 6}  \notag \\*
   &\ \ \ \ + 
   3150 \betasv{6& 0\\10& 6} - \frac{35 \zeta_{5}}{88} \betasv{2\\6}  + 
   1260 \zeta_{5}  \betasv{2\\10} - \frac{1260 \zeta_{5}}{y}  \betasv{3\\10}
    + \frac{945  \zeta_{5}}{4 y^2} \betasv{4\\10}  \notag\\*
    &\ \ \ \
 - \frac{175  \zeta_{9}}{4 y^2} \betasv{0\\6}
 + \frac{ 175  \zeta_{9}}{4 y^3} \betasv{1\\6}
 - \frac{ 525  \zeta_{9}}{64 y^4} \betasv{2\\6}
     - \frac{5 \zeta_{9}}{1728 y} \notag\,,
\\
\FFm{5}{3}{5} &= -9450 \betasv{2& 4\\6& 10} + 
   3780 \betasv{3& 3\\6& 10} + 5670 \betasv{4& 2\\6& 10} + 
   9450 \betasv{4& 2\\10& 6} - 3780 \betasv{5& 1\\10& 6} \notag \\*
   &\ \ \ \  - 
   5670 \betasv{6& 0\\10& 6}  - 2268 \zeta_{5} \betasv{2\\10}  
   - \frac{378  \zeta_{5}}{y}  \betasv{3\\10}+ 
  \frac{ 945 \zeta_{5}}{4 y^2} \betasv{4\\10}  \\*
  &\ \ \ \  + 
  \frac{ 315  \zeta_{9}}{4 y^2} \betasv{0\\6} + 
  \frac{ 105 \zeta_{9}}{8 y^3} \betasv{1\\6} - 
  \frac{ 525  \zeta_{9}}{64 y^4} \betasv{2\\6}+ \frac{ \zeta_{9}}{864 y} \notag\,,
\\
\FFm{7}{3}{5} &= -9450 \betasv{2& 4\\6& 10} - 
   10080 \betasv{3& 3\\6& 10} - 1260 \betasv{4& 2\\6& 10} + 
   9450 \betasv{4& 2\\10& 6} + 10080 \betasv{5& 1\\10& 6}   \notag \\*
   &\ \ \ \ + 
   1260 \betasv{6& 0\\10& 6}+ 504  \zeta_{5}  \betasv{2\\10} + 
   \frac{1008 \zeta_{5}}{y} \betasv{3\\10}  + 
   \frac{945  \zeta_{5}}{4 y^2}  \betasv{4\\10} \notag \\*
   &\ \ \ \ - 
   \frac{35  \zeta_{9}}{2 y^2} \betasv{0\\6} - 
   \frac{35  \zeta_{9}}{y^3} \betasv{1\\6} - 
   \frac{525  \zeta_{9}}{64 y^4} \betasv{2\\6} - \frac{ \zeta_{9}}{8640 y}\, . \notag
\end{align}}%

%%%%%%%%%%%%%%%%%%%%%%%%%%%%%%%%%%%%%%%%%%%%%%%%%%%%%%%%
\section{\texorpdfstring{Integration constants $\overline{\alpha[\ldots]}$ for even functions}{Integration constants alphabar for even functions}}
\label{appalp}
%%%%%%%%%%%%%%%%%%%%%%%%%%%%%%%%%%%%%%%%%%%%%%%%%%%%%%%%

This appendix is dedicated to the combinations $\alpha_{m,k}^{N,j}$ of antiholomorphic integration
constants defined in (\ref{oct24.22}) that enter the even $\FFp{s}{m}{k}$. 

\subsection{\texorpdfstring{Examples at $m\leq 3$}{Examples at m<=3}}
\label{appalp.A}

We shall now gather the remaining instances of the closed formula (\ref{oct24.26})
for the $\alpha_{m,k}^{N,j}$ with $m\leq3$ that are determined from
reality of $\FFp{s}{m}{k}$ with $m{+}k\leq 8$.

At $m=2$, the examples that are not
yet covered by (\ref{exsalpha}) and (\ref{oct24.25}) read
\begin{align}
\alpha^{0,0}_{2, 5}&= 
4\zeta_3\, {\cal E}_0(10, 0^3)  \, ,
& \alpha^{0,0}_{2, 6}&= 
16 \zeta_3\,{\cal E}_0(12, 0^4)  \,,
\notag \\
\alpha^{1,0}_{2, 5}&=
80 \zeta_3\,{\cal E}_0(10, 0^4)  \, ,
&\alpha^{1,0}_{2, 6}&=
480 \zeta_3\,{\cal E}_0(12, 0^5) \,,
\notag \\
\alpha^{2,0}_{2, 5}&=
1440 \zeta_3\,{\cal E}_0(10, 0^5) \, ,
&\alpha^{2,0}_{2, 6}&=
13440\zeta_3\, {\cal E}_0(12, 0^6) \,,
 \label{oct24.31}\\
\alpha^{3,0}_{2, 5}&=
20160\zeta_3\, {\cal E}_0(10, 0^6)  + \frac{4}{3} \zeta_9\,{\cal E}_0(4)  \, ,
&\alpha^{3,0}_{2, 6}&=
322560 \zeta_3\,{\cal E}_0(12, 0^7)  \,,
\notag \\
\alpha^{4,0}_{2, 5}&=
161280\zeta_3\, {\cal E}_0(10, 0^7)  + 
 \frac{32}{3} \zeta_9\,{\cal E}_0(4, 0)  \, , \ \ \ 
 &\alpha^{4,0}_{2, 6}&=
5806080 \zeta_3\,{\cal E}_0(12, 0^8) + 
 \frac{48}{11} \zeta_{11}\,{\cal E}_0(4)  \,,
 \notag \\
&&\alpha^{5,0}_{2, 6}&=
 58060800\zeta_3\, {\cal E}_0(12, 0^9) + 
 \frac{480}{11} \zeta_{11}\,{\cal E}_0(4, 0) \, .
 \notag
\end{align}
At $m=3$, the simplest examples are
\begin{align}
\alpha^{0,0}_{3, 4}&= 
\frac{2}{5} \zeta_5\,{\cal E}_0(8, 0)  \, ,
& \alpha^{0,1}_{3, 4}&=
 0  \,,\notag \\
\alpha^{1,0}_{3, 4}&=
4 \zeta_5\,{\cal E}_0(8, 0^2)  + \frac{2}{7} \zeta_7\,{\cal E}_0(6) \, ,
&\alpha^{1,1}_{3, 4}&=  
\frac{4}{5} \zeta_5\,{\cal E}_0(8, 0^2) \,,
\notag \\
\alpha^{2,0}_{3, 4}&=
48 \zeta_5\,{\cal E}_0(8, 0^3)  + \frac{16}{7} \zeta_7\,{\cal E}_0(6, 0)  \, ,
&\alpha^{2,1}_{3, 4}&=
\frac{96}{5} \zeta_5\,{\cal E}_0(8, 0^3) + \frac{4}{7} \zeta_7\,{\cal E}_0(6, 0)\,,
 \notag \\
\alpha^{3,0}_{3, 4}&=
576\zeta_5\, {\cal E}_0(8, 0^4) + \frac{144}{7} \zeta_7\,{\cal E}_0(6, 0^2)  \, ,
&\alpha_{3,4}^{3,1} &= \frac{1}{2} \alpha_{3,4}^{3,0} \,,  \label{oct24.32} \\
\alpha^{4,0}_{3, 4}&=
4608 \zeta_5\,{\cal E}_0(8, 0^5) + 
 \frac{1152}{7}\zeta_7\, {\cal E}_0(6, 0^3)  \, , \ \ \
&\alpha_{3,4}^{4,1} &= \frac{1}{2} \alpha_{3,4}^{4,0}  \,,
\notag
\end{align}
as well as
\begin{align}
\alpha^{0,0}_{3, 5}&=
\frac{4}{5} \zeta_5\,{\cal E}_0(10, 0^2)  \, ,
&\alpha^{0,1}_{3, 5}&= 
 0 \,, \notag \\
\alpha^{1,0}_{3, 5}&=
\frac{72}{5} \zeta_5\,{\cal E}_0(10, 0^3)  \, ,
&\alpha^{1,1}_{3, 5}&= 
\frac{12}{5} \zeta_5\,{\cal E}_0(10, 0^3)  \,,\notag \\
\alpha^{2,0}_{3, 5}&=
288 \zeta_5\,{\cal E}_0(10, 0^4)  + \frac{4}{9}\zeta_9\, {\cal E}_0(6)  \, ,
&\alpha^{2,1}_{3, 5}&=
96\zeta_5\, {\cal E}_0(10, 0^4) \,,  \label{oct24.33} \\
\alpha^{3,0}_{3, 5}&= 
5760 \zeta_5\,{\cal E}_0(10, 0^5)+ \frac{16}{3} \zeta_9\,{\cal E}_0(6, 0)  \, ,
&\alpha^{3,1}_{3, 5}&=
2592 \zeta_5\,{\cal E}_0(10, 0^5) +\frac{ 4}{3} \zeta_9\,{\cal E}_0(6, 0)  \,,\notag \\
\alpha^{4,0}_{3, 5}&= 
96768\zeta_5\, {\cal E}_0(10, 0^6) + 
 64 \zeta_9\,{\cal E}_0(6, 0^2)  \, ,
 &\alpha_{3,5}^{4,1} &= \frac{1}{2}\alpha_{3,5}^{4,0} \,,  \notag \\
 \alpha^{5,0}_{3, 5}&=
 967680\zeta_5\, {\cal E}_0(10, 0^7) + 
 640 \zeta_9\,{\cal E}_0(6, 0^3) \, ,\ \ \ 
&\alpha_{3,5}^{5,1} &= \frac{1}{2} \alpha_{3,5}^{5,0} \, .
\notag
\end{align}
Note that shuffle relations imply the vanishing of $\alpha_{2,5}^{5,0},\alpha_{2,6}^{6,0},\alpha_{3,4}^{5,0},\alpha_{3,5}^{6,0}$ as well as the combinations $\alpha_{3,4}^{3,0} - 2 \alpha_{3,4}^{3,1},\alpha_{3,4}^{4,0} - 2 \alpha_{3,4}^{4,1}$
and $\alpha_{3,5}^{4,0} - 2 \alpha_{3,5}^{4,1},\alpha_{3,5}^{5,0} - 2 \alpha_{3,5}^{5,1}$.

\subsection{Conjectural closed formula}
\label{appalp.B}

The closed formula (\ref{oct24.26}) for the combinations $\alpha^{N,j}_{m,k}$ of antiholomorphic 
integration constants in (\ref{oct24.22}) is proposed to generalise to 
\begin{align}
A^{N, j}_{m, k} = 
 \frac{ N!}{ ( k {-} m {-} N {+} 2 j {+}  1)!} \bigg\{  &\frac{ 2 \zeta_{2 m - 1}}{2 m {-} 1}
 \frac{(  k {-} m {+} j)!  (N {+} k {-} m {+} 1)! }{(N {-} j)!} {\cal E}_0(2k,0^{N+k-m}) 
 \label{allalph.1} \\
  + &\theta_{N+m-k \geq 0}  
 \frac{  2 \zeta_{2 k - 1}}{2 k {-} 1} \frac{ j!  (N {+} m {-} k {+} 1)! }{  (N {+} m {-} k {-} j)!}
 {\cal E}_0(2 m, 0^{N+m - k }) \bigg\} \,,
  \notag
       \end{align}
with $j \leq m{-}2$ and
\begin{align}
A^{N, j}_{m, k} &= \sum_{\ell=0}^{m-2-j}
 \frac{ (-1)^{\ell}(2 m {-} 2 {-} 2 j {-} 2 \ell) ( 2 m {-} 3 {-} 2 j)!}{\ell! (2 m {-} 2 {-} 2 j {-} \ell)!}\alpha^{N,\ell + j}_{m, k}\,.
  \label{allalph.2}
\end{align}
All instances of this formula with $m{+}k\leq 14$ can be derived from the
reality of $\FFp{s}{m}{k}$, and its validity at higher weight is conjectural. The
step function $\theta_{N+m-k \geq 0}  $ in (\ref{allalph.1}) is defined by
\beq
\theta_{M \geq 0} = \left\{ \begin{array}{cl}
 1 &:\ M \geq 0 \\
 0 &:\ M<0
\end{array} \right.
 \label{allalph.3} 
\eeq
and ensures the absence of negative numbers of zeros in ${\cal E}_0(2m,0^{m-k+N})$.
Moreover, the inverse factor of $( k {-} m {-} N {+} 2 j {+}  1)!$ causes $A^{N, j}_{m, k}$
with $N\geq k{-}m{+}2j{+}2$ to vanish.
One can easily check that (\ref{allalph.1}) and (\ref{allalph.2}) reduce 
to (\ref{oct24.26}) for $m=2,3 $ and $j=0,1$.

%%%%%%%%%%%%%%%%%%%%%%%%%%%%%%%%%%%%%%%%%%%%%%%%%%%%%%%%%%%
\section{\texorpdfstring{Cauchy--Riemann derivatives of $\FFpm{s}{m}{k}$}{Cauchy--Riemann derivatives of Fplusminus(s,m,k)}}
\label{app:CRderiv}
%%%%%%%%%%%%%%%%%%%%%%%%%%%%%%%%%%%%%%%%%%%%%%%%%%%%%%%%%%%

In this appendix, we list representative examples of Cauchy--Riemann derivatives
$\nabla^p$ and $\overline{\nabla}^p$ of $ \FFpm{s}{m}{k}$ for $0<p<s$. As explained in 
section \ref{sec:odd.bsvrep}, their $\beta^{\rm sv}$ representations are fixed from
those of $ \FFpm{s}{m}{k}$ together with their Laurent polynomials and Laplace 
equations. The case $p=s$ removes all irreducible depth-two terms and was treated in sections~\ref{sec:4.1CR} and~\ref{sec:odd.dpt1}, respectively. The ancillary file contains all cases with $m{+}k\leq 14$.

%%%%%%%%%%%%%%%%%%%%%%%%%%%%%%%%%%%%%%
\subsection{\texorpdfstring{Derivatives of even $\FFp{s}{m}{k}$}{Derivatives of even Fplus(s,m,k)}}
\label{app:CRderiv.1}
%%%%%%%%%%%%%%%%%%%%%%%%%%%%%%%%%%%%%%

The first derivatives of $\FFp{2}{2}{2}$ in (\ref{oct24.11}) are given by \cite{Gerken:2020yii}
\begin{align}
\frac{ (\pi \overline{\nabla}) \FFp{2}{2}{2} }{y^2} &=
-144 \betasv{1& 0\\4& 4}  + 
\frac{ 24 \zeta_3}{y} \betasv{0\\4}  - \frac{\zeta_3}{15} - \frac{ \zeta_3^2}{2 y^3} + \frac{5 \zeta_5}{12 y^2}\,,
\notag \\
%+
(\pi  \nabla) \FFp{2}{2}{2} &=
-9 \betasv{2& 1\\4& 4} + 6  \zeta_3 \betasv{1\\4} - \frac{ \zeta_3^2}{ 2 y} + \frac{5 \zeta_5}{12}\, ,
\end{align}
whereas the second derivatives are determined by depth-one data, $(\pi  \nabla)^2 \FFp{2}{2}{2} 
= \frac{1}{2} (\pi \nabla {\rm E}_2)^2$. The analogous derivatives of $ \FFp{3}{2}{3}$
in (\ref{oct24.13}) are \cite{Gerken:2020yii}
\begin{align}
\frac{ (\pi \overline{\nabla})^2 \FFp{3}{2}{3} }{y^4} &=
960 \betasv{0& 1\\4& 6} + 1920 \betasv{1& 0\\4& 6} + 
 2880 \betasv{1& 0\\6& 4}  - \frac{
 320 \zeta_3}{y}  \betasv{0\\6}  - \frac{40 \zeta_3}{y^2}  \betasv{1\\6} \notag \\
 &\ \  - \frac{18  \zeta_5}{y^3} \betasv{0\\4}  - \frac{2 \zeta_3}{189} +
 \frac{ \zeta_5}{30 y^2} + 
\frac{ 3 \zeta_3 \zeta_5}{4 y^5} - \frac{7 \zeta_7}{32 y^4}\,,
\notag \\
% -
 \frac{ (\pi \overline{\nabla}) \FFp{3}{2}{3} }{y^2} &=
 -240 \betasv{1& 1\\4& 6} - 120 \betasv{2& 0\\4& 6} - 
 360 \betasv{2& 0\\6& 4} + 80  \zeta_3 \betasv{0\\6} + 
\frac{ 40  \zeta_3}{y} \betasv{1\\6} \notag \\
&\ \ + \frac{ 9  \zeta_5}{y^2} \betasv{0\\4} - \frac{ \zeta_5}{60 y} - \frac{3 \zeta_3 \zeta_5}{8 y^4}
  + \frac{ 7 \zeta_7}{32 y^3} \,,
  \\
% +
 (\pi  \nabla) \FFp{3}{2}{3} &=
 -\frac{45}{2} \betasv{2& 2\\4& 6} - 15 \betasv{3& 1\\6& 4} - 
 \frac{15}{2} \betasv{4& 0\\6& 4} + 15  \zeta_3 \betasv{2\\6} \notag \\
 &\ \ +  3  \zeta_5 \betasv{0\\4} + \frac{3  \zeta_5}{2 y} \betasv{1\\4} - 
\frac{ 3 \zeta_3 \zeta_5}{8 y^2} + \frac{7 \zeta_7}{32 y}\,,
\notag \\
% ++
 (\pi  \nabla)^2 \FFp{3}{2}{3} &=
 \frac{45}{4} \betasv{2& 3\\4& 6} + \frac{15}{4} \betasv{3& 2\\6& 4} + 
 \frac{15}{2} \betasv{4& 1\\6& 4} - \frac{15}{2} \zeta_3  \betasv{3\\6} \notag \\
 &\ \  -  3  \zeta_5 \betasv{1\\4} - \frac{3 \zeta_5}{8 y} \betasv{2\\4}  + 
\frac{ 3 \zeta_3 \zeta_5}{4 y} - \frac{7 \zeta_7}{32}
\notag
\end{align}
 with higher derivatives determined by
$ (\pi  \nabla)^3 \FFp{3}{2}{3} = \frac{1}{2} (\pi \nabla {\rm E}_2) (\pi \nabla)^2 {\rm E}_3
+ (\Im \tau)^4 {\rm G}_4 \pi \nabla {\rm E}_3$.

{\allowdisplaybreaks 
For $ \FFp{4}{2}{4}$ in (\ref{oct24.13}) we have
\begin{align}
\frac{ (\pi \overline{\nabla})^3 \FFp{4}{2}{4} }{y^6} &=
-80640 \betasv{0& 1\\ 4& 8} - 80640 \betasv{1& 0\\ 4&8} - 
 161280 \betasv{1& 0\\ 8& 4} + \frac{(\pi \overline{\nabla})^2 \EE_2(\pi \overline{\nabla}) \EE_4}{12y^6}\nn\\*
 &\ \
+ \frac{13440 \zeta_3}{y} \betasv{0\\8} + \frac{
 3360\zeta_3}{y^2} \betasv{1\\8}+ \frac{45 \zeta_7}{y^5}\betasv{0\\4}
 - \frac{\zeta_3}{135}  - \frac{\zeta_7}{16 y^4} 
  - \frac{15 \zeta_3\zeta_7}{8 y^7} + \frac{25\zeta_9}{72y^6}  \,, \nn\\
% --
\frac{ (\pi \overline{\nabla})^2 \FFp{4}{2}{4} }{y^4} &=
 3360 \betasv{0& 2\\4& 8} + 13440 \betasv{1& 1\\4& 8} + 
 3360 \betasv{2& 0\\4& 8} + 20160 \betasv{2& 0\\8& 4} - 
 2240  \zeta_3 \betasv{0\\8} \notag \\*
 &\ \ - \frac{2240  \zeta_3}{y} \betasv{1\\8} - \frac{ 140  \zeta_3}{y^2} \betasv{2\\8} - 
\frac{ 45 \zeta_7}{2 y^4}  \betasv{0\\4} + \frac{ \zeta_7}{24 y^3} + \frac{15 \zeta_3 \zeta_7}{ 16 y^6} - \frac{25 \zeta_9}{72 y^5}\,,
\notag \\
% -
 \frac{ (\pi \overline{\nabla}) \FFp{4}{2}{4} }{y^2} &=
-840 \betasv{1& 2\\4& 8} - 840 \betasv{2& 1\\4& 8} - 
 1680 \betasv{3& 0\\8& 4} + 560  \zeta_3 \betasv{1\\8} + 
\frac{ 140  \zeta_3}{y} \betasv{2\\8} \notag \\*
&\ \  + \frac{ 15  \zeta_7}{2 y^3} \betasv{0\\4} - \frac{ \zeta_7}{80 y^2} - \frac{5 \zeta_3 \zeta_7}{ 16 y^5} +
 \frac{25 \zeta_9}{144 y^4}\,,
 \\ 
 (\pi  \nabla) \FFp{4}{2}{4} &=
 -105 \betasv{2& 3\\4& 8} - \frac{105}{2} \betasv{4& 1\\8& 4} - 
 \frac{105}{2} \betasv{5& 0\\8& 4} + 70  \zeta_3 \betasv{3\\8} \notag \\*
 &\ \ + \frac{15\zeta_7}{4 y}  \betasv{0\\4} + 
\frac{ 15 \zeta_7}{16 y^2}  \betasv{1\\4} - 
\frac{ \zeta_7}{480}  - \frac{5 \zeta_3 \zeta_7}{ 16 y^3} 
+ \frac{25 \zeta_9}{144 y^2}\,,
\notag \\ 
 (\pi  \nabla)^2 \FFp{4}{2}{4} &=
\frac{ 315}{4} \betasv{2& 4\\4& 8} + \frac{105}{8} \betasv{4& 2\\8& 4} + 
 \frac{105}{2} \betasv{5& 1\\8& 4} + \frac{105}{8} \betasv{6& 0\\8& 4} - 
\frac{ 105}{2}  \zeta_3 \betasv{4\\8} \notag \\*
&\ \  - \frac{15}{4}  \zeta_7 \betasv{0\\4} - 
\frac{ 15  \zeta_7}{4 y}  \betasv{1\\4}- \frac{15 \zeta_7}{ 64 y^2} \betasv{2\\4} 
 + \frac{15 \zeta_3 \zeta_7}{16 y^2} - \frac{25 \zeta_9}{72 y} \,,\notag\\
 (\pi  \nabla)^3 \FFp{4}{2}{4} &= 
  -\frac{315}{8} \betasv{2 & 5\\ 4&8} - \frac{315}{16} \betasv{5& 2\\ 8 & 4} - 
 \frac{315}{16} \betasv{6 & 1\\8& 4} + \frac{35}{4} \betasv{4\\8} (\pi\nabla)^2\EE_2\nn\\*
 &\ \ 
 +\frac{105\zeta_3}{4} \betasv{5\\8}  + \frac{45\zeta_7}{8} \betasv{1\\4}  + \frac{ 45\zeta_7}{32y} \betasv{2\\4}  - \frac{5\zeta_7}{32y^2} (\pi\nabla)\EE_2- \frac{15 \zeta_3 \zeta_7}{8y}+ \frac{25 \zeta_9}{72}\,. \notag
  \end{align}
The third derivatives feature $(\pi\nabla)^2 \EE_2 = 6 (\Im\tau)^4 \GG_4$ and its complex conjugate. For other functions also derivatives of the (anti-)holomorphic Eisenstein series can arise, but we shall only write out the derivatives to the orders where they do not for simplicity.}

For  $ \FFp{s}{3}{3}$ in (\ref{oct24.11}), the derivatives
which go beyond products of depth one are given by
 %-
 \begin{align}
\frac{ (\pi \overline{\nabla}) \FFp{2}{3}{3} }{y^2} &=
 -2400 \betasv{2& 1\\6& 6} + 800 \betasv{3& 0\\6& 6}  - \frac{80  \zeta_5}{y} \betasv{0\\6} + 
\frac{ 60 \zeta_5}{y^2} \betasv{1\\6}  - 
\frac{ \zeta_5}{189} - \frac{ \zeta_5^2}{8 y^5} + \frac{5 \zeta_7}{ 288 y^2}\,,
\notag \\
(\pi  \nabla) \FFp{2}{3}{3} &=
-150 \betasv{3& 2\\6& 6} + 50 \betasv{4& 1\\6& 6} - 
 20 \zeta_5 \betasv{1\\6} + \frac{15 \zeta_5}{y}  \betasv{2\\6} - \frac{ \zeta_5^2}{8 y^3} + \frac{5 \zeta_7}{288}\,,
 \end{align}
as well as
 {\allowdisplaybreaks
 \begin{align}
\frac{ (\pi \overline{\nabla})^3 \FFp{4}{3}{3} }{y^6} &=
 -76800 \betasv{0& 1\\6& 6} - 268800 \betasv{1& 0\\6& 6} + \frac{1680  \zeta_5}{y^3} \betasv{0\\6}\notag \\*
&\ \ + 
\frac{ 120  \zeta_5}{y^4} \betasv{1\\6} + \frac{
 5 \zeta_5}{126 y^2}  - \frac{27 \zeta_5^2}{16 y^7} + 
\frac{ 35 \zeta_9}{192 y^6}\,,
\notag \\
 \frac{ (\pi \overline{\nabla})^2 \FFp{4}{3}{3} }{y^4} &= 
 19200 \betasv{1& 1\\6& 6} + 24000 \betasv{2& 0\\6& 6}  - \frac{600 \zeta_5}{y^2} \betasv{0\\6} \notag \\*
&\ \  - 
\frac{ 120  \zeta_5}{y^3}\betasv{1\\6}  - \frac{ 5 \zeta_5}{378 y} + \frac{27 \zeta_5^2}{32 y^6} - \frac{ 35 \zeta_9}{192 y^5}\,,
\\ 
 \frac{ (\pi \overline{\nabla}) \FFp{4}{3}{3} }{y^2} &=
 -2400 \betasv{2& 1\\6& 6} - 1200 \betasv{3& 0\\6& 6}  + \frac{120  \zeta_5}{y}  \betasv{0\\6} \notag \\*
&\ \ + \frac{ 60  \zeta_5}{y^2} \betasv{1\\6} + \frac{ \zeta_5}{756} - \frac{9 \zeta_5^2}{32 y^5} + \frac{ 35 \zeta_9}{384 y^4}
\notag
\end{align}}%
and 
\begin{align}
 (\pi  \nabla) \FFp{4}{3}{3} &=
 -150 \betasv{3& 2\\6& 6} - 75 \betasv{4& 1\\6& 6} + 
 30  \zeta_5 \betasv{1\\6} + \frac{15  \zeta_5}{y} \betasv{2\\6} - 
\frac{ 9 \zeta_5^2}{32 y^3} + \frac{35 \zeta_9}{384 y^2}\,,
\notag \\
 (\pi  \nabla)^2 \FFp{4}{3}{3} &=
 75 \betasv{3& 3\\6& 6} + \frac{375}{4} \betasv{4& 2\\6& 6} - 
 \frac{75}{2}  \zeta_5 \betasv{2\\6} - \frac{15  \zeta_5}{ 2 y}  \betasv{3\\6}
 + \frac{27 \zeta_5^2}{32 y^2} - \frac{35 \zeta_9}{192 y}\,,
\\ 
(\pi  \nabla)^3 \FFp{4}{3}{3} &=
-\frac{75}{4} \betasv{3& 4\\6& 6} - \frac{525}{8} \betasv{4& 3\\6& 6} + 
 \frac{105}{4}  \zeta_5 \betasv{3\\6} + \frac{15  \zeta_5}{ 8 y}  \betasv{4\\6}
 - \frac{27 \zeta_5^2}{16 y} + \frac{35 \zeta_9}{192} \, ,
 \notag
 \end{align}
whereas higher derivatives yield products of depth one $(\pi  \nabla)^2 \FFp{2}{3}{3} 
 = \frac{1}{6} (\pi \nabla {\rm E}_3)^2$
as well as $(\pi  \nabla)^4 \FFp{4}{3}{3} = \frac{3}{4} [(\pi \nabla)^2 {\rm E}_3]^2
 + 20 (\Im \tau)^6 {\rm G}_6 \pi \nabla {\rm E}_3$.

%%%%%%%%%%%%%%%%%%%%%%%%%%%%%%%%%%%%%%
\subsection{\texorpdfstring{Derivatives of odd $\FFm{s}{m}{k}$}{Derivatives of odd Fminus(s,m,k)}}
\label{app:CRderiv.2}
%%%%%%%%%%%%%%%%%%%%%%%%%%%%%%%%%%%%%%

The first Cauchy--Riemann derivatives of $\FFm{s}{2}{3}$ in (\ref{odd1st}) are given by
\begin{align}
\frac{ (\pi \overline{\nabla}) \FFm{2}{2}{3} }{y^2} &=
360 \betasv{0& 2\\4& 6} - 720 \betasv{1& 1\\6& 4} - 
 360 \betasv{2& 0\\4& 6} + 720 \betasv{2& 0\\6& 4} + 
 \frac{20}{7} \zeta_3  \betasv{0\\4} \notag \\
 &\ \ + 240 \zeta_3  \betasv{0\\6} - 
\frac{ 15  \zeta_3}{y^2} \betasv{2\\6}  - 
\frac{ 18  \zeta_5}{y^2} \betasv{0\\4} +
 \frac{9  \zeta_5}{ 2 y^3} \betasv{1\\4}
+ \frac{ \zeta_5}{20 y}\,,
\notag\\
%+
(\pi  \nabla) \FFm{2}{2}{3} &=
45 \betasv{1& 3\\4& 6} - 45 \betasv{2& 2\\4& 6} - 
 \frac{45}{2} \betasv{2& 2\\6& 4} + \frac{45}{2} \betasv{4& 0\\6& 4} + 
 \frac{5  \zeta_3}{28}  \betasv{2\\4}  \\
 &\ \  + 30  \zeta_3 \betasv{2\\6} - 
 \frac{15 \zeta_3}{2 y}  \betasv{3\\6} - 9  \zeta_5 \betasv{0\\4} + 
 \frac{9\zeta_5}{16 y^2} \betasv{2\\4} 
 \notag
 \end{align}
and
\begin{align}
\frac{ (\pi \overline{\nabla}) \FFm{4}{2}{3} }{y^2} &=
360 \betasv{0& 2\\4& 6} + 1200 \betasv{1& 1\\4& 6} - 
 720 \betasv{1& 1\\6& 4} + 240 \betasv{2& 0\\4& 6} - 
 1080 \betasv{2& 0\\6& 4} \notag \\
 &\ \  - 160 \zeta_3 \betasv{0\\6}  - 
\frac{ 200  \zeta_3}{y}  \betasv{1\\6} - \frac{15  \zeta_3}{y^2} \betasv{2\\6} + \frac{27  \zeta_5}{y^2} \betasv{0\\4} +
\frac{ 9  \zeta_5}{2 y^3} \betasv{1\\4}-
 \frac{ \zeta_5}{30 y}\,,
\notag \\
(\pi  \nabla) \FFm{4}{2}{3} &=
45 \betasv{1& 3\\4& 6} + \frac{135}{2} \betasv{2& 2\\4& 6} - 
 \frac{45}{2} \betasv{2& 2\\6& 4} - 75 \betasv{3& 1\\6& 4} - 
 15 \betasv{4& 0\\6& 4} - 45  \zeta_3 \betasv{2\\6} \notag \\
 &\ \  -  \frac{15  \zeta_3}{2 y} \betasv{3\\6} + 6  \zeta_5 \betasv{0\\4} + 
\frac{ 15  \zeta_5}{2 y} \betasv{1\\4} + \frac{9  \zeta_5}{ 16 y^2} \betasv{2\\4} \, ,
\end{align}
whereas the second derivatives introduce ${\rm G}_4$ as for instance
seen in (\ref{nabf223}). The first derivatives of $\FFm{s}{2}{4}$ in (\ref{odd4th}) read
{\allowdisplaybreaks
\begin{align}
\frac{ (\pi \overline{\nabla}) \FFm{3}{2}{4} }{y^2} &=
1680 \betasv{0& 3\\4& 8} + 1680 \betasv{1& 2\\4& 8} - 
 3360 \betasv{2& 1\\4& 8} - 5040 \betasv{2& 1\\8& 4} + 
 5040 \betasv{3& 0\\8& 4} \notag \\*
 & + 28 \zeta_3  \betasv{1\\6} + 
 2240  \zeta_3 \betasv{1\\8} - \frac{280  \zeta_3}{y} \betasv{2\\8} - 
\frac{ 70  \zeta_3}{y^2} \betasv{3\\8} \notag \\*
&\ \ - 
\frac{ 45 \zeta_7}{2 y^3} \betasv{0\\4}  + 
\frac{ 45  \zeta_7}{8 y^4} \betasv{1\\4}
+ \frac{ \zeta_7}{16 y^2} \,,
 \\
(\pi  \nabla) \FFm{3}{2}{4} &=
315 \betasv{1& 4\\4& 8} - 315 \betasv{2& 3\\4& 8} - 
 105 \betasv{3& 2\\8& 4} - 105 \betasv{4& 1\\8& 4} + 
 210 \betasv{5& 0\\8& 4} \notag \\*
 &+ \frac{7}{4}  \zeta_3 \betasv{3\\6} + 
 210  \zeta_3 \betasv{3\\8} - \frac{105\zeta_3}{2 y}  \betasv{4\\8}   - \frac{15  \zeta_7}{y} \betasv{0\\4}
 \notag \\*
 &\ \  + 
 \frac{15  \zeta_7}{8 y^2} \betasv{1\\4}+ 
\frac{ 15  \zeta_7}{32 y^3} \betasv{2\\4} +
 \frac{ \zeta_7}{48} \notag
\end{align}}%
and
\begin{align}
\frac{ (\pi \overline{\nabla}) \FFm{5}{2}{4} }{y^2} &=
1680 \betasv{0& 3\\4& 8} + 7560 \betasv{1& 2\\4& 8} + 
 2520 \betasv{2& 1\\4& 8} - 5040 \betasv{2& 1\\8& 4} - 
 6720 \betasv{3& 0\\8& 4}  \notag \\
 &\ \ - 1680  \zeta_3 \betasv{1\\8} - 
\frac{ 1260  \zeta_3}{y} \betasv{2\\8} - 
\frac{ 70 \zeta_3}{y^2}  \betasv{3\\8} + 
\frac{ 30  \zeta_7}{y^3} \betasv{0\\4} + \frac{45 \zeta_7}{ 8 y^4} \betasv{1\\4} 
- \frac{5 \zeta_7}{144 y^2} \,,
\notag \\
(\pi  \nabla) \FFm{5}{2}{4} &=
315 \betasv{1& 4\\4& 8} + 420 \betasv{2& 3\\4& 8} - 
 105 \betasv{3& 2\\8& 4} - \frac{945}{2} \betasv{4& 1\\8& 4} - 
 \frac{315}{2} \betasv{5& 0\\8& 4}   \\
 &\ \  - 280\zeta_3  \betasv{3\\8}  - 
\frac{ 105 \zeta_3}{2 y}  \betasv{4\\8}  + 
\frac{ 45  \zeta_7}{4 y}  \betasv{0\\4}+ \frac{135 \zeta_7}{ 16 y^2}  \betasv{1\\4}
+ \frac{15  \zeta_7}{32 y^3} \betasv{2\\4} - \frac{ \zeta_7}{288}\, ,
\notag
\end{align}
where second derivatives again introduce ${\rm G}_4$.

%%%%%%%%%%%%%%%%%%%%%%%%%%%%%%%%%%%%%%%%%%%%%%%%%%%%%%%%%%%
\section{A more convoluted example of red-herrings}
\label{app:RH}
%%%%%%%%%%%%%%%%%%%%%%%%%%%%%%%%%%%%%%%%%%%%%%%%%%%%%%%%%%%

Following the discussion in section \ref{sec:fold.3}, we want to present here a more convoluted example
of the apparent discrepancy between two representations of alternatively folded seed functions:
The expressions in (\ref{casc.2}) obtained from the Laplace system in step form turn out to yield
the same Poincar\'e sums as their counterparts derived from the inhomogeneous Laplace system (\ref{eq:seedeq}), where we take our formul\ae{} (\ref{eq:genseed}) and (\ref{eqseedR}) extended to $m>k$.

Let us consider the seeds for $\FFp{s}{2}{5}$ and $\Jp{\ell}{2}{5}$ derived from (\ref{eq:genseed}) in the alternative folding:
\begin{align}
y \Re[{\cal E}_0(10, 0^{5}) ]  +\frac{  y^7}{7780033800} +
{\rm rh}_a &=
\frac{1}{168}  \seedpalt{5}{ 2}{ 5} \,,
 \notag \\
y^2 \Re[{\cal E}_0(10, 0^{4}) ] 
 - \frac{ y^7}{740955600 }    +
{\rm rh}_b
    &= -\frac{1}{672 }(20 \seedpalt{5}{ 2}{ 5} + \Jseedpalt{0}{ 2}{ 5}) \,, \label{redherrr}\\
y^3 \Re[{\cal E}_0(10, 0^{3}) ] +\frac{  y^7}{148191120 }  +
{\rm rh}_c&=
\frac{1}{2688}  (180 \seedpalt{5}{ 2}{ 5} + 20 \Jseedpalt{0}{ 2}{ 5} + \Jseedpalt{1}{ 2}{ 5})  \,, \notag
\end{align}
and
\begin{align}
y^4 \Re[{\cal E}_0(10, 0^2) ] -  \frac{ y^7}{49397040}  &= 
-\frac{1}{32256}  (2520 \seedpalt{5}{ 2}{ 5} + 540 \Jseedpalt{0}{ 2}{ 5} + 54 \Jseedpalt{1}{ 2}{ 5} + 
 \Jseedpalt{2}{ 2}{ 5}) \,, 
\\
y^5 \Re[{\cal E}_0(10, 0) ] + \frac{ y^7}{26345088}  &= 
\frac{1}{516096} \ (20160 \seedpalt{5}{ 2}{ 5} + 10080 \Jseedpalt{0}{ 2}{ 5} + 
  1512 \Jseedpalt{1}{ 2}{ 5} + 56 \Jseedpalt{2}{ 2}{ 5} + \Jseedpalt{3}{ 2}{ 5}) \,,
\notag
\\
y^6 \Re[{\cal E}_0(10) ] - \frac{  y^7}{23950080 } &= 
-\frac{1}{10321920}  (100800 \Jseedpalt{0}{ 2}{ 5} {+} 20160 \Jseedpalt{1}{ 2}{ 5} {+} 
  1120 \Jseedpalt{2}{ 2}{ 5} {+} 40 \Jseedpalt{3}{ 2}{ 5} {+} \Jseedpalt{4}{ 2}{ 5}) \,.
   \notag
\end{align}
These do not match exactly with the seeds we would have obtained from the Laplacian 
system in step form discussed in section \ref{sec:fold.2} which would only contain a single
term $y^{6-p} \Re[{\cal E}_0(10, 0^p) ] $ with $ 0 \leq p <6$ per line.
We see that a multitude of red-herrings appears in (\ref{redherrr}), namely 
\begin{align}
{\rm rh}_a & = 3 \Re[{\cal E}_0(10, 0^{6}) ] + \frac{ 15 \Re[{\cal E}_0(10, 0^{7}) ]}{4 y} + 
 \frac{ 15}{8 y^2} \big( \Re[{\cal E}_0(10, 0^8) ]  - \tfrac{ \zeta_9}{9!} \big)\,, \notag \\
 {\rm rh}_b & =- \frac{ 15 \Re[{\cal E}_0(10, 0^{6}) ]}{4 } - \frac{ 45 \Re[{\cal E}_0(10, 0^{7}) ]}{8 y}
 - \frac{  45 }{16 y^2} \big( \Re[{\cal E}_0(10, 0^8) ]  - \tfrac{ \zeta_9}{9!} \big)\,, \label{redhers}\\
 {\rm rh}_c & = \frac{ 15 \Re[{\cal E}_0(10, 0^{6}) ]}{8 } + \frac{ 45 \Re[{\cal E}_0(10, 0^{7}) ]}{16 y} + 
 \frac{ 45 }{32 y^2} \big( \Re[{\cal E}_0(10, 0^8) ]  - \tfrac{ \zeta_9}{9!} \big)\,, \notag
\end{align} 
related by the Laplace system
\begin{align}
\mathcal{O}_1 {\rm rh}_a & ={\rm rh}_b\,, \notag \\
\mathcal{O}_2 {\rm rh}_b & ={\rm rh}_c\,,\\
\mathcal{O}_3 {\rm rh}_c & =0\,. \notag
\end{align}
Following our discussion in section \ref{sec:fold.3}, it follows that ${\rm rh}_c$ being in the kernel of $\mathcal{O}_3$ should  be related to ${\rm E_3}$. 
To see that we can first rewrite the combination of iterated integrals in ${\rm rh}_c$ as
\begin{align}
&\quad \frac{ 15 \Re[{\cal E}_0(10, 0^{6}) ]}{8 } + \frac{ 45 \Re[{\cal E}_0(10, 0^{7}) ]}{16 y} + 
 \frac{ 45 }{32 y^2}  \Re[{\cal E}_0(10, 0^8) ]   \\
 &\notag = -\frac{2}{9!} \frac{15}{8} \sum_{n=1}^\infty  n^2\sigma_{-9}(n) \Big[\sqrt{ n \Im \tau} K_{3-1/2}(2\pi n  \Im \tau) \Big(e^{2\pi i n \Re \tau}+e^{-2\pi i n \Re \tau}\Big) \Big]\,,
\end{align}
while the remaining term can be rewritten trivially as $-\frac{ 45 }{32} \tfrac{ \zeta_9}{9!} y^{1-3}$.

Unsurprisingly this red-herring is comprised of two different terms both of which appear as points (or rather infinite sums thereof) on the {\textit{same}} Poincar\'e orbit (\ref{eq:PSEk}) for ${\rm E}_3$.
Using (\ref{Besselsum}) we can compute the Poincar\'e sum over the Bessel function and then perform the analytically continued Dirichlet series in $n$ to arrive at
\beq
\sum_{\gamma \in B(\ZZ)\backslash {\rm SL}(2,\ZZ)}  \Big[{\rm rh}_c\Big]_\gamma = -\frac{15}{4}\frac{2 \zeta_9}{9!}\frac{1}{\zeta_5} ({\rm E}_3 - {\rm E}_3)= 0\,.
\eeq
Clearly since ${\rm rh}_b = -2 {\rm rh}_c$ we have that ${\rm rh}_b$  is also a red-herring, a very convoluted way to write $0$ as the difference between two identical non-vanishing Poincar\'e sums.
However, we see that we have not quite eliminated all the red-herrings since ${\rm rh}_a$ is not a multiple of ${\rm rh}_c$. Nonetheless we can easily rewrite its Laplace equation as
\beq
\Delta\bigg( {\rm rh}_a - \frac{4}{3} {\rm rh}_c \bigg) = 0\,,
\eeq
hence we expect the combination
\beq
{\rm rh}_a - \frac{4}{3} {\rm rh}_c = \frac{1}{2}  \Re[{\cal E}_0(10, 0^6) ] \,,
\eeq
to generate, upon Poincar\'e summation, an element in the kernel of $\mathcal{O}_1$, i.e.\ ${\rm E}_0$.

Following a reasoning very close to section \ref{sec:fold.3}, we can rewrite this iterated integral as
\beq
\frac{1}{2}\Re[{\cal E}_0(10, 0^6) ]  = -\frac{1}{9!} \sum_{n=1}^\infty n^2\,\sigma_{-9}( n  )\Big[\sqrt{n  \Im \tau} K_{0-1/2}(2\pi n \Im \tau) \Big(e^{2\pi i n  \Re \tau}+e^{-2\pi i n \Re \tau}\Big) \Big]\,.
\eeq
We can perform the Poincar\'e sum over this Bessel function using (\ref{Besselsum}) arriving at\footnote{Similar to
appendix \ref{app:Poincare}, we write the Riemann zeta function as $\zeta(s)$ instead of $\zeta_s$ in order to 
make the various different arguments more legible.}
\begin{equation}
\sum_{\gamma \in B(\ZZ)\backslash {\rm SL}(2,\ZZ)} \bigg[ \frac{1}{2}\Re[{\cal E}_0(10, 0^6) ]  \bigg]_\gamma = - \frac{1}{9!}  \sum_{n=1}^\infty  \frac{  \pi^{2s+1/2}  n^2\,\sigma_{-9}( n  )  \sigma_{2s-1}( n ) }{ n ^{s-1} \cos(\pi s) \Gamma(s{+}1/2)\zeta(2s{-}1)}  \frac{{\rm E}_s(\tau)}{2\zeta(2s)}\,,
\end{equation}
where we kept the Bessel index $s$ as a regulator, taking the limit $s\to 0$ only at the end.

To perform the Dirichlet sum over $n$ we make use of Ramanujan identity
\begin{equation}
\sum_{n=1}^\infty  \frac{\sigma_a(n)\sigma_b(n)}{n^s} = \frac{\zeta(s)\zeta(s{-}a)\zeta(s{-}b)\zeta(s{-}a{-}b)}{\zeta(2s{-}a{-}b)}\,,
\end{equation}
finally obtaining
\begin{align}
\sum_{\gamma \in B(\ZZ)\backslash {\rm SL}(2,\ZZ)} \bigg[ \frac{1}{2}\Re[{\cal E}_0(10, 0^6) ] \bigg]_\gamma &=
-\frac{1}{9!}\lim_{s\to0} \frac{\pi^{2s+1/2}  \zeta(-2{-}s)\zeta(7{-}s)\zeta(s{-}3)\zeta(6{+}s)}{ \cos(\pi s) \Gamma(s{+}1/2)\zeta(2s{-}1) \zeta(4)}  \frac{{\rm E}_s(\tau)}{2\zeta(2s)} \notag \\
&= 0\times {\rm E}_0\,.
\end{align}
In conclusion, the Poincar\'e sum over the remaining red-herring ${\rm rh}_a - \frac{4}{3} {\rm rh}_c$ is indeed proportional to ${\rm E}_0$ as expected from its Laplace equation. However, the proportionality constant, which is effectively given by the Dirichlet series above, vanishes upon analytic continuation.

We see in this more convoluted example that in general, when we compute the alternative folding seeds (\ref{eq:genseedpalt}) starting from the inhomogeneous Laplace system (\ref{eq:seedeq}), 
we will generate a variety of red-herrings at different levels. However, 
these red-herring seeds such as (\ref{redhers}) give rise to vanishing Poincar\'e sums
even though they are obviously non-vanishing functions.  
Hence, the Laplace system written in step form as presented in section \ref{sec:fold.2}
leads to considerably simpler representatives of the seeds, see~\eqref{trunc.seed} 
and (\ref{mineq:genseedmalt}). 

\providecommand{\href}[2]{#2}\begingroup\raggedright\endgroup

\end{document}